\documentclass[12pt]{article}
\usepackage[utf8]{inputenc}
\usepackage[export]{adjustbox}
\usepackage{authblk}
\usepackage{bbm}
\usepackage{pgfpages}
\usepackage{graphicx}
\usepackage{multicol}
\usepackage{csquotes}
\usepackage{multirow}
\usepackage{graphicx} 
\usepackage{booktabs} 
\usepackage{authblk}
\usepackage{caption}
\usepackage{subcaption}
\usepackage{stackengine}
\usepackage{amsmath}
\PassOptionsToPackage{hyphens}{url}\usepackage{hyperref}
\usepackage{comment}
\usepackage[capposition=top]{floatrow}
\usepackage{hyperref}
\hypersetup{
    colorlinks=true,
    linkcolor=blue,
    filecolor=blue,      
    urlcolor=blue,
    citecolor=blue,
}
\usepackage[capposition=top]{floatrow}

\usepackage[spanish,english]{babel}

\usepackage{amssymb}
\usepackage{stackrel}
\usepackage{natbib} 

\usepackage[a4paper, margin=0.75in]{geometry}

\title{Borrowing Constraints in Emerging Markets}

\author[1]{Santiago Camara\footnote{Santiago Camara owes an unsustainable debt of gratitude to Martin Eichenbaum, Giorgio Primiceri, Larry Christiano, Matt Rognlie and Guido Lorenzoni for their guidance and advice. Marios Angeletos and Diego Kanzig provided helpful comments. We would like to thank Luis Camara whose real life experience sparked  interest in this subject. We would like to thank Federico Forte, Juan José Lanzarotti, Marcos Impala and Mariano Bozzi for sharing their real life work experience with me. Yong Cai, Javier Bianchi, Paco Bruera, Juan Sanchez, Rohan Kekre, Per Krusell, Oleg Itskhoki, Kurt Mitman and Javier Garcia Cicco provided interesting comments and insights. We would like to thank Thomas Dreschel, Christoph Thoenissen and Max Croce who kindly shared codes from their papers. We would like to thank Susan Belles for her support.}}

\affil[1]{Northwestern University}

\author[2]{M\'aximo Sangi\'acomo}

\affil[2]{Banco Central de la Rep\'ublica Argentina}

\begin{document}
\date{}
\maketitle
\begin{center}
\textbf{Job Market Paper}
\footnotesize

\normalsize
This version: \today. For the latest version \href{https://scamara91.github.io/JMP/JMP.pdf}{Click here!}
\end{center}

\begin{abstract}
    \footnotesize
     Borrowing constraints are a key component of modern international macroeconomic models. The analysis of Emerging Markets (EM) economies generally assumes collateral borrowing constraints, i.e., firms' access to debt is constrained by the value of their collateralized assets. Using credit registry data from Argentina for the period 1998-2020 we show that less than 15\% of firms' debt is based on the value of collateralized assets, with the remaining 85\% based on firms' cash flows. Exploiting central bank regulations over banks’ capital requirements and credit policies we argue that the most prevalent borrowing constraints is defined in terms of the ratio of their interest payments to a measure of their present and past cash flows, akin to the interest coverage borrowing constraint studied by the corporate finance literature. We claim that this result can be extrapolated to other EMs by showing that firms' interest payments are strongly and positively correlated with their cash flows for a panel of firms from 13 EMs, a novel stylized fact to the literature. Lastly, we argue that EMs exhibit a greater share of interest sensitive borrowing constraints than the US and other Advanced Economies. From a structural point of view, we show that in an otherwise standard small open economy DSGE model, an interest coverage borrowing constraints leads to significantly stronger amplification of foreign interest rate shocks compared to the standard collateral constraint. This greater amplification provides a solution to the Spillover Puzzle of US monetary policy rates by which EMs experience greater negative effects than Advanced Economies after a US interest rate hike. In terms of policy implications, this greater amplification leads to managed exchange rate policy being more costly in the presence of an interest coverage constraint, given their greater interest rate sensitivity, compared to the standard collateral borrowing constraint.

    \medskip
    
    \medskip
    
    \noindent \footnotesize
    \textbf{Keywords:} Borrowing constraints; firm-dynamics; micro-to-macro; collateral constraints; foreign interest rate shocks; open economy monetary policy. \textbf{JEL codes:} F34, E44, G12, G15
\end{abstract}

\newpage
\section{Introduction} \label{sec:introduction}

\begin{figure}[ht]
    \centering
    \includegraphics[width=16cm,height=8cm]{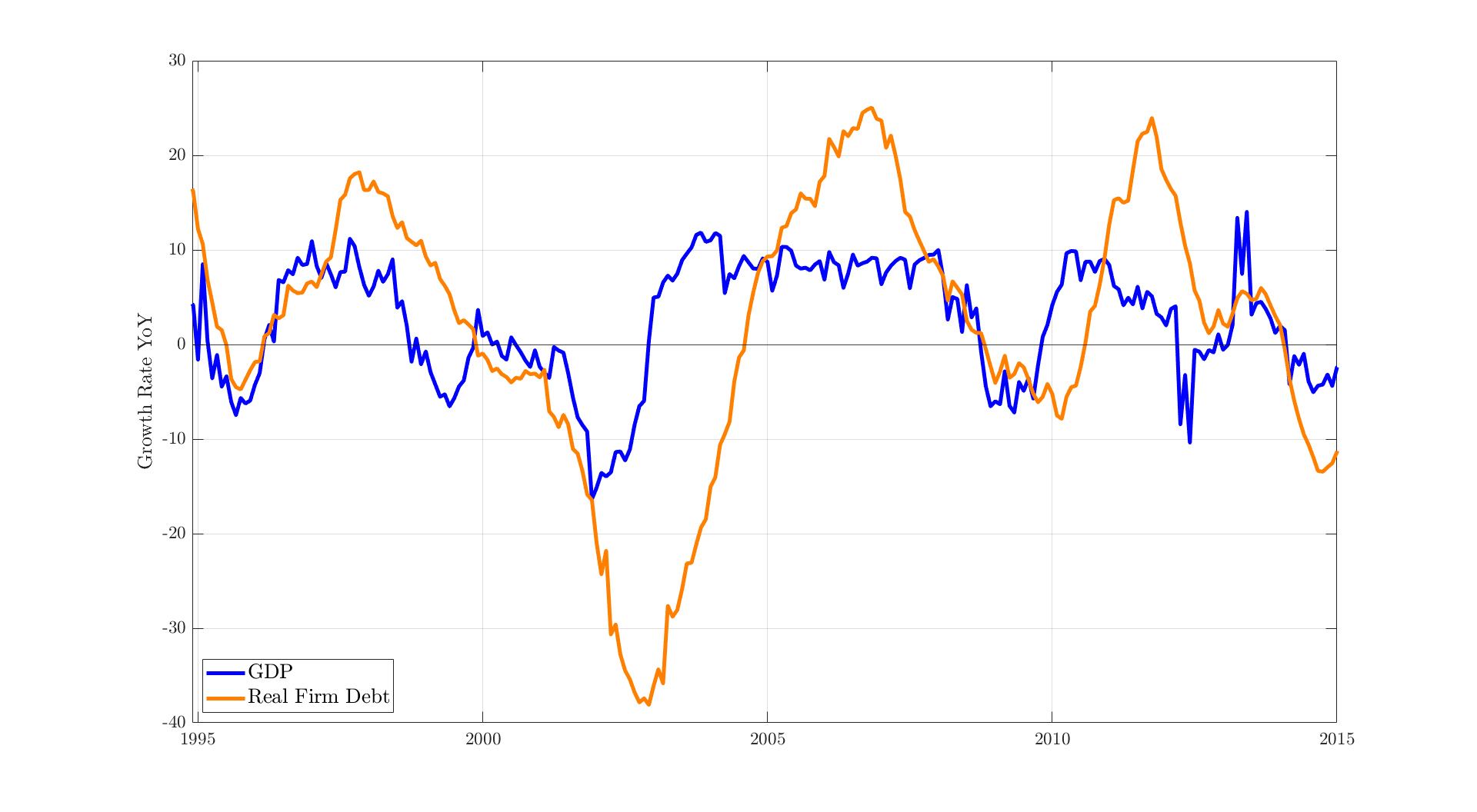}
    \caption{GDP \& Firm Debt}
    \label{fig:Growth_Rates}
    \floatfoot{\textbf{Note:} The figure presents the dynamics of real GDP and real firm debt for the period 1995-2015 at the monthly frequency. Variables are plotted using the year-on-year growth rate.}
\end{figure}

Borrowing constraints have become a key component of international macroeconomic models. By introducing a feedback loop between a firm's debt limit and economic activity, borrowing constraints allow models to match moments from the data such as the pro-cyclical and highly volatile dynamics of firms' debt, as found by \cite{neumeyer2005business}, (see Figure \ref{fig:Growth_Rates}).
Borrowing constraints can amplify models' economic shocks and explain episodes of financial crises, as argued by \cite{mendoza2010sudden}. The key role of borrowing constraints in structural models leads to a crucial empirical question: What determines a firm's debt limit? The benchmark assumption in international macroeconomics is that a firm's access to debt is limited by the re-sale value of their assets, i.e., firms borrow collateral-based debt (\cite{mendoza2010sudden,christiano2011introducing}). This type of borrowing constraint is the solution to a particular limited commitment problem (\cite{townsend1979optimal,hart1994theory,bernanke1999financial}). However, in the presence of imperfect information and adverse selection, borrowing capacity depends on a firm's cash flow from operations (\cite{stiglitz1981credit,holmstrom1997financial}). In this article, we use detailed credit registry data and structural models to empirically investigate the determinants of firms' borrowing constraints and test their aggregate implications. 

First, we exploit the universe of firm-bank credit linkages from Argentina for the period 1998-2020 to document the central role of firm's cash flows, not the liquidation value of collateral, in determining firm's access to bank debt. At the aggregate level, only 15\% of bank debt is backed or based on the specific value of physical or financial assets. In the finance literature, this type of debt is commonly referred as collateral-based lending, as the liquidation value of these assets are the key determinant of banks’ payoffs in case of bankruptcy. This implies that 85\% of aggregate bank debt is not backed by collateralized assets and is based on the value of cash flows from firms’ continuing operations, commonly referred as cash flow-based lending. Additionally, the median firm has 100\% of its bank debt expressed in cash flow-lending. We provide supporting evidence by studying hundreds of hand collected corporate loan debt contracts or ``Obligaciones Negociables''. Overall, the composition of bank and corporate debt suggests that the liquidation value of assets may not be the key determinant of firms' borrowing constraints in Emerging Markets. 

Second, we argue that the most prevalent cash flow-based borrowing constraint firms face is highly sensitive to interest rates. Exploiting regulations by the Argentinean Central Bank (BCRA) over banks' capital requirements and their credit policy we show that firms' bank debt contracts exhibit a borrowing limit in terms of the ratio between a firm's interest payments to their cash flows. This type of debt limit is commonly referred by the finance literature as interest coverage borrowing constraints (see \cite{greenwald2019firm} for example). We argue that the key characteristic of this borrowing constraint is its high sensitivity to interest rates, so that a 100 basis point increase reduces the borrowing capacity of a firm by approximately 20\%. Additionally, by comparing our results with those coming from a recent literature for US and other Advanced Economies' firms, we argue that Argentina shows a greater prevalence of interest sensitive borrowing constraints. In terms of external validity, we argue we can extrapolate our results to other Emerging Markets by highlighting a novel stylized fact: firms' debt interest payments are positively and significantly correlated with their cash flows for a panel of thousands of Emerging Market firms. We later argue that this novel stylized fact can not be replicated by a model with a standard collateral constraint.

Third, we build on our empirical findings by carrying out a structural analysis to study the implications of cash flow-based lending in an open economy. We introduce an interest coverage cash flow-based borrowing constraints into an otherwise standard Small Open Economy model with financial frictions and working capital requirements. We focus our attention in studying the transmission of foreign interest rate shocks, a widely studied economic shock and key driver of business cycles in Emerging Markets (\cite{garcia2010real,camara2021spillovers}). An economy with an interest coverage borrowing constraint has orders of magnitude greater amplification in GDP and firms' debt than an economy with a benchmark collateral-based borrowing constraint. We demonstrate that this greater amplification is driven by the borrowing constraint's interest sensitivity and not the presence of firms' cash flows in it. 

Fourth, we construct a quantitative version of our model where firms are subject to nominal price frictions and can borrow in both domestic and foreign currency. We demonstrate that our benchmark results are also present in the presence of this frictions, with interest coverage constraints still leading to significantly larger amplification than collateral constraints. We claim that this greater amplification has implication over the benefits and costs of exchange rate regimes. We argue that the cost of ``dirty float'' or ``pegged'' exchange rate regimes is substantially larger in the presence of interest coverage constraints than under a standard collateral borrowing constraints. Thus, floating exchange rate regimes may be more beneficial than expected.

Finally, we argue that the higher prevalence of interest sensitive borrowing constraints in EMs provides a straightforward solution to the \textit{Spillover Puzzle}. The \textit{Spillover Puzzle} of US monetary policy is the fact that US monetary policy shocks are associated with significantly larger spillovers in Emerging Markets than in Advanced Economies and/or within the US economy. We document this puzzle using both SVAR and local projections techniques for the period 2004-2016, with Emerging Markets exhibiting, on average, twice as large spillovers from US monetary policy shocks than Advanced Economies. We calibrate our model to match the relative shares of interest coverage and the standard collateral borrowing constraints for both Argentina and the Advanced Economies. Simulating the impact of a 100 basis point increase in the foreign interest rate we find that our model predicts a twice as large drop in real variables for Argentina, in line with our empirical findings.

\noindent
\textbf{Related literature.} This paper relates to five strands of economic literature. First and foremost, this paper contributes to the literature that studies the role of financial frictions in macroeconomics, which goes back to the seminal work of \cite{kiyotaki1997credit} and \cite{bernanke1999financial}. In this paper we follow a recent trend on  the appeal of disciplining macroeconomic models with microeconomic evidence, as suggested by \cite{nakamura2018identification}. We do so by exploiting a detailed credit registry data set to study firms' debt contracts and its quantitative and policy implications. We claim, to the best of our knowledge, that this is the first paper which study the underlying borrowing constraints and their aggregate implications for firms in Emerging Markets.

Second, this paper relates to literature  which analyzes the importance of borrowing constraints in international macroeconomics. Seminal work in this literature are \cite{neumeyer2005business} which highlights the importance working capital constraints to match the volatility of credit in Emerging Markets, and \cite{mendoza2010sudden} which shows that borrowing constraints are key to amplify shocks and match moments of the data around financial crises. The benchmark assumption in international macroeconomics is that firms face collateral borrowing constraints, see \cite{mendoza2010sudden}, \cite{bianchi2010credit} and \cite{korinek2014sudden}. This assumption is present in both quantitative models, see \cite{christiano2011introducing}, and policy analysis see \cite{bianchi2021prudential}. However, there is little to no evidence on whether collateral borrowing constraints are an appropriate representation of the underlying frictions firms face when in credit markets. Exploiting a highly detailed credit registry dataset that allows us to observe the universe of firm-bank linkages, our first empirical contribution to this literature is to show that the vast majority of firms' debt is based on firms' cash flows, with only 15\% of firms' debt being backed by collateralized assets. Furthermore, we argue that the most common borrowing constraint firms face takes the form of an interest coverage constraint, highly sensitive to changes in the interest rate. Our theoretical contribution to this literature is that introducing cash flow-based lending to an otherwise standard open economy model with financial frictions leads to orders of magnitude greater amplification compared to an economy with collateral constraints. We argue that this is particularly the case in response to a foreign interest rate shock, a key driver of business cycles in Emerging Markets (see \cite{neumeyer2005business}, \cite{mendoza2010sudden}, \cite{garcia2010real}, \cite{camara2021FXI}).

Third, this paper contributes to a relatively novel literature which studies the prevalence and importance of cash flow-based borrowing constraints. Based on a comprehensive empirical analysis, \cite{lian2021anatomy} argues that the key constraint on US corporate debt are cash flows measured by firms' earnings. Additionally, \cite{greenwald2019firm} studies the role of interest coverage covenants in the transmission of monetary policy shocks at the firm level. This paper contributes to this literature in a twofold manner. First, we argue that Emerging Markets exhibit a greater prevalence of cash flow-based lending than the US economy and other Advanced Economies. Second, we argue that in an open economy setting, cash flow-based borrowing constraints are of particular importance as firms generate revenues, face costs and borrow in both domestic and foreign currency. 

Fourth, this paper contributes to a literature which studies policy design in open economies. Early examples of this literature find that freely floating exchange rate regimes are optimal, see \cite{gali2005monetary} and \cite{faia2008optimal}. A more recent literature has found that ``dirty float'' exchange rate regimes may be welfare improving in the presence of agent heterogeneity, see \cite{cugat2019emerging}, \cite{auclert2021exchange} and \cite{camara2022tank}, or in the presence of significant financial frictions, see \cite{bianchi2021prudential}. We contribute to this literature by showing that in the presence of interest coverage constraints, ``dirty float'' exchange rate regimes are significantly more costly in terms of economic activity than collateral constraints. Thus, we provide evidence that the cost of interest rate policies to reduce exchange rate volatility may be greater than previously thought.

Finally, this paper contributes to a literature which studies the impact of US monetary policy shocks in EMs using empirical models. Seminal work in this literature are \cite{eichenbaum1995some} and \cite{uribe2006country} which found that US monetary policy shocks have negative spillovers in EMs in terms of an economic recession, an exchange rate depreciation and tighter financial conditions. \cite{mackowiak2007external} suggests that the point estimates of these spillovers are significantly larger for EMs than when estimated for the US economy. We contribute to this literature by providing evidence of significantly larger negative spillovers of US monetary policy shocks in Emerging Markets than in the US economy and/or other Advanced Economies, a empirical fact we denote as the \textit{Spillover Puzzle} of US monetary policy. We claim that the greater prevalence of interest coverage borrowing constraints in EMs, and its greater associated amplification, provide a straightforward solution this puzzle.

\noindent
\textbf{Structure of the paper.} The paper has \ref{sec:conclusions} sections starting with the present introduction. Section \ref{sec:cash_flow_lending} shows the prevalence of cash flow-based lending for Argentinean firms across time and firm characteristics. In Section \ref{sec:interest_sensitive_borrowing_constraints} we exploit BCRA regulations to show that the most common borrowing constraint firms face takes the form of an interest coverage constraint, where the debt limit is established as a ratio of interest payments to cash flows. Section \ref{sec:model_simple} shows that introducing a cash flow-based borrowing constraint in an otherwise standard small open economy leads to orders of magnitude greater amplification than collateral constraints in response to a foreign interest rate shock. Section \ref{sec:nominal_frictions_policy} shows that our results hold in a quantitative model and conveys the policy implication of costly exchange rate pegs. Section \ref{sec:spillover_puzzle} shows that the greater prevalence of interest sensitive borrowing constraints and its associated amplification provides a straightforward solution to the Spillover Puzzle of US monetary policy. Section \ref{sec:conclusions} concludes.

\section{Prevalence of Cash-Flow Based Lending} \label{sec:cash_flow_lending}

In this section of the paper we present our first novel stylized fact, the prevalence of cash flow-based lending among Argentinean firms. In Section \ref{subsec:lending_asset_vs_cash_flow} we describe the core definitions of collateral and cash flow-based lending, we explain the key financial and bankruptcy regulations in place and cover our classification procedure. In Section \ref{subsec:lending_main_results} we report our main results. In Section \ref{subsec:lending_heterogeneity} we provide supporting evidence of our main results by studying the prevalence of cash flow-based lending across firm and credit line characteristics. Relatively young and small firms tend to borrow relatively more through collateral-based credit lines, while older and larger firms borrow almost exclusively cash flow-based. Credit lines associated with capital expenditures exhibit a greater share of collateral-based lending, while credit lines associated with working capital expenditures show a vast prevalence of cash flow-based lending. In Section \ref{subsec:lending_comparison_international} we compare our results with evidence from US firms and argue that our results for Argentina may be extrapolated to other Emerging Market economies.

\subsection{Collateral-Based vs Cash Flow-Based Lending} \label{subsec:lending_asset_vs_cash_flow}

The essence of collateral-based versus cash flow-based firm lending is debt that is issued against the liquidation value of specific (mostly physical) assets versus debt that is issued against the present and future cash flows from the firm's continuing  operations. The different types of backing of firm lending is central for credit-scoring and bank lending in practice.\footnote{Additionally, as argued by \cite{lian2021anatomy} these backings map closely to modelling strategies by the macro-finance literature.}

We follow \cite{lian2021anatomy} and classify collateral-based and cash flow-based debt contracts across four main aspects: (i) general definition, in which we exploit whether or not debt contracts have an associated underlying asset as collateral in case of default, (ii) BCRA regulations on banks' credit policy and monitoring role over firms, (iii) debt structure (type of credit lines) and, iv) default resolution.\footnote{Note that the three main aspects used in this paper do not coincide directly with the aspects considered by \cite{lian2021anatomy}. For instance, the BCRA's ``\textit{Central de Deudores}'' credit-registry data requires banks to identify which debt contracts have an associated asset as collateral. Furthermore, exploiting the type of credit line and additional data from the credit-registry data we are able to narrow down on the type of asset associated with the debt-contract.} In Appendix \ref{subsec:appendix_BCRA_Regulations_ABC} we describe in detail our classification procedure and provides insights and examples.

\noindent
\textbf{Collateral-based lending:}

\begin{itemize}
    \item[(i)] General definition: Collateral-based lending debt contracts are based on the liquidation value of specific assets which include real estate, machinery, equipment, agricultural land, cattle, account receivables by large firms and even sophisticated financial assets. The banks' payoffs in case of default are driven by the liquidation value of these assets.
    
    \item[(ii)] BCRA regulations: The central bank establishes specific regulations on these type of debt contracts such as imposing upper bounds on how much firms can borrow, loan to value limits, according to the different types of assets which used as collateral, and requiring the registry of the asset in different government agencies.\footnote{See Table \ref{tab:bounds_preffered_assets_appendix} in Appendix \ref{subsec:appendix_BCRA_Regulations_ABC}.} The registration of assets used as collateral serves for two distinct reasons. First, the fact these assets are associated to a specific debt-contract and that these tag is public information allows other potential creditors to assess the debtor's credit-worthiness. Second, it allows for certain asset standardization which would increase the assets value in case of default.\footnote{For instance, in Appendix \ref{subsec:appendix_BCRA_Regulations_ABC} we show that machinery used for agriculture and fishery must be registered in detail at different sector-specific governmental agencies. This registration allows banks to better assess the underlying value of assets in the case of default. We later argue that these asset standardization and registration leads to certain sectors having a greater share of collateral-based debt contracts than others.} Lastly, BCRA regulations state that for the case of collateral-based lending, banks are not obliged to assess debtor's repayment capacity nor assess their probability of default using any measure of current or projected cash flows.
    
    \item[(iii)] Debt structure: Exploiting the different type of credit lines, collateral-based debt contracts are usually associated with credit lines specific to the purchase of real estate, machinery or equipment, and those associated with financial leasing.\footnote{Financial leasing credit lines are debt contracts through which a bank acquires a productive asset, which has previously been selected by the tenant, and delivers it to the latter for use in exchange for a fee.} 
    
    \item[(iv)] Default resolution: Creditors have claims against the liquidation value of the specific asset which is associated with the collateral-based debt-contract. In case of default, creditors of collateral-based lending have the highest priority over the associated collateral.\footnote{See \url{http://servicios.infoleg.gob.ar/infolegInternet/anexos/25000-29999/25379/texact.htm} for the full text of the bankruptcy law of Argentina.} However, as is the case for the bankruptcy law of the US, creditors of collateral-based debt contracts cannot seize these assets and interrupt a firms productive process, and must wait for any formal debt renegotiation process (``Concurso de acreedores'' or ``Proceso Preventivo de Crisis'' in Argentina) between the firm and cash flow-based debt creditors to take place. Furthermore, creditors of collateral-based debt contracts have no formal say or vote on these debt negotiations.\footnote{We discuss in detail the different type of debt-renegotiation processes in Section \ref{sec:interest_sensitive_binding_borrowing}.}
\end{itemize}
 
\noindent
\textbf{Cash flow-based lending:}

\begin{itemize}
    \item[(i)] General definition: cash flow-based debt contracts are based on the current and projected value of a firm's cash flows from continuing their productive and commercial operations (i.e. going-concern value as defined by \cite{djankov2008debt}). Consequently, debt contracts and any constraint on the amount borrowed is established as a function of firms' ability to generate sufficient cash flows to pay back capital and interests. In the case of a breach in the contract and a debt renegotiation taking place, the restructured debt will reflect the adjusted cash flows of the continuing operations of the firm.  
    
    \item[(ii)] BCRA regulations: The central bank regulations clearly stipulate that debt contracts not backed by assets must be based on debtors current and projected cash flows.\footnote{In Section \ref{sec:interest_sensitive_borrowing_constraints} and in Appendix \ref{subsec:appendix_BCRA_Regulations_ABC} we describe in detail the different set of BCRA's regulations which sharply describe the monitoring role of banks over firms performance and risk of default based on firms' cash flows.} BCRA regulations impose that banks' assessment over repayment capacity must not involve the value of any of the debtor's assets. Apart from regulations over banks' credit policy and risk-assessment, BCRA regulations impose over banks a monitoring role over their clients, requiring them to scrutinize firms' cash flows at a high frequency and report immediately any breach of contract or unexpected change in repayment capacity.\footnote{In Appendix \ref{subsec:appendix_BCRA_Regulations_monitoring} discusses the regulations which describe banks' monitoring role over firms.} For instance, any breach of a debt-contract is reported to the BCRA's ``Central de Deudores'' and is public information.

    \item[(iii)] Debt structure: Exploiting the different type of credit lines, cash flow-based debt contracts are usually associated with short-term credit lines, discounted documents over potential future cash-flows and the financing of international trade transactions. As argued in Section \ref{subsec:lending_comparison_international}, these type of credit lines are associated with shorter maturities than credit lines generally associated with collateral-based debt contracts. Also, corporate bonds known as `Obligaciones Negociables'' are also written primarily in terms of cash flow-based metrics such as ratios of interest payments to cash flows, instead of being determined by the liquidation value of assets.\footnote{In Section \ref{sec:interest_sensitive_borrowing_constraints} we describe in detail the debt limits in corporate loans known as ``Obligaciones Negociables''.}
    
    \item[(iv)] Default resolution: Creditors of cash flow-based debt contracts have claims against the value of a firm. These creditors play a crucial role in firms debt renegotiation processes such as the ``Concurso de Acreedores''. During a debt renegotiation, cash flow-based creditors are senior creditors and vote on any re-structuring plan proposed by the debtor. In case the debt renegotiation fails (either because of no agreement or the debtor not being able to meet the restructured debt payments), they recover any residual value of the firm after collateral creditors recovered their assets.
\end{itemize}

\noindent
\textbf{Data description \& classification procedures.} The BCRA ``Central de Deudores'' credit registry data set allows us to directly observe: (i) tax identifying number of the firm, (ii) the associated bank, (iii) the amount borrowed, (iv) the type of credit line (which we describe in detail in Section \ref{subsec:lending_heterogeneity}), (v) the credit line's situation and the firm's overall credit riskiness (which we describe in detail in Section \ref{sec:interest_sensitive_borrowing_constraints}), (vi) and whether the credit line has an underlying asset serving as collateral. Thus, we classify these credit lines directly as collateral-based. In Section \ref{sec:interest_sensitive_borrowing_constraints} and in Appendix \ref{subsec:appendix_BCRA_Regulations_ABC} we describe in greater detail the regulations imposed by BCRA on banks which shape collateral-based lending. We classify the credit lines in our data set which do not have an asset as underlying collateral as cash flow-based lending. In Section \ref{sec:interest_sensitive_borrowing_constraints} and Appendixes \ref{subsec:appendix_BCRA_Regulations_ABC} and \ref{subsec:appendix_BCRA_Regulations_monitoring} we describe in greater detail the BCRA regulations which define banks' lending policy as a function of firms' cash flows. 

\subsection{Main Results} \label{subsec:lending_main_results}

We turn to presenting our first novel stylized fact: the prevalence of cash flow-based lending at both the aggregate and firm level in Argentina.

\noindent
\textbf{Aggregate level results.} Table \ref{tab:composition_firm_debt} presents the results of our first novel stylized fact regarding the composition of debt contracts between collateral-based and cash flow-based bank debt.
\begin{table}[ht]
    \centering
    \caption{Composition of Argentinean Firm's Debt \\ \footnotesize Collateral vs Cash Flow-Based}
    \label{tab:composition_firm_debt}
    \begin{tabular}{l l c c c c} 
                  & &  \multicolumn{4}{c}{Share of Debt in \%} \\
        \multicolumn{2}{c}{ }  & \small Total & \small  Firms & \small $L\geq$ 100 & \small $L\geq 500$  \\
        \multicolumn{2}{c}{Category}  & (1) & (2) & (3) & (4) \\
        \hline \hline 
        \multicolumn{2}{l}{Collateral-Based Lending}  & 16.12\% & 15.24\% & 12.94\% & 8.75\% \\
        \multicolumn{2}{l}{Cash Flow-Based Lending} & 83.88\% & 84.76\% & 88.06\% & 91.25\%  \\ \hline \hline
        \floatfoot{\textbf{Note:} This table summarizes the composition of the universe of firm-bank debt from Argentina. The data presented is for the year 2017. The choice reflects the composition by the end of our sample where we have significant complementary firm information. We avoid using data from years 2019 and 2020 so that the information is not distorted by the sovereign and financial crisis which started in March 2018 and the impact of the COVID-19 pandemic. Government loans are not accounted in either category.}
    \end{tabular}
\end{table}
The first column shows that only 16\% of total bank debt in Argentina is collateral-based, while close to 84\% of it is cash flow-based. The second column shows the prevalence of cash flow based debt is slightly greater when focusing only on firms' debt with collateral-based lending explaining close to 15\% of aggregate firms' debt and cash flow-based lending explaining the remaining 85\%.\footnote{In order to differentiate between firms and professional individuals we exploit the two first digits of their tax identification number of ``CUIT''. The CUIT number is granted by the AFIP (``Administraci\'n Federal de Ingresos P\'blicos'') and is made up of 2 digits that indicate the type, 8 digits that are the document number and a last number randomly assigned. We classify as firms observations with first two digits of the CUIT with 30 or higher.} Finally, columns three and four show the decomposition of aggregate firms' debt for firms with employment levels above 100 and 500 workers respectively, with cash flow-based debt explaining between 88\% and 91\% of total firm debt. Consequently, by studying the aggregate composition of bank debt it is clear that cash flow-based debt explains the vast majority of total bank debt. the

\begin{figure}[ht]
    \centering
    \caption{Share of Asset \& Cash Flow-Based Lending \\ \footnotesize Comparison around Sovereign Default Crises }
    \label{fig:evolution_across_time_cash_flow_based}
     \centering
     \begin{subfigure}[b]{0.495\textwidth}
         \centering
         \includegraphics[width=\textwidth]{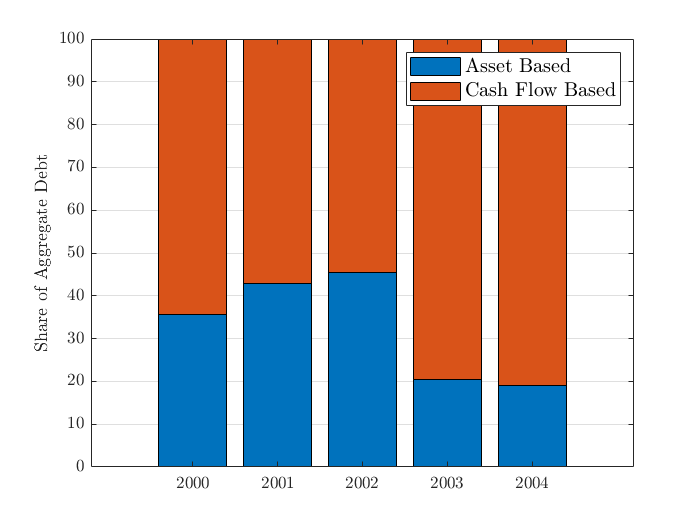}
         \caption{Around 2002 Financial Crisis}
         \label{fig:evolution_cash_flow_around_2002}
     \end{subfigure}
     \hfill
     \begin{subfigure}[b]{0.495\textwidth}
         \centering
         \includegraphics[width=\textwidth]{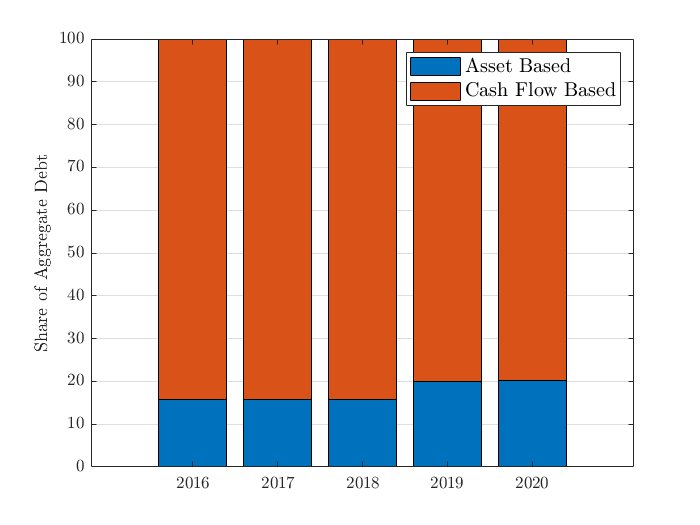}
         \caption{Around 2018 Financial Crisis}
         \label{fig:evolution_cash_flow_around_2018}
     \end{subfigure} 
     \floatfoot{\textbf{Note:} Both figures present a decomposition of aggregate debt into cash flow-based debt (in red) and collateral-based debt (in blue). Figure \ref{fig:evolution_cash_flow_around_2002} presents the results around the 2002 financial crisis with a plus minus two year window. Figure \ref{fig:evolution_cash_flow_around_2018} presents the results around the 2018 financial crisis with a plus minus two year window.}
\end{figure}

Figure \ref{fig:evolution_across_time_cash_flow_based} describes the historical evolution of the decomposition of bank debt contracts between cash flow-based and collateral-based lending.
In particular, Figure \ref{fig:evolution_cash_flow_around_2002} presents the composition around the 2002 Argentinean financial crisis, while Figure \ref{fig:evolution_cash_flow_around_2018} presents the dynamics around the 2018 Argentinean financial crises. The share of cash flow-based debt increased from an average close to 70\% in the early 2000s to more than 80\% by the end of the 2010s. This increase in the share of cash flow-based debt is in line with dynamics of corporate firms in the US in the last two decades, as shown by \cite{lian2021anatomy} and \cite{bianchi2021}. 

\noindent
\textbf{Firm level results.} We turn to analyzing the prevalence of cash flow-based debt contracts at the firm level. Table \ref{tab:composition_firm_debt_share_firms} presents evidence on the share of firms which have cash flow-based debt contracts and their relative importance as a share of their total debt. The first row shows the share of firms which only have cash flow-based debt contracts. This row's first entry shows that close to 81\% of private agents in our sample have 100\% of their debt explained by cash flow-based debt contracts. The first row entry of Columns (2)  show that 80\% of firms have the totality of their debt in cash flow-based debt contracts, while Columns (3) and (4) show that this share increases to 85\% and close to 88\% for firms with a labor force above 100 and 500 workers, respectively. 
\begin{table}[ht]
    \centering
    \caption{Composition of Argentinean Firm's Debt \\ \footnotesize Collateral vs Cash Flow-Based}
    \label{tab:composition_firm_debt_share_firms}
    \begin{tabular}{l c c c c} 
          & \small Total & \small  Firms & \small $L\geq$ 100 & \small $L\geq 500$  \\ 
          & (1) & (2) & (3) & (4) \\ \hline \hline
        \small I. Share of Firms with only cash flow-based Debt  & 80.92\% & 79.86\%	& 85.14\%	& 87.85\% \\
        \small II. Median Share of cash flow-based Debt \footnotesize (\textit{unconditional}) & 100\% & 100\% & 100\% & 100\% \\
        \small III. Mean Share of cash flow-based Debt \footnotesize (\textit{unconditional}) & 89.34\% & 89.53\% & 91.58\% & 92.37\% \\
        \small IV. Mean Share of cash flow-based Debt \footnotesize (\textit{conditional}) & 44.13\% & 48.02\% & 43.32\% & 37.23\%
        \\ \hline \hline
        \floatfoot{\textbf{Note:} This table summarizes the composition of the universe of firm-bank debt from Argentina. The data presented is for the year 2017. In the second row, the ``Mean Share of cash flow-based Debt (\textit{unconditional})'' represents the average or mean share of firms' debt that is categorized as ``cash flow-based'' debt across all firms. In the third row, the ``Mean Share of cash flow-based Debt (\textit{conditional})'' represents the average or mean share of firms' debt that is categorized as ``cash flow-based'' debt across firms which have a positive percentage of their debt as capital-based.}
    \end{tabular}
\end{table}
Cash-flow based debt contracts explain the vast majority of firms' debt. The second and third rows of Table \ref{tab:composition_firm_debt_share_firms} show that the unconditional median and mean shares of cash flow-based debt in total debt is 100\% and 89\% for all the observations in our sample. Columns (2) through (4) shows that the median share of cash flow-based debt is 100\% for all firms and even within firms with a workforce above 100 or 500, while the mean share is 89.5\% for all firms and 91.6\% and 92.4\% for firms with a workforce above 100 or above 500, respectively. Lastly, the bottom row of Table \ref{tab:composition_firm_debt_share_firms} shows that the mean share of cash flow-based debt for observations with a positive amount of collateral-based debt is 44\%. For the case of firms, this conditional mean share is 48\% for all firms, and 43\% and 37\% for firms with a workforce above 100 and above 500 workers, respectively.

\noindent
\textbf{Evidence from corporate bonds.} Corporate bonds explain only a relatively small share of total financing in Argentina and have only recently gained importance in Emerging Markets (see \cite{chang2017bond}). While the market for corporate loans is small (in terms of GDP and in terms of total financing) and heavily skewed toward relatively large firms, it provides publicly available documents on firms debt contracts. This type of debt contracts are the key source of information exploited in a recent literature which studies borrowing constraints for US firms, see \cite{lian2021anatomy} and \cite{greenwald2019firm}. Thus, studying the domestic market for corporate loans provides a comparable dataset with this previous literature.

We focus on corporate loans known as ``Obligaciones Negociables'' or ON. The ONs are fixed-income corporate loans in domestic and foreign currency listed on the Buenos Aires Stock Exchange. The issuance of these corporate bonds are carried out by investment banks in association with commercial banks which carry out the auction.\footnote{Usually these auctions are carried out sequentially, with a first auction for ``competitive investors'' or large investors, such as other investment and commercial banks, and a later auction for ``non-competitive investors'' or smaller firms, which comprises all other potential investors.} These corporate bonds are rated by international rating agencies, such as Fitch or Moody's, and must comply with the scrutiny of the ``Comisi\'on Nacional de Valores'' (CNV) which is the Argentinean counterpart of the US' SEC.  

We hand collected the information on all of the ``Obligaciones Negociables'' currently listed on ``\textit{Bolsas y Mercados Argentinos}'' or BYMA. BYMA is a stock exchange that integrates and represents the main players in the Argentina's stock market. In order to address the needs of a new capital market provided for by the Law 26,831, the Buenos Aires Stock Market S.A. implemented a reorganization under the terms of Article 77 of the Income Tax Law, proceeding with the partial spin-off of its assets to create a new entity: ``Bolsas y Mercados Argentinos S.A.'' “BYMA”, resulting in this being the continuation of the activity of the Buenos Aires Stock Market S.A., with the particularity that in the constitution of the new entity the Buenos Aires Stock Exchange has been incorporated as a shareholder.

These corporate loans represent unsecured debt, with debt re-negotiation triggered by creditors through different clauses specified in the contracts. While we present in detail the different types of technical default events or triggers of debt re-negotiations in Section \ref{sec:interest_sensitive_borrowing_constraints}, Table \ref{tab:ON_ABC_vs_EBC} shows the aggregate composition of debt between asset and cash flow-based debt.
\begin{table}[ht]
    \centering
    \caption{Collateral vs Cash Flow-Based: ``\textit{Obligaciones Negociables}''}
    \label{tab:ON_ABC_vs_EBC}
    \begin{tabular}{l c c c c} 
     Category & Total & Without State Owned & LC & Dollar \\ \hline \hline
Collateral-Based \tiny in \%	&	1.27	&	1.45	&	0.74	&	4.46	\\ 
Cash Flow-Based  \tiny in \%	&	98.73	&	98.55	&	99.26	&	95.54	\\
\\ \hline
Total Debt Value \tiny (in millions of USD)	&	\$63,235 &	 \$55,041 & \$54,265 & \$8,970 	\\
Share of Total ONs \tiny in \%	&	100.00	&	87.04	&	85.82	&	14.18	\\ \hline \hline
        \floatfoot{\textbf{Note:} This table summarizes the composition of the corporate debt market of ``\textit{Obligaciones Negociables}''. The data is sourced for the period August-2022 and reflects all listed bonds at BYMA. The second column removes the partially stated owned firm YPF S.A., which issued ONs during the government's nationalization of the company in 2014.}
    \end{tabular}
\end{table}
In line with the results presented above for bank debt contracts, almost the entirety of corporate debt loans are cash flow-based with only 1.27\% of total debt backed by the value of an asset. This results holds when removing state owned firms from our sample (see Column (2)), and when considering only local currency denominated bonds (see Column (3) under ``LC''), or when considering only dollar denominated bonds (see Column (4) under ``Dollar'').

\noindent
\textbf{Discussion.} The prevalence of cash flow-based over collateral-based debt has deep implications for the analysis of transmission channels and sources of amplifications of financial crises. First, the fact that physical assets, such as real estate and machinery, are not heavily used as collateral for debt contracts suggest that amplification effects through debt-deflationary spirals \`a la Fisher in which borrowing constraints become tighter due to reductions in the price of collateral may not play a predominant role during episodes of financial crises as hypothesized by \cite{mendoza2010sudden}. In addition, the fact that financial assets are rarely used as collateral for firms' bank debt provides evidence that the amplification effect of financial crises through a \textit{fire-sale} of assets such as proposed by \cite{shleifer2011fire} and empirically studied by \cite{duarte2021fire} should not play a significant role in constraining firms' access to credit. This is supported by the lack of an opaque shadow-banking sector in Emerging Markets such as Argentina.\footnote{For details on the study of shadow-banking in Emerging Markets see \cite{ghosh2012chasing} and \cite{cozer2015there}.}

\subsection{Heterogeneity across Credit Lines \& Firm Characteristics} \label{subsec:lending_heterogeneity}

We extend our previous analysis by studying the prevalence of cash flow-based lending across different types of credit lines and across different firm characteristics. We argue that these results validate our findings in Section \ref{subsec:lending_main_results}, and provides insights on the underlying frictions which lead firms and banks to sign the different types of debt contracts.

\noindent
\textbf{Results by type of credit line.} We begin by studying the relative importance of asset and cash flow-based lending across the different types of credit lines in our dataset. Our credit-registry dataset presents detailed information over the different type of debt contracts firms and banks can agree to, with the characteristics of these debt contracts being specified by the BCRA.\footnote{In Section \ref{sec:interest_sensitive_borrowing_constraints} we exploit the details of these regulations to proxy the borrowing constraints and debt covenants firms face.} 

Table \ref{tab:composition_firm_debt_type} presents the share of cash flow-based lending across the different type of credit lines for the year 2017.
\begin{table}[ht]
    \centering
    \caption{Cash Flow-Based Debt across Credit Lines}
    \label{tab:composition_firm_debt_type}
    \small
    \begin{tabular}{l l c c}
 & & \multicolumn{1}{c}{} \\ 
 & & \normalsize \textbf{Share of CFB} & \normalsize \textbf{Share of Total Debt} \\
\multicolumn{2}{l}{ \textbf{Credit line}}  & (1) & (2) \\ \hline \hline
\\
\multicolumn{2}{l}{\textbf{Capital Expenditures}} & \textbf{42.31\%} & \textbf{17.02\%} \\
& Automotive loans &	4.8\% & 0.3\%	\\
 & Machinery \& equipment loans &	38.4\%  & 5.6\%	\\
 & Real estate loans &	43.1\%	 & 5.8\% \\
 & Credits for financial leasing	&	47.9\%  & 5.3\%	\\
\\

\multicolumn{2}{l}{\textbf{Working Capital Expenditures}} &  \textbf{92.24\%} & \textbf{82.98\%} \\
 & Discounted documents	&	89.3\%	 & 42.4\% \\
 & Short term credit lines ($<$30 days) &	92.6\%	 & 16.6\% \\
 & Financing of working capital for exporting	&	96.5\%  & 	18.6\% \\
 & Credit card debt	& 99.5\%	 & 5.5\% \\ \hline \hline
    \end{tabular}
    \floatfoot{\textbf{Note:} This table presents the share of firm-bank debt which is cash flow-based according to different type of debt contracts reported by banks to the Central Bank to meet with current regulation. Data is presented for the year 2017. The reasoning behind this choice of years is to not taint our sample with the effects of financial crises, default or severe financial repression episodes. The types of credit lines are grouped into two sub panels, those associated with capital expenditures and those associated with working capital expenditures. We drop debt contract types which are defined as ``\textit{other loans of ...}'' of different nature.}
\end{table}   
Observing the information on Column (1), the first takeaway from Table \ref{tab:composition_firm_debt_type} is that credit lines associated with capital expenditures exhibit a higher share of collateral-based debt than credit lines associated with working capital expenditures. On the one hand, only 4.8\% of ``Automotive Loans'' credit lines are stipulated in terms of cash flow-based lending. Similarly, credit lines for the purchase of machinery and real estate have between 50\% and 65\% of their lending in collateral-based lending. On the other hand, credit lines associated with working capital expenditures, such as ``Short-term credit lines'' and ``Financing of working capital for exporting'' exhibit more than 90\% of debt in terms of cash flow-based lending.

Observing the information on Column (2), the second takeaway from Table \ref{tab:composition_firm_debt_type} is that firms borrow from banks primarily to finance working capital expenditures. For the year 2017, more than 80\% of firms' bank debt is explained by credit lines associated with working capital expenditures. The remaining 20\% of firms' bank debt is explained by credit lines associated with capital expenditures. In Section \ref{subsec:lending_comparison_international} we argue that this result is completely at odds with those found for US firms.

\noindent
\textbf{Results by firm size.} The relative importance of cash flow-based debt varies across firms' size. In particular, there is a non-monotonic relationship between firms' size and the share of cash flow-based: both relatively small and relatively large firms have cash flow-based debt explaining the vast majority of their debt, whilst medium-sized firms exhibit a relatively lower share of cash flow-based debt in comparison.

\begin{figure}[ht]
    \centering
    \caption{Relationship between Cash Flow-Based Debt \& Firm Size}
    \label{fig:lpoly_cash_flow_size}
     \centering
     \begin{subfigure}[b]{0.495\textwidth}
         \centering
         \includegraphics[width=\textwidth]{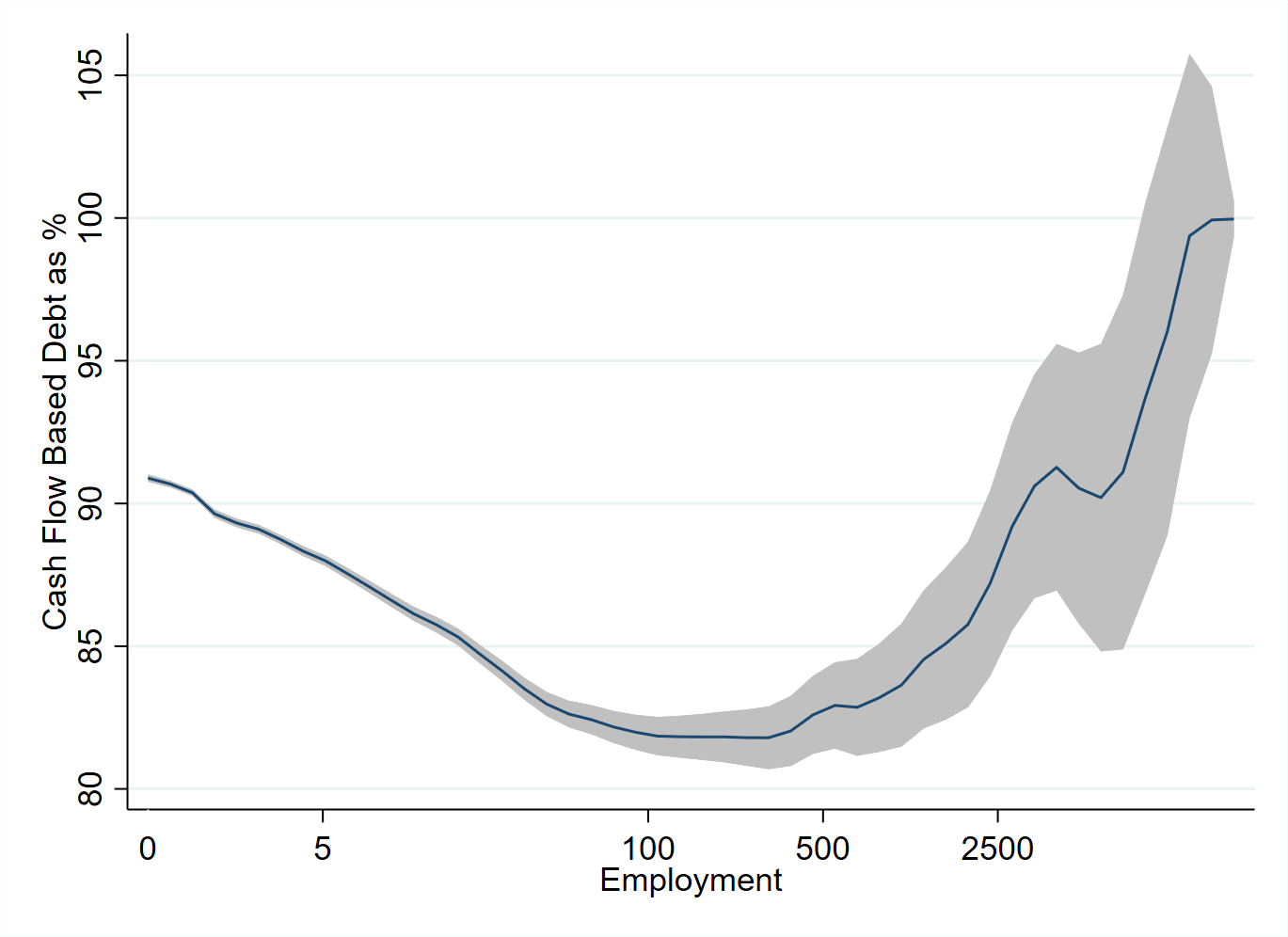}
         \caption{Total Sample}
             \label{fig:Lpoly_Cash_Flow_Empleo}
     \end{subfigure}
     \hfill
     \begin{subfigure}[b]{0.495\textwidth}
         \centering
         \includegraphics[width=\textwidth]{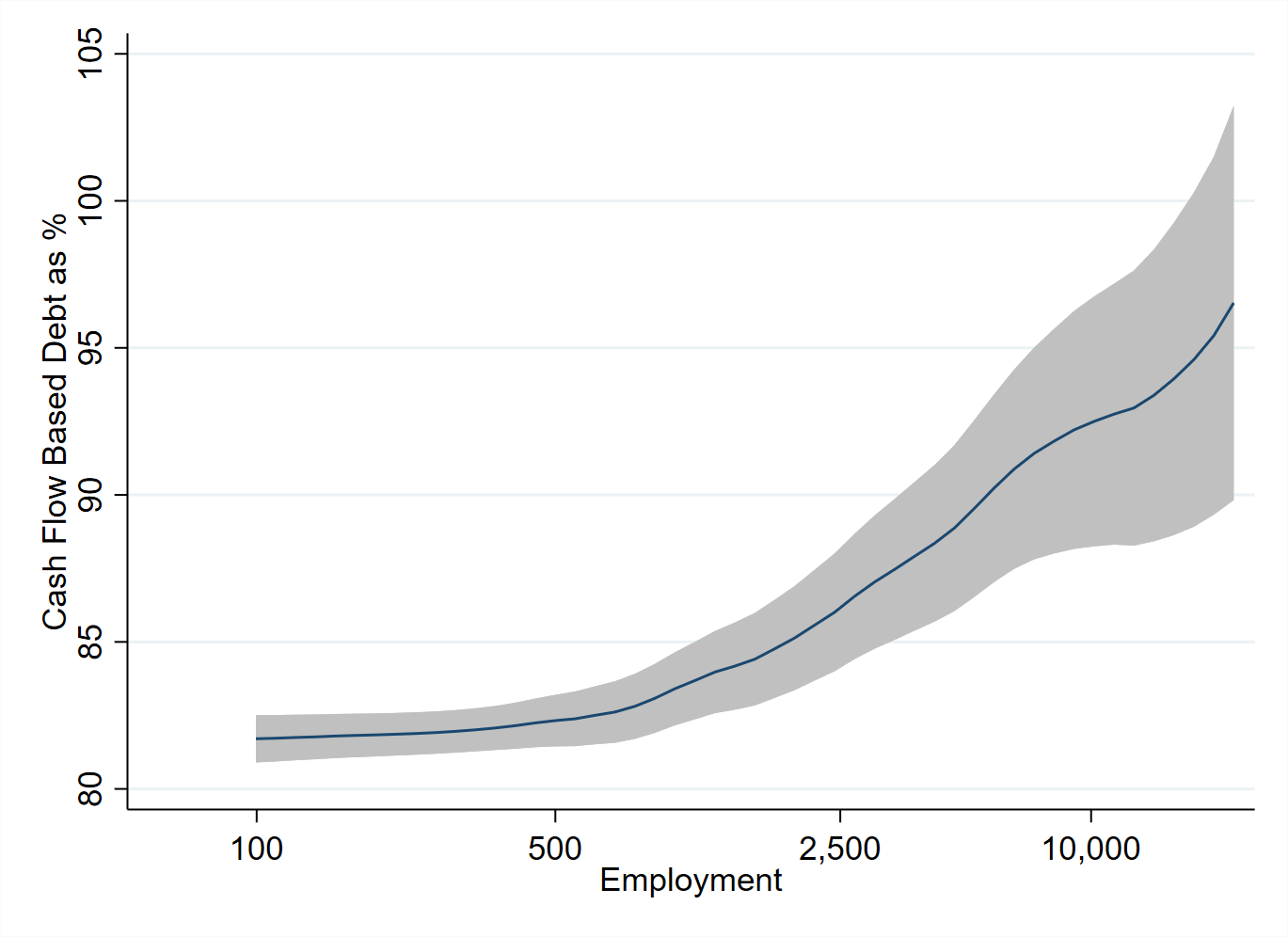}
         \caption{Large Firms}
         \label{fig:Lpoly_Cash_Flow_Empleo_LargeFirms}
     \end{subfigure} 
     \floatfoot{\textbf{Note:} The red full line represents the estimated local polynomial. The grey area represents a 95\% confidence interval. The panel on the right presents evidence for firms with employment levels above 100 and excludes independent contractors which are different ``legal persons'' than what are usually referred as firms.  }
\end{figure}

To study the relationship between cash-based debt contracts and firms' size we fit a kernel-weighted local polynomial smoothing function, presented in Figure \ref{fig:lpoly_cash_flow_size}. 
On the left panel, Figure \ref{fig:Lpoly_Cash_Flow_Empleo} shows the relationship between cash flow-based debt and firms' size, proxied by total employment. The figure exhibits an U-shape relationship between firms' share of cash flow-based lending and firms' employment. The share of cash flow-based lending is decreasing in the level of employment, reaching a trough around 50 employees. After this trough, the level of cash flow-based lending increases with the number of employees. On the right panel, Figure \ref{fig:Lpoly_Cash_Flow_Empleo_LargeFirms} shows an increasing relationship between cash flow-based debt and firms' size for relatively large firms. 

The non-monotonic relationship between cash flow-based lending and firms' employment is not surprising. Relatively smaller firms in Emerging Market economies exhibit relatively high levels of informal employment, and sell a significant share of their output in shadow markets.\footnote{For a greater discussion of the relationship between firm size and informality in Argentina see  \cite{galiani2012modeling} and/or \cite{beccaria2015informality}.} Activity in these informal markets is not reflected in any kind of legal document, such as receipts, accounting books or tax payments. Thus, these firms face significant challenges to access credit in the first place, and explain a relatively minor share of total employment and total bank-debt.\footnote{For references on informality and credit access see \cite{koeda2008informality} and/or \cite{wellalage2016informality}.}$^{,}$\footnote{Firms with levels of employment above 50 employees explain above 70\% of total bank debt (with firms in the top decile of employment explaining 83\% of total bank-debt). Furthermore, while our dataset does not allow us to observe firms' informal employment levels, we believe that total employment is an accurate proxy of firms' size. This is due to labor informality being negatively correlated with formal employment and/or other proxy variables for firm size} This has two effects on firms' borrowing patterns. Small firms borrow little from banks and when they do its mostly short-term through debt contracts which are cash flow-based by their own nature, such as credit-card debt. Relatively larger firms have a greater share of their transactions in formal markets with the accompanied paperwork. Consequently, firms' accounting is more transparent for relatively larger firms, which facilitates both a greater access to total financing and greater amount of cash flow-based debt. 

\noindent
\textbf{Results by firm age.} Several papers have found that a firm's age is correlated with key financial indicators, such as \cite{cloyne2018monetary}.\footnote{The authors find that younger firms in the US show the highest responsiveness to interest rate shocks.} For the case of collateral and cash flow-based lending, \cite{lian2021anatomy} show that relatively younger firms exhibit low to no profits and borrow primarily cash flow-based. The authors argue that younger firms finance capital expenditures through collateral-based lending and older firms, which have already accumulated capital, have positive streams of profits and thus can borrow backed by these cash flows.

\begin{figure}[ht]
    \centering
    \caption{Relationship between Cash Flow-Based Debt \& Firm Age}
    \label{fig:lpoly_age_cash_flow}    
    \includegraphics[width=12cm,height=8cm]{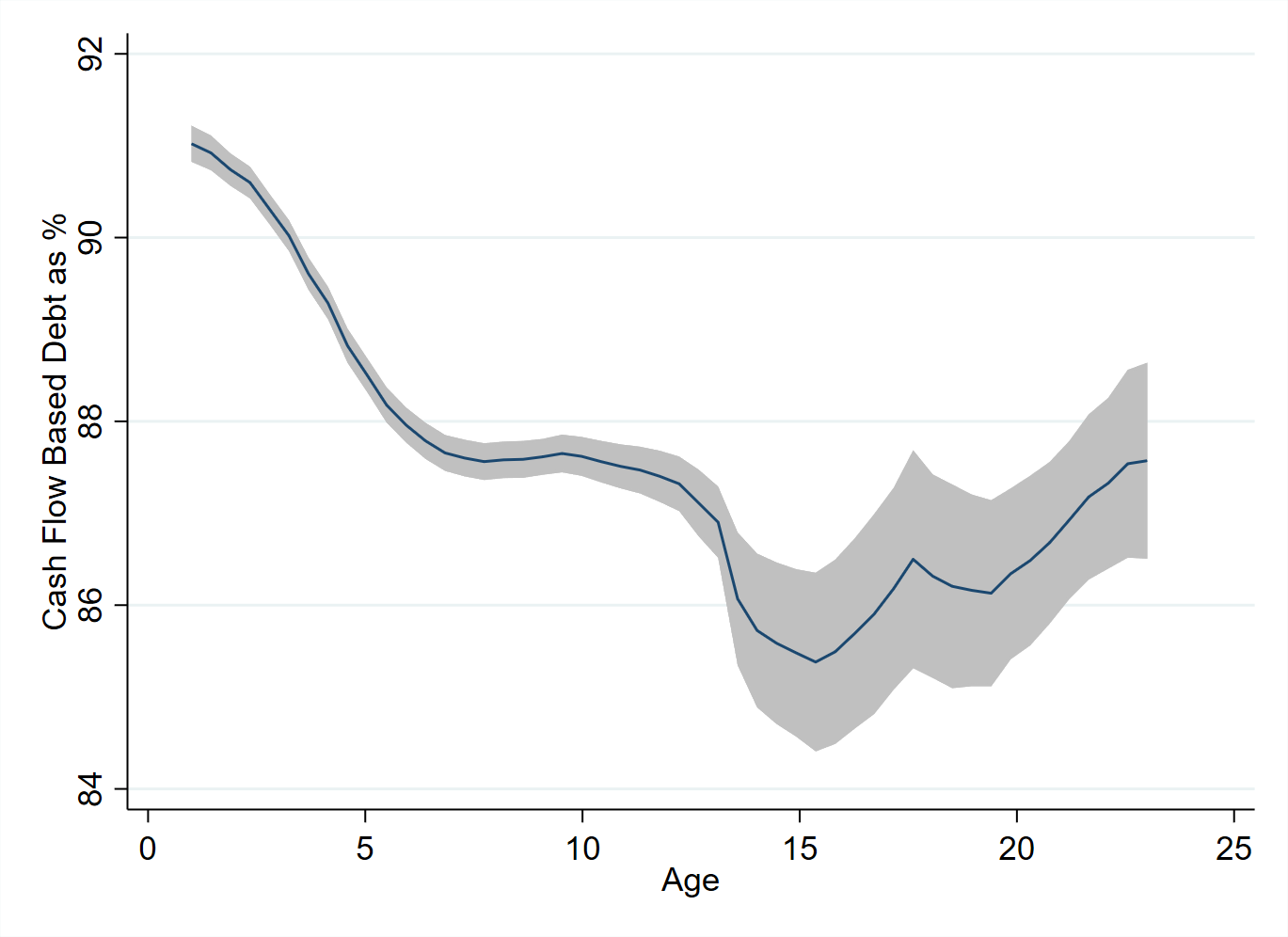}
    \floatfoot{\textbf{Note:} The blue full line represents the estimated local polynomial. The grey area represents a 95\% confidence interval.  }
\end{figure}

Figure \ref{fig:lpoly_age_cash_flow} presents the results for our full sample of firms for the year 2017.\footnote{Our datasets does not provide an indicator of firms' age. We construct our indicator of age by establishing a proxy of firms' ``\textit{birthday}''. To do so, we exploit the length of our datasets. Our international trade data goes back into the year 1994, our credit registry data goes back to 1998 and our firm level employment dataset goes up to the year 2000. We choose the year of the first observation in any of the datasets as the firm's birthday and assign them the age of 1.}
Similar to the results for firms' size, the data suggests a U-shaped relationship between firms' age and their share of cash flow-based lending.\footnote{However, this result is not driven by firms' size as the correlation between firms' size and age is positive, but relatively week at 0.2265.} While cash flow-based lending is prevalent across firms of all ages, relatively younger and older firms borrow cash flow-based, with medium age firms being the group which borrows relatively more collateral-based.

This result may be driven by firms' life cycle in an economy with informality and under-developed credit markets. Relatively young firms have few to no assets to present as collateral and little access to bank debt, let alone capital markets. Consequently, firms' typically borrow through credit card debt and short term lending based on discounted documents or account receivables which act as \textit{screening} devices. As firms mature and build larger credit records, they are able to borrow more which leads them to increase the size of their operation, both in terms of sales and capital accumulation. To carry out these capital expenditures, firms now have a higher access to collateral-based lending. Relatively older firms have scarcer investment opportunities but greater cash flows, which they can use to back their borrowing of working capital. 

\noindent
\textbf{Results by sector of activity.} Next, we turn to analyzing how cash flow-based debt patterns vary across production sectors. For the case of the US, \cite{lian2021anatomy} show that while close to 80\% of firms' debt is cash flow-based, airlines and utilities companies are the exception and exhibit a large share of collateral-based debt. The authors argue that while most non-financial firms have high asset specificity, airlines and utilities companies are special cases where firms have a large amount of standardized, transferable assets (aircraft for airlines and power generators for utilities) that facilitate collateral-based lending. We test whether there are sector differences in cash flow-based bank debt across production sectors in Argentina.

Table \ref{tab:composition_by_sector} shows the decomposition of total firm-bank debt for the main productive sectors of the Argentinean economy in descending order of the importance of collateral-based lending.\footnote{The productive sectors arise from the Argentinean National Account classification.}
\begin{table}[ht]
    \centering
    \caption{Cash Flow-Based Debt across Production Sectors}
    \label{tab:composition_by_sector}
    \footnotesize
    \begin{tabular}{l c c c}
\textbf{Sector}	&	Collateral-Based	&	Cash Flow-Based	&	Share of Total Debt \\ \hline \hline
Fishing	&	42.6\%	&	57.4\%	&	0.1\%	\\
Transportation, storage and communications	&	40.8\%	&	59.2\%	&	4.0\%	\\
Construction	&	26.0\%	&	74.0\%	&	4.0\%	\\
Health and social services	&	25.4\%	&	74.6\%	&	1.2\%	\\
Education services	&	22.5\%	&	77.5\%	&	0.3\%	\\
Hotels and restaurants	&	21.9\%	&	78.1\%	&	0.7\%	\\
Real estate, business and rental activities	&	21.9\%	&	78.1\%	&	3.8\%	\\
Agriculture, livestock, hunting and forestry	&	21.0\%	&	79.0\%	&	12.7\%	\\
Wholesale, retail and repairs	&	14.6\%	&	85.4\%	&	20.7\%	\\
Manufacturing industry	&	10.2\%	&	89.8\%	&	35.4\%	\\
Utilities (Electricity, gas and water supply)	&	8.4\%	&	91.6\%	&	1.7\%	\\
Exploitation of mines and quarries	&	5.1\%	&	94.9\%	&	5.9\%	\\
Financial intermediation	&	4.4\%	&	95.6\%	&	4.6\%	\\
 \hline \hline
    \end{tabular}
    \floatfoot{\textbf{Note:} This table summarizes the composition of the universe of firm-bank debt from Argentina according to firm's self-declared sector of activity for tax-purposes. The table presents the decomposition of total firm-bank debt to the sector. The data presented is for the year 2017. Results are robust and persistent across time. We exclude public administration, national defense and unclassified services. }
\end{table}
Two sectors clearly stand out with collateral-based debt contracts explaining almost half of their total debt: ``Fishing'' and ``Transportation, storage and communications''. This is not surprising as these two industries are special cases where firms have a large amount of standardized and transferable assets which facilitate collateral-based lending. The ``Fishing'' industry is heavily regulated (boat building permits, fishing permits, oceanic fishing permits among others) with a significant degree of over-sight by government institutions (particularly by the Ministry of Agriculture and Fishing). Besides standardized assets such as warehouses and supplies (such as nets), fishing permits are granted at the boat level with ownership transfers stipulated within regulations.\footnote{For additional details on the regulations, permits and transfer of permits/boats see \url{https://www.magyp.gob.ar/sitio/areas/registro_pesca/tramites/}.} Similarly, firms in the ``Transportation, storage and communications'' sector are also intensive in standardized assets such as storage units, warehouses and automotive capital. Consequently, this result is in line the information presented in Table \ref{tab:composition_firm_debt_type}, which showed that the vast majority of the ``Automotive loans'' credit lines are associated with collater-lendingfi. Lastly, it is also noteworthy and not surprising that the ``Agriculture, livestock, hunting and forestry'' sector exhibits a relatively higher share of collateral-based debt. As argued in Appendix \ref{subsec:appendix_BCRA_Regulations_ABC}, collateral-based regulations are stipulated in great detail for the agricultural sector as they establish borrowing constraints as function of the price agricultural land and the value of finely classified cattle stocks.\footnote{Furthermore, Argentinean banks' have branches specific for the Agricultural sector. For instance, ``\textit{Banco Provincia de Buenos Aires}'' the second largest bank in the country and largest country in the province of Buenos Aires which masses the largest part of fertile agricultural land has sector specific credit lines. For instance, credit lines are specified to the different types of crops which have different maturing cycles. For more information on the ``PROCAMPO'' credit lines of this bank can be observed at \url{https://www.bancoprovincia.com.ar/agro/agro_procampo}.} Furthermore, land for agricultural and livestock purposes is usually considered a safe storage of value in Argentina, maintaining its price (or even increasing) during periods of financial stress.\footnote{For instance, \cite{villenamarchetti2003} show that the price in US dollars of the agricultural and live stock land in Argentina increased significantly during 1980/1982 corporate and sovereign debt crisis, the 1989/1990 hyperinflation crisis, and the 2001/2002 sovereign debt crisis.} 

In summary, while collateral-based debt contracts explain a relatively larger share of bank debt for certain sectors, the largest sectors of the economy in terms of total bank debt exhibit significantly high levels of cash flow-based debt. For instance, the ``Wholesale, retail and repairs'' and the ``Manufacturing industry'' sectors have 85.4\% and 89.8\% of their total debt expressed in cash flow-based contracts. 

\noindent
\textbf{Participation in International Trade.} Cash flow based debt is more prevalent for exporting firms compared to non-exporting firms even when controlling for firm size. As suggested by \cite{lian2021anatomy}, Emerging Market economies have less transparent accounting and tax systems which may difficult the emergence of debt contracts specified in terms of financial terms such as EBITDA. Exporting operations provide firms with a verifiable future revenue flow which allow firms to circumvent, at least partially, poor accounting quality and/or other agency costs. 

The BCRA introduced in the year 2014 the ``Financing of working capital for exporting'' or ``Pre-financiaci\'on de exportaciones'' credit line which allows firms to borrow in foreign currency. Under BCRA regulations, borrowing through this credit lines requires firms to present verifiable information of an outstanding import order by a foreign firm. BCRA regulations over this credit line impose clear borrowing limits over this credit lines:
\begin{itemize}
    \item The sum of total bank debt and/or other operations with financial institutions carried out by the exporting firm exporter for this type credit line must not exceed 75\% of the amount (FOB value) of exports.\footnote{For full details on this regulation see Section 2 of the Central Bank's ``Texto ordenado de las normas sobre pol\'iticas de cr\'edito'', available at \url{https://www.bcra.gob.ar/Pdfs/Texord/t-polcre.pdf}, and Section 3.1 of BCRA's ``Document A-3561'', available at     \url{http://www.bcra.gov.ar/pdfs/comytexord/A3561.pdf}.}$^{,}$\footnote{Regulations also specify that banks may ask/use additional verifiable information over firms' export flows of revenue in order to determine any borrowing limit. The regulation also contemplates access to foreign currency for firms which act as direct suppliers of firms which have a verifiable exporting source of foreign currency funds. These firms also must provide evidence of the regular and period exports of their direct customers. In particular, regulation specifies that firms must provide information on their export invoices for the previous 12 months.}
\end{itemize}
Consequently, the inherent paperwork involved in exporting operations provides firms with the necessary documentation banks require to carry out risk evaluations. Overall, these regulations imply that firms can borrow up to some fraction of their present and past values of their exporting revenue.\footnote{In Section \ref{sec:interest_sensitive_borrowing_constraints} we come back to this point when we describe the cash flow-based borrowing constraint BCRA regulations imply.}

\begin{figure}[ht]
    \centering
    \caption{Cash Flow-Based Debt according to Export Performance }
    \label{fig:lpoly_exporter}
     \centering
     \begin{subfigure}[b]{0.495\textwidth}
         \centering
         \includegraphics[height=6cm,width=8cm]{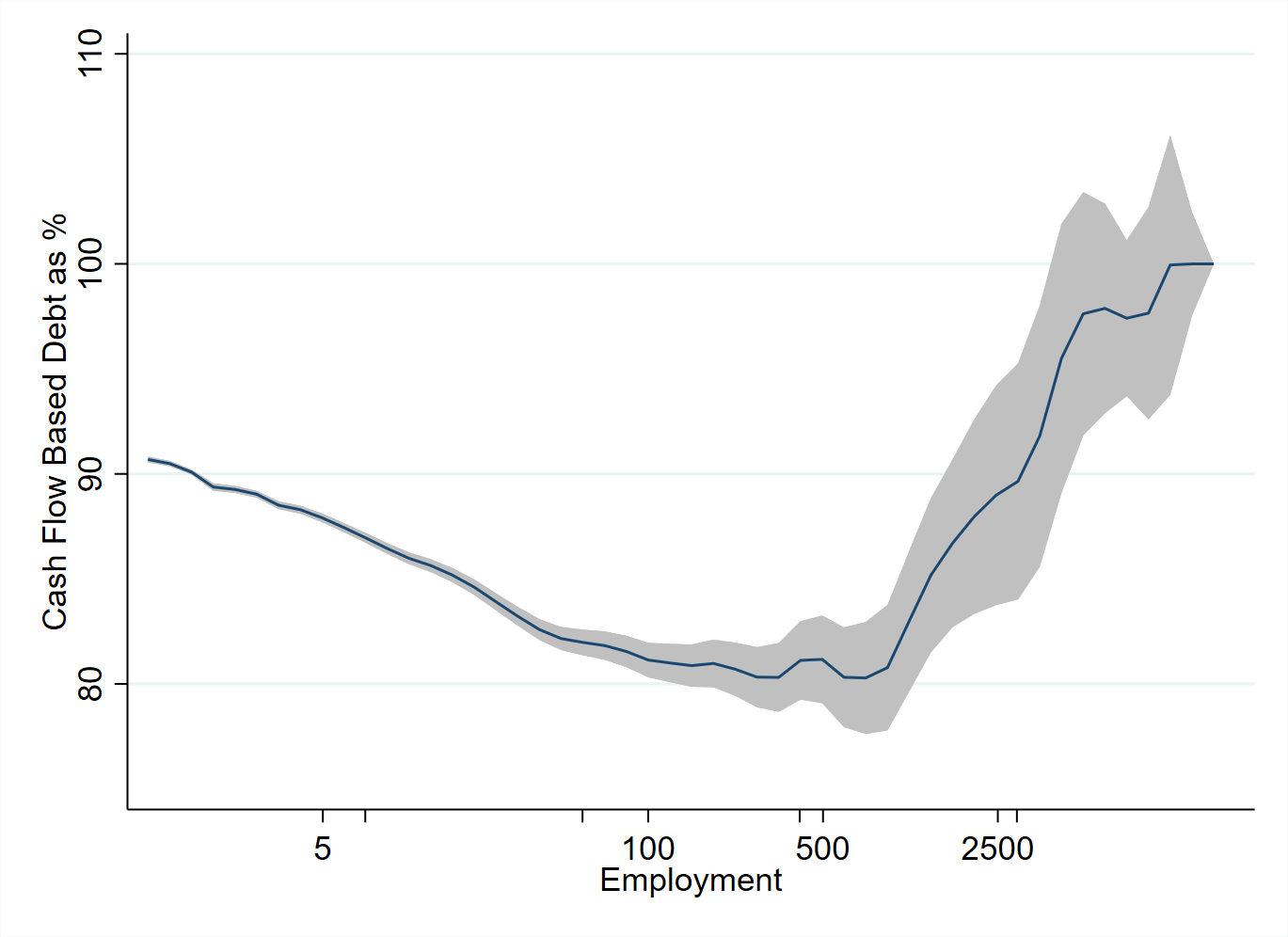}
         \caption{Non Exporters}
         \label{fig:lpoly_empleo_non_exporter}
     \end{subfigure}
     \hfill
     \begin{subfigure}[b]{0.495\textwidth}
         \centering
         \includegraphics[height=6cm,width=8cm]{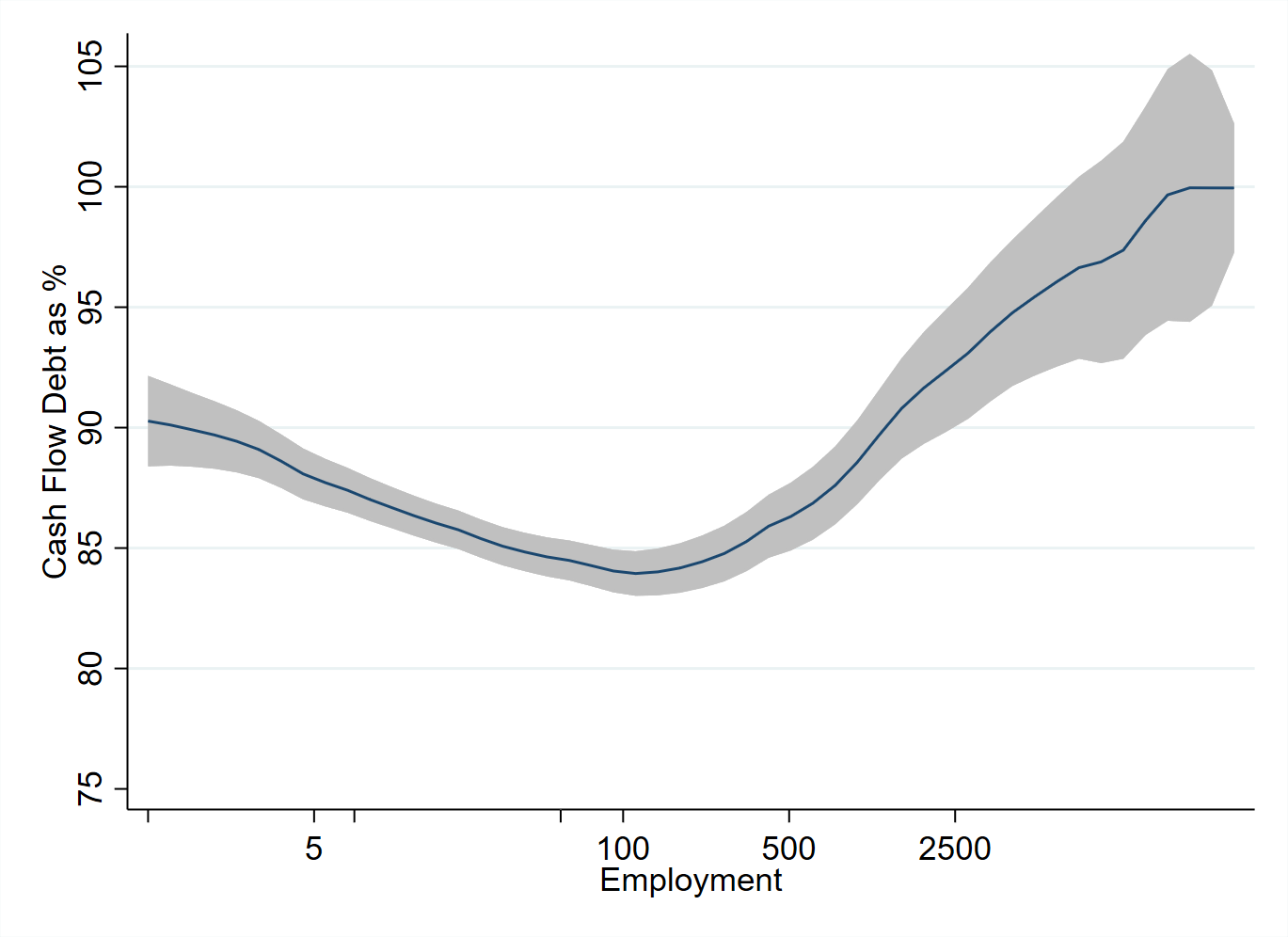}
         \caption{Exporters}
         \label{fig:lpoly_empleo_exporter}
     \end{subfigure} 
     \begin{subfigure}[b]{0.495\textwidth}
         \centering
         \includegraphics[height=6cm,width=8cm]{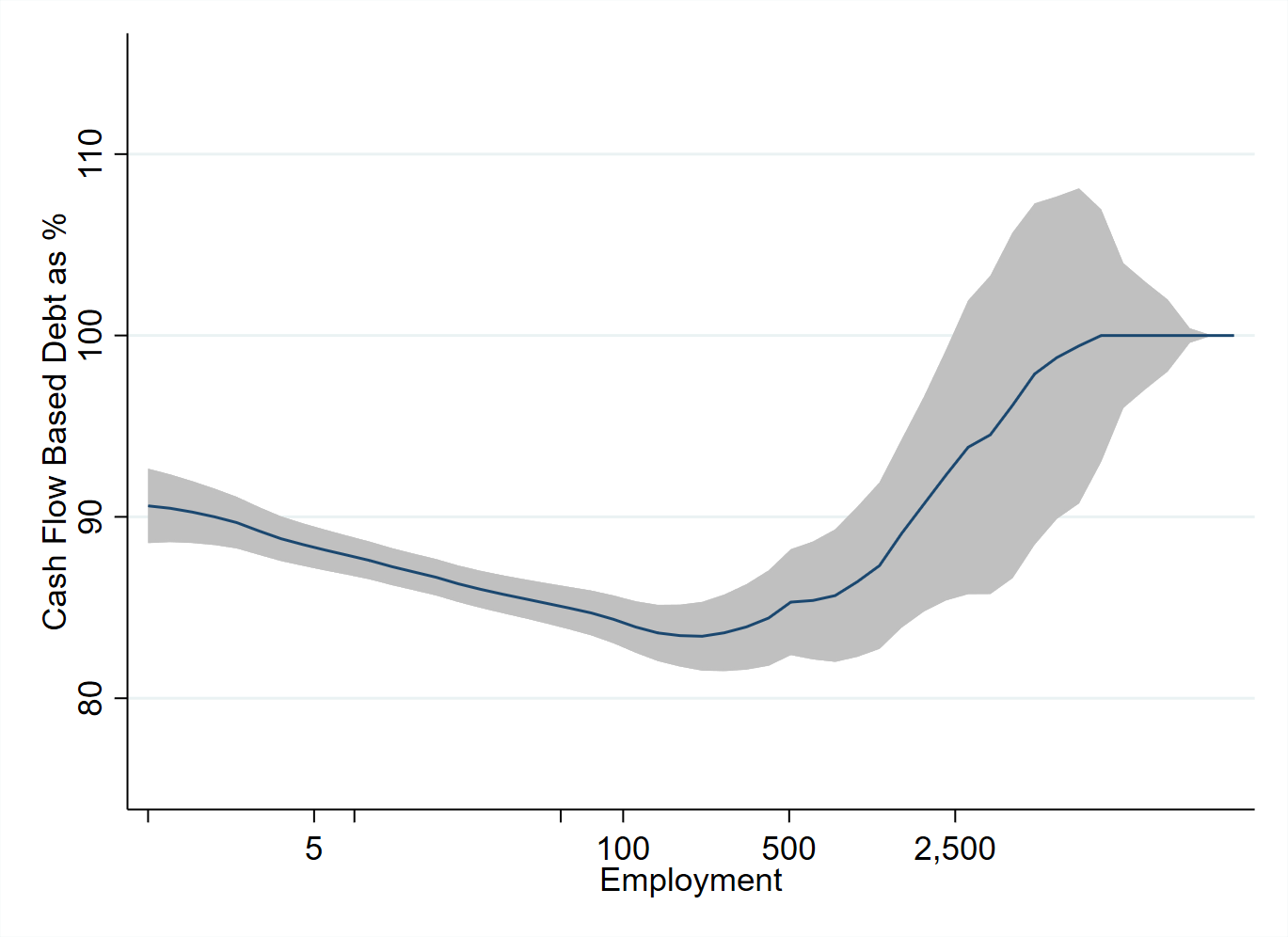}
         \caption{Exporters Below Sector's Median}
         \label{fig:lpoly_empleo_exporter_below_sector_median}
     \end{subfigure}
     \hfill
     \begin{subfigure}[b]{0.495\textwidth}
         \centering
         \includegraphics[height=6cm,width=8cm]{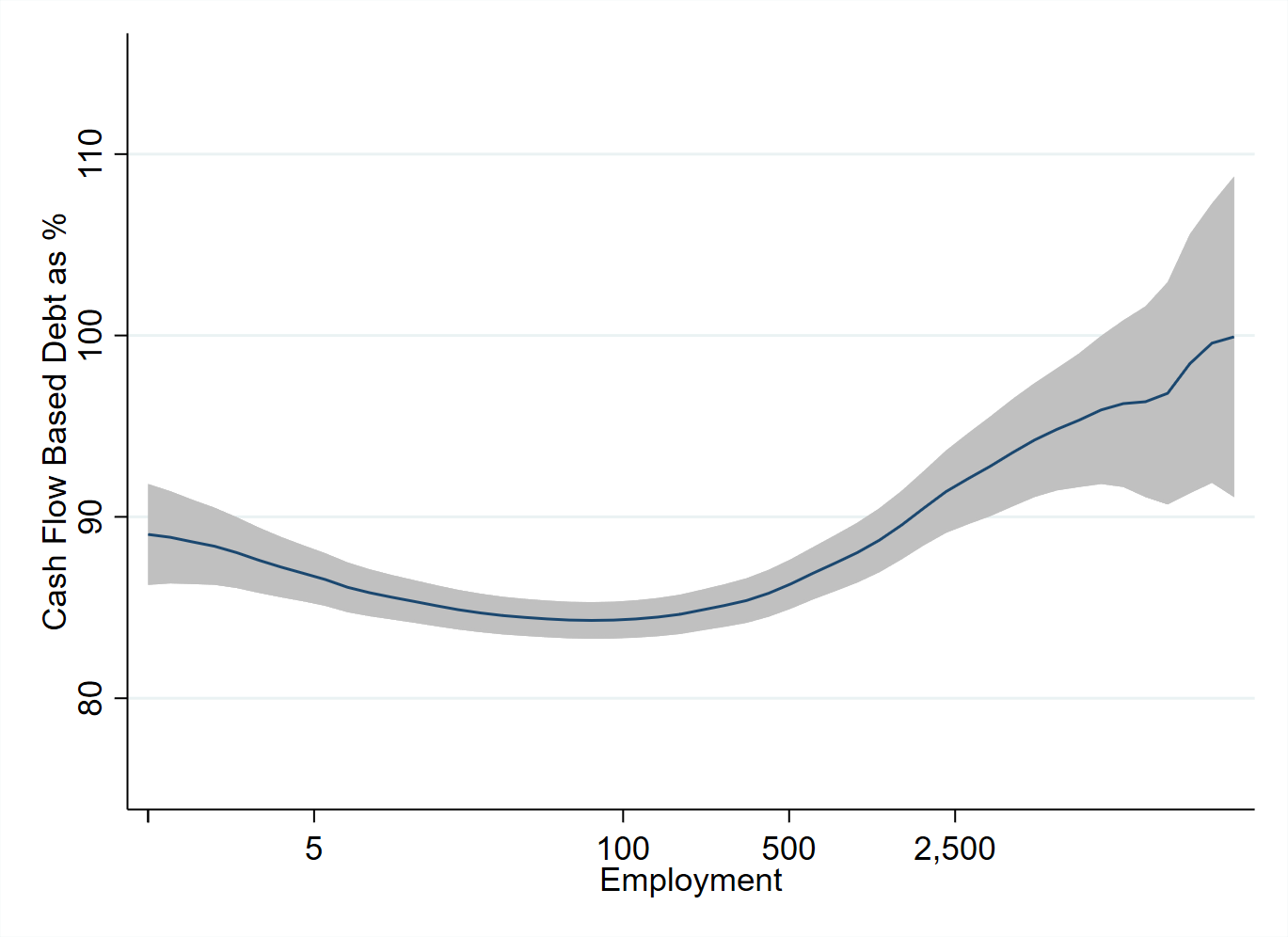}
         \caption{Exporters Above Sector's Median}
         \label{fig:lpoly_empleo_exporter_above_sector_median}
     \end{subfigure}
     \floatfoot{\textbf{Note:} The blue full line represents the estimated local polynomial. The grey area represents a 95\% confidence interval. Figures reflect data from the year 2017. Top-left panel presents data for non-exporting firms, top-right panel presents data for exporting firms. Bottom left and right panels present data for firms below and above the sector specific median of exported value at the firm level.}
\end{figure}

Figure \ref{fig:lpoly_exporter} shows the relationship between the firms' share of cash flow-based debt for four different sub-samples: (i) the top-left panel presents the figure for non exporting firms (i.e. firms which did not exported at all during the year 2017); (ii) the top-right panel presents the figure for exporting firms (i.e. firms which did export during the year 2017); (iii) the bottom left panel presents the figure for the subset of exporting firms which are below the sector-specific median amount of exported value per firm; (iv) the bottom right panel presents the figure for the subset of exporting firms which are above the sector-specific median amount of exported value per firm.
In line with the results presented in Figure \ref{fig:lpoly_cash_flow_size}, there is an U-shaped relationship between firms' share of cash flow-based debt and employment across all four sub-samples. By comparing the figure for non exporting firms (Figure \ref{fig:lpoly_empleo_non_exporter}) with the figure for exporting firms (Figure \ref{fig:lpoly_exporter}), it is clear that the latter have a greater share of cash flow-based debt, particularly for medium levels of employment, 50 to 500 employees. The differences between exporting firms are more subtle. Figures \ref{fig:lpoly_empleo_exporter_below_sector_median} and \ref{fig:lpoly_empleo_exporter_above_sector_median} present the relationship for exporting firms below and above the sector specific median exported value. The 

In Appendix \ref{sec:appendix_complimentary_facts} we present additional evidence through several regression exercises. Overall, the evidence suggests that exporting firms have a greater share of cash flow-based debt than non-exporting firms, in line with the insights provided by Figure \ref{fig:lpoly_exporter}. Furthermore, this relationship holds when controlling for other firm characteristics and sector fixed effects. These results present supporting evidence to the hypothesis that exporting operations, and the intrinsic paperwork involved provides firms with verifiable information about future cash flows. Consequently, exporting firms exhibit a greater share of cash flow-based debt.\footnote{Additionally, as we show in Appendix \ref{sec:appendix_complimentary_facts}, exporting firms also borrow relatively more than firms of the same employment or size. Whether this increase in total bank debt is explained by being able to provide additional verifiable information over future cash-flows, by the nature of exporting activities which are credit intensive, or by other confounding variable, it is not clear and not pursuit in the current paper. }

\subsection{Comparison with International Evidence} \label{subsec:lending_comparison_international}

We conclude this section by comparing the results arising from our datasets of Argentinean firms with evidence from the US and other countries. First, we argue that cash flow-based lending is more prevalent in Argentina than in the US and provide several probable conjectures on why this is the case. Second, we argue that our results for Argentina may be extrapolated to other Emerging Market economies.

\noindent
\textbf{Comparison with evidence from US firms.} There is a significant and recent literature which uses firm and other micro level data to study non-financial firms' borrowing constraints in the US (see \cite{greenwald2018mortgage,greenwald2019firm,drechsel2019earnings,lian2021anatomy,chodorow2022loan}). Thus, a straightforward questions is how do our results for Argentinean firms compare to those from the recent literature which studies US firms? We argue that cash flow-based lending is more prevalent for Argentina firms than for US firms, conditional on certain caveats about data comparability. 

The main limitation to carry out a comparison between our results and those from US firms arise from the differences between our datasets. On the one hand, the literature which studies the importance of cash flow-based lending among US firms relies on balance sheet and supervisory data sets of syndicated loans. These datasets usually comprises relatively large and public companies. On the other hand, our credit-registry dataset contains information on firm-bank links for the universe of Argentinean firms. However, we argue that in spite of this data limitations, there is significant evidence of greater prevalence of cash flow-based lending in Argentina. 

Table \ref{tab:cash_flow_lending_comparison_US} compares our results for Argentinean firms with the findings of \cite{lian2021anatomy} for the US non-financial corporations. 
\begin{table}[ht]
    \centering
    \caption{ Comparison of Cash Flow-Based Lending with US Evidence}
    \label{tab:cash_flow_lending_comparison_US}
    \small
    \begin{tabular}{l c c c || c}
             & \multicolumn{3}{c}{\textbf{Argentina}} & \textbf{US} \\
             &  BCRA - Full  &  BCRA - $L\geq 500$ & Corp. Loans & \cite{lian2021anatomy} \\
             & (1) & (2) & (3) & (4) \\\hline \hline
         Aggregate Share CFB   & 83\% & 91\% & 94\% & 80\% \\
         Mean Share CFB & 89\% & 92\% & 98\% & 85\%  \\
         Firms with CFB    & 100\%  & 100\%  & 100\%   & 62\%  \\
          \hline \hline
    \end{tabular}
    \floatfoot{\textbf{Note:} This table summarizes the composition of the universe of firm-bank debt from Argentina. The data presented is for the year 2017. The choice reflects the composition by the end of our sample where we have significant complementary firm information. Data for the US is sourced directly from \cite{lian2021anatomy}. The acronym CFB represents cash flow-based lending.}
\end{table}
At the aggregate level, the first row of Table \ref{tab:cash_flow_lending_comparison_US} shows that for our full credit-registry sample (first column) cash flow-based debt contracts from Argentinean firms represent around 83\% of total firm-bank debt, roughly greater than that for US firms on the last column. Column (2) shows that when conditioning on firms with 500 or more employees, a relatively more comparable sample, the share of cash flow-based debt becomes significantly greater at 91\% than that present for US firms. This becomes even more evident when comparing the results that arise from Argentinean corporate loans in Column (3), which exhibit a share of cash flow-based lending of almost 94\%. 

Additionally, the second and third rows of Table \ref{tab:cash_flow_lending_comparison_US} show that cash flow-based lending are significantly more prevalent in Argentina than in the US at the firm level. The second row shows that the mean share of cash flow-based lending at the firm level is significantly greater for both our samples of the credit-registry dataset and our sample of corporate loans than for US firms. Lastly, and perhaps more importantly, the third row of Table \ref{tab:cash_flow_lending_comparison_US} shows that cash flow-based lending is significantly more pervasive across Argentinean firms, with all firms in our sample providing some type of cash flow-based credit line. In comparison, only 62\% of US firms exhibit cash flow-based lending.\footnote{Similarly, \cite{lian2021anatomy} report that among small firms on the Compustat dataset the median share of collateral-based lending is 54\% and the median share of cash flow–based lending is about 8\%.}

Table \ref{tab:cash_flow_lending_comparison_across_countries} compares our results with those arising from \cite{lian2021anatomy}'s results for their sample of the ``Rest of the World''.\footnote{Data shown in Table \ref{tab:cash_flow_lending_comparison_across_countries} is constructed using information presented in the online appendix of \cite{lian2021anatomy}.}
\begin{table}[ht]
    \centering
    \caption{Cash-Flow Based Lending across Countries}
    \label{tab:cash_flow_lending_comparison_across_countries}
    \begin{tabular}{l c c | c c}
                        &  \multicolumn{2}{c}{\cite{lian2021anatomy}} & \multicolumn{2}{c}{Argentina}  \\
                         & US & Rest of the World &  BCRA & Corporate Loans  \\
                        & (1) & (2) & (3) & (4) \\ \hline \hline
         Aggregate Share CFB  & 80\% & 54\%-66\% & 83\% & 98\%  \\
         Median Share CFB  & 85\% & 35\%-55\% &   100\% & 100\% \\ \hline \hline
         \floatfoot{\textbf{Note:} The source of the results presented in columns (1) and (2) are sourced from the Online Appendix of \cite{lian2021anatomy}. The authors differentiate between countries with ``Reorganization'' and ``Non-Reorganization'' bankruptcy processes. The left statistic presents the results for latter and the right statistics presents the results for the former.}
    \end{tabular}
\end{table}
Comparing the results on columns (1) and (2) it is clear that US firms exhibit a larger share of cash flow-based lending than countries in the ``Rest of the World'' category. However, this analysis is based on using the ``CompuStat'' dataset which over-represents relatively large firms from relatively wealthier countries. The authors only keep countries with at least 500 firm-level observations. As shown in \cite{dai2012international}, with the exception of China and India, there are no other Emerging Market economies which report more than 500 firm level observations at any year for the period 1990-2010. Comparing the results in column (2) with those in columns (3) and (4), it is clear that Argentina exhibits a larger share of cash flow-based lending than the countries in the ``Rest of the World'' category. Next, we argue that we can extrapolate our results from Argentina to other Emerging Market economies. 

\noindent
\textbf{Conjectures and extrapolation to other Emerging Market economies.} Our dataset does not allow us to disentangle the underlying informational frictions or agency costs that lead to the greater prevalence of cash flow-based lending in Argentina compared to US. We provide four possible conjectures which may explain this result. Furthermore, we argue that these conjectures allow us to extrapolate our results to other Emerging Market economies. Furthermore, we argue that our results for Argentina can be extrapolated to other Emerging Markets which share the similar bankruptcy laws and banking regulations as Argentina.

\begin{itemize}
    \item[1.] Differences in creditors: the role of banks and equity markets
\end{itemize}

First, we argue that a key difference between firms' financing in Argentina (and Emerging Markets as a whole) compared to the US is the significant differences in the role of banks in financing the private sector. Figure \ref{fig:Role_Banks_Credit} presents the share of private sector financing explained by banks for a sample of countries.
\begin{figure}[ht]
    \centering
    \caption{Role of Banks in Private Sector Financing \\ \footnotesize Selected Examples}
    \label{fig:Role_Banks_Credit_Example}    
    \includegraphics[width=9cm,height=6cm]{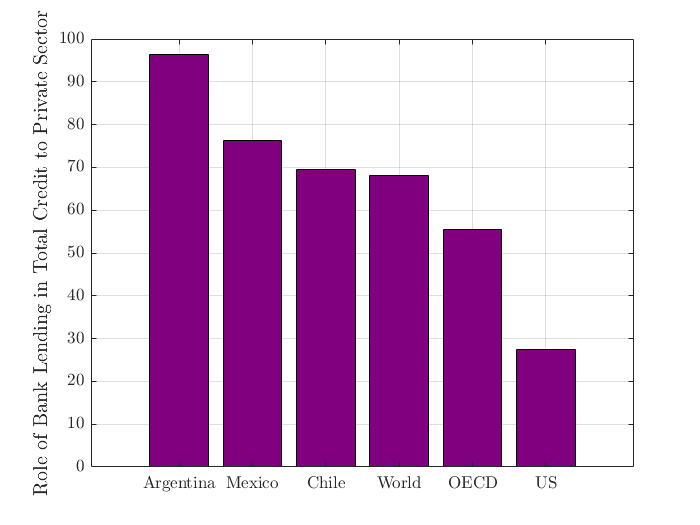}
     \floatfoot{\textbf{Note:} Figure \ref{fig:Lpoly_Ratio_GDP} presents the ratio of ``Domestic credit to private sector by banks (\% of GDP)'' to ``Domestic credit to private sector (\% of GDP)''. All indicators are sourced from the World Bank. }
\end{figure}
For the case of Argentina (on the left of the figure), banks explain more than 90\% of private sector financing. Other Emerging Market economies such as Mexico and Chile exhibit shares of bank financing above 70\%, above the World average of 68\%. These results are in sharp contrast with the result for the US, in which banks only explain 27\% of all private sector financing.  

\begin{figure}[ht]
    \centering
    \caption{Role of Banks \& Equity Markets \\ \footnotesize International Comparison across Income Levels }
    \label{fig:lpoly_international}
     \centering
     \begin{subfigure}[b]{0.495\textwidth}
         \centering
         \includegraphics[width=\textwidth]{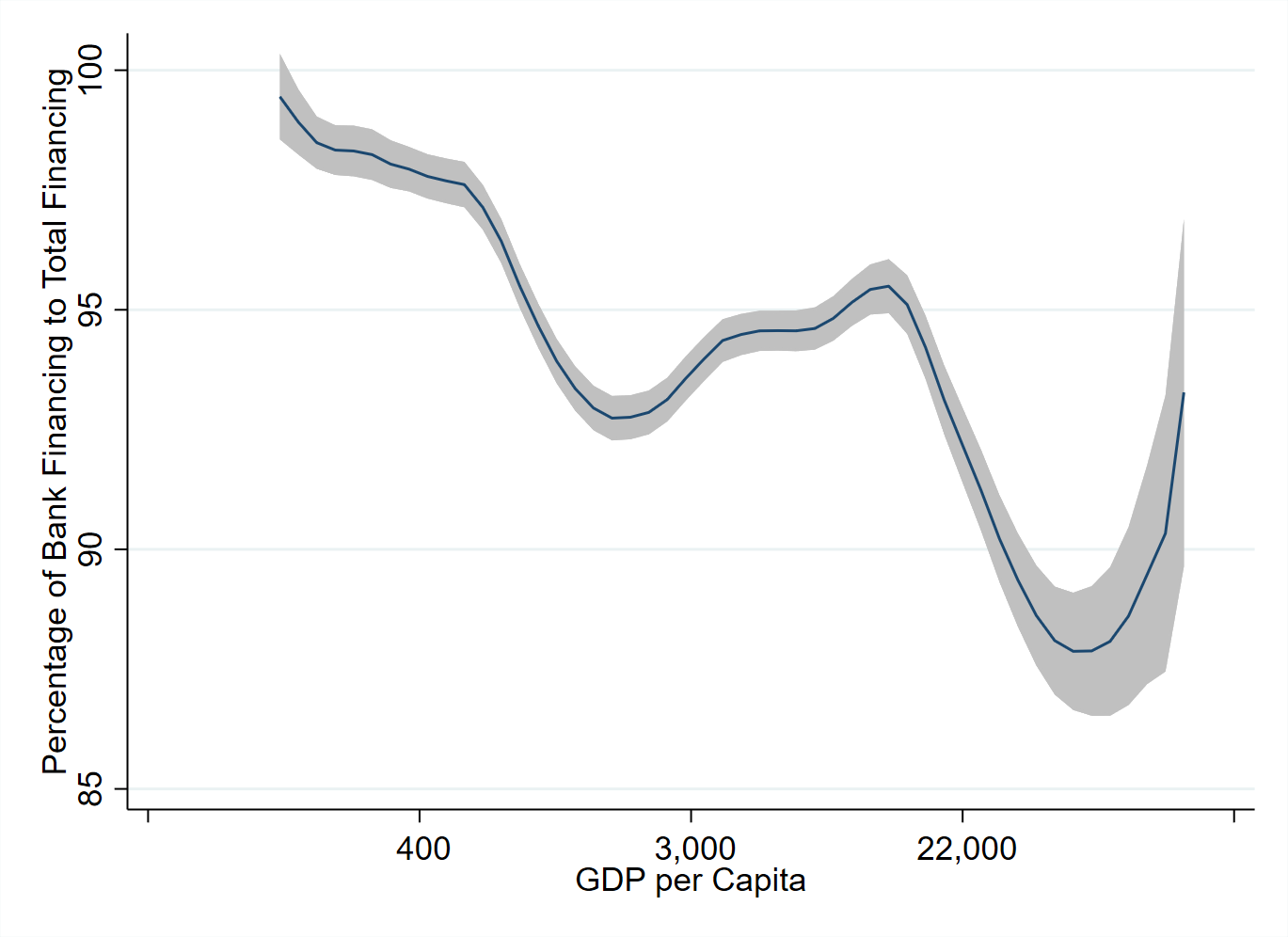}
         \caption{Role of Banks in Total Private Financing}
         \label{fig:Lpoly_Ratio_GDP}
     \end{subfigure}
     \hfill
     \begin{subfigure}[b]{0.495\textwidth}
         \centering
         \includegraphics[width=\textwidth]{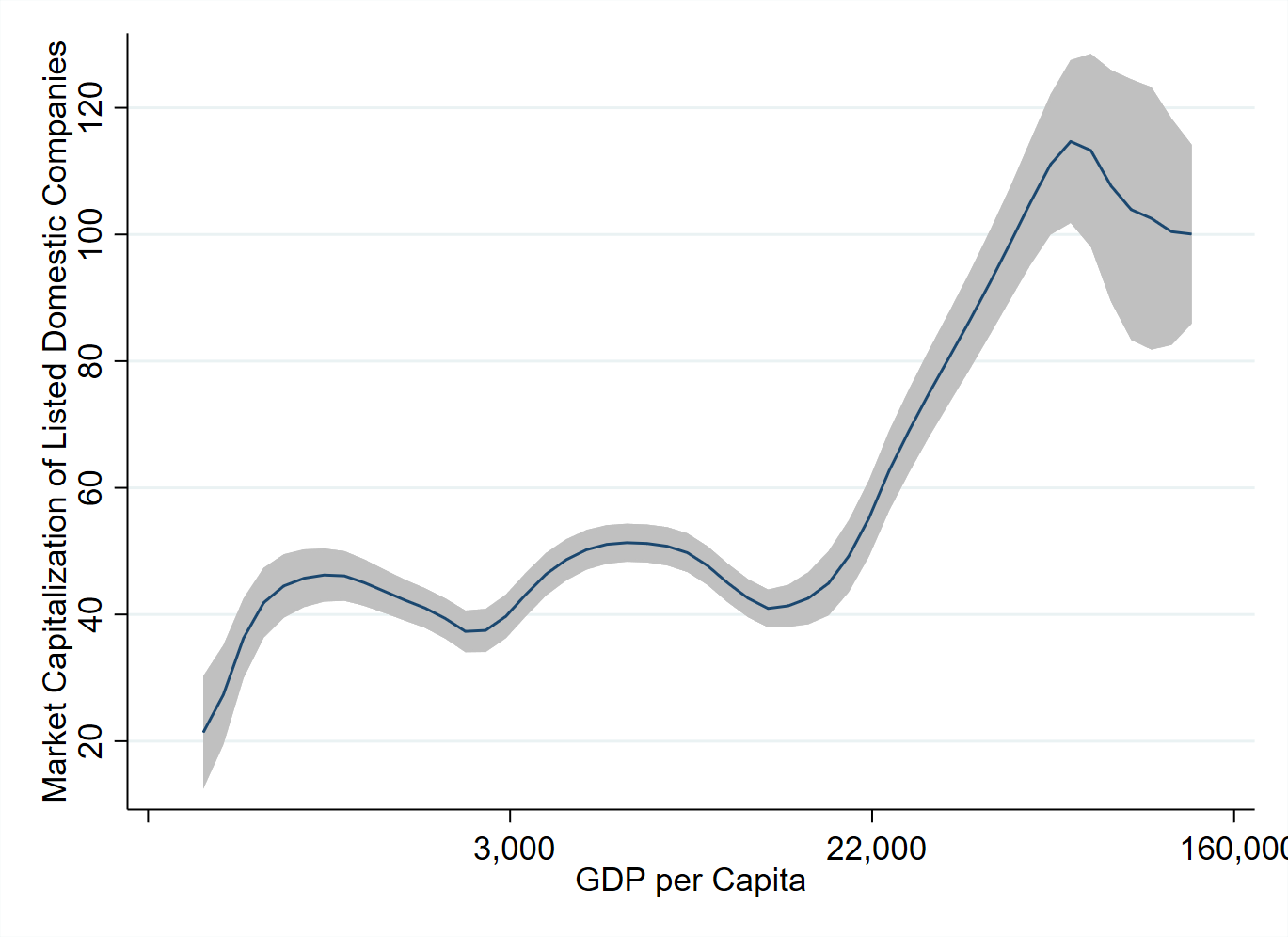}
         \caption{Market Capitalization}
         \label{fig:Lpoly_Market_GDP}
     \end{subfigure} 
     \floatfoot{\textbf{Note:} Figure \ref{fig:Lpoly_Ratio_GDP} presents the ratio of ``Domestic credit to private sector by banks (\% of GDP)'' to ``Domestic credit to private sector (\% of GDP)''. Figure \ref{fig:Lpoly_Market_GDP} presents the ``Market capitalization of listed domestic companies (\% of GDP)''. Data on GDP per capita is constructed using constant dollars for the year 2015. All indicators are sourced from the World Bank. }
\end{figure}

Next, Figure \ref{fig:lpoly_international} shows evidence of a robust and negative relationship between the role of banks and a country's GDP per capita. 
On the left panel, Figure \ref{fig:Lpoly_Ratio_GDP} shows a robust negative relationship between the share of private financing explained by banks and a country's GDP per capita. In line with this decreasing role of banks in private sector financing, on the right panel Figure \ref{fig:Lpoly_Market_GDP} shows an increasing relationship between the market capitalization of listed domestic companies, as a percentage of GDP, and a country's GDP per capita. 

The relative importance of banks and capital markets as sources of financing for firms is a widely and thoroughly studied research topic.\footnote{For the US, \cite{berg2020trends} use Capital IQ data to summarize recent trends in corporate borrowing emphasizing the relative role of banks and capital markets. First, the authors show that balance sheet debt financing by US publicly listed firms is driven by bond financing, exceeding bank loan borrowing by a factor four throughout our sample period. The authors argue that the terms ``bank loan'' and ``term loan'' can be used interchangeably ``based on the fact that ``term loans'' are typically arranged by banks and banks hold a large fraction of term loans''.} Greater preference for market financing as a country grows and financial markets deepen may reflect the high pro-cyclicality of bank credit supply, as suggested by \cite{kashyap1992monetary,kashyap1994credit}.\footnote{In particular, \cite{becker2014cyclicality} find strong evidence of firms substituting bank loans to bonds at times characterized by tight lending conditions (such as high-levels of non-performing loans, etc.).} Emerging Market economies lag in their development of domestic bond and equity markets, despite being a prominent agenda item.\footnote{For Emerging Market policy discussion on the importance of domestic bond and equity markets see \cite{hardie2019financial}.}$^{,}$\footnote{While domestic corporate bond markets have emerged in many countries this development has been far from uniform (see \cite{abraham2020growth,chang2017bond}). The literature suggests that a main cause of less developed domestic bond markets and higher corporate bond yields is sovereign debt. For instance, \cite{augca2012sovereign} shows that an increase in sovereign debt by one standard deviation from its sample mean is associated with 9\% higher corporate loan yield spreads. Similarly, \cite{cavallo2010determinants} shows that sovereign risk remains a significant determinant of corporate risk, and that Emerging Markets' lower transparency and information availability leads to greater financial panics and investor herding behavior.}

There is a large literature which suggests that banks play an important monitoring role over firms. \cite{berg2016total} and \cite{berg2020trends} show that banks increase total borrowing costs and charge firms fees which are key to facilitate ex-ante screening and control ex-post moral hazard. \cite{altman2010bank} analyze banks' monitoring role and show consistent monitoring advantage of bank-loans over bonds by exploiting secondary loan and bond markets close to episodes of firms' default. \cite{hale2009banks} show that banks exploit their informational monopoly over firms and charge greater interest rates by observing the path of interest rates around IPOs. Similarly, \cite{santos2008bank} show that banks exploit their informational advantage over firms by comparing the pricing of loans for bank-dependent borrowers with the pricing of loans for borrowers with access to public debt markets, controlling for risk factors.\footnote{Additional evidence on banks' monitoring role is presented by \cite{altman2004informational}. In particular, the authors examine the informational efficiency of bank loans relative to bonds surrounding loan default dates and bond default dates. Consistent with the hypothesis that banks have a more precise expectations over firms performance, the price reaction of loans is less adverse than that of bonds around loan and bond default dates. Authors conclude that the bank loan market is informationally more efficient than the bond market around default dates. Similarly, \cite{yi2006informational} show that loan and bond ratings are not determined by the same statistical model. The authors study credit spreads incorporating bank loan ratings (among other factors) and find evidence which suggest that bank ratings provide information not reflected in financial information. The authors conclude that ratings may capture idiosyncratic information about recovery rates, as each of the agencies claims, or information about default prospects not available to the market.}$^{,}$\footnote{In an international comparison context, \cite{de2011bank} show that the ratio of bank to bond financing and the debt to equity ratio are lower in the US than in the Euro Area, while the converse is true for the risk premia on bank loans. In line with banks' informational advantage, the authors rationalize these facts through a model with agency costs where heterogeneous firms raise finance through either bank loans or corporate bonds and where banks are more efficient than the market in resolving informational problems. The authors conclude that the Euro Area is characterized by lower availability of public information about corporate credit risk relative to the United States, and when European firms value more than United States firms banks’ information acquisition role.}

The fact that firms in Emerging Market economies rely relatively more in banks than market financing imply that firms are less able to smooth their supply of financing needs and are exposed to greater scrutiny than firms in Advanced Economies. It is straightforward to conclude that greater scrutiny leads to firms providing better and more accurate information on their cash flows which allows banks to assess their default risk.

\begin{itemize}
    \item[2.] Differences in debt maturity
\end{itemize}

Second, we argue that a greater prevalence of cash flow-based lending can be rationalized by differences in debt maturity between Emerging Markets and the US or other Advanced Economies. While cross-country comparisons of corporate debt maturity are hindered by difficulties in data collection and standardization, the literature has reached a consensus that Emerging Market firms exhibit certain ``short-termism'' (see \cite{giannetti2003better,fan2012international}). For instance, for the case of US firms, \cite{berg2020trends} and \cite{chodorow2022loan} use data from both corporate and syndicate loans to show that 90\% of firms' debt value has a maturity above one year and an average maturity above four years for the period 2002-2020. On the other hand, for the case of Emerging Market economies, \cite{cortina2018corporate} uses balance sheet from publicly listed firms in both Emerging Markets and Advanced Economies to show that the median firm from developing countries holds a lower ratio of long-term debt to total liabilities than firms from developed countries (15\% and 19\%, respectively).\footnote{However, \cite{cortina2018corporate} shows that when considering only firms that use corporate bond or syndicated loan markets, the ratio of long-term debt to total liabilities goes up and becomes similar across developing and developed countries. Nevertheless, the authors stress that the patterns that emerge from their analysis applies primarily to a select group of large corporation with Emerging Market economies' firms accessing bond and syndicated loan markets are of size similar having smaller firms than Advanced Economies (see \cite{bento2017misallocation,hsieh2014life}).}$^{,}$\footnote{Furthermore, \cite{demirgucc2015impact} and \cite{ccelik2019corporate} shows that Emerging Market firms' debt maturity has actually decreased since the Great Financial Crisis.} Furthermore, \cite{cortina2018corporate} argues that Emerging Markets' ``short-termism'' is related to domestic bonds, which firms in these countries issue at shorter maturity than firms from Advanced economies countries, and their greater reliance in traditional bank borrowing.\footnote{This fact has been thoroughly studied for the case of differences between government debt maturities across these economies.See \cite{chang1999liquidity,chang2001model,broner2013emerging} for both an empirical and theoretical analysis of Emerging Markets' reliance on short-term sovereign debt and its implications. See \cite{greenwood2015optimal} for an analysis of the US government's debt maturity.}

\begin{figure}[ht]
    \centering
    \caption{Role of Short Term Debt across Income Levels \\ \footnotesize International Comparison}    
    \label{fig:Lpoly_Current_Liabilities}
    \includegraphics[scale=0.20]{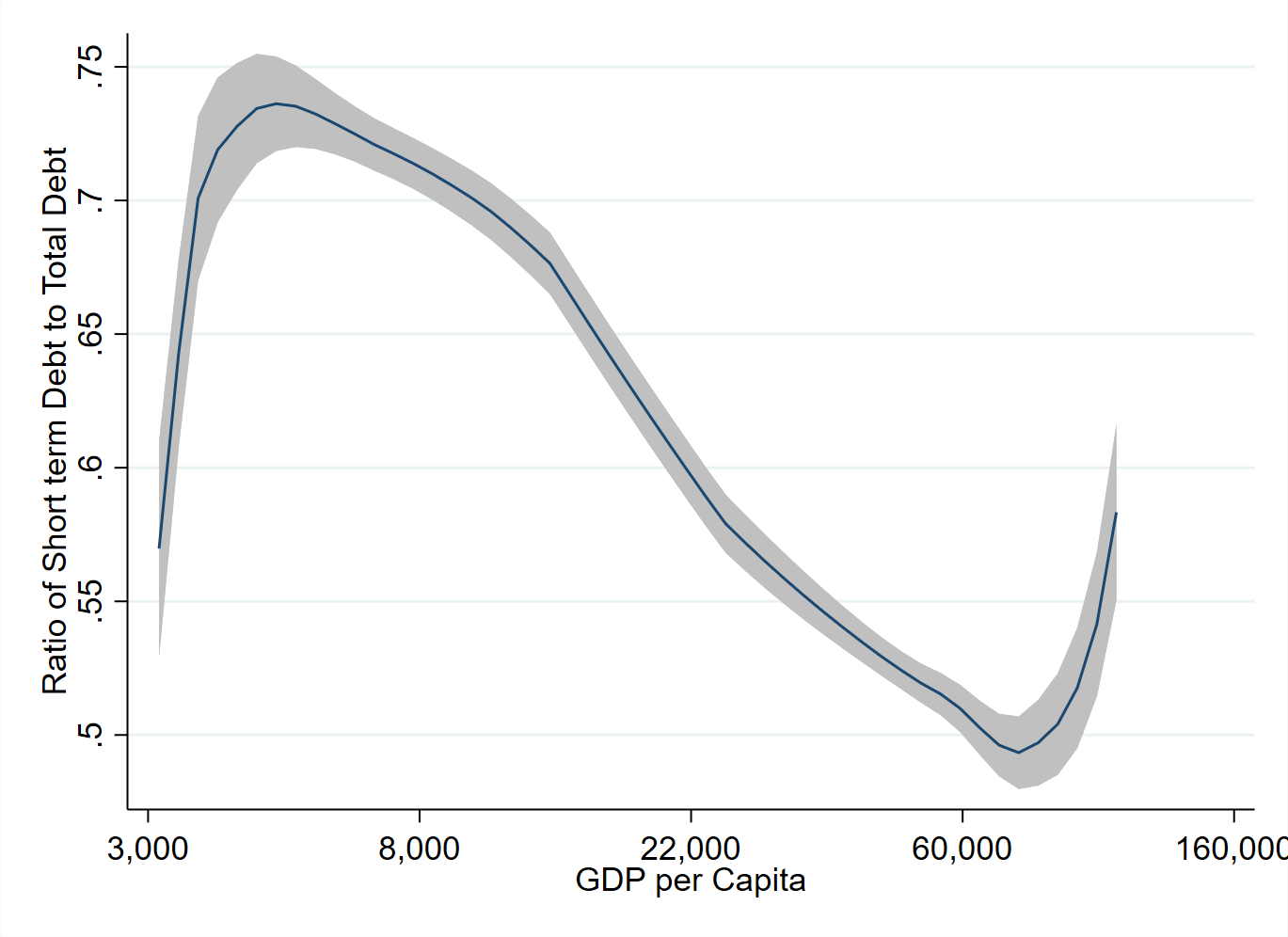}
    \floatfoot{\textbf{Note:} The ratio of short term debt to total debt is computed as the ratio between Orbis variables ``Total Current Liabilities'' and ``Total Liabilities'' multiplied by 100. Data on GDP per capita is constructed using constant dollars of the year 2015 and is sourced from the World Bank.}
\end{figure}

We show that there is a negative relationship between firms' debt maturity and their country's income per capita. To do so, we construct a panel of  16,564 firms across 28 countries (both Emerging Market and Advanced Economies) for the period 2013-2017 through the Orbis dataset.\footnote{Access to the Orbis dataset was granted through Northwestern University's Library. See \url{https://www.google.com/search?q=northwestern+university+orbis+library&rlz=1C1CHBD_enUS920US920&oq=north&aqs=chrome.0.69i59l2j69i57j69i61l3j69i65l2.890j0j7&sourceid=chrome&ie=UTF-8}.}$^,$\footnote{In Appendix \ref{subsec:appendix_additional_data_description_orbis} we describe in detail the panel dataset constructed and present summary statistics.} Note that within this panel of firms the representation is towards both relatively richer countries and towards relatively larger firms within each country, a common feature of standardized international financial and balance sheet datasets. Nevertheless, Figure \ref{fig:Lpoly_Current_Liabilities} shows evidence of a strong and negative relationship between firms' ratio of short term debt (under one year) to total debt and countries' GDP per capita. 
On the one hand, the figure suggest that firms in countries categorized as ``Middle Income'' by the World Bank ($\approx $US\$ 5,100 in the year 2017) exhibit a ratio of short term to total debt close to 75\%. On the other hand, the figure suggests that firms in countries categorized as ``High Income'' by the World Bank ($\approx $US\$41,100 in the year 2017) exhibit a ratio of short term debt to total debt close to 50\%. The fact that firms in Emerging Market economies borrow at relatively shorter term/maturity expose them to fast changing borrowing conditions and a greater exposure to roll-over and liquidity risk.\footnote{Table \ref{tab:ratio_current_gdp} in Appendix \ref{sec:appendix_international_comparison} shows that this result holds when controlling for firm characteristics and sector and country fixed effects. In addition, Table \ref{tab:ratio_current_gdp} in Appendix \ref{sec:appendix_international_comparison} shows that even when controlling for countries' level of income per capita, firms in the US exhibit a ratio of current to total liabilities 10\% lower.}$^{,}$\footnote{There is a literature which studies in detail firms' debt maturity in detail. \cite{berg2020trends} show that for US publicly listed firms the average maturity for the period 2002-2019 is 4.5 years. Furthermore, the authors find that firms tend to extend maturity during economic expansions and can only borrow at relatively shorter maturities during recessions or episodes of financial distress. Additionally, in line with Stylized Fact 2, \cite{nguyen2022debt} show that firms that have a greater preference over bank loans with respect to market financing have 70 months shorter maturity on average than firms with low preference for bank loans. }

\begin{itemize}
    \item[3.] Differences in credit lines and financing needs
\end{itemize}

Third, we argue that the greater prevalence of cash flow-based lending in Argentina can be explained by differences in financing needs across countries. To this end, we show that Emerging Market's firms' short run borrowing is explained by firms borrowing primarily for working capital needs. The difference in underlying financing needs is particularly sharp when comparing the evidence from Argentina and the US' firms.

\begin{figure}[ht]
    \centering
    \caption{International Comparison in Financing Needs \\ \footnotesize Working Capital vs Capital Expenditures }
    \label{fig:international_comparison_wk_ke}
     \centering
     \begin{subfigure}[b]{0.495\textwidth}
         \centering
         \includegraphics[width=\textwidth]{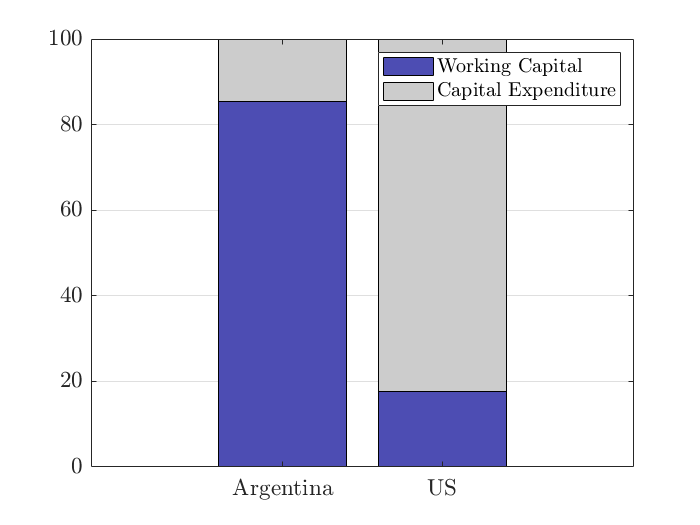}
        \caption{Argentina vs US}
    \label{fig:wk_vs_ke_main}
     \end{subfigure}
     \hfill
     \begin{subfigure}[b]{0.495\textwidth}
         \centering
         \includegraphics[width=\textwidth]{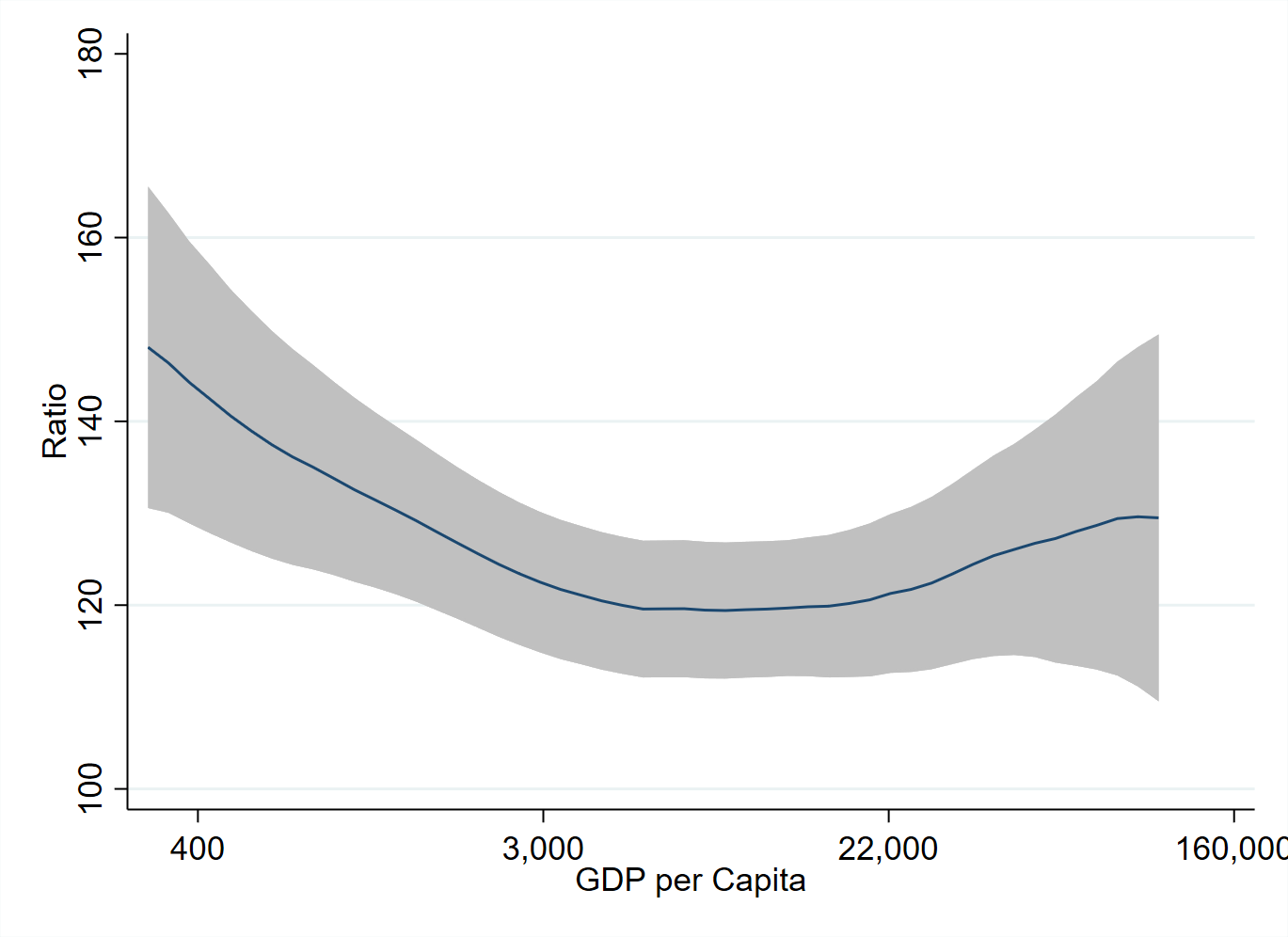}
         \caption{International Comparison Capitalization}
         \label{fig:Lpoly_Firms_Ratio_main}
     \end{subfigure} 
     \floatfoot{\textbf{Note:} On the left panel, Figure \ref{fig:wk_vs_ke_main}, the source of data for firms in Argentina is the Credit-Registry dataset from BCRA. The classification of the different type of credit lines is described in Appendix \ref{sec:appendix_stylized_facts}. Data for firms in the US is sourced from the Quarterly Financial Report by the BLS. The share of ``Working Capital'' debt is computed as the ratio of ``Net Working Capital'' debt to ``Total Liabilities''. For both countries, the figure reflects evidence from the year 2017. Figure \ref{fig:Lpoly_Firms_Ratio_main} presents the ratio of the share of firms which rely in banks for working capital financing to the share of firms which rely in banks for capital expenditure needs. Data on GDP per capita is sourced from the World Bank at constant 2015 USD. Given that waves of the WBES occurred from 2002 to 2020, we compute the average GDP per capita per country for said period.}
\end{figure}
On the left panel, Figure \ref{fig:wk_vs_ke_main}, shows that, on the one hand, firms in Argentina borrow primarily to cover working capital needs, with 82\% of total bank debt explained by credit lines associated with working capital needs.\footnote{In Appendix \ref{sec:appendix_stylized_facts} we show in detail the classification of bank debt into working capital or capital expenditures. } On the other hand, firms in the US exhibit the opposite pattern with working capital needs explaining only 18\% of firms' total debt.


\section{Prevalence of Interest Sensitive Borrowing Constraints} \label{sec:interest_sensitive_borrowing_constraints}

In this section, we present evidence that the standard cash flow-based borrowing constraint is highly sensitive to interest rates. This constraint stipulates limits on a firm's debt payments based on measures of cash-flows. In Section \ref{subsec:interest_sensitive_bank_regulations} we show that BCRA regulations impose clear guidelines over both cash flow and collateral-based borrowing constraints. In Section \ref{subsec:interest_sensitive_corporate} we present supporting evidence by studying the debt covenants of hundreds of corporate loans. Section \ref{subsec:interest_sensitive_corporate} presents evidence that violating borrowing constraints lead to significant reductions in firms' credit around the 2001-2002 sovereign default crisis. 

\subsection{Evidence from BCRA Regulations} \label{subsec:interest_sensitive_bank_regulations}

We start by showing that BCRA regulations over banks shape the underlying borrowing constraints firms face. First, we show that these regulations establish detailed limits on firms' collateralized borrowing based on the different types of asset. Second, we show that BCRA regulations require banks to monitor firms and assess their risk of default based on the relationship between debt payments and cash flows. 

\noindent
\textbf{Collateral based borrowing constraints.} BCRA regulations heavily regulate collateral-based lending. BCRA regulations contemplate two distinct classes of assets which can be used as collateral: ``Type A'' assets and ``Type B'' assets. On the one hand, ``Type A'' assets are comprised of the transfer or surety of rights with respect to financial assets, equity or other financial documents of any nature that, reliably instrumented, ensure that the bank or lending entity will be able to dispose of the funds by way of cancellation of the obligation contracted by the client, without the need to previously require the debtor to sell the assets or repay the debt. 

On the other hand, type B ``preferred assets'' are comprised of assets or rights over assets or commitments of third parties that, reliably instrumented, ensure that the bank or lending entity will be able to dispose of the funds by way of cancellation of the obligation contracted by the client, previously complying with the procedures established for the execution of these assets. Once again, the regulation implies that the comprised assets should efficiently liquidated in transparent and liquid markets. This type of assets comprises claims over real estate (built property land lots and construction trusts), claims over automotive vehicles and agricultural, road and industrial machinery (to the extent that they are registered in the pertinent national registry of automotive property and have a market that allows obtaining a reference value), or a fixed pledge with registration on bovine cattle; financial leasing contracts with respect to the purchase of capital goods and/or other machinery.\footnote{Under financial leasing contracts for firms' purchase of capital goods and/or other machinery, the Argentinean firm finances the purchase of a productive good and keeps claims over the good while the borrowing firms makes use of it for production.} A private agent which borrows backed by these assets must present documentation which prove ownership and provide significant detail on assets. A subset of this documentation is the registration of assets, such as real estate, automotive vehicles, agricultural machinery and cattle stock under public registration entities.\footnote{For instance, real estate ownership and use must be registered in province specific registration entities. For the case of the capital city see \url{https://www.argentina.gob.ar/justicia/propiedadinmueble}. For automotive vehicles (including individual and company owned vehicles) there is a national registry which could be accessed at \url{https://www.dnrpa.gov.ar/portal_dnrpa/}. This registry also incldues agricultural machinery, see \url{http://servicios.infoleg.gob.ar/infolegInternet/anexos/45000-49999/46096/norma.htm}. Cattle stock and other agricultural assets must be declared for both registration purposes and for phytosanitary conditions/regulations. For more information on the registry of cattle stocks see \url{https://www.argentina.gob.ar/senasa/programas-sanitarios/cadenaanimal/bovinos-y-bubalinos/bovinos-y-bubalinos-produccion-primaria/registros-y-habilitaciones/bovinos-y-bubalinos-produccion-primaria/identificacion-animal} and \url{https://www.argentina.gob.ar/senasa/programas-sanitarios/cadenaanimal/bovinos-y-bubalinos/bovinos-y-bubalinos-produccion-primaria/registros-y-habilitaciones/bovinos-y-bubalinos-produccion-primaria/identificacion-animal}. For the registry of fishery machinery and equipment see \url{https://www.argentina.gob.ar/palabras-clave/registro-de-la-pesca}.}

The Central Bank's regulation stipulates constraints on how much firms can borrow given the ``preferred assets'' that back debt contracts.\footnote{These constraints on the amount firms can borrow as a function of the value of ``preferred assets'' are usually referred within the Central Bank's regulation as ``m\'argenes de cobertura''.}  Table \ref{tab:bounds_preffered_assets} shows bounds as a percentage of asset values for several types of both Type ``A'' and Type ``B'' preferred assets debt contracts.
\begin{table}[ht]
    \centering
    \caption{Borrowing Limits over Collateral-Based Debt}
    \label{tab:bounds_preffered_assets}
    \footnotesize
    \begin{tabular}{l l c}
                         &  &  Bound as \% of Asset Value \\ \hline \hline
    \multicolumn{3}{c}{ }  \\
    \multicolumn{2}{l}{\underline{Type ``A'' assets}} &  \\
     & Cash or Highly-Liquid Domestic Currency assets & 100\% \\
     & Cash or Highly-Liquid Foreign Currency assets & 80\% \\
     & Gold & 80\% \\
     & Sovereign Bonds & 75\% \\
     & Central Bank Liabilities & 100\% \\
     & Private Equity Claims    & 70\% \\     
    \multicolumn{3}{c}{ }  \\
    \multicolumn{2}{l}{\underline{Type ``B'' assets}} &  \\     & Real Estate  & 50\%-100\% \\   
       & Automotive vehicles    \& agricultural machinery & 60\%-75\% \\
       & Road \& industrial machinery        & 60\% \\
       & Cattle stock                        & 60\% \\
    \multicolumn{3}{c}{ }  \\
     \hline \hline 
    \end{tabular}
    \floatfoot{\textbf{Note:} The table presents bounds on a subset of assets which are usually considered as collateral of private agents'. Consequently, the table is not exhaustive. All bounds are computed as a fraction of current market value of the assets. Asset types with bounds on the value which can back debt contracts represent heterogeneity in bounds within asset types. For instance, how much an agent can borrow backed by real estate depends on whether the property is used as living place or not, whether the property is an empty lot, a lot with construction built on it and/or an agricultural lot.}
\end{table}
The top panel shows the bounds on Type ``A'' assets. How much a private agent can borrow is higher for financial assets in domestic currency than for financial assets in foreign currency. Additionally, private agents can pledge a higher fraction of more liquid assets, such as cash or central bank liabilities, than of relatively less liquid assets such as sovereign bonds and/or private equity. The bottom panel shows private agents' bounds for several Type ``B'' assets. Depending on the type of asset and the debtor's credit-worthiness (more on this in the next section) these debt contracts exhibit bounds between 60\% and 100\% of the market value of the Type ``B'' of preferred assets.

Bank debt contracts backed by collateralized assets impose the following borrowing constraint
\begin{align} \label{eq:borrowing_constraint_text}
    B_t \leq \theta^{LEV} \times p_{k,t} k_t
\end{align}
where $B^{LEV}_t$ denotes the amount of debt, $p_{k,t} k_t$ represents the value of the asset used as collateral and $\theta^{LEV} \in (0,1)$ is parameter that governs the tightness of the constraint. The borrowing constraint in Equation \ref{eq:borrowing_constraint_text} is the standard collateral-based borrowing constraint for international macroeconomic models, such as \cite{mendoza2010sudden}. Under this benchmark borrowing constraint, a drop in the price of capital, $p_{k,t}$, leads to a tighter borrowing constraint, which reduces a firm's ability to accumulate capital, $k_t$. This drop in capital accumulation leads to an even tighter borrowing constraint, amplifying the original shock. This Fisherian deflation that reduces credit and the price and quantity of collateral assets is the key ingredient of the benchmark international macroeconomic models.

\noindent
\textbf{Cash flow based borrowing constraints.} BCRA regulations over banks' credit policy and risk assessment shape firms' cash flow-based borrowing constraints. Two BCRA regulations play a crucial part in shaping these borrowing constraints: (i) regulations over banks' monitoring role over firms performance, (ii) regulations over banks' balance sheet exposure to borrowers' risk and risk assessment. Next, we describe these two set of regulations and argue that the implied borrowing constraints are highly sensitive to the interest rates.

We start by describing the BCRA regulations that bestow on banks a monitoring role over firms' status and/or performance. Banks must update firms' status at the Central Bank of Argentina's ``\textit{Central de Deudores}'' at a monthly frequency. As we described in Section \ref{sec:cash_flow_lending}, this implies that banks must report firms' borrowed amounts by type of credit line, whether the debt contract is in normal situation, if any stipulation in the debt contract has been violated and if so the status of the debt contract.\footnote{This information is easily accessible to the public and can be accessed through the following website
\url{http://www.bcra.gob.ar/BCRAyVos/Situacion_Crediticia.asp}. Furthermore, this dataset provides information on whether firms have recently issued any non-sufficient-fund checks.} 

Banks monitoring role over firms is embedded on the BCRA's regulations on banks' credit policy and debtor classification (or ``\textit{Clasificaci\'on de Deudores}'').\footnote{See \url{https://www.bcra.gob.ar/Pdfs/Texord/t-cladeu.pdf} for the full text.} The regulations stipulate that banks must re-evaluate and re-asses firms' credit status and risk of default  periodically according to its total amount borrowed.\footnote{See Section 3.2 of BCRA's ``\textit{Clasificaci\'on de Deudores}'': The classification of debtors must be carried out with a frequency that takes into account their importance –considering all financing included–, and in all cases the analysis carried out must be documented.} These bank assessments focus on firms' cash flow projections and keeping adequate debt payment to cash-flow ratios. Under normal circumstances, assessments are carried out quarterly or semi-annually depending on firms borrowing as a share of banks total lending.\footnote{A firm's status and re-assessment must be carried out quarterly if its total borrowing is equal to or higher than 5\% of the bank's net-worth or equal to or higher than 5\% of the bank's total assets. If a firm's borrowing is lower than 5\% of the bank's net-worth or total assets then its credit-status and re-assessment must be carried out semi-annually or twice a year.} However, banks must carry out mandatory and immediate firm re-assessments if: (i) the firm violates any condition of the original debt contract or of any debt renegotiation (such as being late on any payment for a period of time greater than 30 days), (ii) the firm's status with another bank at the BCRA's ``\textit{Central de deudores}'' deteriorates to a status below the current one at the bank.\footnote{See Section 6.4 ``Mandatory reconsideration of debtor classification'' or ``\textit{Reconsideraci\'on obligatoria de clasificaci\'on de deudores}''.}  Consequently, firms are subject to constant monitoring and scrutiny by banks which amplify the impact of any change in firms' financing needs.  

BCRA's regulation stipulate that banks must classify firm's risk of default over three main categories. The first category comprises firms which are considered under a ``normal or low-risk (of default)'' situation.
\begin{itemize}
    \item  ``Low-risk.'' Firm's debt structure to cash flow relationship is \textit{adequate}, with cash-flows being able to repay the capital and interests of its obligations with the bank. 
\end{itemize}  
Even more, the regulation stipulate that a debtor's situation could be classified as ``low risk'' if their main sector of activity is projected to report increases in cash-flows.\footnote{Note that, in line with the results presented in Section \ref{sec:cash_flow_lending}, the BCRA's regulations on firms' debt capacity focus on their cash-flows and specify a classification which relates the relationship of installment payments to cash-flows. Furthermore, in Appendix \ref{subsec:appendix_BCRA_Regulations_ABC} we show that BCRA regulations stipulate that banks should not assess the value of a firm's assets as a proxy to repay its debt.} Firm's in a ``low-risk'' situation meets the following condition
\begin{align} \label{eq:EBC_normal_situation}
    r_t \times B_t +  \vartheta_t \times B_t  \leq \theta^{\text{Low Risk}} \times \pi_t
\end{align}
where $r_t \times B_t$ represents firm's payment of debt interests, represents the interest ,$\vartheta_t \times B_t$ represents the firm's payment of maturing debt in period $t$, parameter $\theta^{\text{Low Risk}} > 0$ represents the maximum ratio of debt payments to cash-flows the bank believes to be ``adequate''. Note that this borrowing constraint is sensitive to the interest rate, $r_t$. To gauge the interest sensitive of this borrowing constraint, consider an example in which the net interest rate is $r_t = 5\%$, the share of debt which matures is $\vartheta_t = 25\%$, consistent with a quarterly calibration and debt maturity equal to a year. A decrease in the interest rate from 5\% to 4\% increases the borrowing limit imposed by Equation \ref{eq:EBC_normal_situation} increases by 4\%. Note that this is a direct change in the borrowing limit assuming that firms' cash flows $\pi_t$ remain unchanged. In the limit case $\vartheta_t$ is equal to 1, the condition in Equation \ref{eq:EBC_normal_situation} imposes a debt limit as the ratio between a firm's total debt plus interest payments to cash flows. 

The second category of default risk assessment comprises firms which face ``medium risk of default'', and/or are ``reporting problems'' in meeting their full debt payment obligations and/or under 
\begin{itemize}
    \item ``Medium risk''. The analysis of the debtor's cash flow shows that struggles meeting financial commitments, with cash flows only able to meet interest payments.
\end{itemize}
Furthermore, firms under a situation of ``medium risk'' of default show projections of progressive deterioration of their cash-flows and a high sensitivity to minor and foreseeable modifications of its own variables and/or of the economic environment, further weakening its payment possibilities. Firms in a situation of ``medium risk'' are either in technical default (as defined above), or banks project that they will in the short and foreseeable future. Firms which are not under a ``normal situation'' but do not fall under a situation of ``medium risk'' of default must meet the following borrowing condition
\begin{align} \label{eq:EBC_medium_risk}
    r_t \times B_t \leq \theta^{\text{Medium Risk}} \times \pi_t
\end{align}
where $\theta^{\text{Medium Risk}} > 0$ represents the maximum interest payment to firm's cash-flow the bank analyst believe to be adequate. The borrowing limit imposed by Equation \ref{eq:EBC_medium_risk} is significantly sensitive to changes in the interest rate. Firms in this situation can still re-finance their debt lines as long as they present valid information on their past, present and projected cash flows. Following the example constructed above for the case of a ``low-risk'' firm, a decrease in the interest rate from $5\%$ to $4\%$ leads to a direct decrease in the borrowing limit implied by Equation \ref{eq:EBC_medium_risk} of 25\%. 

Finally, the last category of the risk-assessment classification is comprised of firms which show a ``high risk of insolvency'' and/or show a ``high risk of defaulting'' or not meeting any debt payments
\begin{itemize}
    \item ``High-risk of insolvency (and/or default)''. The analysis of the debtor's cash flow shows that it is highly unlikely that it will be able to meet all of his financial commitments. The debtor's cash of funds is manifestly insufficient, not enough to cover the payment of interest. 
\end{itemize}
Moreover, firms classified into this category should, under the banks' analysis, exhibit projected cash-flows which imply that they will also have severe difficulties in ``complying with possible refinancing agreements'' or ``debt/liability re-structuring''. Firms in this situation have clearly breached their original debt contract, incurred in a technical default event and not satisfying the borrowing conditions described by either Equation \ref{eq:EBC_normal_situation} or Equation \ref{eq:EBC_medium_risk}. Banks consider these loans ``\textit{irrecuperables}'' or ``unrecoverable'' and will be written-off their balance sheets as a loss. 

Banks' monitoring regulations are accompanied by macro-prudential policy regulations over banks' exposure to lending risk. In particular, banks compliance with the monitoring and informational requirements are related to BCRA regulations over banks' capital requirements over the riskiness of their credit portfolio. These regulations are described in documents ``\textit{Lineamientos para la gesti\'on de riesgos en las entidades financieras}'' or ``Guidelines for risk management in financial entities'' and ``\textit{Capitales m\'inimos de las entidades financieras}'' or ``Minimum capital requirements of financial institutions''.\footnote{See \url{https://www.bcra.gob.ar/pdfs/texord/texord_viejos/v-lingeef_19-07-01.pdf} and \url{http://www.bcra.gov.ar/Pdfs/Texord/t-capmin.pdf} for the full text on this regulations.} These BCRA regulations impose capital requirements depending on the riskiness of the assets in their credit portfolio.

Banks' capital requirements stipulated by macro-prudential policies are based on firms' relationship between interest payments and their cash flows. In particular, if a firm is categorized under the risk status ``\textit{Medium risk of default}'', as described in Equation \ref{eq:EBC_medium_risk} the lending bank's capital requirements increase in an amount proportional in an amount equal to all of its exposure to said firm, i.e., not only the exposure to the specific debt-contract breached.\footnote{See  BCRA's ``Policy on minimum capital requirements of financial institutions'', Section 2 on ``Capital Requirements due to Credit Risk'', bullet point ``2.5.7.''} This increase in the bank's capital ranges from 50\% to 150\% depending on the bank's projection over the share of interest payments to be recovered from the firm.\footnote{See  BCRA's ``Policy on minimum capital requirements of financial institutions'', Section 2 on ``Capital Requirements due to Credit Risk'', bullet point ``2.6.11.3''.} 

In summary, the BCRA imposes several regulations over banks' cash flow-based credit policy and risk assessment which are based on firms' relationship between their cash flows and their interest payments. Furthermore, there is nothing in the regulations that stipulate that risk assessment should be carried out by analyzing the relationship between a firm's total debt and their cash flows. Given that firms' are able to re-finance their bank debt contracts even in a situation of ``Medium risk'', it is reasonable to assume that the BCRA regulations impose a debt limit of the form given by Equation \ref{eq:EBC_medium_risk}.

\noindent
\textbf{Evidence from a debt contract example.} We use details from an actual loan contract application to illustrate these features.\footnote{The original and full document of the loan application can be found at \url{https://banco.bind.com.ar/images/documentos/comex/CE003-solicitud-de-prestamo-en-moneda-extranjera-para-la-prefinanciacion-financiacion-de-exportaciones.pdf}.} In August 2015, the ``Banco Industrial S.A.'' or BIND stipulated the following conditions as ``events of technical default'' or ``\textit{eventos de incumplimiento}'':
\begin{enumerate}
    \item Failing to pay any installment and/or services of interests on the debt,

    \item Debtor not complying in full with any other credit or obligation towards the bank,
    
    \item Debtor not complying in full with any other credit or obligation towards other financial situation such that it leads to a third party filing a lawsuit against the debtor,
    
    \item Debtor refusing to supply the whatever information or allow the verification that the bank or BCRA estimate necessary,\footnote{Furthermore, the debtor must comply with any information requirement that if, carried out, would show that the data contained in the application and its annexes are inaccurate or imply that the debtor has use funds for any other destination than the one expressly consigned in the original application.} 
    
    \item Or if any other circumstance occurs such that, in the bank's opinion, affects the moral or commercial solvency of the debtor.
\end{enumerate}
Apart from the enumerate events, banks can declare technical default in case of significant revaluing of collateral (as argued in Section \ref{sec:cash_flow_lending}), or in the case of ownership transfer or events \textit{force majeure}.\footnote{These events comprise the sale, total or partial of the corporation; the merger, transformation or liquidation of the Debtor; the death or disability of the debtor (in the case of natural persons); debtor's filing for insolvency or bankruptcy or if it is requested by third parties, or is subject to any other compulsory asset or liability restructuring procedure; substantial regulatory changes that modify the fundamental market conditions under which the contract was granted.} The first and second conditions for technical default imply that firms face a borrowing constraint in form of an interest coverage constraint as in Equation \ref{eq:EBC_medium_risk}. 

The last three conditions for technical default imply that the bank is continuously monitoring and evaluating a firm's commercial solvency. While this may seem like an arbitrary condition for a debt contract, it falls directly under the monitoring role the BCRA bestows on banks. In particular, BCRA's regulations on banks' credit-policy and debtor classification (or ``\textit{Clasificaci\'on de Deudores}'') stipulate that banks must re-evaluate and re-asses firms' credit status and repayment capacity periodically according to its total amount borrowed.\footnote{See \url{https://www.bcra.gob.ar/Pdfs/Texord/t-cladeu.pdf} for the full text.}$^{,}$\footnote{See Section 3.2 of BCRA's ``\textit{Clasificaci\'on de Deudores}'': The classification of debtors must be carried out with a frequency that takes into account their importance –considering all financing included–, and in all cases the analysis carried out must be documented.} While we describe in depth the BCRA's regulations on banks monitoring role and assessment of firms' default risk in Appendix \ref{subsec:appendix_BCRA_Regulations_monitoring}, we present below key details, relate them with the events of technical default in the example debt contract and how they translate into borrowing limits.

\noindent
\textbf{Decomposition of firm's debt according to borrowing constraints.} Next, we present a decomposition of firms' debt according to borrowing constraints commonly used in macro-finance models. To this end, we build on Table \ref{tab:composition_firm_debt} by classifying collateral and cash flow-based lending into three different types of borrowing constraints considered by the macro-finance literature.

First, we classify all collateral-based lending into a standard collateral borrowing constraint as in Equation \ref{eq:borrowing_constraint_text}. This is straightforward as debt contracts are specified in terms of the liquidation value of specific assets provided as collateral by firms. This is in line with the literature which studies the micro-foundations of collateral-based lending as \cite{townsend1979optimal},\cite{bernanke1999financial} and \cite{jermann2012macroeconomic}.

Second, we decompose cash flow-based lending into two standard borrowing constraints considered by the literature: (i) ``Debt to Cash Flow'' constraint, and a (ii) ``Interest Coverage'' constraint. As argued above, BCRA regulations stipulate a firm's risk-assessment in terms of the relationship between interest payments, not total debt, to cash flows. Given that these regulations apply to all cash flow-based lending, we could classify 100\% of it to an interest coverage borrowing constraint as described by Equation \ref{eq:EBC_medium_risk}. However, as stressed in Section \ref{subsec:lending_heterogeneity} that bank loans to finance working capital is associated with specific BCRA regulations which limit a firm's debt as a function of their past and present export flows. While the same capital requirement regulations apply to this credit line, the characteristics of this credit line imply a borrowing constraint closer to that of a ``Debt to Cash Flow'' than to an ``Interest Coverage'' constraint. Thus, we follow a conservative approach and classify the ``Financing of working capital for exporting'' and all other unclassified cash flow-based lending as ``Debt to Cash Flow'' borrowing constraints.

Table \ref{tab:collateral_vs_cash_flow_borrowing_constraints} presents the results of our decomposition. 
\begin{table}[ht]
    \centering
    \caption{Collateral vs. Cash Flow Lending}
    \label{tab:collateral_vs_cash_flow_borrowing_constraints}
    \begin{tabular}{l l c c c} 
        \multicolumn{2}{c}{ }  & BCRA - Year 2000 & BCRA - Year 2017 \\
        \multicolumn{2}{c}{ }  & (1) & (2) \\ \hline \hline
        \multicolumn{2}{l}{Collateral-Based} & 35.5\% & 15.9\%\\
        \\
        \multicolumn{2}{l}{Cash Flow-Based} & 64.5\% & 84.1\% \\
        & \small Debt to Cash Flow & \small 14.9\% & \small 25.8\% \\
        & \small Interest Coverage & \small 49.6\% & \small 58.4\% \\  \hline \hline
 \floatfoot{\textbf{Note:} The table presents the decomposition of total firm bank debt into two categories: ``Collateral-Based'' and ``Cash Flow-Based'' lending. Furthermore, it decomposes the latter into two sub-categories: (i) ``Debt to Cash Flow'' lending, associated with a borrowing constraint of the form $B_t \leq \theta \times \pi_t$, and (ii) ``Interest Coverage'' lending, associated with a borrowing constraint of the form $r_t \times B_t \leq \theta \times \pi_t$ as described in Equation \ref{eq:EBC_medium_risk}.}
    \end{tabular}
\end{table}
The first result arising from this table is that collateral borrowing constraints only explain a minor share of total firm borrowing, close to 36\% in the year 2000 and less than 16\% by the year 2017. This result is straightforward given the one to one mapping between collateral-based lending and collateral borrowing constraints. The second result is that the vast majority of cash flow-based lending is explained by loans subject to ``Interest Coverage'' borrowing constraints. Even under our conservative classification strategy, close to 70\% of firms' debt is subject to the interest sensitive borrowing constraints, with the remaining 30\% subject to a ``Debt to Cash Flow'' borrowing constraint. 

In summary, the BCRA's regulations described above over banks' monitoring role and risk-assessment suggest that the main borrowing constraint firms face in Argentina is a ``Interest Coverage'' borrowing constraint. This is, the most prevalent borrowing constraint in Argentina is one in which a firm's debt limit is determined by the relationship between its interest payments and its cash flows. 

\noindent
\textbf{Aggregate evidence of interest sensitive borrowing constraints.} We provide supporting evidence of the prevalence of ``Interest Coverage'' or interest sensitive borrowing constraints in Argentina by studying the aggregate dynamics of GDP, firm debt and interest payments. In particular, we argue that while interest rates are counter-cyclical and lead the business cycle, interest payments follow the cycle and are highly correlated with total firm debt.

\begin{figure}[ht]
    \centering
    \caption{GDP, Firm Debt, Interest Rates \& Payments}
    \label{fig:Growth_Rates_Expanded}
     \centering
     \begin{subfigure}[b]{0.95\textwidth}
         \centering
         \includegraphics[height=4cm,width=12cm]{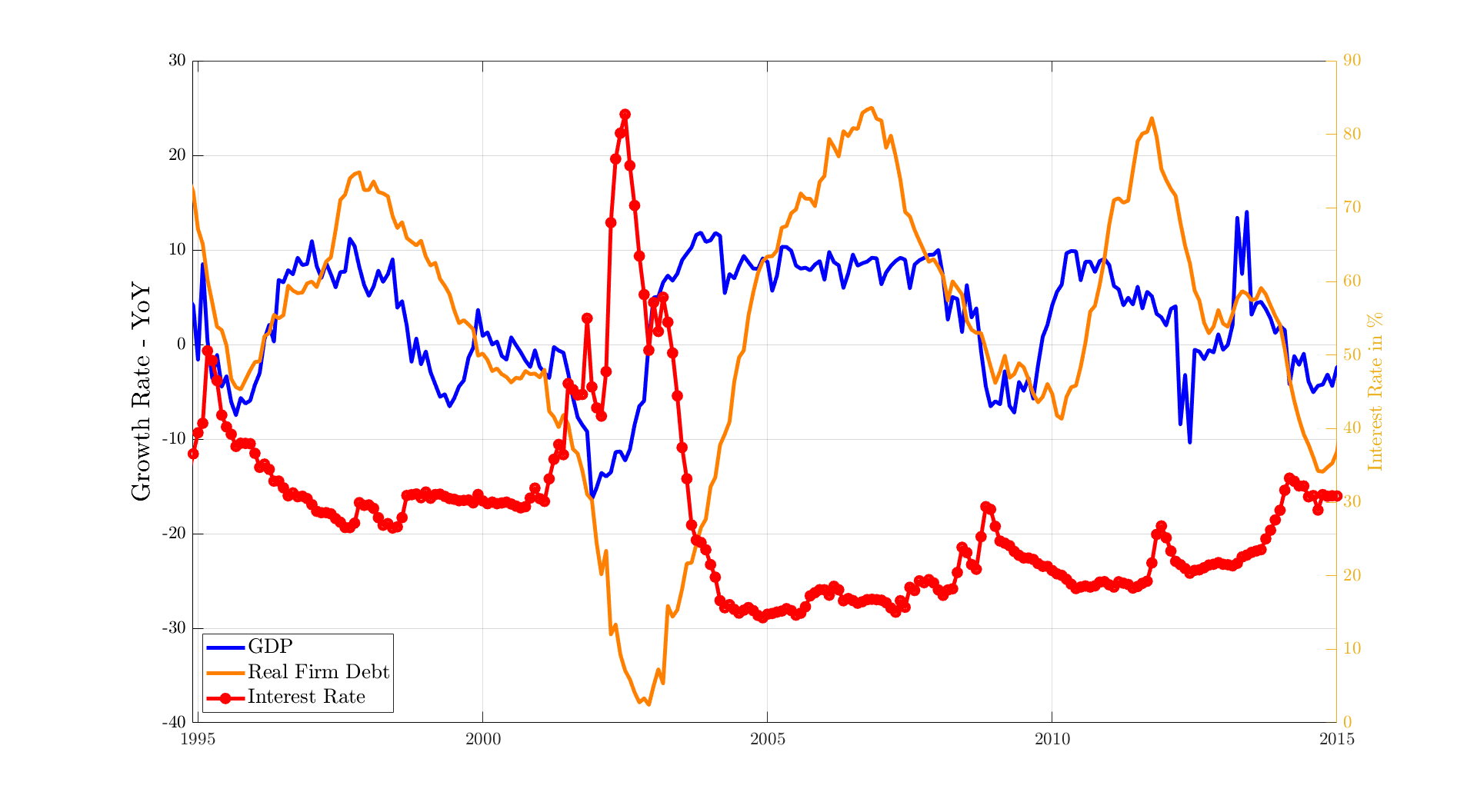}
         \caption{GDP, Firm Debt \& Interest Rates}
         \label{fig:Growth_Rates_Rate}
     \end{subfigure}
     \hfill \\
     \begin{subfigure}[b]{0.95\textwidth}
         \centering
         \includegraphics[height=4cm,width=12cm]{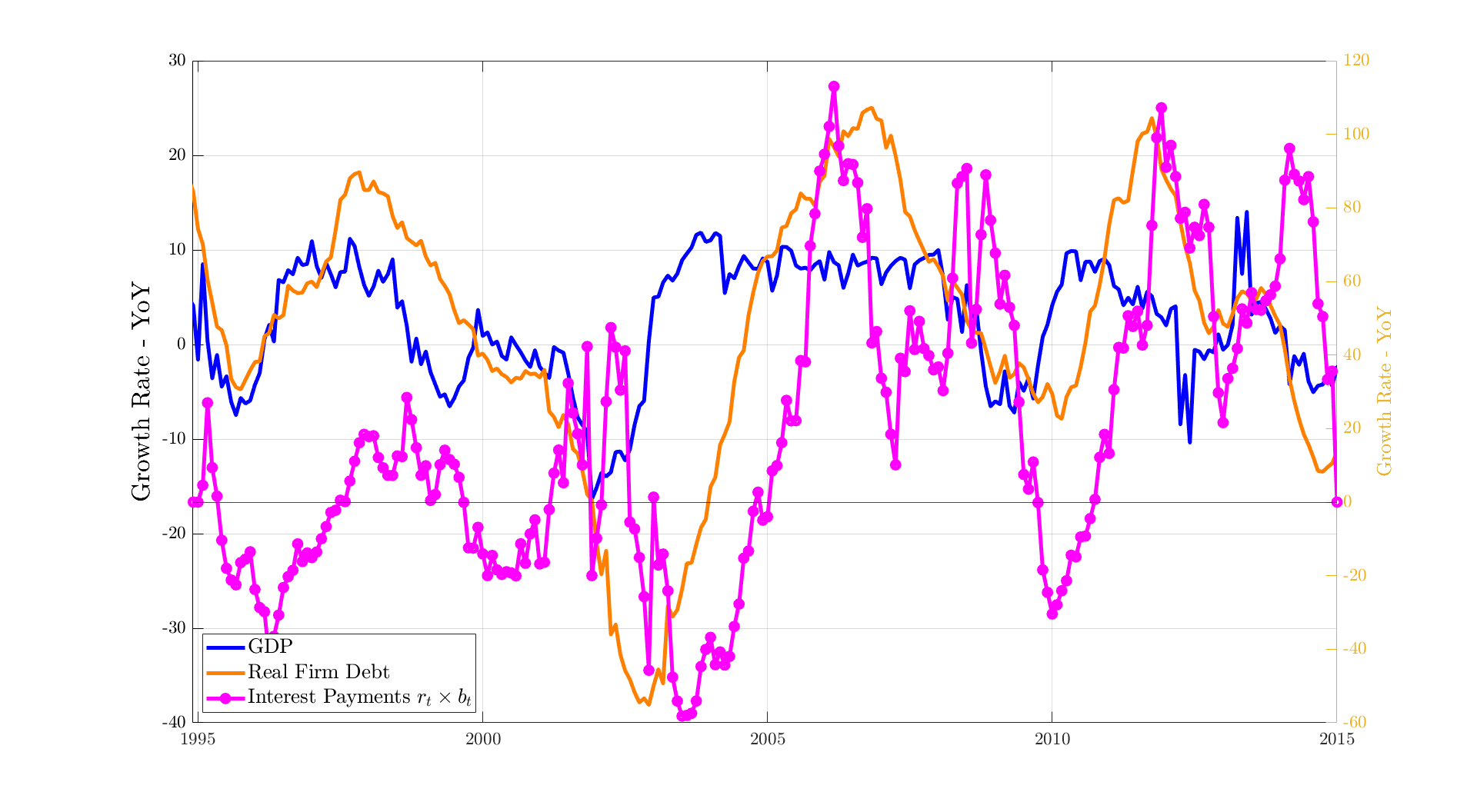}
         \caption{GDP, Firm Debt \& Interest Payments}
         \label{fig:Growth_Rates_Payments}
     \end{subfigure} 
     \begin{subfigure}[b]{0.95\textwidth}
         \centering
         \includegraphics[height=4cm,width=12cm]{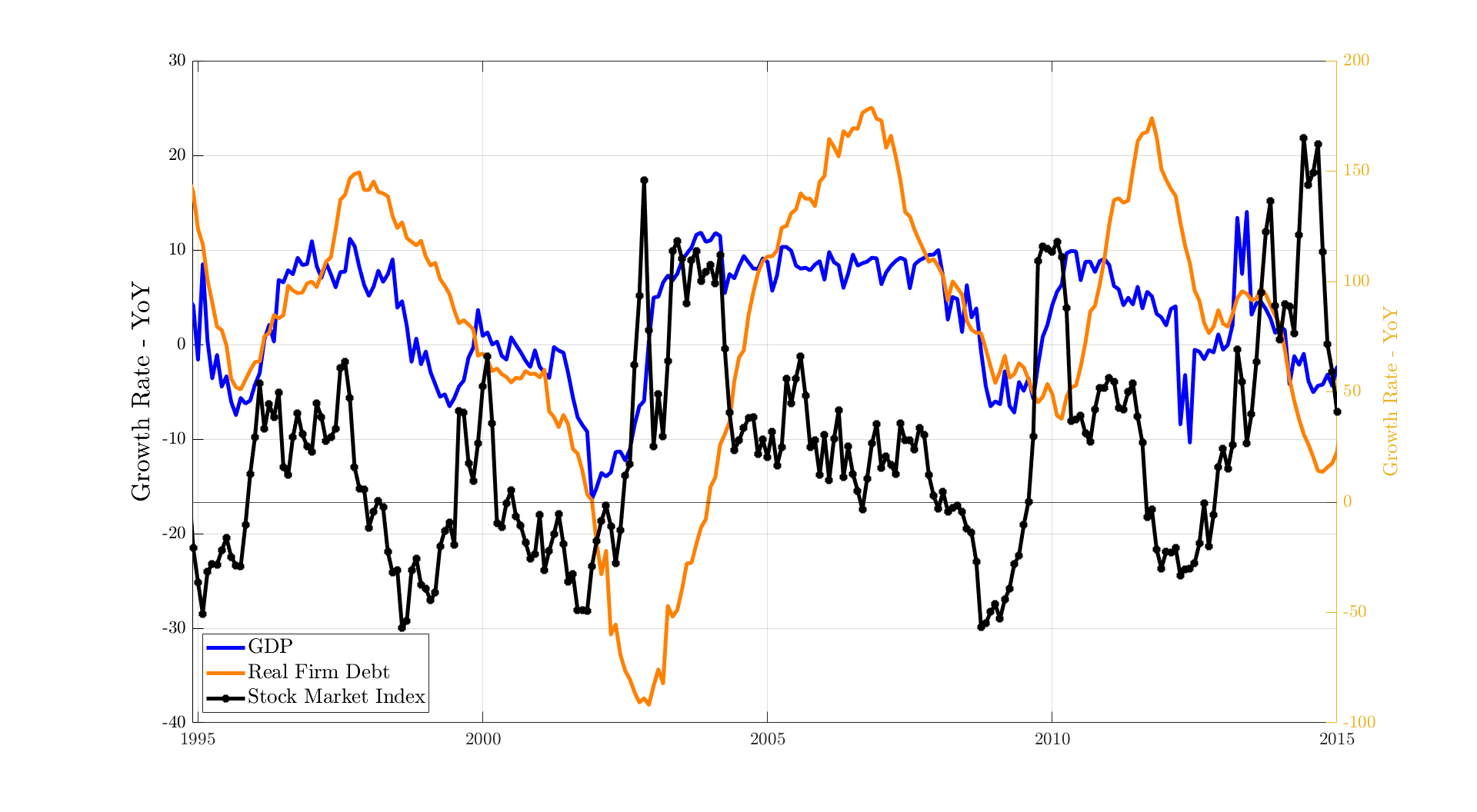}
         \caption{GDP, Firm Debt \& the Stock Market}
         \label{fig:Growth_Rates_StockMarket}
     \end{subfigure} 
    \floatfoot{\textbf{Note:} The figure presents the dynamics of real GDP, real firm debt, interest payments and the average nominal rate for the period 1995-2015 at the monthly frequency. The first two Variables are plotted using the year-on-year growth rate on the left y-axis. The nominal interest rate is plotted on the right y-axis.}
\end{figure}

Interest rates are highly counter-cyclical in Argentina, a well established stylized fact in the international macroeconomic literature, see \cite{neumeyer2005business}. Figure \ref{fig:Growth_Rates_Expanded}, builds on Figure \ref{fig:Growth_Rates} from Section \ref{sec:introduction} by plotting the average interest rate firms face, aggregate interest payments and the stock market. On the top panel, Figure \ref{fig:Growth_Rates_Rate}, shows a strong and negative correlation between the growth rate of real activity (-0.5311), of real firm debt (-0.6398) with interest rates. 
Interest rates lead the business cycles. For instance, the correlation between the average interest rate and the six month ahead growth rate of GDP is even stronger at -0.7185. 

Interest rate payments are strongly correlated with firm debt and the business cycle. The correlation between firms' interest payments and firms' debt is 0.4117 for the full sample and increases to 0.6142 if we exclude the years 2001 and 2002. At the monthly frequency, there is a negative correlation between contemporaneous GDP growth rate and interest payments of -0.1167. However, this correlation becomes quantitative larger and positive when comparing GDP growth rate and interest payments one year ahead, at  0.4823. Similarly, the correlation between GDP growth rate and firm credit becomes quantitative larger when comparing with firm credit 6 or 12 months ahead, at 0.6531. These correlations suggest that both interest rates and interest payments are counter-cyclical for Argentina. Additionally, while interest rates lead the cycle, firm debt and interest payments lag the cycle.

Lastly, on the bottom panel, Figure \ref{fig:Growth_Rates_StockMarket} presents the dynamics of GDP, firm debt and the stock market. The stock market has been used as a proxy for the price of capital by the macro-finance literature, see \cite{christiano2014risk}. The stock market is pro-cyclical and leads the GDP growth rate (contemporaneous and six months ahead correlation of 0.3166 and 0.4407). However, the correlation between the stock market and firms' debt is either negative or non-significantly different from zero across time horizons (contemporaneous, six months ahead and six months lagged of -0.1837, 0.0674 and -0.3857). 

Overall, Figure \ref{fig:Growth_Rates_Expanded} presents a novel stylized fact for an Emerging Market economy like Argentina, that interest payments are counter-cyclical and strongly correlated with firms' debt. This strong correlation is in line with the interest coverage borrowing constraint implied by the BCRA regulations. Additionally, the negative correlation between the stock market and firms' debt provides evidence against firms facing collateral-based constraints. In Section \ref{subsec:interest_sensitive_comparison} we show that these stylized facts are also present for a panel of Emerging Market economies when using both aggregate and firm level data.

\subsection{Evidence from Corporate Bonds} \label{subsec:interest_sensitive_corporate}

We provide further details over the borrowing constraints Argentinean firms face by presenting details on the episodes of technical default in the market for corporate bonds ``Obligaci\'on Negociable'' (described in Section \ref{sec:cash_flow_lending}). Alike the case of bank contracts, we present details from an actual debt contract. In April $11^{th}$ 2022, the firm ``Empresa Distribuidora y Comercializadora Norte S.A.'' or EDENOR, issued a ON corporate bond at a 9.75\% annual interest rate in US dollars for a total amount of US\$ 120 million dollars, maturing in May $12^{th}$ 2025. Lenders provided the full amount upfront and had no additional commitments. The debt contract stipulates semi-annual interest payments and only one capital payment at maturity.\footnote{This is the typical ON corporate bond payment profile. Weighted by debt's face value, 93\% of ON's corporate bonds make only one capital payment at maturity and make interest payments either quarterly or semi-annually.} 

Edenor's ON corporate debt contract specifies particular conditions under which the firm breaches the contract and is under technical default. Below we present the four key conditions for technical default
\begin{itemize}
    \item[1.] Non-compliance with the principal payment of any of the ON upon becoming due, either at maturity, by redemption, expiration of its term, pre-cancellation or in any other case, and such non-compliance remains in force for a period of five consecutive days;
    
    \item[2.] Non-payment of interests when they become due, either at maturity, by redemption, expiration of their term, pre-cancellation or in any other case, and said non-compliance remains in force for a period 30 calendar days;`
    
    \item[3.] If there had been a revocation, cancellation, rescission or suspension for more than 20 calendar days of the Edenor's concession contracts as an utility energy supplier;
    
    \item[4.] Any failure on the part of Edenor to duly observe or comply with any of Edenor's commitments or agreements within the framework of the Trust Agreement of the ONs (except those referred to in points (a) and (b ) precedents) for a period of more than 60 calendar days after the date on which written notification is sent in this regard demanding that Edenor correct it, sent to Edenor by the Trustee or the holders of at least 25\% of the total nominal value of the ONs;
\end{itemize}
The lack of capital payments until maturity implies that Edenor only faces the constraint implied by condition ``1.'' only every time the ON corporate debt contract matures. Thus, Edenor regularly faces the need of sufficient cash-flows to pay interest payments to meet constraint ``2.''. In case Edenor loses its concession contract as an utility energy supplier (condition ``3.''), it is evident that it won't be able to generate the necessary cash-flows to repay any debt payments.

Edenor's ON Trust Agreement imposes a debt limit over the firm's total liabilities. This debt limit is stipulated such that the firm cannot violate, at any time, either of two conditions:
\begin{itemize}
    \item Debt ratio index constraint: ratio between all debt obligations to the EBITDA corresponding to the last four consecutive economic quarters must not surpass 3.75.
    
    \item Interest coverage ratio index constraint: ratio between the EBITDA corresponding to the last four consecutive economic quarters to interest payments must not surpass 2.\footnote{In terms of the ON debt contract, interest payments includes not only the payment of interests on debt and the payment of any punitive interest and other additional amounts given certain conditions of technical default.}
\end{itemize}
Note that the ``Debt ratio index'' and the ``Interest coverage ratio index'' imply the borrowing constraints presented in Equations \ref{eq:EBC_normal_situation} and \ref{eq:EBC_medium_risk}, respectively. Violation of any of these debt-limits is considered a technical default event, accelerating the repayment of all interests and capital. In line with the evidence presented in Sections  \ref{sec:cash_flow_lending} and \ref{subsec:interest_sensitive_bank_regulations}, the debt limits are stipulated in terms of the firm's cash-flow and not in terms of the liquidation value of its capital.\footnote{The ON debt contract does imply a clause in which a technical default event is triggered if a judge declares the sale of the firms' assets to repay the debt to a third party given insufficient cash-flows. However, the clause does not imply any relationship between the firm's debt and the value of the capital. Furthermore, the lack of sufficient cash-flow to pay the debt to the third party would probably trigger any of the cash flow-based debt limits stipulated as technical default or as an actual debt limit.} Note that these two debt limits may be violated by changes in economic and financial conditions which are not directly under the borrower's control. For instance, a decline in earnings could cause an involuntary violation of the debt limit by the firm. Declines in earnings can be caused by domestically originated shocks, such as demand shocks, or internationally originated shocks, such as exchange rate depreciations, foreign financial shocks and commodity price shocks. 

Lastly, ON corporate bonds contain a key clause in which creditors can call for an ``Extraordinary assembly of ON holders'' or ``\textit{Asamblea extraordinaria de tenedores de ON}''. These extraordinary assemblies provide creditors with the opportunity to vote over key decisions of firms' liabilities such as declaring an overall debt re-negotiation o ``Concurso de Preventivo de Acreedores''. Each ON bond comes with specific metrics creditors must meet to call these extraordinary assemblies and to vote over specific matters. For the case of Edenor
\begin{itemize}
    \item An Extraordinary Assembly may be called by the Trustee or by the Issuer of the ONs at the request of the holders of that represent at least 5\% of the nominal value. 
    
    \item To be able to vote at an Extraordinary Assembly, a person must be (i) the holder of one or more ON corporate bonds and must sign up three days before the registration date.
    
    \item Quorum in Extraordinary Assemblies convened to adopt any resolution will be constituted by persons who own or represent at least 60\% of the total face value of the outstanding New Notes.
    
    \item Any resolution tending to modify or amend or waive the fulfillment of any provision of the ON will be validly adopted if it is approved by the majority of the persons with the right to vote present or represented at the Extraordinary Assembly.
\end{itemize}
This debt re-negotiation mechanisms embedded within ON corporate bonds implies that creditors play a recurrent monitoring role over firms. This is in line with the monitoring role banks have over firms according to BCRA's regulations. Furthermore, this debt re-negotiation mechanisms allow firms to restructure their liabilities and debt payments as a function of the dynamics of their cash-flows, allowing them to remain active.\footnote{A prime example of this debt re-negotiation process is the case of ``Roch S.A.'', a leading independent operators in the exploitation, exploration and commercialization of gas and oil in Argentina. See the company website at \url{https://www.roch.com.ar/en/home/}. On May $16^{th}$ 2022, the creditors called for an Extraordinary Assembly, voted in favor of an overall debt re-negotiation. See the full document over the meeting in \url{https://ws.bolsar.info/descarga/pdf/408120.pdf}. This allowed the firm to avoid default and bankruptcy and carried out a successful liability re-structuring by July $8^{th}$ and resume operations as usual, see \url{https://www.iprofesional.com/negocios/366273-una-petrolera-local-logra-evitar-el-fantasma-del-default}.}


\subsection{Comparison with International Evidence} \label{subsec:interest_sensitive_comparison}

We conclude this section by comparing the results arising from our datasets of Argentinean firms with evidence from the US and other countries. First, we argue that interest sensitive borrowing constraints are more prevalent in Argentina than in the US and other Advanced Economies. Second, we argue that our results for Argentina may be extrapolated to other Emerging Markets. We do so by presenting a novel stylized facts to the literature, that for firms in Emerging Markets interest payments are significantly correlated to their cash flows.

\noindent
\textbf{Comparison with evidence from firms in the US and other Advanced Economies.} As in Section \ref{subsec:lending_comparison_international}, we face the question on how our results for Argentinean firms compare to those from the recent literature which studies US firms? We argue that interest sensitive borrowing constraints are more prevalent for Argentina firms than for US firms. 

First, Table \ref{tab:comparison_borrowing_constraints} presents a comparison of our results for both US and UK firms, sourced from \cite{drechsel2019earnings} and \cite{bahaj2019employment} respectively. For US firms, the share of collateral-based borrowing constraints is close to 20\%, similar to the findings of \cite{lian2021anatomy}. For UK firms, the share of collateral-based borrowing constraints is significantly greater, explaining 50\% of firms' debt.\footnote{Note, \cite{bahaj2019employment} find that relatively smaller and younger firms exhibit a larger share of collateral-based constraints, while relatively larger firms borrow primarily cash flow-based. }
\begin{table}[ht]
    \centering
    \caption{Collateral vs Cash Flow Borrowing Constraints \\ \footnotesize International Comparison}
    \label{tab:comparison_borrowing_constraints}
    \begin{tabular}{l l c | c | c }
                       & & US & UK & Argentina \\
                       & & \cite{drechsel2019earnings} & \cite{ bahaj2019employment} & BCRA  \\ 
                       & & (1) & (2) & (3) \\ \hline \hline
\multicolumn{2}{l}{Collateral-Based} & 21.3\% & 50\% & 16\% \\ 
 & & & & \\
\multicolumn{2}{l}{Cash Flow-Based}  & 47\% - 61\% & 50\% & 84\% \\
    & \small Debt to Cash Flow & \small 60.5\% & - & \small 26\% \\
    & \small Interest Coverage & \small 46.7\% & - & \small  58\% \\ \hline \hline
    \end{tabular}
    \floatfoot{\textbf{Note:} Data on column (1) are sourced from Table 1 of \cite{drechsel2019earnings}. This figures present the frequency of each type of debt covenant across financial debt contracts. Data on column (2) are sourced from Table 2 of \cite{bahaj2019employment}. These figures express the share of total firm debt which is explained by the different types of borrowing constraints. The authors show that at least 50\% of firms debt is collateral-based. We interpret their finding as at most 50\% of UK firms' debt is cash flow-based. This allows us to provide an international comparison of our results.}
\end{table}
For US firms, cash flow debt covenants are present in 47\% to 61\% of financial debt contracts. On the one hand, the most common type of cash flow debt covenant is the ``Debt to Cash Flow'' debt covenant, present in 60.5\% of all debt contracts, also in line with the findings by \cite{lian2021anatomy}. On the other hand, interest sensitive debt covenants are present in almost 47\% of all debt contracts. Overall, given our results for Argentina in column (3), it is clear that cash flow-based lending is more prevalent in Emerging Markets than in the US and in the UK, and potentially other Advanced Economies. Furthermore, comparing the results in columns (1) and (3), it is clear that Argentina exhibits a larger share of interest sensitive cash flow-based borrowing constraints. 

Evidence for the US is sourced from large and mostly public firms exploiting debt contracts which face multiple debt covenants.\footnote{For evidence on the multiplicity of debt covenants for US firms, see \cite{greenwald2019firm}.} To carry out a more accurate comparison we leverage on the work by \cite{greenwald2019firm} which carries out a statistical analysis to recover the share of firms which face each of the cash flow-based borrowing constraints. Table \ref{tab:comparison_interest_sensitive} compares the prevalence of interest sensitive borrowing constraints in Argentina and the US.
\begin{table}[ht]
    \centering
    \caption{Prevalence of Interest Sensitive Borrowing Constraints \\ \footnotesize Argentina vs US }
    \label{tab:comparison_interest_sensitive}
    \begin{tabular}{l c | c }
                        & US - \cite{greenwald2019firm} & Argentina - BCRA  \\
                        & (1) & (2) \\ \hline \hline
    Debt to Cash Flow & 70\% & 31\% \\
    Interest Coverage & 30\% & 69\% \\ \hline \hline
    \end{tabular}
    \floatfoot{\textbf{Note:} Data on the column (1) are sourced from Figure 3 of \cite{greenwald2019firm}. We calculated the average share of firms which face interest coverage borrowing constraints across the author's sample.}
\end{table}
In particular, we calculated the average share of firms which face interest coverage borrowing constraints across the author's sample during a period of time comparable to ours.\footnote{Note that \cite{greenwald2019firm} sample ends in 2008. The author argues that extending the sample would only lead to a reduction in the relative share of firms which face interest sensitive borrowing constraints.}$^{,}$\footnote{In \cite{greenwald2019firm}, the author stress the prevalence of simultaneous cash flow-based borrowing constraints and the time-varying share of binding interest sensitive constraints. We believe that this time-varying share is also present in Argentina. However, we leave it for future work.} The share of debt subject to interest coverage borrowing constraints in Argentina, column (2), is more than twice as large as in the US, column (1). This greater prevalence implies that the vast majority of debt in Argentina is highly sensitive to interest rates. In summary, the evidence in Tables \ref{tab:comparison_borrowing_constraints} and \ref{tab:comparison_interest_sensitive} suggests that Argentina exhibits a greater share of interest sensitive cash flow-based borrowing constraints than the US and other Advanced Economies.  

\noindent
\textbf{Extrapolation to other Emerging Market economies.} We provide evidence of external validity of our results for Emerging Markets by highlighting a novel stylized facts: firms interest payments over their debts is significantly correlated with their observable measures of their cash flows.

We construct a panel of Emerging Markets firms using Orbis.\footnote{See \ref{subsec:appendix_additional_data_description_orbis} for the data set description and coverage.} This allows us to study the dynamics of cash flows, interest payments and other financial firm level variables for more than 500 firms from 10 Emerging Markets for the year 2000 to 2020. This dataset suffers from the same limitations as those used for the study of US firms: only few firms are part of this data set, and the representation is highly skewed towards relatively large and public firms. Given these limitations, we believe it provides significant evidence of the presence of interest sensitive cash flow-based constraints.

\begin{figure}[ht]
    \centering
    \caption{Correlations between Firm Interest Payments \& Cash Flows \\ \footnotesize Aggregate Level Analysis }
    \label{fig:correlation_aggregate_level}
     \centering
     \begin{subfigure}[b]{0.495\textwidth}
         \centering
         \includegraphics[height=5cm,width=8cm]{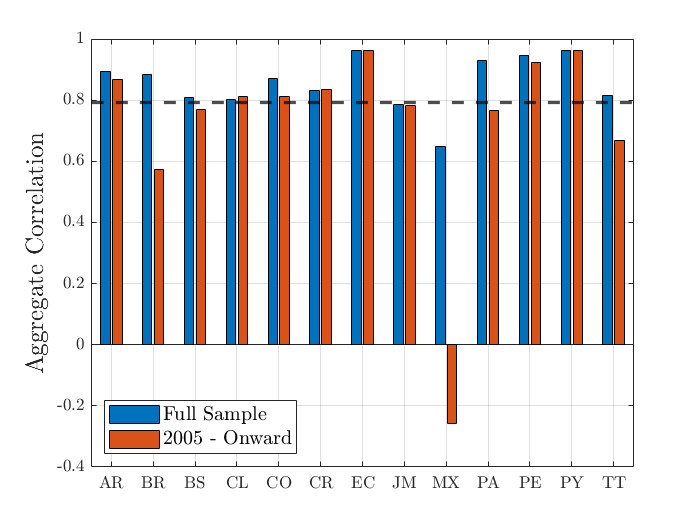}
         \caption{Interest Payments to Cash Flows}
         \label{fig:Correlation_Interest_EBITDA}
     \end{subfigure}
     \hfill
     \begin{subfigure}[b]{0.495\textwidth}
         \centering
         \includegraphics[height=5cm,width=8cm]{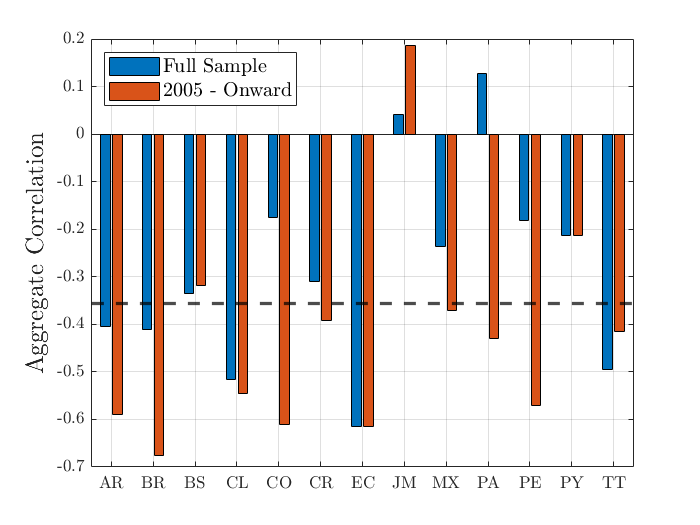}
         \caption{Interest Payments to GDP Growth Rate}
         \label{fig:Correlation_Interest_GDP}
     \end{subfigure} 
    \floatfoot{\textbf{Note:} The figure presents the aggregate correlation between firms' interest payments and cash flows. To compute aggregate measures of these variables we sum over all the firms in our dataset. The countries in our sample are Argentina, Brazil, Barbados, Chile, Colombia, Costa Rica, Ecuador, Jamaica, Mexico, Panama, Peru, Paraguay, Trinidad y Tobago. The dashed black line represents the average correlation across countries. The blue bars represent the aggregate correlation for the full length of sample for each country, which differs across countries due to coverage differences. The red bar represents the aggregate correlation from the year 2005 onward.}
\end{figure}

Figure \ref{fig:correlation_aggregate_level} presents the correlation between firms' interest payments and their cash flows and the correlation between firms' interest payments and GDP at the aggregate level.\footnote{Aggregate measures are computed by summing across firms within a country.} On the left panel, Figure \ref{fig:Correlation_Interest_EBITDA} shows that at the aggregate level, there is a strong correlation, close to 0.80 on average, between firms' interest payments and their cash flows. 
This strong and positive correlation at the aggregate level is in line with firms facing an interest coverage borrowing constraint, as suggested by the BCRA regulations in Section \ref{subsec:interest_sensitive_bank_regulations}. On the right panel, Figure \ref{fig:Correlation_Interest_GDP} shows the a strong and negative correlation between firms' interest payments and the GDP growth rate. Both of these results are in line with the evidence presented in Figures \ref{fig:Growth_Rates_Expanded} for Argentina at the aggregate level.

\begin{figure}[ht]
    \centering    
    \caption{Correlations between Firm Interest Payments \& Cash Flows \\ \footnotesize Firm Level Analysis }
    \label{fig:Correlation_Firm_Level_WA_Median}
\includegraphics[width=10cm,height=6cm]{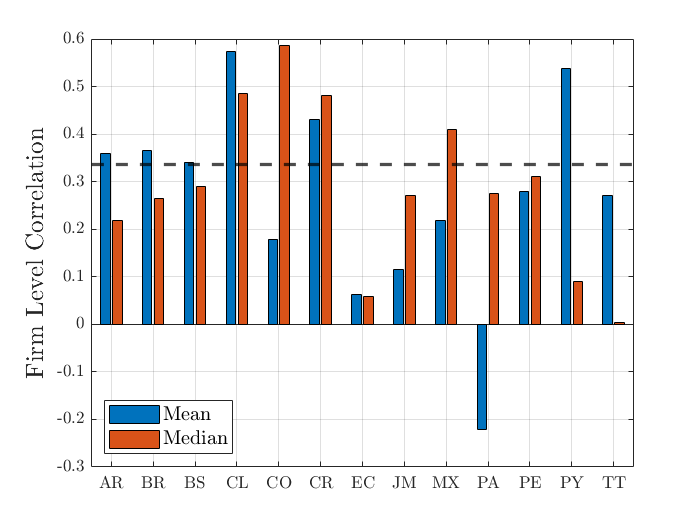}
\floatfoot{\textbf{Note:} The figure presents the mean (blue bar) and median (red bar) firm level correlation between interest payments and cash flows at the annual level. The mean is computed as a weighted average of the firm level correlation using firms' cash flows as weights. The dashed black line presents the weighted average across countries.}
\end{figure}

Lastly, we present evidence of a positive and significant correlation between firms' interest payments and their cash flows at the firm level. Figure \ref{fig:Correlation_Firm_Level_WA_Median} presents the mean and median firm level correlation between interest payments and cash flows at the annual level.
The correlation is positive and significant across most of the countries in our sample. Note that the mean firm level correlation is roughly half of the country level correlation (0.35 vs. 0.75), but still significantly positive. In Appendix \ref{sec:appendix_stylized_facts} we show that the average firm level correlation is significantly different from zero.

In summary, we provided evidence of the presence of interest sensitive coverage borrowing constraints for a panel of Emerging Market economies. We do this by analyzing the aggregate and firm level correlation between interest payments and cash flows. At the aggregate level there is a strong and positive correlation between firms' interest payments and cash flows and a strong and negative correlation between firms' interest payments and the growth rate of GDP. This is in line with the dynamics presented in Figure \ref{fig:Growth_Rates_Expanded} for Argentina. Additionally, we presented evidence that there is a positive and significant firm level correlation between interest payments and cash flows. Overall, these correlations are in line with the presence of interest coverage borrowing constraints for Emerging Markets, providing external validity of our firm level results for Argentina. In Section \ref{subsec:nominal_frictions_matching_aggregate} we build on this empirical results by arguing that models with collateral-based constraints are not able to match the correlations between interest payments, cash flows and the business cycle.

\section{A Real Model with Cash Flow Borrowing Constraints} \label{sec:model_simple}

In this section we lay out a simple real model in which firms face the borrowing constraints presented in Section \ref{sec:interest_sensitive_borrowing_constraints}. The model demonstrates the key economic mechanisms at work and provides the benchmark for comparison for our extended versions of the model. In Section \ref{sec:nominal_frictions_policy} extends the model into a quantitative framework which allows for nominal frictions and both domestic and foreign currency debt. In Section \ref{sec:spillover_puzzle}, we extend the model to allow for firms to face both cash flow and collateral-based constraints.

\subsection{Model Description} \label{subsec:model_simple_environment_rbc}

Our structural framework builds on the work of \cite{mendoza2010sudden} and \cite{garcia2010real} and introduces financial frictions following the approach of \cite{jermann2012macroeconomic}. 

\noindent
\textbf{Model environment.} Time is discrete, indexed by $t$, and continues infinitely. The economy is populated by three types of agents: (i) a continuum household, (ii) a continuum of firms and (iii) a fiscal authority. The firms produces by hiring labor and accumulating capital, subject to a borrowing constraint. The households' problem is standard, it consumes, supplies labor, saves in internationally traded bonds and shares issued by the firm and receive dividend payments. The fiscal authority has a passive role, only levying lump sum taxes.

Each firm produces using a final standard good through the following production technology
\begin{align}
    Y_t = a_t K^{\alpha}_{t-1} M^{\eta}_t \left(X_t h_t\right)^{1-\alpha-\eta} \label{eq:model_simple_rbc_production}
\end{align}
where $Y_t$ represents output in period $t$, $K_{t}$ denotes capital chosen in period $t-1$ and determined in period $t$, $h_t$ denotes hours worked in period $t$, $M_t$ represents imported inputs, $\{\alpha, \eta\} \in (0,1)$ represents the capital and imported input elasticity subject to $\alpha+\eta \leq 1$,  and $a_t$ and $X_t$ represents productivity shocks. We follow \cite{garcia2010real} and interpret these shocks as encompassing both exogenous variations in technology and other disturbances such as terms of trade shocks. 

We assume that $a_t$ follows an auto-regressive process in logs, while $X_t$ is a permanent shock, with its growth rate $g_t \equiv X_t/X_{t-1}$ following an auto-regressive process. These productivity processes can be characterized by the following expressions.
\begin{align*}
    \ln a_{t+1} &= \rho_a \ln a_t + \epsilon^{a}_{t+1}; \qquad \epsilon^{a}_t \sim \mathcal{N} \left(0,\sigma^{2}_a\right) \\
    \ln \left(\frac{g_{t+1}}{\bar{g}}\right) &= \rho_g \ln \left(\frac{g_{t}}{\bar{g}}\right) + \epsilon^{g}_{t+1}; \qquad \epsilon^{g}_t \sim \mathcal{N} \left(0,\sigma^{2}_g\right), \quad g_t \equiv \frac{X_t}{X_{t-1}}
\end{align*}
where $\{\rho_a,\rho_g\} \in [0,1)$. We use uppercase letters to denote variables which in equilibrium contain a trend, and use lowercase letters to denote variables which do not contain a trend in equilibrium.

The representative firm owns and accumulate capital subject to the following law of motion
\begin{align} 
    K_t = \left(1-\delta \right) K_{t-1} + I_t \left[1-\frac{\phi_i}{2} \left(\frac{I_{t}}{I_{t-1}}- \bar{g}\right)^{2} \right] \label{eq:model_simple_rbc_LOC}
\end{align}
where $I_t$ is investment, $\delta$ is the depreciation rate of capital, $\phi_i$  is a parameter which governs the curvature of the investment adjustment cost. Note that as the economy exhibits trend growth rate, the adjustment costs of investments exhibit the deterministic component of trend productivity growth rate $\bar{g}$ inside the parenthesis. This specification of investment adjustment costs is in line with that studied in \cite{christiano2005nominal,christiano2011introducing}, and introduces a time-varying price of capital. This arises from the standard result that adjustment costs lead to variations in the value of capital inside the firm relative to its replacement value, i.e., affecting the ratio known as \textit{Tobin's Q} (see \cite{hayashi1982tobin}).

Firms use equity and debt. Debt, denoted by $B_t$, is preferred to equity, pecking order, because of its tax advantage. This is a standard assumption in the macro-finance literature, as in \cite{hennessy2005debt}. We assume that given an interest rate $r_t$, the effective interest rate firms face is
\begin{align} \label{eq:real_tax_advantage}
    \tilde{R}_t = 1 +r_t\left(1-\tau\right)
\end{align}
where $\tau$ represents the tax benefit. Every period, the firm must also decide on the amount of equity payouts to its shareholders, denoted by $D_t$. We come back to description of firm's decision over equity payouts. 

In addition to the intertemporal debt, $B_t$, firms are subject to working capital constraint, given the vast importance of this type of financing argued in Section \ref{sec:cash_flow_lending} and \ref{sec:interest_sensitive_borrowing_constraints}. We assume that firms must pay a fraction $\phi$ of the wage bill and imported intermediate inputs in advanced and must finance it at interest rate $\tilde{R_t}$. Thus, we denote the firm's borrowing for working capital needs as
\begin{align}
    L_{t} = \phi \left(W_t h_t + p^{M}_t M_t \right)
\end{align}
where $W_t$ is the wage rate of labor, $p^{M}_t$ is the relative price of imported inputs. This working capital constraint can be rationalized through firms necessity to cover cash flow mismatches between the payments necessary to hire labor and purchase imported intermediate goods, and the realization of revenues from production. We assume that this working capital constraint is repaid within period $t$.

We can write the firm's flow of funds constraint as
\begin{align} \label{eq:model_simple_rbc_flow_of_funds}
    Y_t + \frac{B_{t+1}}{R_t} = \left(1-\phi\right)\left(W_t h_t+p^{M}_t M_t\right) + I_t + D_t + L_t \tilde{R}_t + B_t
\end{align}
where the left hand side represents the sources of funds, production and issue of new debt; and the right hand side represents the uses of funds, hiring labor, purchasing imported inputs, investment, payments of dividends, paying working capital and intertemporal debt. 

We assume that firms face a borrowing constraint which establishes a debt limit $\Bar{B}_t$. This debt limit will be endogenous in our analysis and depend on both aggregate and firm level variables. We assume that the borrowing constraint applies to both intertemporal and working capital debt and is expressed in terms of the amount lent by the firms creditors
\begin{align} \label{eq:real_model_generic_borrowing_constraint}
    \Bar{B}_t \geq \frac{B_{t+1}}{1+r_t} + \tilde{R}_t L_t  
\end{align}
Higher debt, either inter temporal and/or for working capital, makes the borrowing constraint tighter. Given the tax advantage over debt in Equation \ref{eq:real_tax_advantage}, firms borrow up to its constraint in every period. 

To see more clearly how changes in the borrowing constraints affect firms' financing and production decisions it is useful to think about an unexpected shock which reduces the debt limit $\bar{B}_t$. At the beginning of period $t$, a firm only has the obligation to repay its debt $B_t$ issued in period $t-1$. The firm must decide how much intertemporal debt to issue, $B_{t+1}$, how much labor to hire and inputs to purchase, and its equity payouts $D_t$. If at the pre-shock starting point the firm is borrowing up to its limit in Equation \ref{eq:real_model_generic_borrowing_constraint}, and the firm wishes to keep its production scale unchanged, i.e., do not reduce $L_t$ in order to reduce which $h_t$ or $M_t$, a reduction in $\bar{B}_t$ requires a reduction in the equity payout $D_t$. In other words, in face of a tighter borrowing constraint the firm is forced to increase its equity and reduce the amount of newly issued intertemporal borrowing. However, if the firm faces adjustment costs in its equity payouts $D_t$, it must reduce either hire less labor, purchase a lower amount of imported inputs, or both. Consequently, whether a tighter borrowing constraint affects a firm's production decision depends on the flexibility of firms to adjust its financial structure, i.e., the relative shares of debt and equity. 

We formalize the rigidities affecting the substitution between debt and equity by following the approach of \cite{jermann2012macroeconomic}. This approach assumes that a firm's payout is subject to quadratic adjustment costs which we define in detrended terms as
\begin{align}
    \Psi \left(d_t\right) = d_t + \psi \left(d_t - \bar{d}\right)^2 \label{eq:model_simple_rbc_dividends}
\end{align}
where $\bar{d}$ is the steady state value of dividend payout deflated by technology, and $\psi \geq 0$ is a parameter that governs the adjustment costs. This adjustment costs are meant to capture managers preference over dividends smoothing, which has been documented for US firms (see \cite{lintner1956distribution}), and for Emerging Markets (see \cite{jeong2013determinants,al2014dividend,al2017corporate,ahmed2020corporate}). This preference for dividend smoothing could derive from agency problems or other financial frictions. We abstract from explicitly modelling these underlying frictions and take them as a given.

Lastly, it is useful to define a firm's cash flow as their operational profits, which we denote by $\pi_t$
\begin{align}
    \pi_t \equiv Y_t - \left(W_t n_t + P^{M}_t M_t \right)\label{eq:model_simple_rbc_cash_flow}
\end{align}
This definition of cash-flow corresponds to the accounting definition of \textit{EBITDA}, i.e., sales net of overhead and labor costs, but without subtracting investment, interest payments or taxes. We also study a specification of the firm's cash flow which considers the interest payments on working capital borrowing, i.e., $\tilde{\pi}_t \equiv Y_t - \left(1-\phi+\phi \tilde{R}_t\right) \left(W_t n_t + P^{M}_t M_t\right)$. Our benchmark results are robust and amplified under this specification.

\noindent
\textbf{Borrowing constraints.} Next, we describe the functional form for the debt limit $\bar{B}_t$ in Equation \ref{eq:real_model_generic_borrowing_constraint}. We assume that firms are subject to a borrowing constraints on top of their working capital constraint. In line with our empirical work we consider three possible borrowing constraints: (i) a standard collateral or leverage (LEV), (ii) a cash flow to debt borrowing constraint (DC), (iii) a cash flow to interest payments or interest coverage borrowing constraint (IC). We assume that these borrowing limits are given by the following expressions
\begin{align}
    B^{LEV}_t &\leq \theta^{LEV} p_{k,t} K_t \left(1-\delta\right) \label{eq:model_simple_RBC_LEV} \\ 
    B^{DC}_t &\leq \theta^{DC} \varphi\left(L\right) \pi_t \label{eq:model_simple_RBC_DC} \\
    \tilde{r}_tB^{IC}_t &\leq \theta^{IC} \varphi\left(L\right) \pi_t \label{eq:model_simple_RBC_IC}
\end{align}
where $\varphi\left(L\right)$ is a lag of operator and parameters $\theta \in \{\theta^{LEV},\theta^{DC},\theta^{IC}\}$ capture the exogenous tightness of these borrowing constraints. As argued in Sections \ref{sec:cash_flow_lending} and \ref{sec:interest_sensitive_borrowing_constraints}, both the borrowing constraints implied by BCRA regulations and those implied by corporate loan debt covenants are not established in terms of the current period cash flows, but generally in terms of observable lagged cash flows in the near past. 

We assume that a firm's collateral is evaluated at market value. In our current structural model, the price of capital is determined in equilibrium as the Lagrange multiplier of the firm's capital accumulation equation. Given the investment adjustment costs in Equation \ref{eq:model_simple_rbc_LOC}, the price of capital appreciates with an increase in investment demand and depreciates with a decrease in investment demand.

\noindent
\textbf{Representative firm's optimization problem.} The objective function of the firm is to maximize the expected discounted stream of the dividends paid to its owners,
\begin{align}
    \max \mathbb{E}_0 \sum^{\infty}_{t=0} \Lambda_t \times d_t
\end{align}
subject to Equations \ref{eq:model_simple_rbc_production}, \ref{eq:model_simple_rbc_LOC}, \ref{eq:model_simple_rbc_dividends}, \ref{eq:model_simple_rbc_flow_of_funds}, \ref{eq:model_simple_rbc_cash_flow}, and one of the borrowing constraints given by Equations \ref{eq:model_simple_RBC_LEV} to \ref{eq:model_simple_RBC_IC}. The term $\Lambda_t$ in the objective function is the firm owners' stochastic discount factor between periods $0$ and $t$. 

In Appendix \ref{sec:appendix_details_simple_model} we present and describe in full the intermediate firm's optimality conditions. However, it is useful to describe how financial frictions affect the firm's input purchases and investment decisions. First, financial frictions introduce two wedges into firms input purchase decisions. Equation \ref{eq:labor_decision_collateral} shows this for a firm subject to a collateral borrowing constraint
\begin{align}
    MPL_t &= w_t \left[ \overbrace{\left(1-\phi+\phi \tilde{R}_t\right)}^{\text{Working Capital Wedge}} + \overbrace{\mu_t\times\Psi'_{d,t}\times\phi}^{\text{Borrowing Constraint Wedge}}  \right] \label{eq:labor_decision_collateral}
\end{align}
where $MPL_t$ represents the marginal product of labor. The first term in the square brackets on the right hand side of Equation \ref{eq:labor_decision_collateral} represents a working capital wedge, created by the increase in the cost of hiring labor due to fraction of inputs which must be paid in advanced and financed at interest rate $\tilde{R}_t$. An interest rate $\tilde{R}_t$ hike exacerbates the negative effects of the wedge reducing firm's optimal labor choice, as argued by \cite{neumeyer1998currencies}. 

The overall debt limit implied by the borrowing constraint introduces a second wedge on firms' optimal labor choice. This is represented by the second term in the square brackets on the right hand side of Equation \ref{eq:labor_decision_collateral}. This wedge is comprised of three terms: the Lagrange multiplier on firm's borrowing constraint, $\mu_t$, the derivative of the firm's dividend adjustment cost function, $\Psi'_{d,t}$, and the share of inputs which must be paid in advanced. First, a reduction in a firm's debt limit increases $\mu_t$ and leads to a reduction in the optimal labor choice. Second, greater dividend adjustment costs, implying a high degree of imperfect substitutability between debt and equity, lead to a greater wedge. The intuition behind this result is that a firm can, at least partially, offset a tighter borrowing constraint by reducing dividend payouts. However, this offset is limited by the dividend adjustments costs. As dividend adjustment costs increase, a tighter borrowing constraint leads to a greater reduction in a firm's access to finance, including both debt and equity. Lastly, a greater share of inputs which must be financed in advanced, $\phi$, leads to a greater share of the total debt limit being used to finance working capital, exacerbating the overall wedge.

The intermediate good firm's investment decision is also affected by financial frictions, but in a less direct way. Equation \ref{eq:investment_decision} presents the firm's optimal investment decision 
\begin{align} \label{eq:investment_decision}
    \frac{1}{\Psi'_{d,t}} = Q_{t} \Phi'_{i_t} + \mathbb{E}_t Q_{t+1} \Phi'_{i_{t+1}}
\end{align}
where $Q_t$ represents the price of capital in period $t$ and the terms $\Phi'_{i_t}$ and $\Phi'_{i_{t+1}}$ represents the derivatives of the investment adjustment costs with respect to $I_t$ in periods $t$ and $t+1$. Financial frictions affect the firm's investment decisions by altering the shadow price of one additional unit of investment. This shadow price is represented by the left hand side of Equation \ref{eq:investment_decision}. The optimal investment decision balances the cost of an additional investment unit, given by one fewer unit of dividend payout $1/\Psi'_{d,t}$, and the benefit of an additional investment unit given by relaxing the firm's constraint over its law of motion of capital. Note that investment expenditures are not subject to the working capital requirement. In consequence, financial frictions affect investment decisions only through the imperfect substitutability between debt and equity.

\noindent
\textbf{Household, fiscal authority and equilibrium.} The representative household maximizes the following life time utility function
\begin{align*}
    \mathbb{E}_0 \sum^{\infty}_{t=0} \beta^{t} \frac{\left(C_t - \zeta \omega^{-1} X_{t-1} h^{\omega}_t \right)-1}{1-\sigma}
\end{align*}
where $\beta$ represent a discount factor, $h_t$ represents hours worked. The household consumes and decides how much to save in risk-free bonds, $B_t$, and in shares of the representative firm, $s_t$. Given the firm's tax advantage over debt, in equilibrium firms are debtors and households save. Additionally, we do not allow foreigners to own shares of the representative firm, which leads to $s_t = 1$ $\forall$ periods $t$. 

We assume that the fiscal authority plays a passive role. The firm runs a budget balance with only lump sum taxes over the household as a source of tax revenue. These lump sum taxes, denoted by $T_t$, which is the household takes as given, is used to finance the firms' tax benefit enjoyed by firms when borrowing
\begin{align*}
    T_t = \frac{B^{f}}{\tilde{R}_t}-\frac{B^{f}}{1+r_t}
\end{align*}

We assume that the economy as a whole faces a debt-elastic interest rate premium as in \cite{schmitt2003closing} in order to induce independence of the deterministic steady state from initial conditions. In particular, the interest rate agents in the economy face is assumed as the sum of the world interest rate $r^{*}>0$ and a country premium that is a decreasing function of a detrended measure of net foreign assets, and an exogenous shock
\begin{align}
    r_t = r^{*} + \psi \left(e^{NFA_{t+1}/X_t - \bar{NFA}}-1\right) + e^{\mu_t-1} -1 
\end{align}
where $NFA_t$ is given by
\begin{align}
    NFA_t &= \left(B_t - B^{f}_t\right)
\end{align}
and $\bar{NFA}$ represents the steady state value of net foreign assets. We follow \cite{garcia2010real} and assume that the exogenous shock term $e^{\mu_t-1}$ is characterized by a stochastic shock which follows an auto-regressive process of order 1 given by
\begin{align} \label{eq:foreign_interest_rate_shock}
    \ln \mu_{t+1} = \rho_{\mu} \ln \mu_{t}+\epsilon^{\mu}_t, \qquad \epsilon^{\mu}_t \sim \mathcal{N} \left(0,\sigma^{2}_{\mu}\right)
\end{align}

\noindent
\textbf{Calibration and solution.} We parametrize our model following a two prong approach. First, there is a set of parameters which we calibrate by choosing standard values in the literature of international macroeconomics. There is a set of parameters which we calibrate to match certain aggregate moments of the Argentinean data.

We begin by presenting in Table \ref{tab:real_model_standard_parameters} the parametrization of the set of parameters that are chosen to match the international macroeconomic literature.
\begin{table}[ht]
    \centering
    \caption{Parametrization: Real Model \\ \footnotesize Standard Values}
    \label{tab:real_model_standard_parameters}
    \small
    \begin{tabular}{c l l} \hline \hline
    \textbf{Parameter} & \textbf{Value} & \textbf{Details / target}  \\ \hline
    $\phi_i$  & Investment Adjustment Cost & $2$ - \footnotesize \cite{christiano2011introducing} \\
    $\alpha$ & Capital Elasticity & $0.32$ - \footnotesize \cite{garcia2010real,mendoza2010sudden} \\
    $\eta$ & Imported Input Elasticity & $0.10$ - \footnotesize \cite{mendoza2010sudden} \\
    $\delta$ & Capital depreciation rate  & $0.1255/4$ - \footnotesize \cite{garcia2010real}  \\
    $\beta$  & Household discount factor   & $0.9852$ - \footnotesize \cite{thoenissen2014news} $\rightarrow$ $R = 6\%$ \\    
    $\sigma$  & Household risk aversion   & $2$ - \footnotesize \cite{garcia2010real,mendoza2010sudden} \\     
    $\phi$ & Working capital parameter & $0.25$ - \footnotesize \cite{mendoza2010sudden} \\     
    $\tau$   & Tax advantage on debt & $0.35$ - \footnotesize Argentinean Data \&  \cite{thoenissen2014news} \\
    $\psi$   & Dividend adjustment cost    & $0.20$ - \footnotesize \cite{jermann2012macroeconomic} \\ \hline \hline
    \end{tabular}
    \floatfoot{\textbf{Note:} See Appendix \ref{sec:appendix_details_simple_model} for additional details on the calibration of the model and for a full list of model parameters.}
\end{table}
The capital share in production, $\alpha$, and the capital depreciation rate, $\delta$, are sourced from \cite{garcia2010real}. The household discount factor, $\beta$, is set to match an annual interest rate of $6\%$, consistent with \cite{thoenissen2014news}. We set the fraction of the wage bill and imported intermediate inputs that must be pay in advanced, $\phi_{WK} = 0.25$, in line with the estimates of \cite{mendoza2010sudden} and other values in the literature. Note that this value is significantly lower than those studied by \cite{neumeyer2005business} which choose a value of $\phi_{WK} = 1$ as their main parametrization. This parameter choice is intended to reflect the importance of working capital constraints within a broader overall borrowing constraint without overemphasizing its role as a transmission channel of economic shocks. The parameter that governs the degree of debt subsidy $\tau$ is set equal to $\tau = 0.35$ following \cite{thoenissen2014news}. Lastly, we parametrize $\psi$ which governs the dividend adjustment cost equal to $0.2$. This value is standard in the literature which study financial frictions for US firms, see \cite{jermann2012macroeconomic} and \cite{drechsel2019earnings}. This parameter value has also been used by the international macroeconomic literature which builds on \cite{jermann2012macroeconomic}, as exemplified by \cite{thoenissen2014news}, and thus we consider it a useful initial benchmark. However, we will later emphasize that this parameter plays a crucial role in determining the tightness of borrowing constraints and impact of interest rate shocks. We show that for higher values of $\psi$, implying greater financial frictions in Emerging Markets than for US firms, then the quantitative implications of interest coverage borrowing constraints are exacerbated.

Next, Table \ref{tab:real_model_calibrated_parameters} presents the results of our set of parameters which are calibrated to match Argentinean data. We seek to the hours worked at 20\% of the households total hours, as in \cite{garcia2010real}; a balanced net-exports to GDP ratio and a private sector credit to GDP of 35\%, the average value for Argentina for the period 1993-2020.\footnote{Assuming balanced net exports to GDP is close to the observed level in the data, and in line with values assumed in the literature. For example, \cite{garcia2010real} calibrates the model to steady state trade balance of 0.25\% of GDP.}$^{,}$\footnote{The average ratio of private sector credit to GDP is sourced from the World Bank \url{https://data.worldbank.org/indicator/FS.AST.PRVT.GD.ZS}. We carry out robustness checks to match higher ratios of private sector credit to GDP. We find that the amplification of interest coverage constraints to leverage constraints is exacerbated for higher ratios.}   
\begin{table}[ht]
    \centering
    \caption{Parametrization: Real Model \\ \footnotesize Calibrated Values}
    \label{tab:real_model_calibrated_parameters}
    \small
    \begin{tabular}{c l l} \hline \hline
    \textbf{Parameter} & \textbf{Value} & \textbf{Details / target}  \\ \hline
    $\zeta^{LEV}$  & Dis-utility of Labor Lev. Const. & 1.9354 - \footnotesize Match 20\% hours worked \\
    $\zeta^{DC}$   & Dis-utility of Labor DC Const. & 1.9348 - \footnotesize Match 20\% hours worked \\
    $\zeta^{IC}$   & Dis-utility of Labor IC. Const. & 1.9348 - \footnotesize Match 20\% hours worked \\  
    \\
    $\theta^{LEV}$  & Tightness Leverage Const. & 0.2279 - \footnotesize Match 35\% of Credit to GDP \\
    $\theta^{DC}$   & Tightness DC Const. & 4.2261 - \footnotesize Match 35\% of Credit to GDP \\
    $\theta^{IC}$   & Tightness IC Const. & 0.0955 - \footnotesize Match 35\% of Credit to GDP \\     
     \hline \hline
    \end{tabular}
    \floatfoot{\textbf{Note:} See Appendix \ref{sec:appendix_details_simple_model} for additional details on the calibration of the model and for a full list of model parameters.}
\end{table}
We match these aggregate moments for each of the economies with borrowing constraint from Equations \ref{eq:model_simple_RBC_LEV} to \ref{eq:model_simple_RBC_IC}, by choosing a set of parameters $\{\zeta, \bar{NFA}, \theta \}$. From the first panel of Table \ref{tab:real_model_calibrated_parameters}, it is re-assuring that the resulting parameters of the dis-utility of labor are close across economies with different borrowing constraints, and close the calibrated value of 2.24 used in \cite{garcia2010real}. Next, it is worthy to describe the resulting values of the tightness parameters. For the leverage constraint, the parameter $\theta^{LEV} = 0.2279$ is close to the calibrated value used by \cite{mendoza2010sudden} and \cite{mendoza2020fipit} of 0.20. Thus, we believe our benchmark parametrization allows us to carry out a quantitative comparison between the standard leverage constraint and the two cash flow-based borrowing constraints. The calibrated value of $\theta^{DC} = 4.2261$ is smaller than those found by \cite{greenwald2019firm}, 8.613, and \cite{drechsel2019earnings}, $4 \times 4.6$, for US firms. This is expected as the ratio of private credit to GDP in Argentina is 10 times smaller than the same ratio for the US economy. Similarly, the estimated tightness parameter for the interest coverage constraint, $\theta^{IC} = 0.0955$, is smaller than that found by by \cite{greenwald2019firm}, 0.154.

\subsection{Counterfactual Analysis} \label{subsec:model_simple_IRF}

We study the implications of cash flow-based borrowing constraints under the light of a foreign interest rate shock, $R^{*}_t$. The impact of US monetary policy shock is a classic question in international macroeconomics, going back to \cite{fleming1962domestic}, \cite{mundell1963capital}, \cite{dornbusch1976expectations} and \cite{frenkel1983monetary}. Furthermore, it is still an active research topic both empirically (see \cite{degasperi2020global,camara2021spillovers}) and structurally (see \cite{kalemli2019us,camara2021FXI}). We argue that the introduction of interest sensitive cash flow-based borrowing constraint leads to a significantly greater amplification of foreign interest rate shocks compared to the benchmark leverage constraint. This greater amplification is driven by the borrowing constraint's interest sensitivity and not by the introduction of firms' cash flow. This amplification is significantly exacerbated with an increase in the magnitude of working capital constraints, $\phi$, and in the frictions which govern the substitutability of debt and dividend payouts, $\psi$.

\noindent
\textbf{Benchmark results.} We assume that an exogenous 100 annual basis point increase in the foreign interest rate shock occurs modeled through a positive realization of $\epsilon^{\mu}_t$ in Equation \ref{eq:foreign_interest_rate_shock}. We assume that this exogenous shock has a persistence of $\rho_{\mu} = 0.75$, which is in line with the observed path of the Federal Reserve's Federal Funds Rate after a positive interest rate shock. 

\begin{figure}[ht]
    \centering
    \includegraphics[width=15cm,height=10cm]{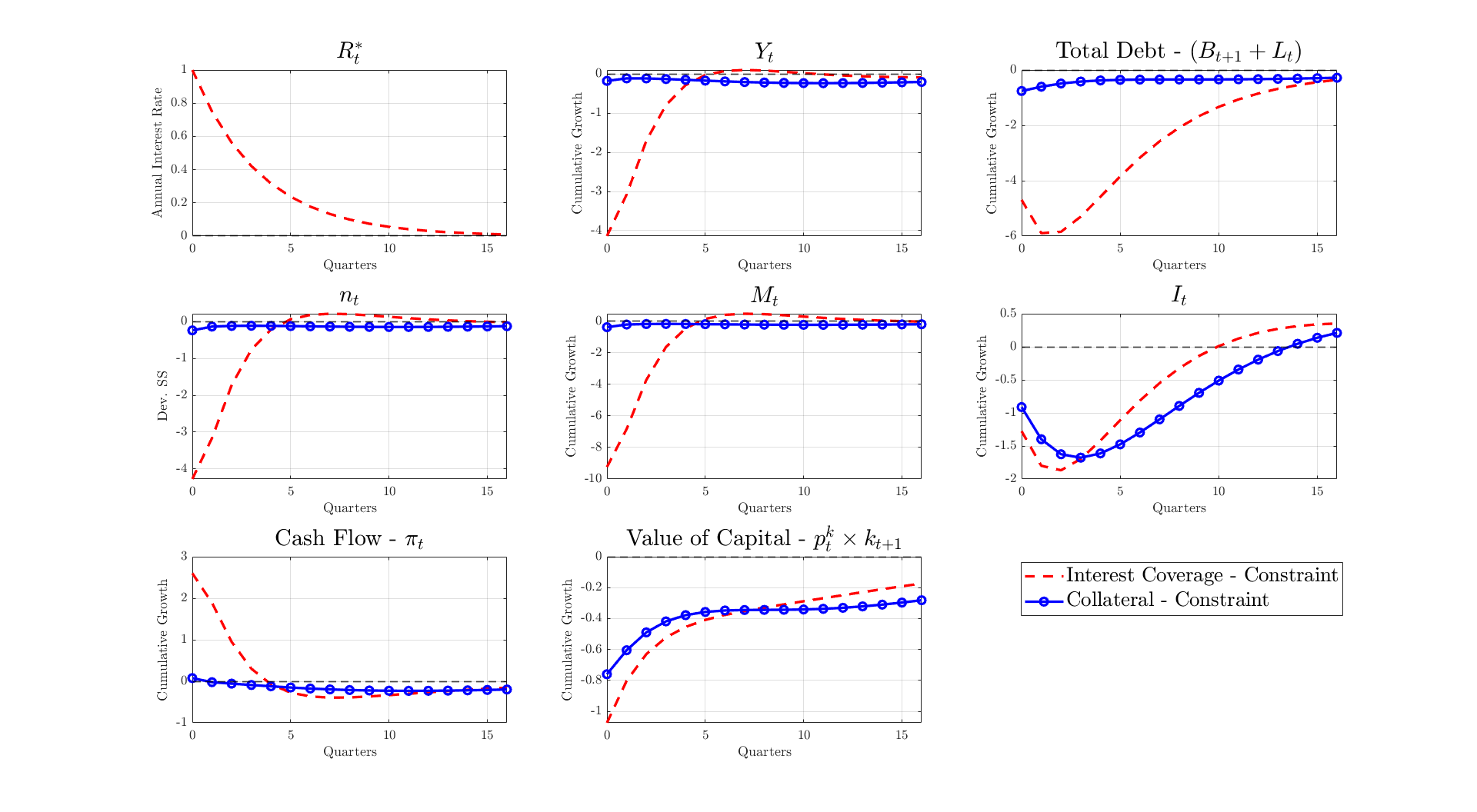}
    \caption{Collateral \& Interest Coverage Borrowing Constraints \\ \footnotesize IRF Analysis - $R^{*}$ Shock}
    \label{fig:Rshock_rho_75_Draft}
    \floatfoot{\textbf{Note:} The figure is comprised of 8 panels ordered in three rows and three columns. In the text we refer to each panel following a $(x,y)$ convention, where $x$ and $y$ represent the row and column. Each panel presents the dynamics of an economy with a collateral borrowing constraint (blue dotted line), and an economy with a interest sensitive cash flow borrowing constraint (red dashed line).}
\end{figure}

Figure \ref{fig:Rshock_rho_75_Draft} presents the impulse responses to a foreign interest rate shock, see Panel $(1,1)$, under a collateral borrowing constraint (dotted blue line) and under an interest coverage cash flow constraint (dashed red line).
The first conclusion that arises from this figure is that an economy with an interest sensitive borrowing constraint (IC) exhibits a significantly greater amplification on real variables in response to an increase in the foreign interest rate than an economy with a collateral borrowing constraint. To see this, Panel $(1,2)$ shows the impact of the foreign interest rate shock in production $Y_t$. The drop in production is around 40 times larger under the interest coverage constraint compared to the collateral constraint. The drop in production $Y_t$ is significantly larger for the interest coverage constraint for the first year after the shock. Panel $(1,3)$ shows that greater drop in production under the interest coverage constraint is associated with a greater drop in total debt compared to the collateral borrowing constraint. 

To understand how the greater drop in firms' debt leads to a greater drop in production it is key to understand the dynamics of firms' employment, input and investment decisions. On the one hand, Panels $(2,1)$ and $(2,2)$ show that the greater reduction in firms' debt under an interest coverage constraint leads to a significantly larger and sharper drop in input purchases compared to a collateral borrowing constraint. On impact, the drop in employment and imported inputs is 40 and 90 times larger under the interest coverage constraint. On the other hand, Panel $(2,3)$ shows that the drop on impact in investment is only 50\% larger under an interest coverage constraint compared to a collateral borrowing constraint. Additionally, the recovery is faster under the collateral borrowing constraint. The greater impact on input purchases is in line with the presence of both working capital and borrowing constraint in the optimal purchase choice, with the optimal investment decision is only affected indirectly by financial frictions. Lastly, Panel $(3,2)$ shows that the dynamics of the value of capital is relatively close across the two economies. 

The second conclusion that arises from Figure \ref{fig:Rshock_rho_75_Draft} is that the key driver of the greater amplification of foreign interest rate shocks is the inherent interest sensitivity of the interest coverage borrowing constraint. Panel $(3,1)$ shows the dynamics of firms' cash flow, as defined in Equation \ref{eq:model_simple_rbc_cash_flow}. Note that the economy under an interest coverage borrowing constraint exhibits an initial increase in firms' cash flow. This initial increase is caused by the significant drop in real wages, employment and imported inputs, triggered by firms' drop in input demand.\footnote{This result is caused by the lack of price and/or wage frictions. In Section \ref{sec:nominal_frictions_policy} we introduce nominal price frictions and study how this frictions matter for the transmission of shocks. However, we do not consider wage frictions in our current analysis. Nevertheless, note that introducing wage frictions would lead to lower increase in cash flows and/or decrease in cash flows, exacerbating the impact on firms' access to debt.} Thus, the reduction in firm's total debt shown in Panel $(1,3)$ is driven by the borrowing constraint's interest sensitivity. 

\begin{figure}[ht]
    \centering
    \includegraphics[width=15cm,height=10cm]{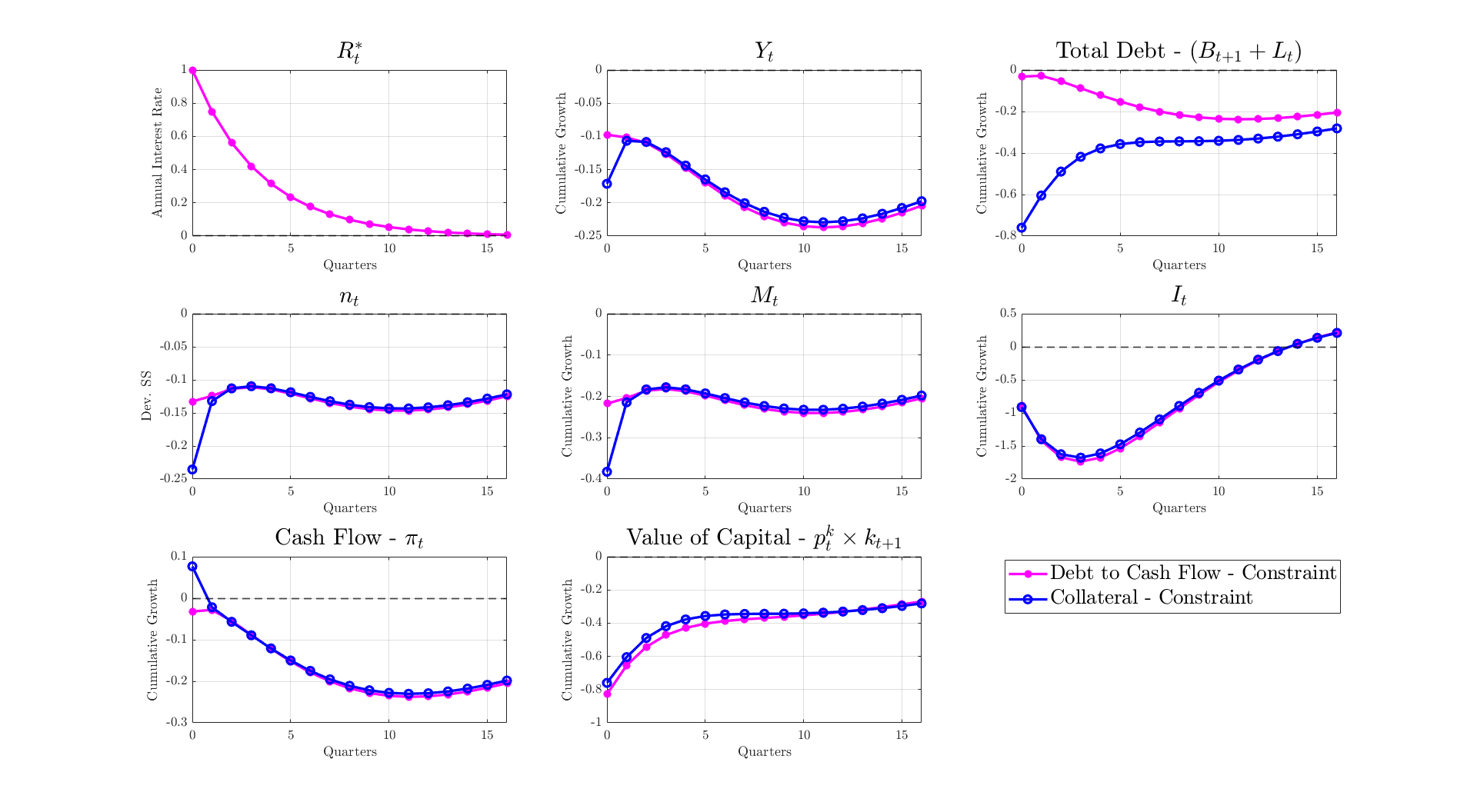}
    \caption{Collateral \& Debt to Cash Flow Borrowing Constraints \\ \footnotesize IRF Analysis - $R^{*}$ Shock}
    \label{fig:Rshock_rho_75_DC_vs_AB_Draft}
    \floatfoot{\textbf{Note:} The figure is comprised of 8 panels ordered in three rows and three columns. In the text we refer to each panel following a $(x,y)$ convention, where $x$ and $y$ represent the row and column. Each panel presents the dynamics of an economy with a collateral borrowing constraint (blue dotted line), and an economy with a debt to cash flow borrowing constraint (magenta dotted line).}
\end{figure}

In order to highlight the importance of the borrowing constraint's interest sensitive, Figure \ref{fig:Rshock_rho_75_DC_vs_AB_Draft} presents the impulse responses under a collateral and a debt to cash flow borrowing constraint.
Under a debt to cash flow borrowing constraint the reduction on impact in firm's debt is significantly smaller than under a collateral borrowing constraint. Consequently, the impact on the firm's input purchases is twice as large under a collateral borrowing constraint. Altogether, the dynamics are remarkably similar across both constraints. This is not surprising given that in this benchmark model without any nominal friction there is a tight relationship between a firm's cash flow and the value of capital.\footnote{For instance, \cite{greenwald2019firm} in a similar structural model finds that debt to cash flow and collateral borrowing constraints yield quantitatively similar impulse response functions in response to a discount factor shock.} Thus, the sharp reduction in the firm's debt under an interest coverage constraint shown in Figure \ref{fig:Rshock_rho_75_Draft} can be attributed to the interest coverage constraint's interest sensitivity.

Overall, in response to a foreign interest rate shock, an interest sensitive borrowing constraint leads to an orders of magnitude greater amplification of production, debt and input purchases than under a standard collateral borrowing constraint. This greater amplification is driven by the interest coverage constraint's interest sensitivity. The interest rate hike decreases the firm's debt limit, which leads to a reduction in the firm's working capital loans, associated with a sharp fall in the hiring of labor and purchase of intermediate inputs. Given that financial frictions only affect a firm's investment decisions indirectly, the impact on investment and the price of capital is similar under an interest coverage and collateral borrowing constraints.

\noindent
\textbf{Sensitivity to working capital needs.} The main transmission channel of a foreign interest rate shock $R^{*}_t$ in our model is through tighter borrowing and working capital constraints. Thus, parameter $\phi$ which governs the importance of working capital needs plays a crucial part. Next, we test the quantitative importance of changes in the benchmark calibration of parameter $\phi$.

\begin{figure}[ht]
    \centering
    \includegraphics[width=15cm,height=7cm]{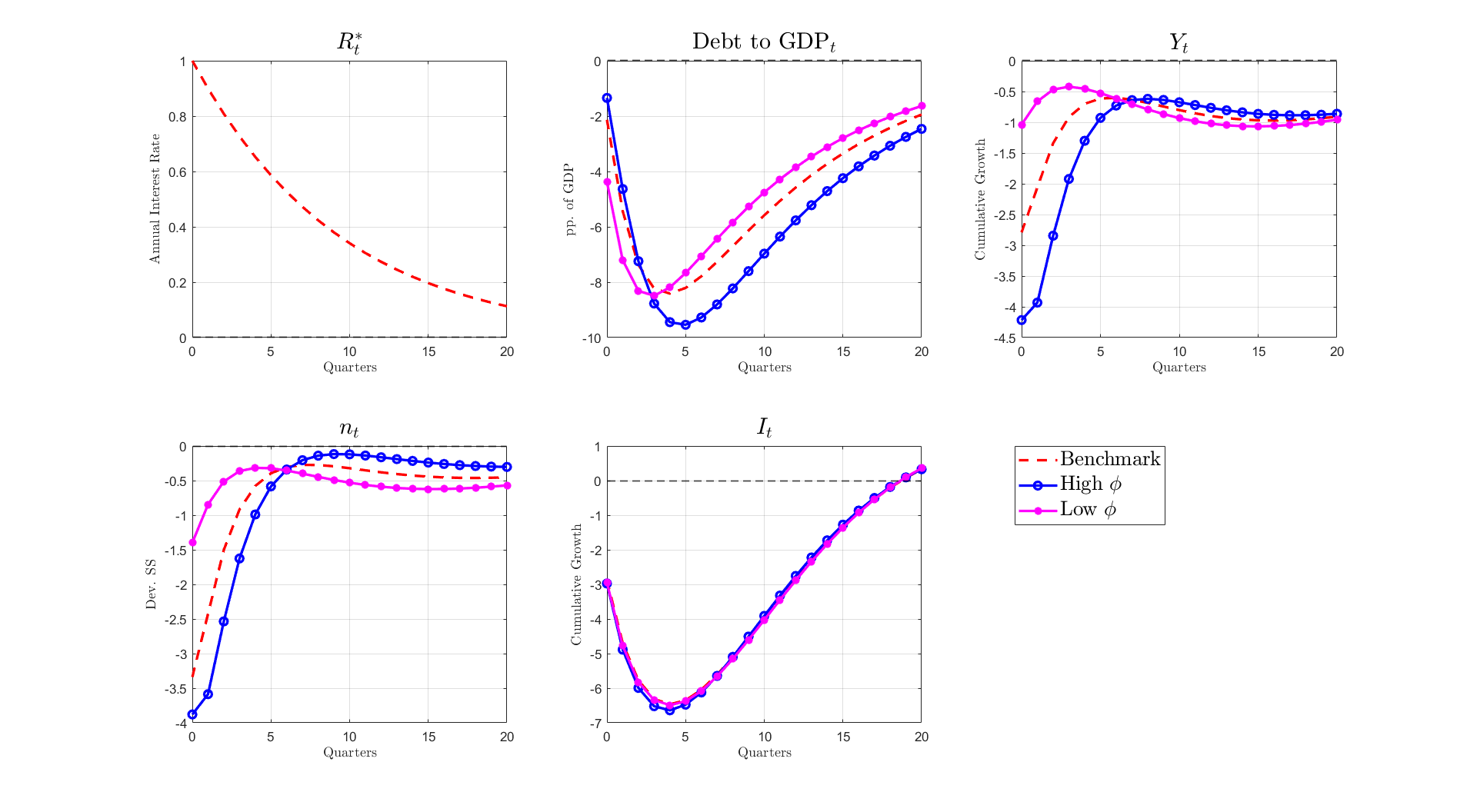}
    \caption{Collateral \& Interest Sensitive Borrowing Constraints \\ \footnotesize Sensitivity to Working Capital $\phi$}
    \label{fig:Real_Model_WK_Sensitivity}
    \floatfoot{\textbf{Note:} The figure is comprised of 5 panels ordered in three rows and three columns. In the text we refer to each panel following a $(x,y)$ convention, where $x$ and $y$ represent the row and column. Each panel presents the dynamics of an economy with interest sensitive cash flow borrowing constraints with different values of parameter $\phi$. The red dashed line presents the results under the benchmark calibrated value of $\phi = 0.25$. The blue dotted line presents the results under a calibration of $\phi = 0.5$, and the dotted magenta line presents the results under a calibration of $\phi = 0.05$. For these two extreme cases with different values of $\phi$, parameters $\theta^{IC}$ and $\zeta^{IC}$ are re calibrated to match the aggregate moments in Table \ref{tab:real_model_calibrated_parameters}.}
\end{figure}

Figure \ref{fig:Real_Model_WK_Sensitivity} shows the quantitative importance of parameter $\phi$ by comparing the impulse response functions of a foreign interest rate shocks for three values $\phi$: (i) the benchmark value (red dashed line), (ii) a high value of $\phi=0.5$ (blue dotted line), (iii) a low value of $\phi=0.05$ (magenta dotted line). 
The results are straightforward and intuitive as greater working capital requirements lead to tighter borrowing constraints and greater impacts on labor demand and total production. Panel $(1,2)$ shows that the drop in the debt to GDP ratio is only significantly greater under the dynamics of a high value of $\phi$. Panel $(1,3)$ shows that doubling the working capital requirements leads to a 50\% greater impact on $Y_t$. Similarly, almost shutting off working capital requirements with a value of $\phi = 0.05$ leads to a significantly lower impact on $Y_t$. The last row of Figure \ref{fig:Real_Model_WK_Sensitivity} highlights the transmission channel of foreign interest rate shocks through working capital constraints. On the one hand, Panel $(2,1)$ shows that greater values of $\phi$ leads to a significantly greater impact on labor demand, $n_t$, as working capital requirements directly affect the wedge on firms' labor demand. On the other hand, Panel $(2,2)$ shows that increasing the value of $\phi$ does not affect significantly investment dynamics which are not subject to working capital requirements.

\noindent
\textbf{Sensitivity to dividend adjustment costs.} The frictions in adjusting dividends plays a crucial role in our model. The introduction of quadratic adjustment costs in Equation \ref{eq:model_simple_rbc_dividends} causes an imperfect substitution between debt and equity, leading to tighter borrowing constraints having a real effect on firm's output. In order to gauge the quantitative importance of the frictions in dividend payouts, we compute the impulse response functions of a foreign interest rate shock for different values of parameter $\psi$ which governs the adjustment costs of dividends.

\begin{figure}[ht]
    \centering
    \includegraphics[width=15cm,height=7cm]{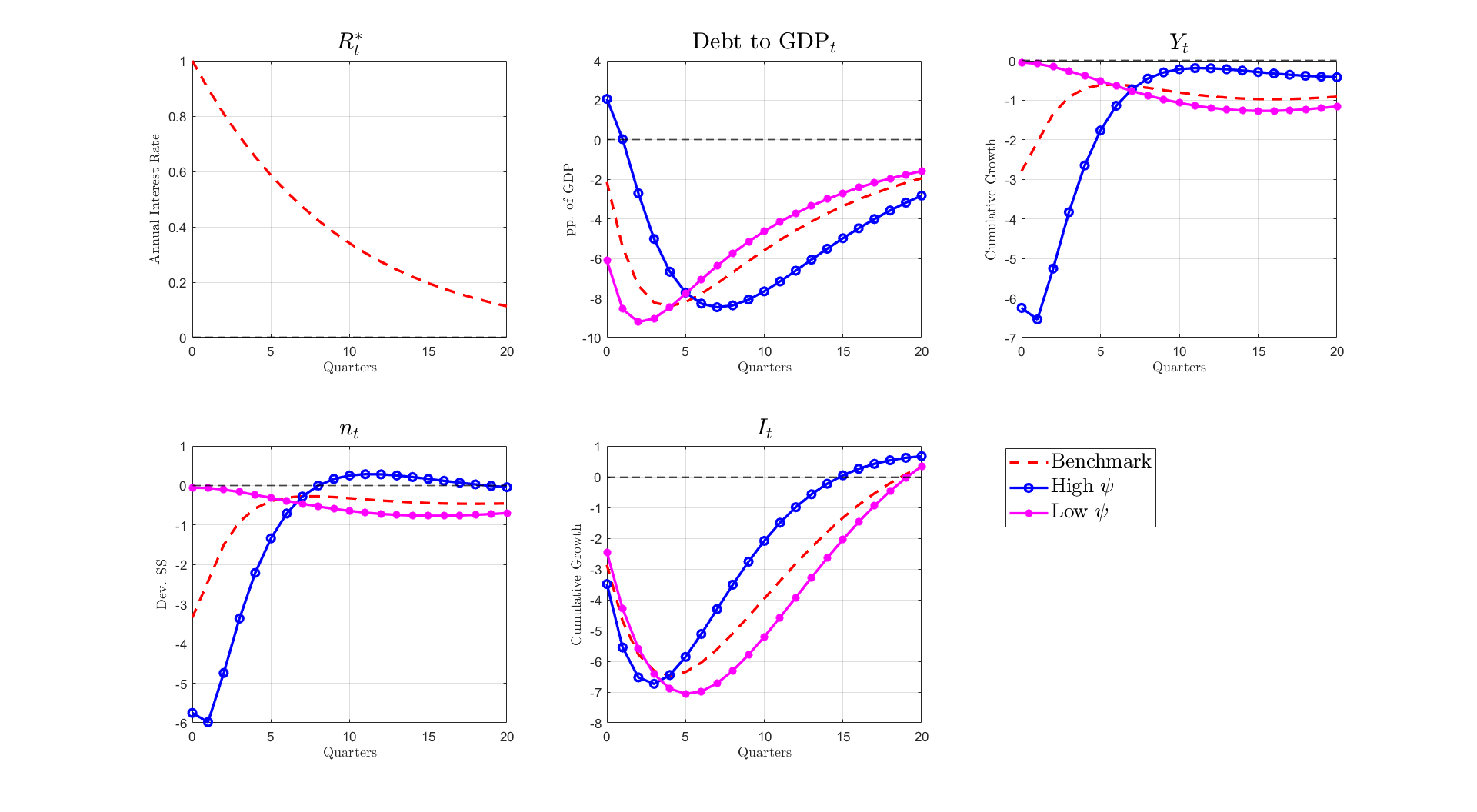}
    \caption{Collateral \& Interest Sensitive Borrowing Constraints \\ \footnotesize Sensitivity to Dividend Frictions $\psi$}
    \label{fig:Real_Model_PSI_Sensitivity}
    \floatfoot{\textbf{Note:} The figure is comprised of 5 panels ordered in three rows and three columns. In the text we refer to each panel following a $(x,y)$ convention, where $x$ and $y$ represent the row and column. Each panel presents the dynamics of an economy with interest sensitive cash flow borrowing constraints with different values of parameter $\psi$. The red dashed line presents the results under the benchmark calibrated value of $\phi = 0.25$. The blue dotted line presents the results under a calibration of $\phi = 1$, and the dotted magenta line presents the results under a calibration of $\phi = 0.005$. Note that different values of parameter $\psi$ do not affect the calibration of the model as it not affects its non-stochastic steady state.}
\end{figure}

Figure \ref{fig:Real_Model_PSI_Sensitivity} shows the quantitative importance of parameter $\psi$ by comparing the impulse response functions of a foreign interest rate shocks for three values $\psi$: (i) the benchmark value (red dashed line), (ii) a high value of $\psi=1$ (blue dotted line), (iii) a low value of $\phi=0.005$ (magenta dotted line). 
Once more, results are straightforward and intuitive as greater dividend frictions lead to an increase impact of tightening borrowing constraints over firms' input demand. Panel $(1,2)$ shows that increasing the value of $\psi$ leads to a greater and more persistent decrease in the ratio of debt to GDP. This greater drop in firms' debt is associated with a greater drop in production $Y_t$ and labor demand $n_t$, see Panels $(1,3)$ and $(2,1)$ respectively. In particular, increasing the adjustment cost parameter $\psi$ from 0.2 to 1, doubles the impact of a foreign interest rate shock on production and labor demand. Unlike an increase in working capital requirements, $\phi$, an increase in the dividend adjustment costs affects investment choices as seen in Panel $(2,2)$.  While a greater value of $\psi$ leads to a relatively larger initial impact on firms' investment after an interest rate shock, this change in parametrization also leads to a significantly faster recovery.


\section{Nominal Frictions \& Policy Implications} \label{sec:nominal_frictions_policy}

In this section, we augment our model by introducing nominal price frictions. Beyond analysing the importance of collateral and cash flow borrowing constraints in presence of nominal frictions, the framework allows us to consider two key economic features of interest in Emerging Market economies. The first feature is that it allows to consider for cash flows and debt in both domestic and foreign currency. The second feature is that it allows us to consider the effects of monetary policy of economies under different borrowing constraints. Section \ref{subsec:nominal_frictions_policy_description} describes the additional features of the model. Section \ref{subsec:nominal_frictions_policy_IRF} carries out an impulse response analysis to study the importance of differences in borrowing constraints across different economic shocks. Section \ref{subsec:nominal_frictions_matching_aggregate} shows how an economy with an interest sensitive borrowing constraint is able to match aggregate moments of the data presented in Section \ref{sec:interest_sensitive_borrowing_constraints}. Lastly, Section \ref{subsec:nominal_frictions_policy_implications} shows the policy implications of interest sensitive borrowing constraints, emphasizing the costs of policies which limit exchange rate volatility.

\subsection{Model Description} \label{subsec:nominal_frictions_policy_description}

In this section we describe the main added features to the model in Section \ref{sec:model_simple} which allow us to introduce nominal price frictions. We leave the full description of the model to Appendix \ref{sec:appendix_details_simple_model_nk}. The economy is comprised of four agents: (i) firms which produce a domestic homogeneous good, (ii) intermediate firms alike those of Section \ref{sec:model_simple}, (iii) firms which produce a variety of final goods, (iv) a government comprised of both fiscal and monetary authorities. We calibrate the model such that it replicates aggregate moments from an average Emerging Market economy.

\noindent
\textbf{Domestic homogeneous good production.}  Production of the domestic homogeneous good is carried out aggregating different varieties of intermediate goods
\begin{align*}
    Y_t = \left[\int^{1}_{0} y^{\frac{1}{\eta}}_{i,t} di \right]^{\eta}, \quad \eta>1
\end{align*}
The optimal behavior of the final good producing firm leads to the following demand curve:
\begin{align*}
    p_{i,t} = P_t Y^{\frac{\eta-1}{\eta}}_t y^{\frac{1-\eta}{\eta}}_{i,t}
\end{align*}
The aggregate price index of the final good becomes
\begin{align*}
    P_t = \left(\int^{1}_0 p^{\frac{1}{1-\eta}}_{i,t}\right)^{1-\eta}
\end{align*}
It is useful to define the intermediate firms' real demand and real revenue as a function of aggregate variables $Y_t$, $P_t$ and the quantities $Y_{i,t}$. Real demand, i.e., in terms of the domestic homogeneous good:
\begin{align*}
    D_{i,t} = \frac{p_{i,t}}{P_t} = Y^{\frac{\eta-1}{\eta}}_t y^{\frac{1-\eta}{\eta}}_{i,t}
\end{align*}
Real revenues, i.e., in terms of the domestic homogeneous good:
\begin{align*}
    F_{i,t} = \frac{p_{i,t} y_{i,t}}{P_t} = Y^{\frac{\eta-1}{\eta}}_t y^{\frac{1}{\eta}}_{i,t}
\end{align*}

\noindent
\textbf{Intermediate Good Firms.} We describe the additional features introduced to the intermediate good firm presented in Section \ref{sec:model_simple}. We assume that there is a continuum of size 1 of intermediate good firms. First, we assume that firms face Rotemberg price adjustment cost of the form
\begin{align}
    \Upsilon_{p} \left(p_{i,t},p_{i,t-1},Y_t\right) = \phi_{p}\left(\frac{p_{i,t}}{p_{i,t-1}} - 1\right)^2 Y_t
\end{align}
where $p_{i,t}$ and $p_{i,t-1}$ are the prices firm $i$ charge in period $t$ and $t-1$ respectively. Firms internalize the effects that their price and input purchase decisions affect the demand they face for their products and, consequently, their mark-ups.

Second, we allow for firms to borrow in both domestic and foreign currency. We assume that firm finances a fraction $\phi_b$ of total debt in domestic currency and finances a fraction $1-\phi_b$ of total debt is financed in terms of foreign currency. This implies that
\begin{align*}
    B^{\text{peso}}_t &= \phi_b B_t \\
    S_t B^{\text{dollar}}_t &= \left(1-\phi_b\right) B_t \\
\end{align*}
where $B^{\text{peso}}_t$ represents the amount of domestic currency debt which is financed at interest rate
\begin{align*}
    \tilde{R}_{d,t} = 1 + \left(R_{d,t}-1\right)\times\tau
\end{align*}
and $B^{\text{dollar}}_t$ represents the amounts of dollars borrowed at the non-state contingent rate
\begin{align*}
    \tilde{R}^{*}_{t} = 1 + \left(R^{*}_{t}-1\right)\times\tau
\end{align*}
where $\tau$ represents a tax advantage for firms financing through debt, as in Section \ref{sec:model_simple}. Note that while interest rates paid in period $t+1$ are fixed in period $t$ are non-state contingent, the effective interest rate paid is
\begin{align*}
    \tilde{R}_{t} = s_t \left(1-\phi_b\right) \tilde{R}^{*}_{t-1} + \phi_b \tilde{R}_{d,t-1}
\end{align*}

Third, we introduce a debt smoothing technology. This is motivated by the fact that banks evaluate firms riskiness taking into account both present and past firm indicators of overall risk of default, such as interest and capital payments, and their history of revenues and cash flows. An alternative motivation is presented by \cite{greenwald2019firm} which argues that syndicated loans in the US introduce debt covenants which are generally written at the annual rather than the quarterly frequency. Note that the corporate loan used as an example in Section \ref{subsec:interest_sensitive_corporate} exhibited debt covenants written at the annual frequency.

We choose the functional form introduced by \cite{greenwald2019firm} which assumes that the overall debt limit is a moving average of the current debt limits limit defined by the interest coverage or collateral borrowing constraints as defined in Section \ref{sec:model_simple},
\begin{align} \label{eq:debt_smoothing}
    B^{*}_t = \left(1-\rho_B\right) \bar{B}_t + \rho_B \pi^{-1}_t B^{*}_{t-1}
\end{align}
where $B^{*}_t$ is the actual debt limit the firm faces in period $t$ and $\bar{B}_t$ is the debt limit specified by the interest coverage or collateral constraint. Parameter $\rho_B$ reflects the debt limit's weight on the previous periods debt limit. This implies that the term $1-\rho_B$ reflects how sensitive a firm's debt limit is to changes in current economic conditions. 

\noindent
\textbf{Final good producers.} There are three final goods in the economy which are produced combining the domestic and foreign homogeneous goods: (i) consumption goods, (ii) investment goods, (iii) export goods.

Consumption goods purchased and consumed by domestic households. They are produced by a representative, competitive firm using the following production function:
\begin{align}
    C_t = \left[\left(1-\omega_c\right)^{\frac{1}{\eta_c}} \left(C_{d,t}\right)^{\frac{\eta_c-1}{\eta_c}}  + \omega_c^{\frac{1}{\eta_c}} \left(C_{m,t}\right)^{\frac{\eta_c-1}{\eta_c}} \right]^{\frac{\eta_c}{\eta_c-1}}
\end{align}
where $C_{d,t}$ is a domestic homogeneous output good with price $P_t$, $C_{m,t}$ is an foreign good with price $P^{m}_t = S_t P^{f}_t$, $C_t$ is the final consumption good and $\eta_c$ is the elasticity of substitution between domestic and foreign goods. Note that we are assuming that the domestic currency price of the foreign good is equal to its foreign currency price times the nominal exchange rate. This set up leads to the following demand functions
\begin{align}
    C_{m,t} &= \omega_c \left(\frac{P^{c}_t}{P^{m}_t} \right)^{\eta_c} C_t \\
    C_{d,t} &= \left(1-\omega_c\right) \left(\frac{P^{c}_t}{P_t} \right)^{\eta_c} C_t
\end{align}
Parameter $(1-\omega_c)$ represents the degree of home bias, with $\omega_c = 0$ implying that the consumption good is a one to one transformation from the domestic homogeneous good. In our calibrated exercise, we assume the presence of home bias, i.e., $1-\omega_c > 0.5$.

There is a large number of identical investment good producing firms. In order to produce investment goods, $I_t$, the firm combines domestic and foreign investment goods using the following production function:
\begin{align}
    I_t = \left[ \gamma^{\frac{1}{\nu_I}}_I I^{\frac{\nu_I-1}{\nu_I}}_{d,t} + \left(1-\gamma_I\right)^{\frac{1}{\nu_I}} I^{\frac{\nu_I-1}{\nu_I}}_{m,t} \right]^{\frac{\nu_I}{\nu_I-1}}
\end{align}
The input demand functions for inputs are given by
\begin{align}
    I_{d,t} &= \gamma_I I_t \left(\frac{P_t}{P^{I}_t} \right)^{-\nu_I} \\
    I_{m,t} &= \left(1-\gamma_I\right) I_t \left(\frac{S_tP^{f}_t}{P^{I}_t} \right)^{-\nu_I}
\end{align}
Parameter $\gamma_I$ represents the degree of home bias, with $\gamma_I = 1$ implying that the investment good is a one to one transformation from the domestic homogeneous good. In our calibrated exercise, we assume that investment goods rely significantly in foreign goods. $\gamma_I<0.5$.\footnote{There is significant evidence that Emerging Market economies rely heavily in imports for both investment goods and intermediate inputs. See \cite{eaton2001trade} for a reference which highlights this fact in the global economy and \cite{camara2022granular} for the case of Argentina.}

Lastly, we assume that the domestic homogeneous good can be directly exported. We denote the amount of the good exported by $X_t$. Foreign demand for the domestic good
\begin{align*}
    X_t = \left(\frac{P^{x}_t}{P^{f}_t}\right)^{-\eta_f} Y^{f}_t
\end{align*}
where $Y^{f}_t$ is a foreign demand shifter, $P^{f}_t$ is the foreign currency price of the foreign good, $P^{x}_t$ is the foreign currency price of the export good. 

We assume that there is a perfectly competitive exporter which purchases the domestic homogeneous good at price $P_t$. It sells the good at dollar price $P^{x}_t$, which translates into domestic currency units, $S_t P^{x}_t$. Competition implies that price, $S_t P^{x}_t$, equals marginal cost, $P_t$, so that
\begin{align} \label{eq:benchmark_export_pricing}
    S_t P^{x}_t = P_t
\end{align}
We can express the foreign demand as
\begin{align} \label{eq:benchmark_export_demand}
    X_t &= \left(p^{x}_t\right)^{-\eta_f} Y^{f}_t
\end{align}
where $p^{x}_t = P^{x}_t/P^{f}_t$.

Note that the export pricing assumption given by \ref{eq:benchmark_export_pricing}, accompanied with the demand function given by Equation \ref{eq:benchmark_export_demand} imply that producer currency pricing. Consequently, a nominal exchange rate depreciation leads to an immediate drop in the foreign currency price of exports goods and increase in exported quantities. We believe that this a suitable starting assumption as it is the benchmark case for most of the New Keynesian literature in small open economies. However, given the increasing evidence of pricing to market techniques and sticky dollar export prices.\footnote{See \cite{gopinath2010currency} and \cite{gopinath2020dominant} as examples of a literature which provides evidence of sticky dollar export prices.}

\noindent
\textbf{Government policy.} Under this framework, government policy is comprised of a fiscal authority, as described in Section \ref{sec:model_simple}, and a monetary authority which sets the nominal interest rate $R_{d,t}$. We assume that the monetary authority follows an interest rate rule specified as
\begin{align} \label{eq:taylor_rule_benchmark}
    \log \left(\frac{R_{d,t}}{R_d}\right) = \rho_R \log \left(\frac{R_{d,t-1}}{R_d}\right) + \left(1-\rho_R\right)\left[r_{\pi} \log \left(\frac{\pi^{c}_t}{\bar{\pi}^{c}} \right) \right]
\end{align}
where $\bar{R}_{d}$ is the steady state value of the interest rate, $\rho_R$ is an auto-regressive term on past values of the interest rate, $\pi^{c}_t / \bar{\pi}^{c}$ is the deviation of the gross consumption good inflation rate from the monetary authority's target, and $r_{\phi}$ represents the reaction coefficient of the policy rate to this deviation.\footnote{Note that we assumed that }

\noindent
\textbf{Households.} The representative household's problem remains relatively unchanged from that in Section \ref{sec:model_simple}, other than allowing for savings in both domestic and foreign currency deposits or assets. In order to do so, we introduce a term in the utility function which depends on their holdings of foreign currency deposits. The utility function is given by 
\begin{align*}
    &\mathbb{E}_0 \sum^{\infty}_{t=0} \beta^{t} \bigg\{\frac{\left[c_t - \theta \omega^{-1} L^{\omega}_t \right]^{1-\sigma}-1}{1-\sigma} + h_t \left(\frac{S_t D_t}{P^{c}_t} \right) \bigg\}
\end{align*}
where function $h_t \left(\frac{S_t D_t}{P^{c}_t} \right)$ is given by
\begin{align*}
    h_t \left(\frac{S_t D_t}{P^{c}_t} \right) = -\frac{1}{2} \gamma \left(\frac{S_t D^{*}_t}{P^{c}_t} - \Upsilon^{*}\right)^2
\end{align*}
where $\Upsilon^{*}$ is the household's target of foreign currency deposits. Overall, the function $h_t$ introduces a non-pecuniary cost in terms of the deviations of a household's foreign currency deposits from their target. This leads to an endogenous violation of the uncovered interest rate parity which secures stationary dynamics and a unique non-stochastic steady state.

\noindent
\textbf{Market clearing conditions.} Lastly, we describe the good and financial market clearing conditions. First, the supply of domestic currency or peso loans comes from households, $D_t$. The demand for pesos comes from intermediate good firms, $B^{\text{peso}}_t$. We assume that foreigners do not participate in the local currency market, so that market clearing implies:
\begin{align*}
    D_t = B_t
\end{align*}
The supply of dollars comes from households $D^{*}_t$ and foreigners $F^{o}_t$. The demand for dollar financing comes from the domestic intermediate good firms, $B^{\text{dollar}}_t$. Market clearing implies
\begin{align*}
    D^{*}_t + F^{o}_t = B^{\text{dollar}}_t
\end{align*}
Recall that this are \textit{dollar} terms. 

The balance of payments imposes the condition that, expressed in dollars, the trade surplus equals the net accumulation of foreign assets:
\begin{align*}
    \frac{P_t}{S_t} X_t - P^{f}_t \left(I_{m,t} + C_{m,t}  \right) = D^{*}_t - R^{*}_{t-1} D^{*}_{t-1} - \left(B^{\text{dollar}}_t - B^{\text{dollar}}_{t-1} R^{*}_{t-1}  \right)= - \left( F^{o}_t - R^{*}_{t-1} F^{o}_{t-1}  \right)
\end{align*}
The first equality is given by the balance of payments and the second equality is given by the dollar market clearing.

The demand for exports, expressed in dollars, is:
\begin{align*}
    \frac{P_t X_t}{S_t} &= \frac{P_t}{S_t} \left(\frac{P^{x}_t}{P^{f}_t}\right)^{-\eta_f} Y^{f}_t  \\
    &= \frac{P_t}{S_t} \left(\frac{S_t P^{f}_t}{ S_t P^{x}_t}\right)^{\eta_f} Y^{f}_t \\
    &= \left(p^{c}_t q_t \right)^{\eta_f} Y^{f}_t
\end{align*}
where $q_t$ is this economy's real exchange rate. We can express the balance of payments in terms of the domestic homogeneous good as
\begin{align*}
    \left(p^{c}_t q_t \right)^{\eta_f} Y^{f}_t - p^{m}_t \left[ \left(1-\gamma_I\right) \left(\frac{p_{i,t}}{p^{m}_t} \right)^{\nu_I} i_t + \omega_c \left(\frac{p^{c}_t}{p_{m,t}} \right)^{\eta_c} c_t + m_t \right] &= - \frac{S_t \left(F^{o}_t-R^{*}_{t-1}F^{o}_{t-1}\right)}{P_t}
\end{align*}
Market clearing in the market for the domestic homogeneous good is given by 
\begin{align*}
    y_t &= \gamma_I \left(p^{i}_t\right)^{\nu_I} i_t + \left(1-\omega_c\right) \left(p^{c}_t\right)^{\eta_c} c_t + x_t + \Psi \left(D^{E}_t\right) + \Upsilon^{P}_t \left(p_{i,t-1},p_{i,t},y_t\right)
\end{align*}
which includes the demand of final goods and the dividend and price adjustment costs. It is useful to define GDP in this economy as the sum of components of aggregate demand net of imports
\begin{align*}
    GDP_t = \left(p_{t}^{x}\right)^{\eta_{f}}y_{t}^{f}+p_{t}^{c}c_{t}+p_{t}^{I}i_{t}-p_{t}^{m}\left(\left(1-\gamma_{I}\right)\left(\frac{p_{I,t}}{p_{t}^{m}}\right)^{\nu_{I}}i_{t}+\omega_{c}\left(\frac{p_{t}^{c}}{p_{t}^{m}}\right)^{\eta_{c}}c_{t}+m_t\right)
\end{align*}

\noindent
\textbf{Calibration.} Table \ref{tab:nk_model_calibrated_parameters} summarizes the parametrization of the additional parameters introduced in our augmented model with nominal frictions. To keep our analysis in line with that presented in Section \ref{sec:model_simple} we keep the underlying parametrization and calibration strategy the same as in our real model. Thus, Table \ref{tab:nk_model_calibrated_parameters} highlights the value of parameters that govern the production of final goods, additional parameters on the novel features of the intermediate good producing firms and parameters that govern the monetary policy interest rate rule.

The top panel of Table \ref{tab:nk_model_calibrated_parameters} provides the value of parameters that govern the production function of the final goods in the economy. Overall, we assume values which are standard in the literature, such as \cite{gertler2007external} and \cite{christiano2011introducing}. We assume that consumption goods exhibit home bias while investment goods exhibit foreign bias. 
\begin{table}[ht]
    \centering
    \caption{Parametrization: Nominal Frictions Model}
    \label{tab:nk_model_calibrated_parameters}
    \small
    \begin{tabular}{c l l} \hline \hline
    \textbf{Parameter} & \textbf{Value} & \textbf{Details / target}  \\ \hline
    $\eta_c$  & Cons. Good CES Parameter & 1.50 - \footnotesize \cite{gertler2007external,camara2021FXI} \\
    $1-\omega_c$  & Cons. Good Home Bias & 0.70 - \footnotesize \cite{christiano2011introducing,camara2021FXI} \\
    $\nu_I$  & Inv. Good Home CES Parameter & 1.50 - \footnotesize \cite{christiano2011introducing,camara2021FXI} \\
    $\gamma_I$  & Inv. Good Home Bias & 0.25 - \footnotesize \cite{christiano2011introducing,camara2022granular} \\
    $\eta_f$  & Export Elasticity to ToT & 0.25 - \footnotesize - In line with micro data \\
    \\
    $\eta$  & Intermediate Good Mark Ups & 1.13 - \footnotesize \cite{drechsel2019earnings} \\
    $\phi_p$  & Price Adjustment Cost & 77 - \footnotesize \cite{drechsel2019earnings}  \\
    $1-\phi_b$  & Share of FC Debt & 0.25 - \footnotesize \cite{christiano2021financial} \& Match Data \\
    $\rho_B$  & Debt Smoothing Parameter & 0.75 - \footnotesize \cite{greenwald2019firm}  \\
    \\
    $\rho_R$  & Taylor Rule AR Parameter & 0.75 - \footnotesize \cite{de2022monetary} \\
    $r_{\pi}$  & Taylor Rule Inflation Parameter & 1.50 - \footnotesize \cite{gertler2007external,de2022monetary} \\
    $\bar{\pi}^{c}$  & Taylor Rule Inflation Target & 1 - \footnotesize Benchmark assumption \\
     \hline \hline
    \end{tabular}
    \floatfoot{\textbf{Note:} This table builds on the parametrization of the model introduced in Section \ref{sec:model_simple} in Table \ref{tab:real_model_standard_parameters}. Parameterized values presented in Table \ref{tab:real_model_standard_parameters} are kept the same. For instance, parameters over firms' production functions, working capital requirements remain constant. We calibrate parameters $\theta^{k}$ and $\theta^{IC}$ to match a 35\% private sector credit to GDP. We also calibrate the dis-utility of labor to match household hours of work at 0.20. Lastly, we calibrate foreign demand shifter $Y^f$ to match a ratio of exports to GDP in steady state of 25\%. }
\end{table}
We assume that a low exchange rate elasticity of quantities exported, at $\eta_f = 0.25$. This result is in line with both aggregate and firm level empirical evidence, see \cite{ahmed2009chinese}, \cite{alessandria2013export}, \cite{ahmed2015depreciations}, \cite{alessandria2015export}, \cite{fontagne2018international} and \cite{kohn2020financial}. This low exchange rate elasticity can be taken as a reduced form way to achieve the same results as a relatively novel literature which introduces sticky foreign currency pricing, such as \cite{camara2021spillovers}, or other sluggishness in export dynamics, such as \cite{gertler2007external,garcia2015dealing}.

The middle panel describes the parametrization of the novel features of intermediate good producing firms. We choose a value of $\eta = 1.13$, which governs firms' mark-ups in steady state. There is little empirical evidence which studies mark-ups in small open economies, so we take this value from \cite{drechsel2019earnings}, a standard value in the literature for US firms. Similarly, we assume a price adjustment cost parameter $\phi_p$ equal to 77, which is in line with the parametrization chosen by \cite{drechsel2019earnings}.\footnote{Note that Argentina has exhibited inflation rates well above the two digit threshold for at least 15 of the last 30 years. As argued above, the motivation of the calibration is to match aggregate moments of an average or typical Emerging Market economy. Furthermore, given that our benchmark results in Section \ref{sec:model_simple} are constructed for a real economy with no nominal frictions, we can conclude that price rigidities are not the key driver of our results.} This parametrization implies a price duration between 2 and 3 quarters. Lastly, we parametrize the share of foreign currency debt equal to $0.25$. First, this value is in line with the median share of credit and deposit dollarization in Emerging Markets presented by \cite{christiano2021financial}. Second, this result is in line with the average share of foreign currency debt in Argentina during the period 1994-2001. We choose a debt smoothing parameter of 0.75 which is line with the debt limit reflecting a weighted average of last four quarters. This parameter value is in line with the parametrization of \cite{greenwald2019firm} for US firms. In our counterfactual analysis in Section \ref{subsec:nominal_frictions_policy_IRF} we argue that this assumption provides a better quantitative fit and, in any case, stacks the cards against interest coverage constraints.

The bottom panel of Table \ref{tab:nk_model_calibrated_parameters} presents the parametrization of the monetary authority's interest rate rule. We assume that the interest rate rule is relatively sluggish with an AR coefficient of 0.75. This sluggishness of interest rates is in line with empirical evidence from Emerging Market's policy rates, see \cite{de2022monetary}. The interest rate rule's inflation coefficient is set to 1.5. This value and the interest rate sluggishness is in line with parameter values used and estimated in the literature, see \cite{gertler2007external}, \cite{christiano2011introducing} and \cite{camara2021FXI}. Lastly, we assume that the gross inflation target is equal to 1, i.e., zero net inflation rate in steady state. This assumption is not an accurate approximation of the Argentinean economy given the country's recurrent high inflation episodes. As stated before, the main goal of this parametrization exercise is to reflect an average Emerging Market economy and not the Argentinean economy. Furthermore, our results from Section \ref{sec:model_simple} are obtained without any price friction. Thus, we are confident that the results presented in the following sections are not driven by our assumptions over the value of the inflation target, $\bar{\pi}^{c}$ or the degree of nominal price frictions, $\phi_p$.

\subsection{Counterfactual Analysis} \label{subsec:nominal_frictions_policy_IRF}

We study the implications of cash flow-based borrowing constraints under the light of a foreign interest rate shock, $R^{*}_t$, in the presence of nominal frictions and domestic and foreign currency debt. We argue that the benchmark results presented in our real model in Section \ref{sec:model_simple} are also present in a more quantitative model. The presence of interest coverage or interest sensitive borrowing constraints leads to a greater amplification of a foreign interest rate shock than a collateral-based constraint. This result emerges even if the share of foreign currency debt is relatively low.

\noindent
\textbf{Benchmark results.} We assume an exogenous 100 annual basis point increase in the foreign interest rate shock, $R^{*}_t$, with an auto-regressive parameter $\rho_{R^{*}}$ equal to 0.75 as in Section \ref{sec:model_simple}. 

\begin{figure}[hbtp]
    \centering
    \caption{Collateral \& Interest Coverage Borrowing Constraints \\ \footnotesize $R^{*}$ Shock - Nominal Frictions }
\includegraphics[width=16cm,height=12cm]{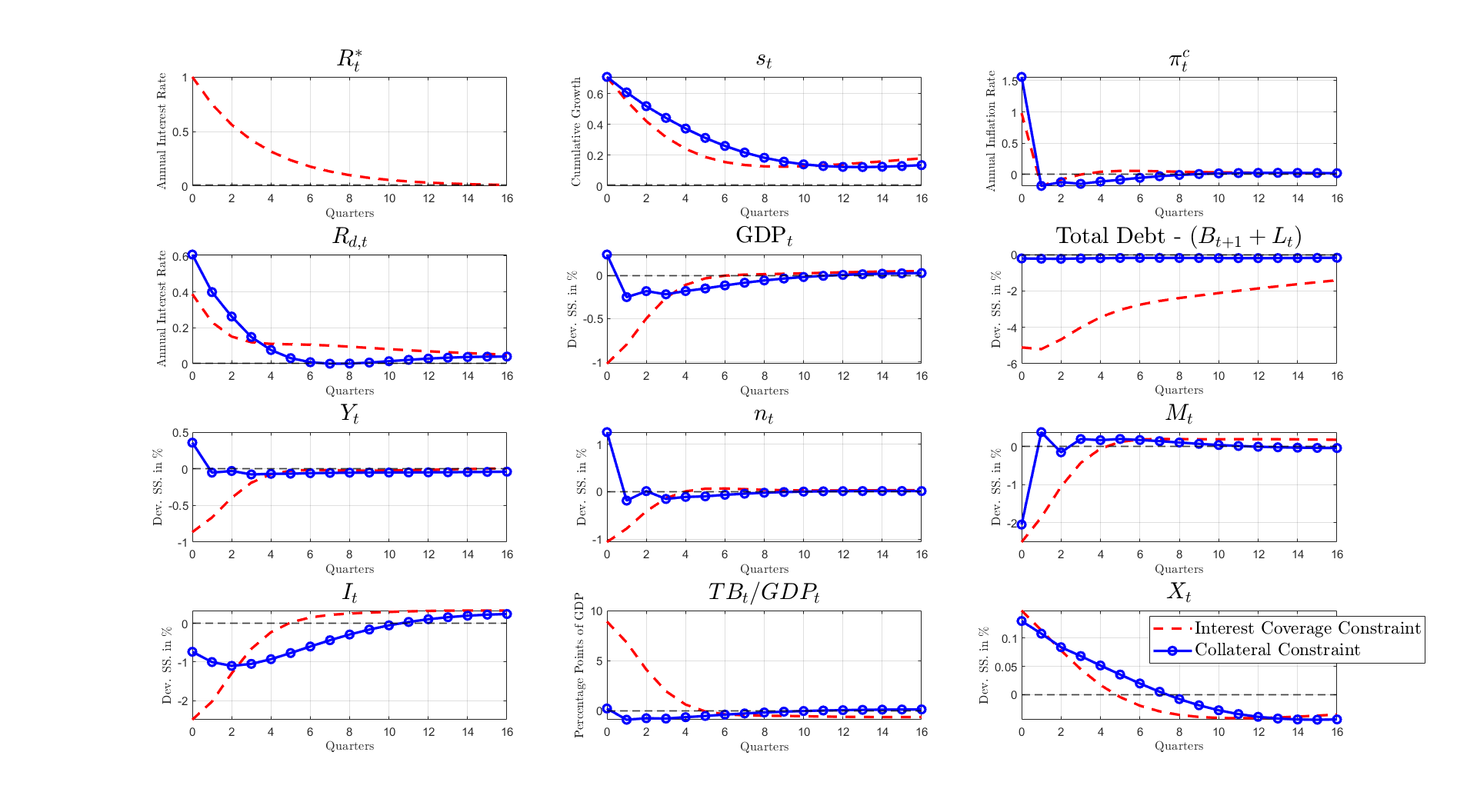}
    \label{fig:Benchmark_Graph_IC_Collateral}
    \floatfoot{\textbf{Note:} The figure is comprised of 12 panels ordered in four rows and three columns. In the text we refer to each panel following a $(x,y)$ convention, where $x$ and $y$ represent the row and column. Each panel presents the dynamics of an economy with a collateral borrowing constraint (blue dotted line), and an economy with a interest sensitive cash flow borrowing constraint (red dashed line).}
\end{figure}

Figure \ref{fig:Benchmark_Graph_IC_Collateral} presents the impulse response dynamics of an economy with a standard collateral borrowing constraint (blue dotted line) and of an economy with an interest coverage borrowing constraint. 
The first conclusion from this figure is that an economy with an interest coverage constraint leads to significantly greater amplification on real variables in response to an increase in the foreign interest rate than an economy with a collateral borrowing constraint. In order to understand this greater amplification we explain the transmission mechanism of a foreign interest rate shock in an economy with nominal frictions.

An increase in the foreign interest rate, $R^{*}_t$, affects the small open economy through the uncovered interest rate parity condition, putting pressure on the nominal exchange rate. Panel $(1,2)$ shows that the nominal exchange rate depreciates on impact close to 0.7, roughly the same across the two economies. This depreciation of the nominal exchange rate impacts the consumer price inflation rate, see Panel $(1,3)$. However, the degree of exchange rate pass-through into prices is almost half as large for the economy with an interest coverage borrowing constraint. The monetary authority reacts to the increase in the inflation rate leading to an increase in the policy rate, see Panel $(2,1)$. The fact that the interest coverage borrowing constraint economy exhibits a lower exchange rate pass-through implies that the increase in the policy rate is lower compared to an economy with a collateral borrowing constraint (40 vs 60 basis point). Note that the increase in the domestic rate is less than half of the increase in the foreign interest rate. \textit{A priori}, this decoupling of domestic and foreign rates weakens the transmission channel of an interest coverage borrowing constraint. Panel $(2,2)$ shows that the impact on GDP is significantly greater under an interest coverage constraint than under a collateral constraint. While an economy with a  collateral borrowing constraint predicts an increase in GDP on impact close to 0.25\%, an economy with an interest coverage borrowing constraints exhibits a drop of 1\%. The drop in GDP in an economy with an interest coverage borrowing constraint is significantly larger than under a collateral constraint for the first 4 quarters to later converge back to steady state. 

Once again, the larger impact on real variables under an interest coverage borrowing constraint is driven by the greater impact in firms' access to debt under an interest coverage compared to a collateral borrowing constraint, as shown in Panel $(2,3)$. The tighter borrowing limit reduces firms' ability to finance working capital hindering intermediate good firms' production, as shown in Panel $(3,1)$. This drop in production is around 8 times larger under an interest coverage compared to a collateral borrowing constraint; the greater drop in the simple model is around 40 times. The reduction in working capital requirements leads to a grater and sharper reduction in input purchases, see Panels $(3,2)$ and $(3,3)$. Similar to the results in Section \ref{sec:model_simple}, the impact on investment is larger under an interest coverage constraint, but the recovery is faster under a collateral constraint. 

\begin{figure}[ht]
    \centering
    \caption{Collateral \& Interest Coverage Borrowing Constraints \\ \footnotesize $R^{*}$ Shock - Nom. Frictions - \textit{Cont.}}
\includegraphics[width=12cm,height=4cm]{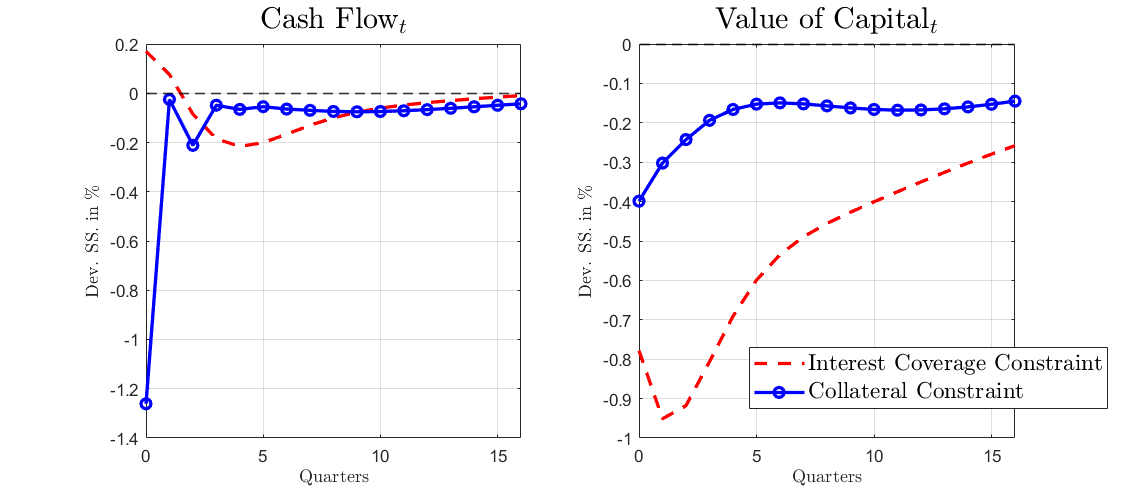}
    \label{fig:Benchmark_Graph_IC_Collateral_CashFlows}
    \floatfoot{\textbf{Note:} The figure is comprised of 2 panels ordered in one row and two columns. In the text we refer to each panel following a $(x,y)$ convention, where $x$ and $y$ represent the row and column. Each panel presents the dynamics of an economy with a collateral borrowing constraint (blue dotted line), and an economy with a interest sensitive cash flow borrowing constraint (red dashed line).}
\end{figure}

Next, we show evidence that the interest sensitivity of the interest coverage borrowing constraint is the main driver of the greater amplification. Figure \ref{fig:Benchmark_Graph_IC_Collateral_CashFlows} presents the impulse response functions of firms' cash flows and value of capital after the foreign interest rate shock.
Panel $(1,1)$ shows that firms' cash flows increase on impact under an interest coverage constraint, while cash flows decrease under a collateral borrowing constraint. However, unlike the results of the model in Section \ref{sec:model_simple}, cash flows fall significantly after the initial two periods. While the collateral borrowing constraint exhibits a larger drop in cash flows on impact, the recovery is faster. Lastly, the drop in the value of capital is significantly larger in the economy with an interest coverage borrowing constraint.

In summary, an interest coverage borrowing constraint leads to a significantly larger amplification of a foreign interest rate shock than a standard collateral borrowing constraint, even in the presence of nominal frictions. Furthermore, this greater amplification is present even when the share of foreign currency debt is only 25\%. This relatively low share of foreign currency debt leads to a decoupling of domestic and international rates. Regardless, once again, an interest coverage borrowing constraint leads to a greater drop in firms' debt limit, which leads to a reduction in firms' input purchases and overall production orders of magnitude greater than in an economy with a  collateral borrowing constraint. 

\subsection{Borrowing Constraints \& Data Moments} \label{subsec:nominal_frictions_matching_aggregate}

In Sections \ref{sec:cash_flow_lending} and \ref{sec:interest_sensitive_borrowing_constraints} we presented the novel stylized fact that at the aggregate and firm level interest payments are counter-cyclical in Emerging Market economies. Additionally, we showed that for a panel of Emerging Market firms interest payments are strongly correlated with cash flows. In this section, we show that interest sensitive borrowing constraints are key to match this novel stylized fact. Furthermore, we show that interest sensitive borrowing constraints are able to match other aggregate moments in the data, such as counter-cyclical interest rates. 

\begin{figure}[htbp]
    \centering
    \caption{Model implied Correlations \\ \footnotesize Interest Coverage Borrowing Constraint}
    \label{fig:correlations_IC}
     \centering
     \begin{subfigure}[b]{0.495\textwidth}
         \centering
         \includegraphics[height=5cm,width=8cm]{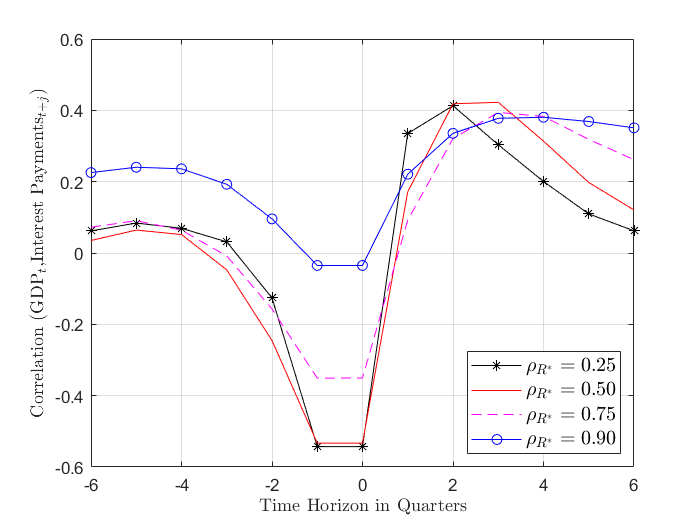}
         \caption{GDP \& Interest Payments}
         \label{fig:IC_Corr_GDP_IP}
     \end{subfigure}
     \hfill
     \begin{subfigure}[b]{0.495\textwidth}
         \centering
         \includegraphics[height=5cm,width=8cm]{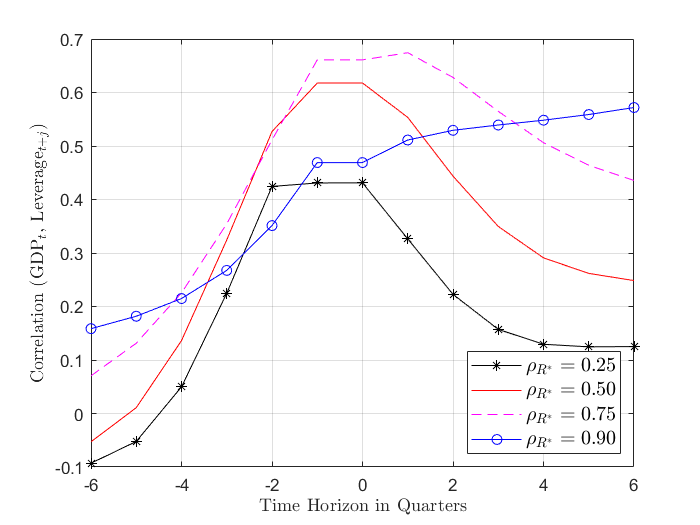}
         \caption{GDP \& Leverage}
         \label{fig:IC_Corr_GDP_Leverage}
     \end{subfigure} 
     \hfill \\
     \begin{subfigure}[b]{0.495\textwidth}
         \centering
         \includegraphics[height=5cm,width=8cm]{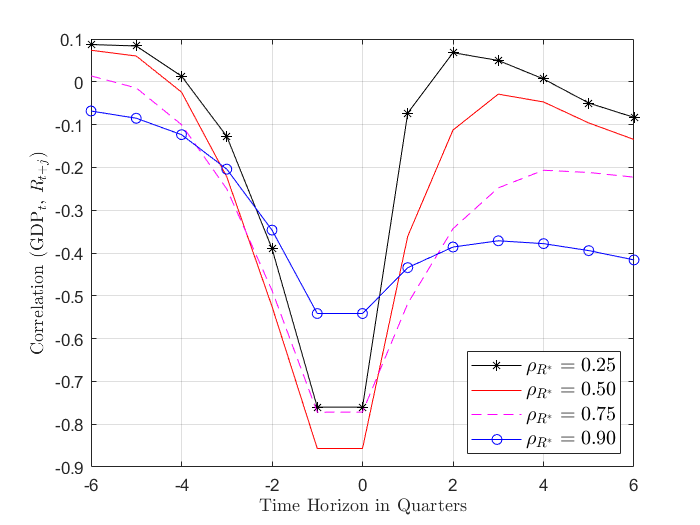}
         \caption{GDP \& Interest Rate}
         \label{fig:IC_Corr_GDP_R}
     \end{subfigure}
     \hfill
     \begin{subfigure}[b]{0.495\textwidth}
         \centering
         \includegraphics[height=5cm,width=8cm]{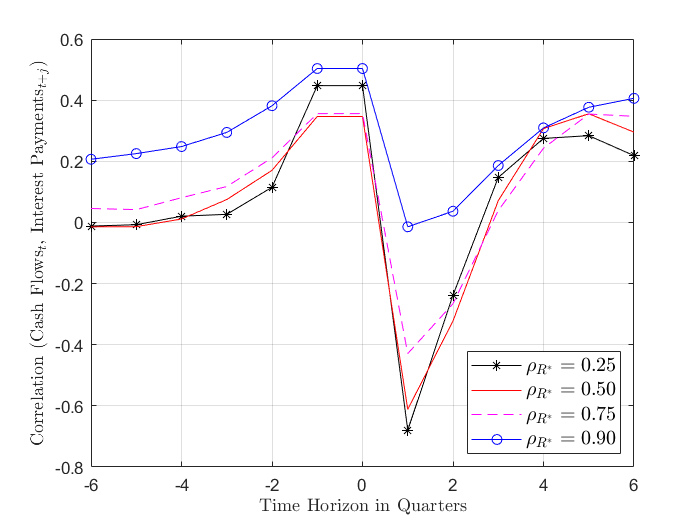}
         \caption{Cash Flows \& Interest Payments}
         \label{fig:IC_Corr_EBITDA_IP}
     \end{subfigure}
\end{figure}

\begin{figure}[htbp]
    \centering
    \caption{Model implied Correlations \\ \footnotesize Collateral Borrowing Constraint}
    \label{fig:correlations_AB}
     \centering
     \begin{subfigure}[b]{0.495\textwidth}
         \centering
         \includegraphics[height=5cm,width=8cm]{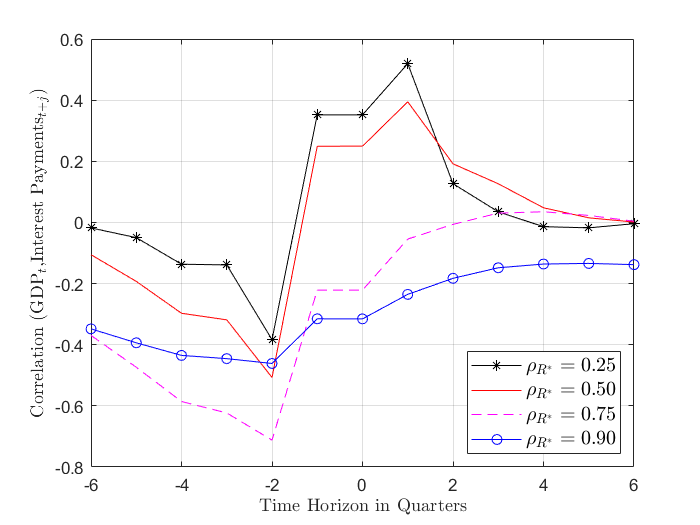}
         \caption{GDP \& Interest Payments}
         \label{fig:AB_Corr_GDP_IP}
     \end{subfigure}
     \hfill
     \begin{subfigure}[b]{0.495\textwidth}
         \centering
         \includegraphics[height=5cm,width=8cm]{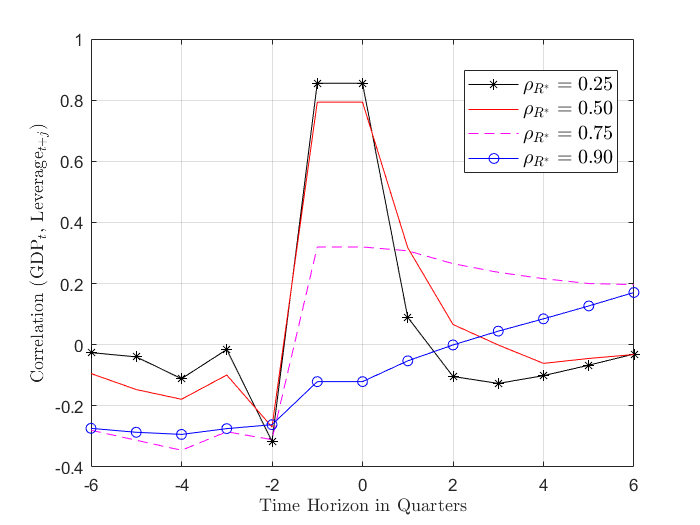}
         \caption{GDP \& Leverage}
         \label{fig:AB_Corr_GDP_Leverage}
     \end{subfigure} 
     \hfill \\
     \begin{subfigure}[b]{0.495\textwidth}
         \centering
         \includegraphics[height=5cm,width=8cm]{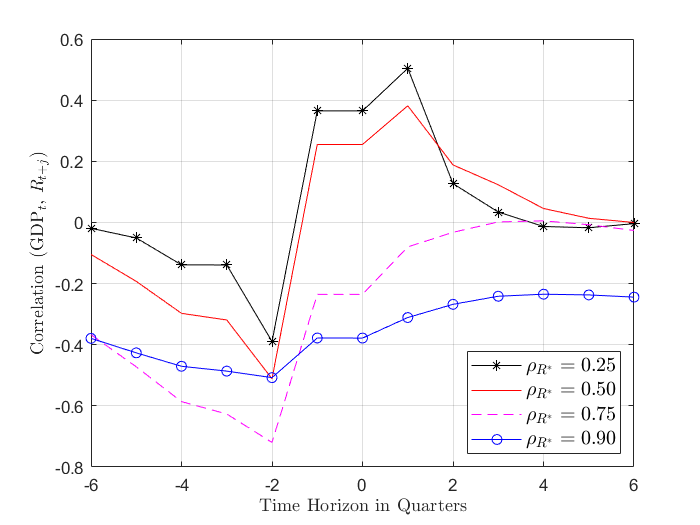}
         \caption{GDP \& Interest Rate}
         \label{fig:AB_Corr_GDP_R}
     \end{subfigure}
     \hfill
     \begin{subfigure}[b]{0.495\textwidth}
         \centering
         \includegraphics[height=5cm,width=8cm]{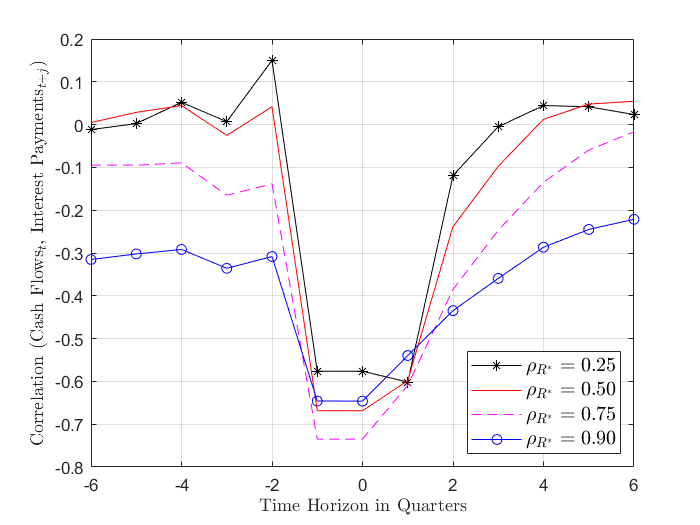}
         \caption{Cash Flows \& Interest Payments}
         \label{fig:AB_Corr_EBITDA_IP}
     \end{subfigure}
\end{figure}

\noindent
\textbf{Aggregate moments using benchmark calibration.} We study the model implied correlations between different aggregate and firm level moments. In order to do this, we simulate the model for 10,000 periods subject only to a foreign interest rate shock $R^{*}_t$ with different degrees of persistence, $\rho_{R^{*}}$. Figures \ref{fig:correlations_IC} and \ref{fig:correlations_AB} present the results under an interest coverage and collateral borrowing constraint respectively.


First, Figure \ref{fig:correlations_IC} describes the correlations for an economy subject to an interest coverage constraint. On the top left panel, Figure \ref{fig:IC_Corr_GDP_IP} shows that under an interest coverage borrowing constraint the correlation between GDP and firms' interest payments exhibits a U-shape. On impact, the correlation between GDP and firms' interest payment is negative and quantitatively large. However, this correlation weakens when considering time periods away from the contemporaneous relationship. On the top right panel, Figure \ref{fig:IC_Corr_GDP_Leverage} presents a strong and positive correlation between GDP and firms' leverage. This is driven by a foreign interest rate shock causing both GDP and firms' leverage to fall simultaneously. This pro-cyclical relationship between GDP and leverage is at odds with the empirical evidence for Emerging Markets, as presented by \cite{fernandez2015interest}. This result is driven by the nature of the foreign interest rate shock and by the fact that borrowing constraints are binding in this model. On the bottom left panel, Figure \ref{fig:IC_Corr_GDP_R} shows that the interest coverage borrowing constraint predicts a U-shaped and negative correlation between GDP and interest rates. This is line with the empirical findings presented in Figure \ref{fig:Growth_Rates_Expanded} in Section \ref{sec:cash_flow_lending} and with the evidence presented by \cite{neumeyer2005business}. Lastly, on the right bottom panel, Figure \ref{fig:IC_Corr_EBITDA_IP} shows the correlations between firms' cash flows and their interest payments. There is a positive correlation between these variables across most time horizons. The correlation weakens and even turns negative considering the relationship two periods ahead, but turns positive in the following periods. Overall, an economy with an interest coverage constraint is able to match the novel stylized fact presented in Section \ref{sec:interest_sensitive_borrowing_constraints}.\footnote{While this result is engineered by the interest coverage borrowing constraint $r_t \times b_t \leq \Phi \left(L\right) \pi_t $, note that this result is also present when introducing the debt smoothing technology in Equation \ref{eq:debt_smoothing}.}

Figure \ref{fig:correlations_AB} presents the results for an economy under a collateral borrowing constraint. On the top left panel, Figure \ref{fig:AB_Corr_GDP_IP} shows a mixed relationship between GDP and firms' interest payments. For relatively low levels of persistence $\rho_{R^{*}}$, there is a strong and positive correlation between GDP and interest payments. For the case of $\rho_{R^{*}} = 0.90$, the correlation becomes weakly negative and completely flat across time horizons. This is at odds with the strong counter-cyclicality found in Figure \ref{fig:Growth_Rates_Expanded} in Section \ref{sec:interest_sensitive_borrowing_constraints}. On the right panel, Figure \ref{fig:AB_Corr_GDP_Leverage} shows a pro cyclical relationship between GDP and leverage, alike the results presented under an interest coverage borrowing constraint. On the bottom left panel, Figure \ref{fig:AB_Corr_GDP_R} shows that under a collateral borrowing constraint, there is a negative correlation between GDP and interest rates, with interest rates leading the economic downturn. However, this correlation is significantly lower than the one implied by an economy under an interest coverage constraint. Lastly, on the bottom right panel, Figure \ref{fig:AB_Corr_EBITDA_IP} shows a strong and negative correlation between a firms' cash flows and their interest payments. The collateral borrowing constraint predicts a U-shaped relationship between these variables, centered around the contemporaneous horizon. Once again, this result is against the empirical evidence presented in Figure \ref{fig:Growth_Rates_Expanded} in Section \ref{sec:interest_sensitive_borrowing_constraints}.

In summary, we simulated two economies, one subject to an interest coverage constraint and another one subject to a collateral constraint, using our benchmark calibration. On the one hand, we showed that an interest coverage constraint is able to match the counter cyclicality of interest rates and interest payments. Furthermore, an economy with an interest coverage borrowing constraint is able to match the strong and positive correlation between firms' cash flows and interest payments, the novel stylized fact presented in Section \ref{sec:interest_sensitive_borrowing_constraints}. On the other hand, we showed that under a collateral borrowing constraint the model is not able to match the counter cyclicality of interest payments or the strong correlation between firms' cash flows and interest payments. Additionally, the correlation between GDP and the interest rate is significantly weakened compared to that implied by an interest coverage borrowing constraint. We take these results as further evidence that interest coverage borrowing constraints are prevalent in Emerging Market economies. 

\noindent
\textbf{Introduction of additional economic shocks.} 

\subsection{Policy Implications: Exchange Rate Regimes} \label{subsec:nominal_frictions_policy_implications}

In this section, we study the policy implications of interest sensitive borrowing constraints. We show that monetary policy regimes which seek to limit the volatility of nominal exchange rates are significantly more costly in economies with interest sensitive borrowing constraints than in economies with the benchmark collateral borrowing constraints. A monetary authority which reacts to a foreign interest rate hike with an increase in the domestic monetary policy rate amplifies the tightening of interest sensitive borrowing constraints, a direct channel not present in an economy with collateral-based constraints. This result suggests that nominal exchange rate pegs may be even more costly than what a previous literature has argued.

To consider different exchange rate regimes we augment the Taylor rule presented in Equation \ref{eq:taylor_rule_benchmark} in Section \ref{subsec:nominal_frictions_policy_description}, with the following specification
\begin{align}
    \log \left(\frac{R_{d,t}}{\bar{R}_d}\right) = \rho_R \log \left(\frac{R_{d,t-1}}{\bar{R}_d}\right) + \left(1-\rho_R\right) \left[\phi_{\pi} \log \left(\frac{\pi_t}{\bar{\pi}}\right) + \phi_s \log \left(\frac{S_t}{\bar{S}}\right) \right]
\end{align}
where the additional terms inside the square brackets, $\phi_s \log \left(\frac{S_t}{\bar{S}}\right)$ is comprised of the nominal exchange rate deviations from its steady state value, $\log \left(\frac{S_t}{\bar{S}}\right)$ and parameter $\phi_s \geq 0$. The literature usually denotes parameter $\phi_s$ as representing the monetary authority's degree of ``fear of floating'', see \cite{cugat2019emerging,camara2022tank}. On the one hand, when $\phi_s = 0$ the central bank allows the nominal exchange rate to float freely. On the other hand, with $\phi_s >0$, the monetary authority reacts to a depreciation of the nominal exchange rate with an interest rate hike. In the extreme case of $\phi_s \rightarrow \infty$ the monetary authority keeps the nominal exchange rate fixed.

Nominal exchange rate regimes have been the subject of economic research for more than a century, see \cite{della2003new,della2007straining}. For instance, \cite{schmitt2016downward} shows that, in a representative agent economy, an exchange rate regime which stabilizes the  non-tradable inflation can reproduce the flexible-price allocation, which is Pareto optimal. This regime corresponds to a floating exchange rate that depreciates when the relative price of non-tradable goods goes down in the flexible-price allocation. In a setting with heterogeneous households, \cite{cugat2019emerging}, \cite{auclert2021exchange} and \cite{camara2022tank} show that agents may exhibit preference over less flexible exchange rate regimes as it stabilizes the price of tradable goods. 

There is a significant empirical literature that argues that monetary authorities in both AE and EM economies exhibit certain degree of ``fear of floating''. For instance, \cite{calvo2002fear}, who coined the term, find that even if countries announce to be following a floating exchange rate regime, they limit the fluctuations in the nominal exchange rate. \cite{lubik2007central} test whether central banks in Advanced Economies react to the nominal exchange rate. The authors find that Canada and England do, but Australia and New Zealand do not.

\noindent
\textbf{Impulse response analysis.} We study the aggregate implications of the fear of floating behavior through the lens of an impulse response analysis after an exogenous 100 basis point increase in the foreign interest rate $R^{*}_t$. We study two economies, one with an interest coverage borrowing constraint and one with a collateral borrowing constraint. Within each economy we consider three cases: (i) the benchmark case presented in Section \ref{subsec:nominal_frictions_policy_IRF}, (ii) a case with a middle value of $r_S$ which is parametrized to match a ``dirty floating'' exchange rate regime, (iii) a case with a high value of $r_S$ which is parametrized to match an ``exchange rate peg''. 
\begin{figure}[htbp]
    \centering
    \caption{Implications of Fear of Floating across Borrowing Constraints}
    \label{fig:fear_of_floating}
     \centering
     \begin{subfigure}[b]{0.95\textwidth}
         \centering
         \includegraphics[width=\textwidth]{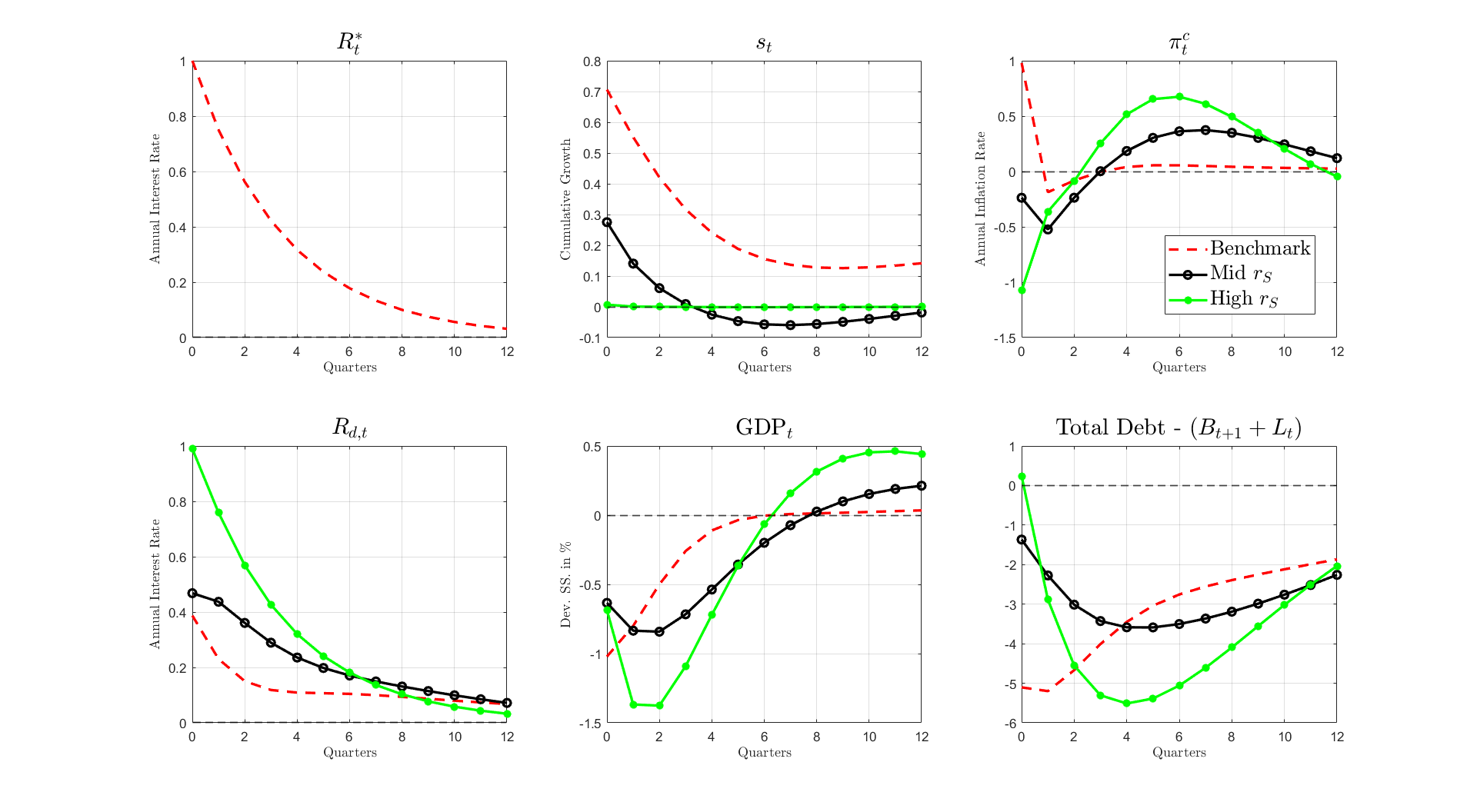}
         \caption{Interest Coverage Borrowing Constraint}
         \label{fig:Fear_of_Floating_IC}
     \end{subfigure}
     \hfill \\
     \begin{subfigure}[b]{0.95\textwidth}
         \centering
         \includegraphics[width=\textwidth]{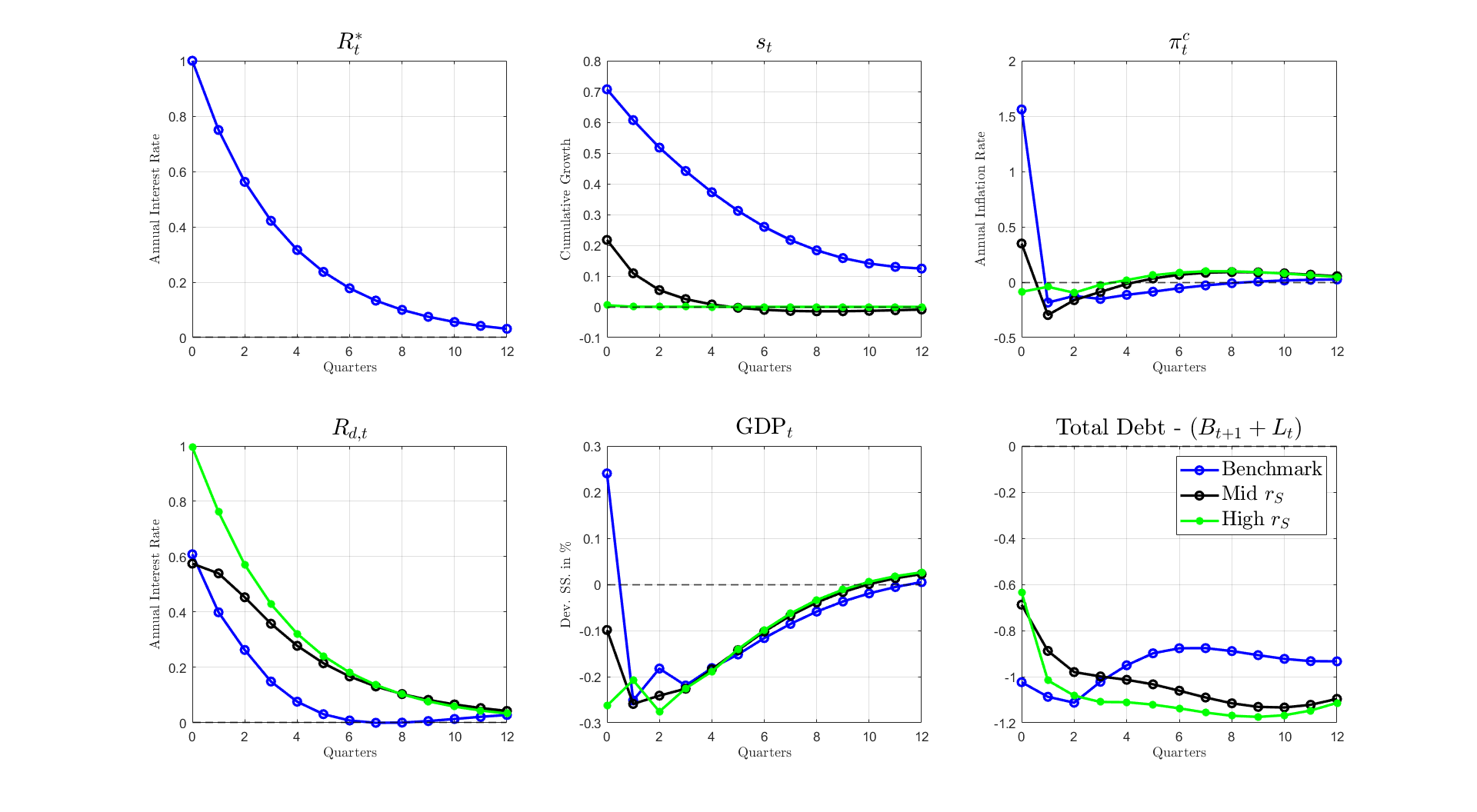}
         \caption{Collateral Borrowing Constraint}
         \label{fig:Fear_of_Floating_AB}
    \end{subfigure}
    \floatfoot{\textbf{Note:} The figure is comprised of 2 sub-figures. Each sub-figure is comprised of 6 panels ordered in two row and three columns. In the text we refer to each panel following a $(x,y)$ convention, where $x$ and $y$ represent the row and column. The ``mid $r_S$'' economy uses a parameter value equal to $r_S = 2$. The ``High $r_S$'' economy uses a parameter value equal to $r_S = 200$.}
\end{figure}
Figure \ref{fig:fear_of_floating} presents the dynamics for the two economies under the three exchange rate regimes.

The first result from Figure \ref{fig:fear_of_floating} is that monetary policy regimes which aim to reduce the volatility of the nominal exchange rate leads to a greater impact on real variables of a foreign interest rate shock. This a standard result in a canonical, representative agent New Keynesian open economy model, see \cite{gali2005monetary}. ``Dirty floating'' or ``exchange rate pegs'' exhibit lower nominal exchange rate depreciations and lower pass through to the domestic inflation rate as expected, see Panels $(1,2)$ and $(1,3)$ on Figures \ref{fig:Fear_of_Floating_IC} and \ref{fig:Fear_of_Floating_AB}. This lower exchange rate depreciation is explained by a greater response of the policy interest rate, see Panel $(2,1)$. The higher domestic interest rate reduces firms' financing cost and any increase in exports. 

The second, and novel result, from Figure \ref{fig:fear_of_floating} is that monetary policy regimes which aim to reduce the volatility of the nominal exchange rate are significantly more costly in terms of real activity in an economy with an interest coverage constraint than in an economy with a collateral constraint. To see this result we compare Panels $(2,3)$ across Figures \ref{fig:Fear_of_Floating_IC} and \ref{fig:Fear_of_Floating_AB}. On the one hand, Figure \ref{fig:Fear_of_Floating_AB} shows that, while the ``dirty floating'' and ``exchange rate peg'' regimes remove the increase on impact of GDP generated by exports reacting to the exchange rate depreciation, the dynamics of GDP are remarkably similar with the one under the benchmark monetary policy regime. On the other hand, while Figure \ref{fig:Fear_of_Floating_IC} shows that in an economy with an interest coverage constraint the drop in GDP is close on impact, the drop in GDP is more than two times larger under an ``exchange rate peg'' regime compared to the benchmark monetary policy regime. Even in the ``dirty floating'' case, the drop in GDP is close to 50\% larger two quarters after the initial shock. The greater impact of a ``dirty floating'' exchange rate regime under an interest coverage compared to a collateral constraint is causes by the interest sensitivity of the former constraint. On the one hand, Panel $(2,3)$ of Figure \ref{fig:Fear_of_Floating_IC} shows that an ``exchange rate peg'' leads to a significantly larger drop in firms' debt two quarters after the initial shock. On the other hand, Panel $(2,3)$ of Figure \ref{fig:Fear_of_Floating_AB} shows that the reduction in firms' debt is roughly in the same order of magnitude across all exchange rate regimes.

In summary, the negative effects of monetary policy regimes which seek to reduce the volatility of the nominal exchange rate are exacerbated in the presence of an interest coverage constraint. On the one hand, in an economy with a collateral borrowing constraint a ``dirty float'' exchange rate regime increases the drop in GDP in response to a foreign interest rate shock. However, this greater drop in GDP only occurs during the first two periods. On the other hand, in an economy with an interest coverage borrowing constraint a ``dirty float'' and an ``exchange rate peg'' increases the drop in GDP by 50\% to 100\%. In consequence, the central bank's cost of exhibiting a fear of floating behavior is significantly augmented in the presence of an interest coverage constraint compared to a collateral constraint.

\section{Borrowing Constraints \& the Spillover Puzzle} \label{sec:spillover_puzzle}

In this section we argue that differences in economies underlying borrowing constraints matter for the transmission of US monetary policy shocks. In Section \ref{subsec:spillover_puzzle_empirical}
we show that Emerging Market economies experience significantly greater negative spillovers of US monetary policy shocks than the US and other Advanced Economies, a phenomenon we denote the \textit{Spillover Puzzle}. We do this through both panel SVAR and Local projection techniques. In Section \ref{subsec:spillover_puzzle_quantitative} we calibrate the model presented in Section \ref{sec:model_simple} to match the share of collateral and cash flow constraints in Argentina, the US and other Advanced Economies, and show that the predicted greater amplification is in line with those predicted by empirical models. The greater prevalence of interest sensitive cash flow-based borrowing constraints and its associated greater amplification of foreign interest rate shocks, provide a straightforward rationale for the greater impact of US monetary policy shocks on Emerging Markets.

\subsection{Empirical Evidence on the Spillover Puzzle} \label{subsec:spillover_puzzle_empirical}

The quantitative larger impact of US monetary policy shocks in Emerging Markets, compared to Advanced economies is a well documented fact in international macroeconomics. For instance, \cite{reinhart2001growth} argue that the US' interest rate cycle can have amplified consequences in Emerging Market economies, highlighting the negative impact of the Federal Reserve's interest rate hike in the early 1980s on Latin America's debt crisis. It has been suggested that financial crises in Emerging Markets, such as \textit{Sudden Stop} and \textit{Sovereign Default} episodes, have been triggered by Federal Reserve interest rate hikes, see \cite{eichengreen2016managing}. The presence of quantitative larger negative spillovers of US monetary policy shocks in Emerging Markets than in the US economy was first suggested by \cite{mackowiak2007external}, and in subsequent papers such as \cite{degasperi2020global} and \cite{camara2021spillovers}).  Next, we use SVAR and local projection techniques to present robust evidence of the greater impact of US monetary policy shocks \textit{outside} the US economy than \textit{within} the US economy, and particularly in Emerging Markets compared to Advanced Economies.

\noindent
\textbf{SVAR evidence.} We estimate a panel SVAR for both Emerging Market (8) and Advanced Economies (7). The model specification is a pooled panel SVAR as presented by \cite{canova2013panel} and used in \cite{camara2021spillovers}.\footnote{This type of model considers the dynamics of several countries simultaneously, but assuming that the dynamic coefficients are homogeneous across units, and coefficients are time-invariant. In this framework, this implies that country $i$'s variables only depend on structural shocks  and the lagged values of country $i$'s variables}. Given that the model is relatively standard, we leave the description of the its details for Appendix \ref{subsec:appendix_details_empirical_SVAR_model}.

The sample of countries in our dataset is presented in Table \ref{tab:panel_description}.
\begin{table}[ht]
    \centering
    \begin{tabular}{c|c}
        Emerging Markets &  Advanced Economies \\ \hline \hline
        Argentina &  Australia \\
        Brazil & Canada \\
        Chile & Japan \\
        Indonesia & South Korea \\
        Mexico & United Kingdom \\
        Peru   & USA \\
        South Africa & \\
        Turkey \\ \hline \hline
    \end{tabular}
    \caption{Countries in Empirical Analysis Sample}
    \label{tab:panel_description}
\end{table}
On our benchmark assumption we include 5 macroeconomic and financial variables: (i) industrial production, (ii) nominal exchange rate with respect to the US dollar, (iii) consumer price index, (iv) domestic lending rates, (v) equity index. To this set of variables we add the identified US monetary policy shocks constructed by \cite{jarocinski2020deconstructing}. Identification of a US monetary policy shock is achieved by ordering the structural shock first in a Choleski sense.\footnote{This allows for the structural shock to impact all other variables in the model on impact.} The model is estimated at the monthly frequency. 

\begin{figure}[ht]
    \centering
    \caption{Spillover Puzzle of US Monetary Policy \\ \footnotesize Evidence from a Panel SVAR Model}
    \label{fig:SVAR_Benchmark}
    \includegraphics[width=18cm,height=12cm]{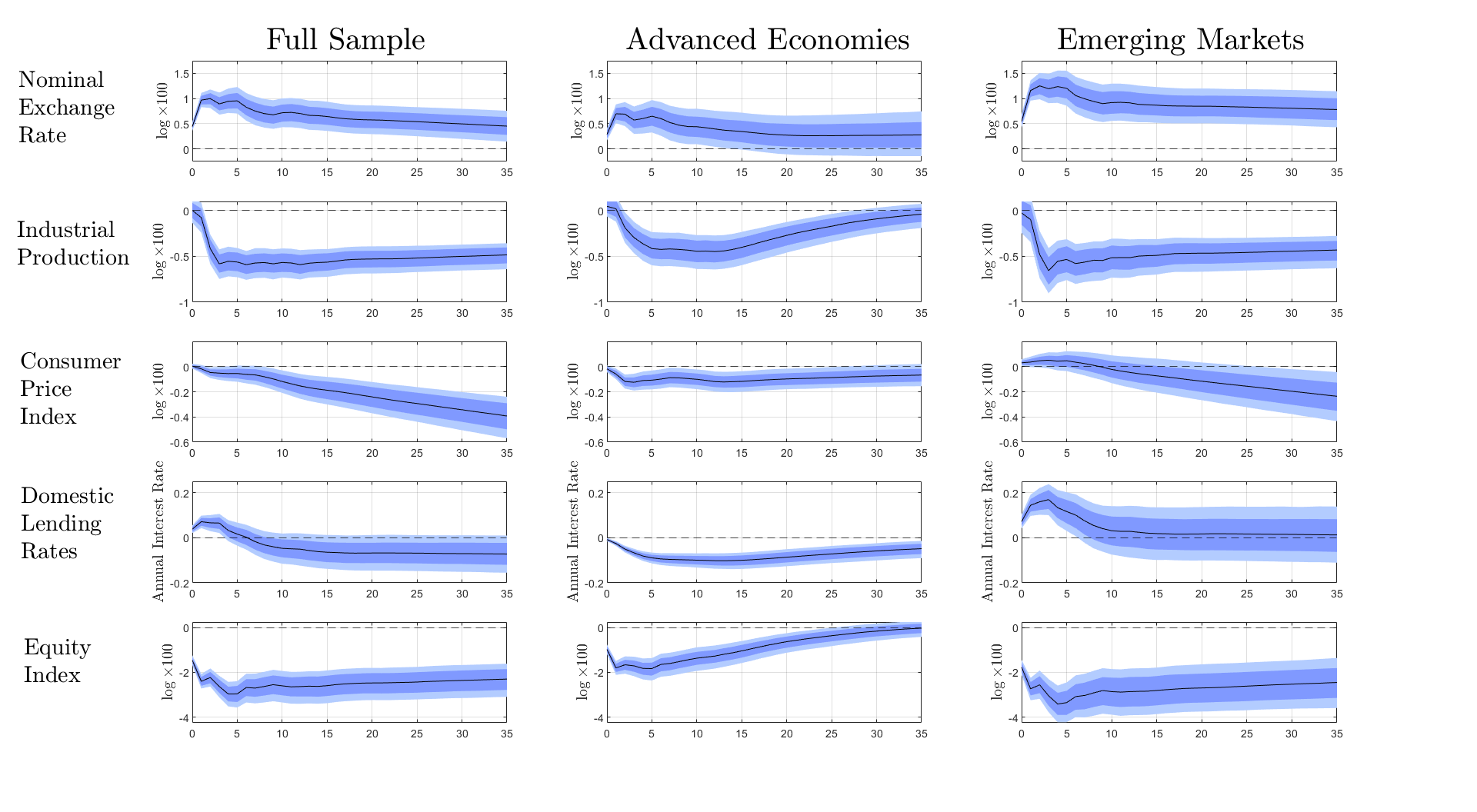}
    \floatfoot{\textbf{Note:} The black solid line represents the median impulse response function. The dark shaded area represents the 16 and 84 percentiles. The light shaded are represents the 5 and 95 percentiles. There are 15 panels with figures, organized in five rows and three columns. Rows represent variables: (i) nominal exchange rate, (ii) industrial production, (iii) consumer price index, (iv) domestic lending rate, (v) equity index. Columns represent the ``Full Sample'', ``Advanced Economies'' and ``Emerging Markets'', as defined by Table \ref{tab:panel_description}. For every variable, the scale vertical axis is set to allow for comparison across samples. In the text, when referring to Panel $(i,j)$, $i$ refers to the row and $j$ to the column of the figure.}
\end{figure}

Figure \ref{fig:SVAR_Benchmark} presents our first empirical result concerning the Spillover Puzzle of US monetary policy.
Each column presents the impulse response functions of a one standard deviation monetary policy shock for our ``Full Sample'', for a panel of ``Advanced Economies'' and for a panel of ``Emerging Markets''. The first result from Figure \ref{fig:SVAR_Benchmark} shows that US monetary policy shocks have negative and statistically and quantitatively significant spillovers. Panel $(1,1)$ shows that a one standard deviation shock leads to a depreciation of the nominal exchange rate on impact, which remains persistently above its pre-shock level, even three years after the initial shock. Panel $(2,1)$ shows that industrial production drops below its pre-shock level between two and three months after the initial shock, exhibiting only a mild recovery. Panel $(3,1)$ shows that the consumer price index does not react significantly during the first 10 months after the shock, to later show a persistent decline. Panel $(4,1)$ shows that domestic lending rates exhibits an increase by $3.8$ basis points. This is relatively in line with the increase of $5$ and $4.5$ basis point increase in the US treasury and excess bond premium respectively, found in the US found by \cite{jarocinski2020deconstructing}.\footnote{See Figure 2 in the paper.} Lastly, Panel $(5,1)$ shows that a US monetary policy leads to a persistent decline in the equity index, which remains between 2\% and 3\% below its pre-shock level.\footnote{Note that the domestic equity index is expressed in domestic currency. Consequently, this measure of the equity index is susceptible to exchange rate risk. However, note that the drop in the equity index, shown in the fifth row of Figure \ref{fig:SVAR_Benchmark}, is significantly greater than the exchange rate depreciation shown in the first row of Figure \ref{fig:SVAR_Benchmark}.}

The second result emerging from Figure \ref{fig:SVAR_Benchmark} is that the impact of a US monetary policy shock is significantly quantitatively larger for Emerging Markets compared to Advanced Economies, across all variables in our empirical specification. To see this we compare the presented in the middle and right columns. Panels $(1,2)$ and $(1,3)$ show that Emerging Markets exhibit a depreciation of the nominal exchange rate between 70\% and 100\% greater than Advanced Economies. Comparing Panels $(2,2)$ and $(2,3)$ show that industrial production exhibit a sharper and more persistent drop in industrial production than Advanced Economies. In terms of the consumer price level, while Advanced Economies exhibit a persistent drop, shown in Panel $(3,2)$, Emerging Markets exhibit a significant increase during the first 6 months after the initial shock, to later exhibit a persistent decline.\footnote{Emerging Markets exhibit a greater dependence of imported consumer and input goods than Advanced Economies, increasing their exposure to exchange rate pass through. See \cite{camara2021spillovers}.} Similarly, Panel $(4,2)$ shows that domestic lending rates decrease in Advanced Economies while Panel $(4,3)$ shows that they increase in Emerging Markets for 10 months after the initial shock. Furthermore, this increase is quantitatively larger than the increase in interest rates in the US, as argued above. Finally, Panels $(5,2)$ and $(5,3)$ show that the drop in the equity index in Emerging Markets is twice as large as in Advanced Economies. Overall, US monetary policy shocks exhibit a significantly larger impact in macroeconomic and financial variables in Emerging Markets compared to Advanced Economies. Thus, the evidence emerging from our panel SVAR model suggests the presence of the US monetary policy Spillover Puzzle. 

\noindent
\textbf{Local projection evidence.} The SVAR evidence presented above shows that after an identified US monetary policy shock the impulse response functions of Emerging Markets is greater in magnitude than for the panel of Advanced Economies. We provide further evidence that stresses the greater negative spillovers of US monetary policy shocks in EM economies through a local projection exercise, \'a la \cite{jorda2005estimation}. The advantage of this methodology is that it allows us to statistically test differences in responses across country groups.

We estimate the following econometric specification
\begin{align} \label{eq:LP_specification_US}
    \ln y_{i,t+h} = \beta^{MP}_{h} i^{\text{MP}}_t + \gamma^{INT}_{h} \mathbbm{1} \left[i \neq \text{US} \right] \times i^{\text{MP}}_t 
    + \sum^{J_y}_{j=1} \delta^{j}_i y_{i,t-j} + \sum^{J_x}_{j=1} \alpha^{j}_i x_{i,t-j} + \gamma t + \mu_i + \epsilon_{i,t} 
\end{align}
where $\ln y_{i,t+h}$ represents one of the variables in our SVAR specification in country $i$ in period $t+h$, $i^{\text{MP}}_t$ is the identified monetary policy shock, $\mathbbm{1} \left[i \neq \text{US} \right]$ is an indicator variable which takes the value of $1$ if country $i$ is not the US, $y_{i,t-j}$ and $x_{i,t-j}$ are lagged values of the dependent variable and control variables from country $i$, the term $\gamma t$ represents a time trend, $\mu_i$ represents a country fixed effect, and $\epsilon_{i,t}$ is a stochastic disturbance. Constants $J_y$ and $J_x$ represent the number of lagged values of the dependent and control variables used in the regression, chosen to $J_y = J_x = 1$.\footnote{We select the number of lags by computing the Schwarz's Bayesian information criterion (SBIC) selection statistic for each country separately.  The optimal number of lags is $J_y = J_x = 1 $ for all but one country in our sample. The exception is Indonesia, for which the SBIC statistic suggest the choice of 2 lags. As a robustness check, we computed the Hannan and Quinn information criterion statistic which also suggest using one lag for the vast majority of countries in our sample.}${,}$. The model is estimated at the monthly frequency, with our full sample of EM and AE economies, including the US economy. Standard errors are clustered at the country and time level.\footnote{Clustering the standard errors at the time level is of key importance as the US monetary policy shock occurs simultaneously for all countries in the sample.} The key parameters of interests are $\beta^{MP}_h$ and $\beta^{INT}_h$. On the one hand, $\beta^{MP}_h$ represents the impact of a US monetary policy shock in the US economy in period $h$. On the other hand, $\beta^{MP}_h+\beta^{INT}_h$ represents the impact of a US monetary policy shock in countries other than the US economy. 

\begin{figure}[ht]
    \centering
    \caption{Spillover Puzzle of US Monetary Policy \\ \footnotesize LP Results - Outside \& Within the US }
    \label{fig:All_Variables}
    \includegraphics[width=18cm,height=10cm]{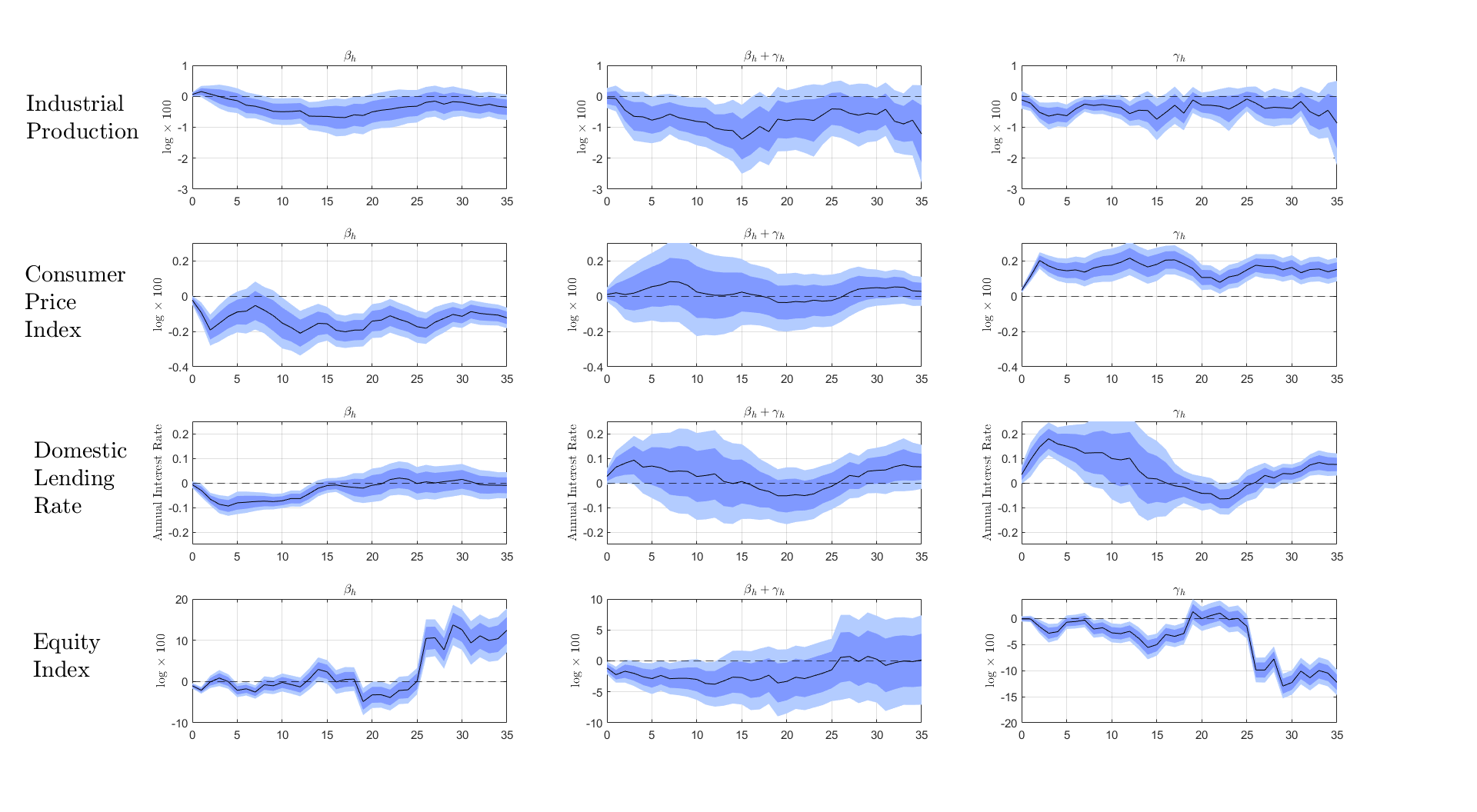}
    \floatfoot{\textbf{Note:} The black line represents the point estimate of the parameter of interest. The dark blue are represents 68\% confidence intervals. The light blue are represents 90\% confidence intervals. Standard errors are clustered at the country and time dimension. There are 12 panels with figures, organized in four rows and three columns. Rows represent variables: (i) industrial production, (ii) consumer price index, (iii) domestic lending rate, (iv) equity index. The left column presents the estimates of parameter $\beta_h$, the middle column presents the estimates of the sum of coefficients $\beta_h + \gamma_h$, while the right column presents the estimates of parameter $\gamma_h$. For every variable, the scale vertical axis is set to allow for comparison across samples. In the text, when referring to Panel $(i,j)$, $i$ refers to the row and $j$ to the column of the figure.}
\end{figure}

Figure \ref{fig:All_Variables} presents shows the impact of a US monetary policy shock in the US and the Rest of the World, by estimating Equation \ref{eq:LP_specification_US}. 
The left column shows the estimates of parameter $\beta_h$ which measures the impact of a US monetary policy shock in the US economy. The middle column shows presents the estimates of $\beta_h+\gamma_h$ which measures the total impact of a US monetary policy shock in countries other than the US. Lastly, the right column presents the marginal impact of a US monetary policy shock on countries other than the US. In particular, Panels $(1,1)$ and $(1,2)$ of the first row of Figure \ref{fig:All_Variables} show that the impact of a US monetary policy shock is twice as large outside the US than within the US economy. Furthermore, Panel $(1,3)$ shows that this difference is statistically significant across different time horizons $h$. Hence, Figure \ref{fig:All_Variables} provides evidence that the greater impact of US interest rates in Emerging Markets shown through a SVAR model, is statistically significant.\footnote{While the impact on the consumer price and the equity index is significantly larger for Emerging Market economies, the impact on lending rates is in line with estimates for the US economy.}

Next, we show that outside of the US economy, the impact of US monetary policy shocks is economically and statistically larger in magnitude for Emerging Market economies than for Advanced Economies. To do so, we estimate the following econometric specification
\begin{align} \label{eq:LP_specification_EM}
    \ln y_{i,t+h} = \beta^{MP-EM}_{h} i^{\text{MP}}_t + \gamma^{INT-EM}_{h} \mathbbm{1} \left[i = \text{EM} \right] \times i^{\text{MP}}_t 
    + \sum^{J_y}_{j=1} \delta^{j}_i y_{i,t-j} + \sum^{J_x}_{j=1} \alpha^{j}_i x_{i,t-j} + \gamma t + \mu_i + \epsilon_{i,t} 
\end{align}
The difference with the econometric specification in Equation \ref{eq:LP_specification_US} is that the indicator function which took the value of 1 if country $i$ was not the US economy, now takes the value of 1 if country $i$ is an EM economy, 
$\mathbbm{1} \left[i \neq \text{US} \right]$. We restrict our sample to countries other than the US.\footnote{Dropping the US economy from our panel allows us to show that the results arising from estimating Equation \ref{eq:LP_specification_EM} is not driven by the US economy. Including the US economy does not affect the results qualitatively or quantitatively.} Under this econometric specification, parameter $\beta^{MP}_{h}$ measures the impact of a US monetary policy shock in Advanced Economies in period $h$, while $\beta^{MP}_{h}+\beta^{INT-EM}_{h}$ represents the impact of a US monetary policy shock in Emerging Market economies in period $h$. 

\begin{figure}[ht]
    \centering
    \caption{Spillover Puzzle of US Monetary Policy \\ \footnotesize Local Projection Results - Adv. vs EM.}
    \label{fig:All_Variables_AE}
    \includegraphics[width=18cm,height=10cm]{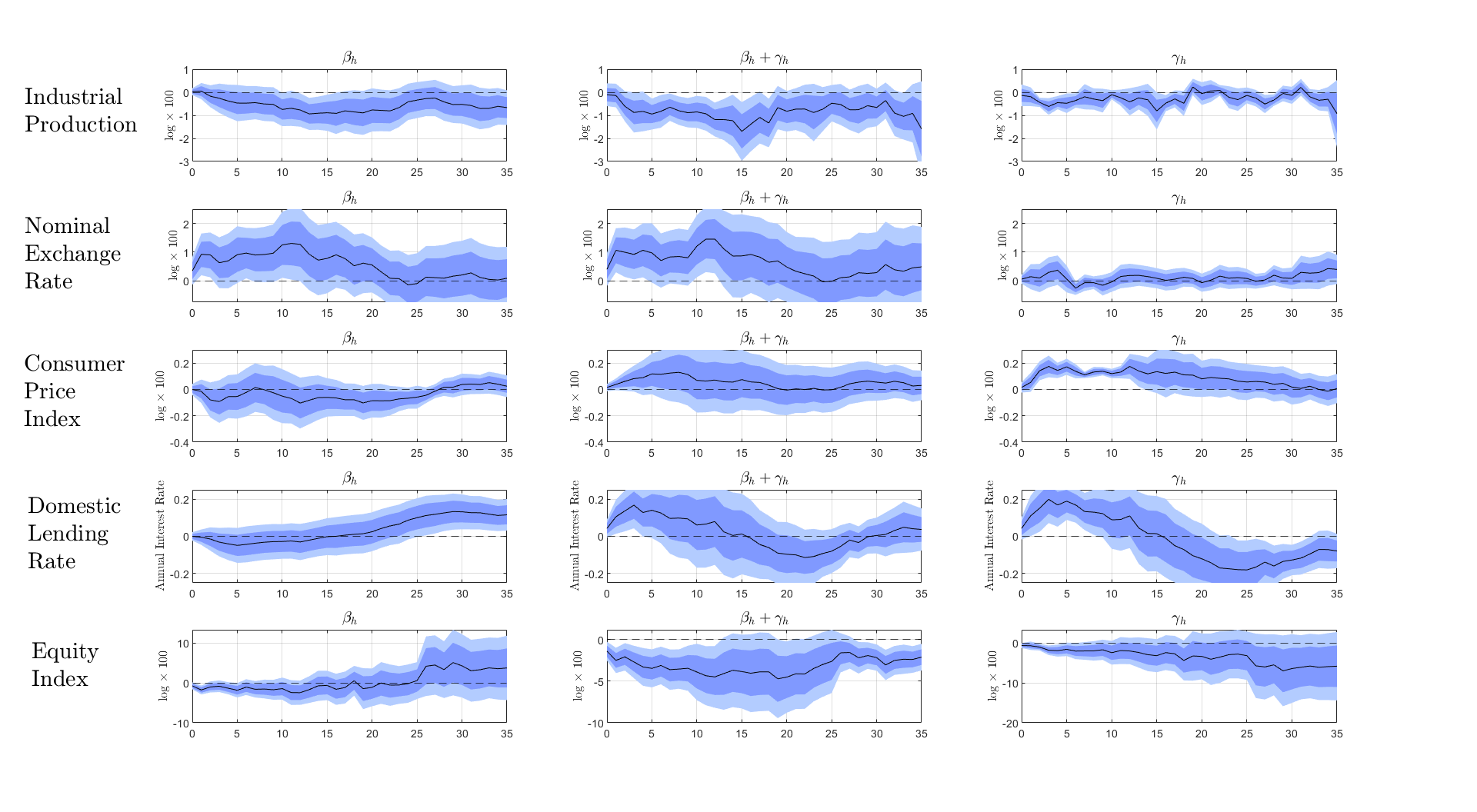}
    \floatfoot{\textbf{Note:} The black line represents the point estimate of the parameter of interest. The dark blue are represents 68\% confidence intervals. The light blue are represents 90\% confidence intervals. Standard errors are clustered at the country and time dimension. There are 15 panels with figures, organized in five rows and three columns. Rows represent variables: (i) industrial production, (ii) nominal exchange rate, (iii) consumer price index, (iv) domestic lending rate, (v) equity index. The left column presents the estimates of parameter $\beta_h$, the middle column presents the estimates of the sum of coefficients $\beta_h + \gamma_h$, while the right column presents the estimates of parameter $\gamma_h$. For every variable, the scale vertical axis is set to allow for comparison across samples. In the text, when referring to Panel $(i,j)$, $i$ refers to the row and $j$ to the column of the figure.}
\end{figure}

Figure \ref{fig:All_Variables_AE} presents shows the impact of a US monetary policy shock in Advanced Economies and in Emerging Markets, by estimating Equation \ref{eq:LP_specification_EM}. 
On the first row, comparing the plots on Panels $(1,1)$ and $(1,2)$ it is clear that the impact of a US monetary policy shock is significantly greater in Emerging Markets than in Advanced Economies (roughly twice as big). Furthermore, the impact transmits into Emerging Markets faster, reaching a level 1\% below pre-shock levels 2 periods after the initial shock, and is more persistent across time. Panel $(1,3)$ shows that this greater impact industrial production is statistically significant for most periods $h$. These results suggest that the \textit{Spillover Puzzle} of US monetary policy is particularly present for Emerging Market economies. 

The resulting dynamics for the rest of the variables in the model are roughly in line with those arising from the SVAR model. While Advanced Economies exhibit a drop in the consumer price index, Emerging Markets exhibit a moderate increase. In terms of domestic interest rates, Emerging Markets exhibit a mild increase roughly in line with those found by \cite{jarocinski2020deconstructing}. Finally, the impact on Emerging Markets' equity indexes is significantly larger, in line with the greater drop in economic activity. 

In summary, US monetary policy shocks have negative and quantitatively large spillovers over both Advanced and Emerging Market economies. The impact of US monetary policy shocks is quantitatively larger outside the US economy than within the US economy. Furthermore, this greater impact is particularly larger in Emerging Market economies than in Advanced Economies other than the US. This result emerges from both SVAR and local projection models. 

\subsection{Quantitative Exercises on the Spillover Puzzle} \label{subsec:spillover_puzzle_quantitative}

We propose a novel explanation to the Spillover Puzzle as thoroughly characterized above. Specifically, we argue that differences in the prevalence of cash flow and collateral-based borrowing constraints across economies provide a natural and straightforward solution. To this end, we combine the greater share of interest sensitive cash flow-based lending in Emerging Markets, documented in Sections \ref{sec:cash_flow_lending} and \ref{sec:interest_sensitive_borrowing_constraints}, with the greater amplification of foreign interest rate shocks predicted by the model presented in Section \ref{sec:model_simple}. In particular, we extend on our model introduced in Section \ref{sec:model_simple} and allow for firms to face a debt limit represented by a weighted average of cash flow and collateral-based borrowing constraints. We calibrate the relative weights of each type of borrowing constraints to match the empirical moments presented for Argentina, US and the UK. We show that the greater amplification is in line with that predicted by the local projection regressions presented in the Figures \ref{fig:All_Variables} and \ref{fig:All_Variables_AE}.

\noindent
\textbf{Quantitative exercise Argentina vs. US.} Table \ref{tab:comparison_US} presents the decomposition of cash flow-based lending into ``Debt to Cash Flow'' and ``Interest Coverage'' constraints for both the US (leveraging on the work of \cite{greenwald2019firm}) and Argentina.
\begin{table}[ht]
    \centering
    \caption{Cash-Flow Based Lending in Argentina \& the US Countries}
    \label{tab:comparison_US}
    \begin{tabular}{l c | c }
                        & US - \footnotesize \cite{greenwald2019firm} & Argentina   \\ \hline \hline
    Debt to Cash Flow & 70\% & 31\% \\
    Interest Coverage & 30\% & 69\% \\ \hline \hline
    \end{tabular}
\end{table}
Results show that there is a greater prevalence of ``Interest Coverage'' constraints in Argentina compared to the US. Our results suggest that 70\% of cash flow-based lending in Argentina is represented by interest sensitive borrowing constraints, with the remaining 30\% represented by cash flow-based borrowing constraints. For the case of US firms, \cite{greenwald2019firm} suggests that on average, 70\% of firms' cash flow-based lending is in the form of ``Debt to Cash Flow'' borrowing constraints, and the remaining 30\% in terms of ``Interest Coverage'' borrowing constraints.

We allow firms to face both ``Interest Coverage'' and ``Debt to Cash Flow'' based lending simultaneously to better match the observed data. We follow the approach introduced by \cite{drechsel2019earnings} which establishes a firm's debt limit as a weighted average of the two constraints
\begin{align}
    \bar{b}_t \leq \omega b^{IC}_t + \left(1-\omega\right) b^{DC}_t
\end{align}
where $\omega \in \left[0,1\right]$ represents the relative importance of cash flow-based borrowing constraints compared to collateral-based constraints.
\begin{table}[ht]
    \centering
    \caption{Quantitative Exercise Argentina \& US \\ \footnotesize Calibration}
    \label{tab:comparison_US_calibration}
    \begin{tabular}{c c c c}
    Parameter   & Description  & Argentina & US - \footnotesize \cite{greenwald2019firm} \\ \hline \hline
    $\omega$      & Share of IC & 0.70      & 0.30  \\
    $\theta^{IC}$ & Tightness $\theta^{IC}$ & 0.1225    & 0.154 \\
    $\theta^{DC}$ & Tightness $\theta^{DC}$ & 4.1437    & 8.613 \\ \hline \hline
    \end{tabular}
\end{table}
Table \ref{tab:comparison_US_calibration} presents the calibration of the share of each type of borrowing constraint and the tightness of each constraint for both Argentina and the US.

\begin{figure}[ht]
    \centering
    \caption{\textit{Spillover Puzzle}: Argentina \& US \\ \footnotesize Quantitative Exercise }
    \label{fig:IRFs_Argentina_vs_USA}
    \includegraphics[width=15cm,height=8cm]{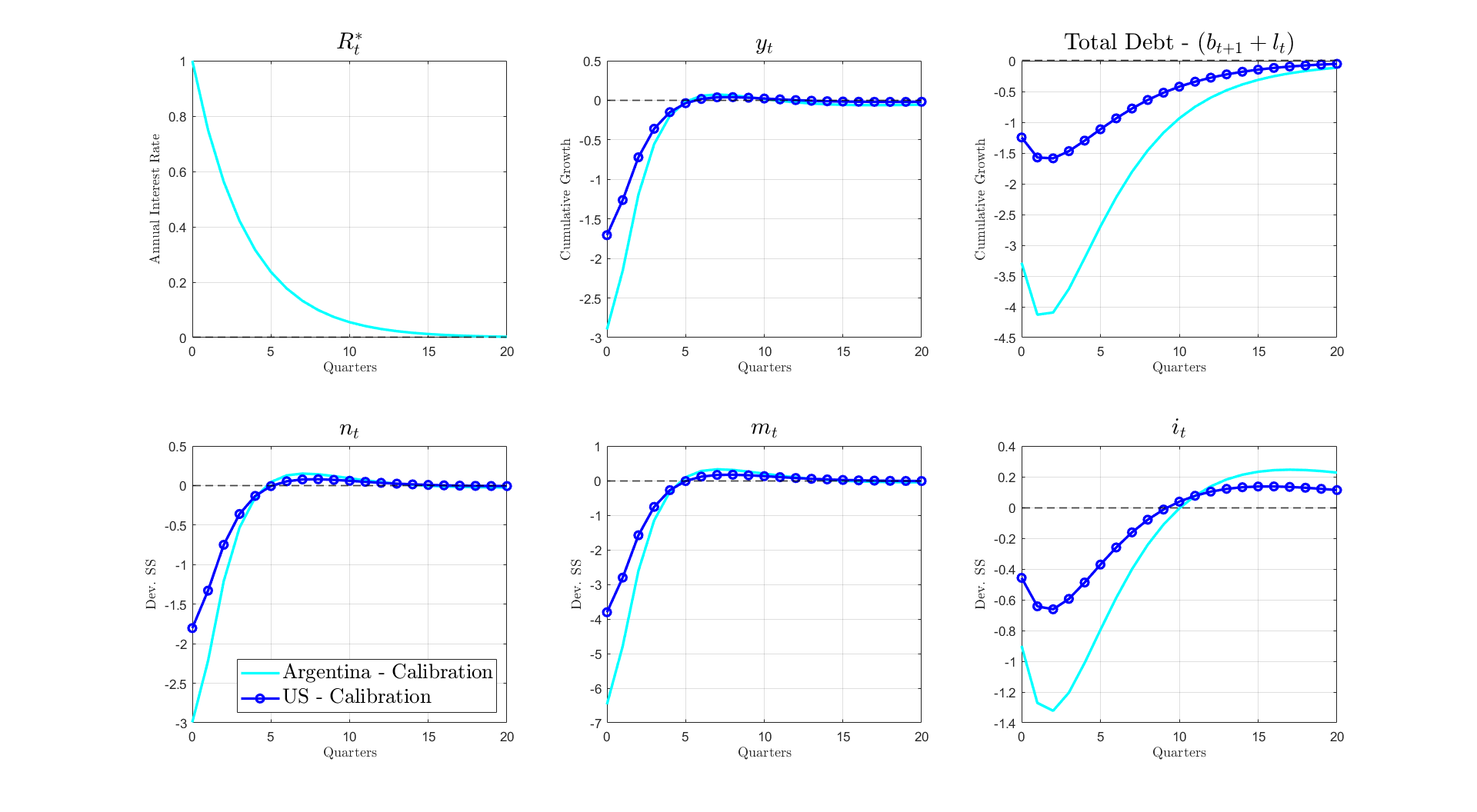}
    \floatfoot{\textbf{Note:} }
\end{figure}

Figure \ref{fig:IRFs_Argentina_vs_USA} presents the impulse response functions of a foreign interest rate shock for two economies calibrated according to the parameter values in Table \ref{tab:comparison_US_calibration}. Across variables, we observe that under this calibration, the impact on real variables is twice as large in the economy with a greater share of ``Interest Coverage'' borrowing constraints. This is quantitatively in line with the results arising from the local projection results presented in Figure \ref{fig:All_Variables}. Consequently, by allowing for heterogeneity in borrowing constraints, we provide a straightforward solution to the fact that US monetary policy shocks lead to greater spillovers in Emerging Markets than within the US economy. 

\noindent
\textbf{Quantitative exercise Argentina vs. UK.}

\section{Conclusions} \label{sec:conclusions}

In this article, we provided empirical evidence on the determinants of firms' borrowing constraints in Emerging Markets and studied their aggregate implications. Empirically, exploiting a highly detailed credit registry data set from Argentina we showed that the vast majority of firms' debt is cash flow-based lending and only a small share is collateral-based. We showed that firms primarily borrow cash flow-based to finance working capital needs with collateral-based lending explaining a relatively larger share of the financing of capital expenditures. Furthermore, we showed that firms in sectors of the economy with a greater share of standardized assets exhibit a greater prevalence of collateral-based lending. We show that firms' share of cash flow-based lending exhibits a U-shaped relationship with their levels of employment and age. 

Second, we exploit the details of firms' debt contracts and Central Bank regulations to argue that the most common type of borrowing constraint firms face is determined by a relationship between their interest payments and a measure of their cash flows. This is in line with an interest coverage borrowing constraint. We show that Central Bank regulations impose a monitoring and risk assessment role on banks over firms based on the relationship between their interest payments to their cash flows. We show supporting empirical evidence of this borrowing constraint by finding at the macro and micro level that firms' cash flows are pro-cyclical. Furthermore, this novel stylized fact is present for Argentina and other Emerging Markets, providing external validity of our results. 

Third, we test the aggregate implications of this prevalence of interest coverage borrowing constraints. We do this by constructing a relatively standard small open economy DSGE model where firms are subject to working capital and cash flow-based borrowing constraints. In a counterfactual analysis where firms' borrowing constraints are always binding, a foreign interest rate shock leads to orders of magnitude greater amplification in an economy with interest coverage borrowing constraint compared to an economy with a benchmark ``collateral'' borrowing constraint. This result is driven by the interest sensitivity of the borrowing constraint and not by a sharp reduction of firms' cash flows. Additionally, we show that this result is also present in a more quantitative setting where firms are subject to nominal frictions and where firms can borrow in both domestic and foreign currency debt. Lastly, we argue that the cost of interest rate policies designed to limit exchange rate volatility are significantly larger in the presence of interest coverage borrowing constraints compared to the benchmark collateral-based constraint.

Finally, we argue that the greater prevalence of interest sensitive borrowing constraints in Emerging Markets provides a solution to the \textit{Spillover Puzzle} of US monetary policy. This is a well documented puzzle in international macroeconomics through which US monetary policy shocks have significantly larger spillovers in Emerging Markets in the US economy or in other Advanced economies. After showing the presence of significantly larger spillovers in Emerging Markets using state of the art identified US monetary policy shocks, we argue that augmenting our models to reflect the relative share of cash flow and collateral-based borrowing constraints can provide a solution. Through two calibrated exercises we show that the greater spillovers founds in Emerging Markets is in line with the predicted greater amplification coming from our structural model.

\newpage
\bibliography{main.bib}

\newpage
\appendix

\section{Additional Data Set Description} \label{sec:appendix_data_description}


\subsection{Employment Data} \label{subsec:appendix_additional_data_description_employment}

In this Appendix we present additional details on our employment dataset. We start by present summary statistics about firm's employment. Argentinean firms can be constituted as independent contractors or ``\textit{Persona f\'isica}'' or constituted as firms or legal independent organizations or ``\textit{Persona jur\'idica}''. Table \ref{tab:data_employment_types} presents details on the levels of employment per type of firm and the number of firms. While the number of independent contractors is slightly greater than the number of independent organizations, the vast majority of employment is explained by the latter.
\begin{table}[ht]
    \centering
    \caption{Employment by Type of Firm}
    \footnotesize
    \label{tab:data_employment_types}
    \begin{tabular}{l c c}
                            & Employment & Number of Firms \\ \hline \hline
    Contractors               & 1,056,786 & 361,954 \\ 
    Independent Organizations & 9,809,750 & 299,573 \\ \hline \hline
    \end{tabular}
    \floatfoot{\textbf{Note:} The data presented in this table is constructed using data for the year 2017. This is in line with data used for Tables in Section \ref{sec:cash_flow_lending}.}
\end{table}

Next, Table \ref{tab:data_employment_summary_statistics} presents summary statistics about the distribution of the number of employees per firm.
\begin{table}[ht]
    \centering
    \caption{Employment per Firm by Type of Firm}
    \label{tab:data_employment_summary_statistics}
    \footnotesize
    \begin{tabular}{l c c c c c c}
	&	Mean	&	p25	&	p50	&	p75	&	p90	&	p95	\\ \hline \hline
Total Sample	&	16.43	&	1	&	2	&	5	&	14	&	31	\\
Contractors	&	2.92	&	1	&	2	&	3	&	6	&	9	\\
Independent Organizations	&	32.75	&	1	&	4	&	11	&	32	&	67	\\ \hline \hline
    \end{tabular}
    \floatfoot{\textbf{Note:} The data presented in this table is constructed using data for the year 2017. This is in line with data used for Tables in Section \ref{sec:cash_flow_lending}.}
\end{table}
The mean level of employment per firm is 16.43 for the full sample, 2.92 for the sample of contractors, and 32.75 for independent organizations. The distribution is highly skewed towards higher values with the median level of employment per firm being 2 for the full sample, 2 for contractors and 4 for independent organizations. 

Next, we turn to describing the role of small, medium and large firms in explaining aggregate employment. We categorize firms by their level of employments into 7 different groups: (i) under 10 employees, (ii) between 10 and 49, (iii) between 50 and 249, (iv) between 250 and 499, (v) between 500 and 999, (vi) between 1,000 and 9,999, (vii) above 10,000. Table \ref{tab:data_employment_shares} presents data on the level of employment and the number of firms in each category. The upper panel of the table presents the results for our full sample while the bottom panel of the table presents the results excluding public administration and publicly provided education and health systems. 
\begin{table}[ht]
    \centering
    \caption{Employment per Firm}
    \footnotesize
    \label{tab:data_employment_shares}
    \begin{tabular}{l c c c c}
    \multicolumn{5}{c}{Excluding Public Administration} \\
Category	&	Employment	&	Number of Firms	&	Share of Employment	&	Share of Firms	\\ \hline \hline
$<10$	&	 1,361,877 	&	 561,175 	&	12.53	&	84.83	\\
$\in [10,50)$	&	 1,540,164 	&	 79,417 	&	14.17	&	12.01	\\
$\in [50,250)$	&	 1,735,432 	&	 17,387 	&	15.97	&	2.63	\\
$\in [250,500)$	&	 647,076 	&	 1,865 	&	5.95	&	0.28	\\
$\in [500,1,000)$	&	 609,212 	&	 887 	&	5.61	&	0.13	\\
$\in [1,000,10,000)$	&	 1,723,327 	&	 716 	&	15.86	&	0.11	\\
$\geq 10,000$	&	 3,249,448 	&	 80 	&	29.90	&	0.01	\\ \hline
\textbf{Total}	&	 \textbf{10,866,536} 	&	 \textbf{661,527} 	&	\textbf{100}	&	\textbf{100}	\\ \hline \hline
    \end{tabular}
    \footnotesize
    \begin{tabular}{l c c c c}
    \\
    \\
    \multicolumn{5}{c}{Excluding Public Administration} \\
Category	&	Employment	&	Number of Firms	&	Share of Employment	&	Share of Firms	\\ \hline \hline
$<10$	&	1,182,386	&	464,599	&	18.74	&	83.44	\\
$\in [50,250)$	&	1,427,645	&	73,662	&	22.63	&	13.23	\\
$\in [250,500)$	&	1,578,864	&	15,879	&	25.03	&	2.85	\\
$\in [500,1,000)$	&	538,676	&	1,561	&	8.54	&	0.28	\\
$\in [500,1,000)$	&	461,180	&	676	&	7.31	&	0.12	\\
$\in [1,000,10,000)$&	945,077	&	431	&	14.98	&	0.08	\\
$\geq 10,000$	&	175,278  	&	12	&	2.78	&	0.00	\\ \hline
\textbf{Total}	&	 \textbf{6,309,105} 	&	 \textbf{556,820} 	&	\textbf{100}	&	\textbf{100}	\\
 \hline \hline
    \end{tabular}
\end{table}
More than half of aggregate private employment is explained by firms with less than 500 workers. However, firms with less than 500 workers explain almost 95\% of all firms. This implies that a significantly small number of firms explains the top 50\% of aggregate employment. 

\noindent
\textbf{Firms' age.} Our datasets do not provide direct information over a firms' age. However, we construct an indirect measure of firms' age. First, our international trade data set covers the period 1994-2019. Thus, if we observe their tax ID identifying number in 1994, we know that in year 2017 the firm is at least 23 years old. Our employment data set starts in the year 2001. If we observe a firm for the first time in our employment dataset in the year 2001, we cannot determine whether the firm was born in the year 2001 or was already alive before said year. Thus, to those firms, we can say that in the year 2017 they were at least 16 years old. This is but an imperfect measure, by no means perfect.

Table \ref{tab:data_employment_by_age} presents data on mean and median employment and the number of firms by three categories of age: (i) below 5 years of age, (ii) between 5 and 9 years, (iii) above 10 years. \begin{table}[ht]
    \centering
    \caption{Employment per Firm by Age}
    \footnotesize
    \label{tab:data_employment_by_age}
    \begin{tabular}{l c c c}
\multicolumn{4}{c}{Total Sample} \\ \hline
Age in Years	&	Mean Employment	&	Median Employment	&	Number of Firms	\\ \hline \hline
Under 5	&	4.40	&	2	&	178,484	\\
5 to 9	&	13.81	&	2	&	148,652	\\
Above 10	&	44.62	&	3	&	80,595	\\ \hline \hline
    \end{tabular}
    \begin{tabular}{l c c c}
    \\
\multicolumn{4}{c}{Independent Organizations} \\ \hline
Age in Years	&	Mean Employment	&	Median Employment	&	Number of Firms	\\ \hline \hline
Under 5	&	8.36	&	3	&	63,288	\\
5 to 9	&	30.25	&	4	&	60,140	\\
Above 10	&	88.39	&	7	&	39,294	\\ \hline \hline
    \end{tabular}
    \begin{tabular}{l c c c}
    \\
\multicolumn{4}{c}{Independent Organizations excluding Public Administration} \\ \hline
Age in Years	&	Mean Employment	&	Median Employment	&	Number of Firms	\\ \hline \hline
Under 5	&	8.36	&	3	&	63,288	\\
5 to 9	&	30.25	&	4	&	60,140	\\
Above 10	&	88.39	&	7	&	39,294	\\ \hline \hline
    \end{tabular}
\end{table}
The top panel presents information for the full sample, the middle sample presents information for independent organization firms, while the bottom panel presents information for the sample of independent organizations excluding public administration and services of health and education provided by the government. In line with literature in firm dynamics and industrial organization, such as \cite{evans1987relationship} and \cite{hansen1992innovation}, firm size proxied by employment and firm age are positively correlated across the three samples.

\newpage
\subsection{Orbis Data} \label{subsec:appendix_additional_data_description_orbis}

In this Appendix we present additional details on the data sourced from Orbis about Argentinean firms. This data set is accessed through the ``Northwestern Library'' between April 1$^{st}$ and April $14^{th}$ of the year 2022. 

The Orbis dataset provides additional financial data which supplements the credit-registry, firm and customs level data used in the paper. In particular, we use data for the following variables
\begin{itemize}
    \item Total Assets
    
    \item Current Liabilities
    
    \item Non-Current Liabilities    
    
    \item Net-Income

    \item Cash-Flows
    
    \item Sales \& Cost of Producing Goods
    
    \item EBIT

    \item EBITDA
    
    \item Operating Turnover

    \item Interest Payments
\end{itemize}
The coverage is significantly shorter than that of the rest of the datasets used across the paper. For the different variables, coverage for the period 2013-2021 we have information for a range of 100 to 600 firms. Furthermore, the coverage of reported firms  is skewed towards relatively larger firms. 
\begin{table}[ht] 
    \centering
    \caption{Coverage of Orbis Data Set \\ \footnotesize Number of Firms which Report ... }
    \label{tab:coverage_orbis}
    \footnotesize
    \begin{tabular}{l c c c c c c c c c}
Variable	&	2013	&	2014	&	2015	&	2016	&	2017	&	2018	&	2019	&	2020	&	2021	\\ \hline \hline
Assets	&	354	&	477	&	593	&	634	&	639	&	594	&	622	&	606	&	442	\\
Current Liabilities	&	121	&	235	&	344	&	379	&	381	&	337	&	362	&	346	&	215	\\
Non Current Liabilities	&	121	&	235	&	344	&	379	&	381	&	337	&	362	&	346	&	215	\\
Cash Flow	&	111	&	101	&	113	&	126	&	153	&	156	&	159	&	155	&	111	\\
EBIT	&	121	&	234	&	343	&	379	&	380	&	336	&	362	&	346	&	215	\\
Net Income	&	356	&	478	&	596	&	639	&	645	&	597	&	628	&	612	&	445	\\
Operating Turnover	&	355	&	473	&	592	&	1,484	&	1,181	&	12,124	&	619	&	868	&	445	\\
P over L	&	354	&	476	&	592	&	635	&	638	&	593	&	622	&	606	&	442	\\
Sales	&	120	&	137	&	170	&	203	&	235	&	200	&	209	&	209	&	143	\\
Solvency	&	354	&	476	&	590	&	633	&	637	&	593	&	620	&	602	&	441	\\ \hline \hline
    \end{tabular}
\end{table}
Table \ref{tab:coverage_orbis} presents the number of firms reporting information for each of the variables considered for our purposes.

First, we provide supporting evidence of the results presented in Section \ref{sec:cash_flow_lending}. In Appendix \ref{sec:appendix_international_comparison} we show that Argentina's private sector financing is primarily covered by banks. Furthermore, in Appendix \ref{subsec:appendix_international_comparison_US} we show that while Argentinean firms tend to borrow short term, show an opposite pattern, issuing debt both from banks and through corporate bonds at maturities usually greater than one year. A possible concern is that relatively larger Argentinean firms which face lower costs of issuing corporate bonds may reflect debt patterns more consistent with the patterns which arise from the US economy. Table \ref{tab:orbis_ratio_non_current} presents moments of firms' distribution of non-current liabilities over total liabilities.
\begin{table}[ht]
    \centering
    \caption{Ratio of Non-Current Liabilities over Total Liabilities \\ \footnotesize Expressed in Percent \%}
    \label{tab:orbis_ratio_non_current}
    \footnotesize
    \begin{tabular}{l c c c c c}
	&	Mean	&	p25	&	Median	&	p75	&	p90	\\ \hline \hline
2013	&	31.90	&	11.19	&	27.47	&	47.94	&	67.86	\\
2014	&	26.28	&	3.67	&	18.94	&	44.45	&	68.01	\\
2015	&	24.08	&	1.61	&	13.17	&	40.03	&	67.61	\\
2016	&	23.59	&	0.86	&	12.86	&	38.51	&	66.95	\\
2017	&	25.11	&	1.08	&	15.79	&	43.81	&	68.54	\\
2018	&	30.48	&	2.77	&	23.92	&	52.35	&	74.12	\\
2019	&	30.55	&	1.47	&	25.04	&	52.75	&	75.19	\\
2020	&	30.33	&	1.28	&	21.52	&	52.93	&	76.17	\\
2021	&	33.72	&	3.77	&	30.55	&	56.39	&	77.08	\\ \hline
\textbf{Total}	&	27.90	&	1.78	&	19.60	&	48.45	&	71.15	\\
\hline \hline
    \end{tabular}
    \floatfoot{\textbf{Note:} The ratio is computed as the amount of non-current liabilities divided by the sum of current liabilities plus non-current liabilities. Results are presented as percentages over total liabilities. }
\end{table}
In line with the results presented in Section \ref{sec:cash_flow_lending}, the vast majority of firms' liabilities is short term or current liabilities, exhibited by the mean and median ratio of non-current liabilities to total liabilities being equal to 27.90\% and 19.60\% respectively.

\newpage
\section{International Comparison} \label{sec:appendix_international_comparison}

In this Appendix of the paper we present a comparison of the results for Argentinean firms presented above with aggregate results from other countries.

First, we present evidence of the important role of bank financing in Argentina. To do so, we compare the role of bank financing with respect to the total sources of financing of the private sector.
\begin{figure}[ht]
    \centering
    \caption{Role of Banks in Financing the Private Sector \\ \footnotesize Selected Countries}
    \label{fig:Role_Banks_Credit}
    \includegraphics[width=16cm,height=8cm]{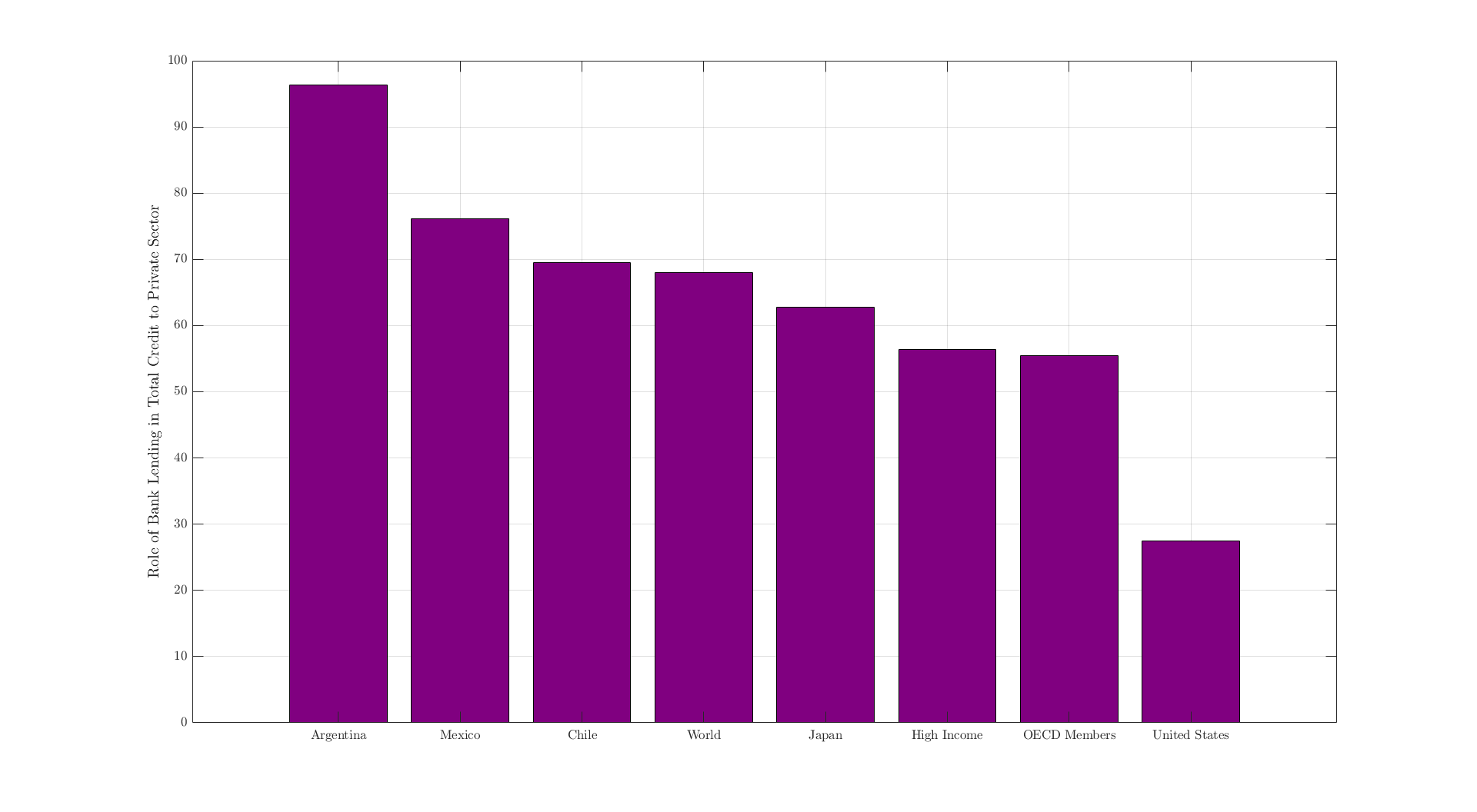}
    \floatfoot{\textbf{Note:} The figure above presents the share of bank financing over the total private sector financing. To do so, we use data from the World Bank's. In particular, we compute the ratio of the indicator ``Domestic credit to private sector by banks (\% of GDP)'' and the indicator ``Domestic credit to private sector (\% of GDP)''.}
\end{figure}
Figure \ref{fig:Role_Banks_Credit} presents a comparison across countries. On the first row from the left, for the case of Argentina, banks financing explains more than 96\% of total private sector financing. Naturally, Argentina is one of the countries in which banking plays the most important source of private sector financing. On the last row from the left, for the case of the US, banks only represent 25\% of total financing of the private sector. Consequently, the US economy is the country in which banking financing plays the smallest role of private sector financing. Countries with similar levels of income per capita, such as Chile and Mexico, exhibit a relatively lower importance of banking in private sector financing but still high (69\% and 76\% respectively) at least compared to High-Income and OECD members (which exhibit an average of 56.4\% 55.51\% respectively). 

Second, we provide an additional piece of evidence on the importance of bank financing in Argentina. To do so, we compare the ``Market capitalization of listed domestic companies as a \% of GDP'' across countries.
\begin{figure}[ht]
    \centering
    \caption{Market Capitalization of Listed Domestic Companies \\ \footnotesize Selected Countries}
    \label{fig:market_capitalization}
    \includegraphics[width=16cm,height=8cm]{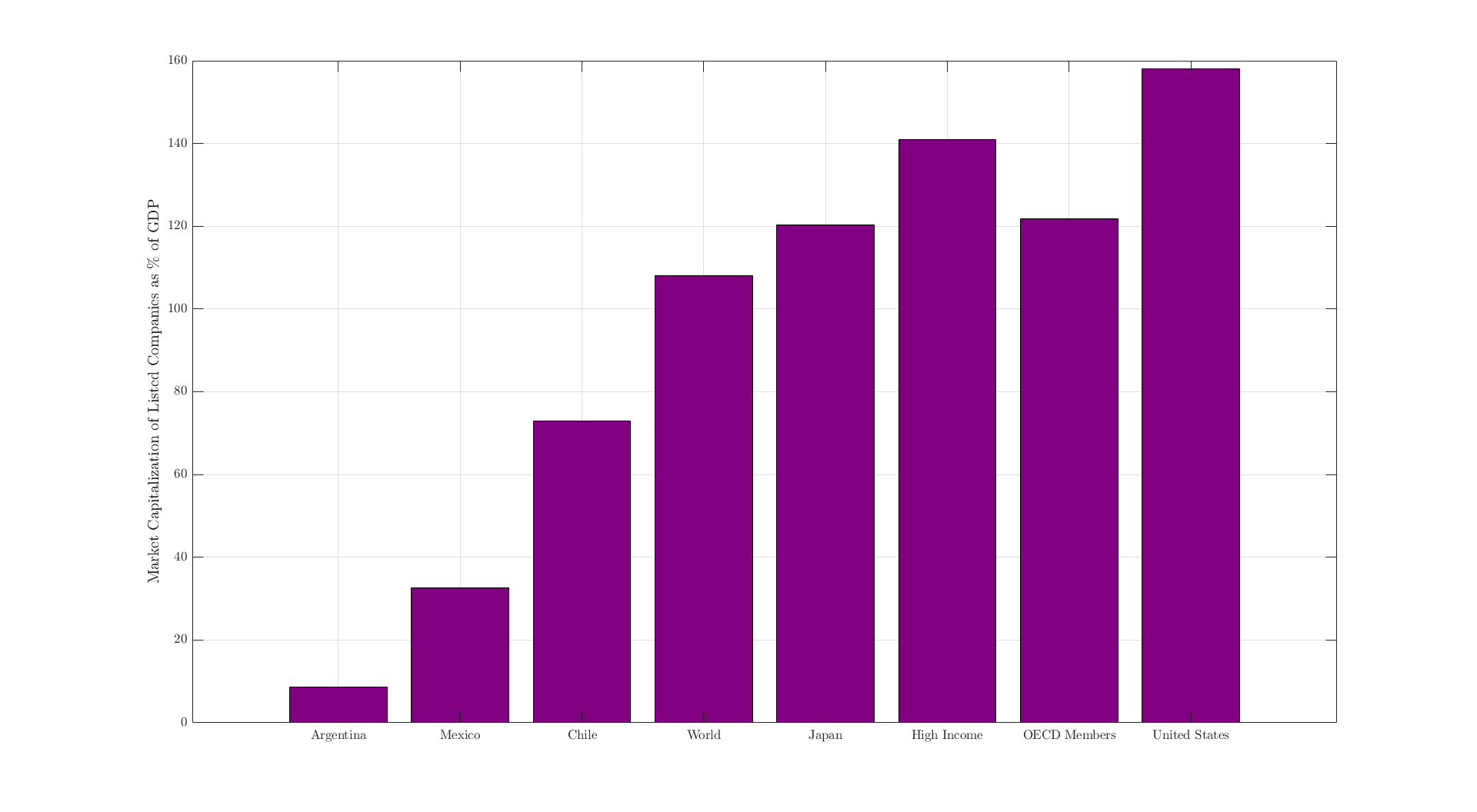}
    \floatfoot{\textbf{Note:} The figure above presents the ``Market capitalization of listed domestic companies as a \% of GDP'' for a selected sample of countries. The sample choice matches that of Figure \ref{fig:Role_Banks_Credit}. Data reflect average values for the year 2019.}
\end{figure}
Publicly listed companies fund themselves through issuing equity and, in practice, have an easier/faster access to corporate bond markets. Thus, an economy which exhibits a greater market value of their listed domestic companies is more likely to have access to financing sources other than domestic banks. Figure \ref{fig:market_capitalization} presents the results of this comparison. For the case of Argentina, the market capitalization of publicly listed companies is 8\% of GDP. This is between 4 and 10 times lower than the value for Mexico and Chile, 32.58\% and 72.94\% respectively. The comparison is even starker for the World average of 108.12\% and that of the US economy 158.12\%. Thus, in line with the results presented in Figure \ref{fig:Role_Banks_Credit}, Argentinean firms rely almost exclusively in bank to cover their financing needs.

\newpage
\subsection{Comparison with US Firms} \label{subsec:appendix_international_comparison_US}

In this subsection of the Appendix, we present comparison of the firm level results for Argentina with aggregate results coming from the Quarterly Financial Report of the US Census Bureau. 

\begin{figure}[ht]
    \centering
    \caption{Composition of US Firms' Liabilities}
    \label{fig:us_comparison}
     \centering
     \begin{subfigure}[b]{0.495\textwidth}
         \centering
         \includegraphics[width=\textwidth]{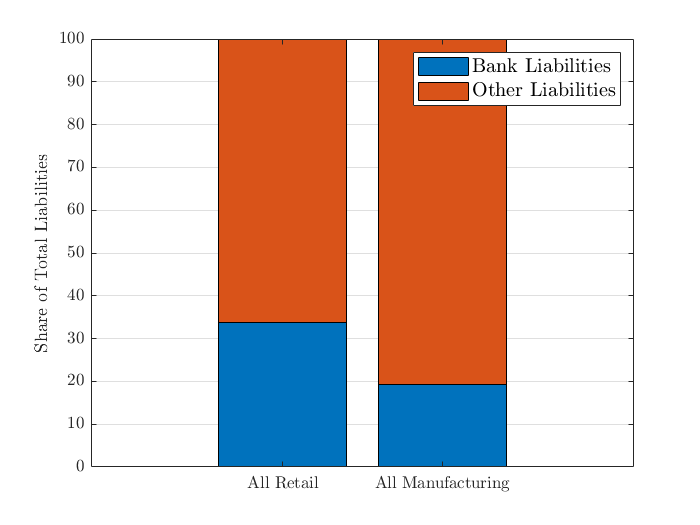}
         \caption{Bank vs Other Debt}
         \label{fig:banks_vs_not}
     \end{subfigure}
     \hfill
     \begin{subfigure}[b]{0.495\textwidth}
         \centering
         \includegraphics[width=\textwidth]{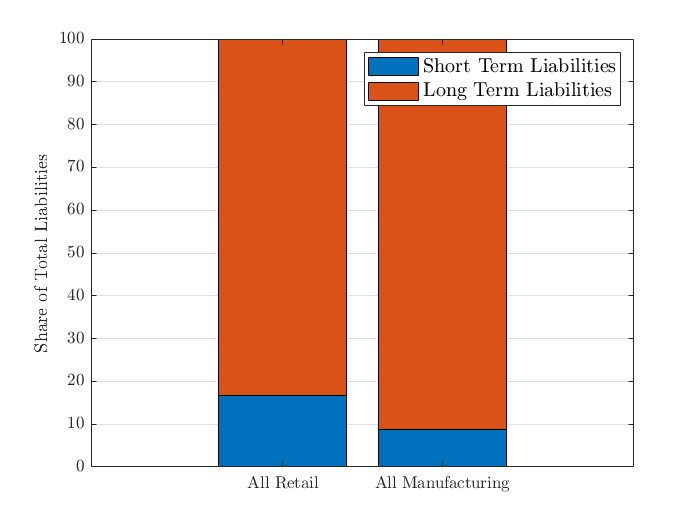}
         \caption{Short vs Long Debt}
         \label{fig:short_vs_long}
     \end{subfigure} \\ 
     \begin{subfigure}[b]{0.495\textwidth}
         \centering
         \includegraphics[width=\textwidth]{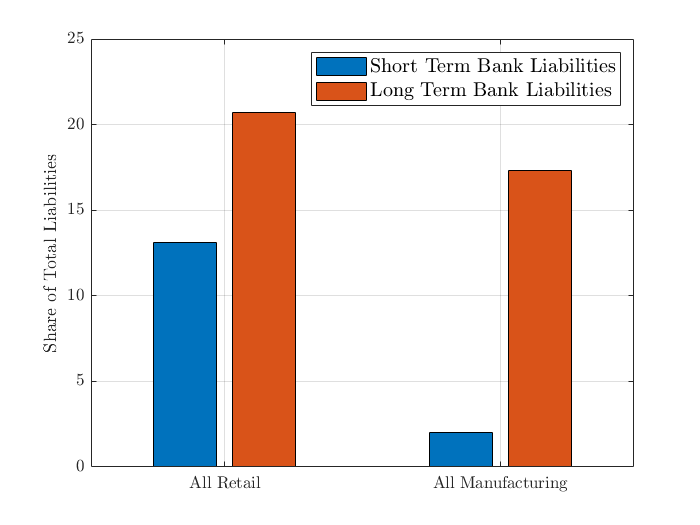}
         \caption{Short vs Long Bank Debt}
         \label{fig:bank_short_vs_long}
     \end{subfigure}
     \hfill
     \begin{subfigure}[b]{0.495\textwidth}
         \centering
         \includegraphics[width=\textwidth]{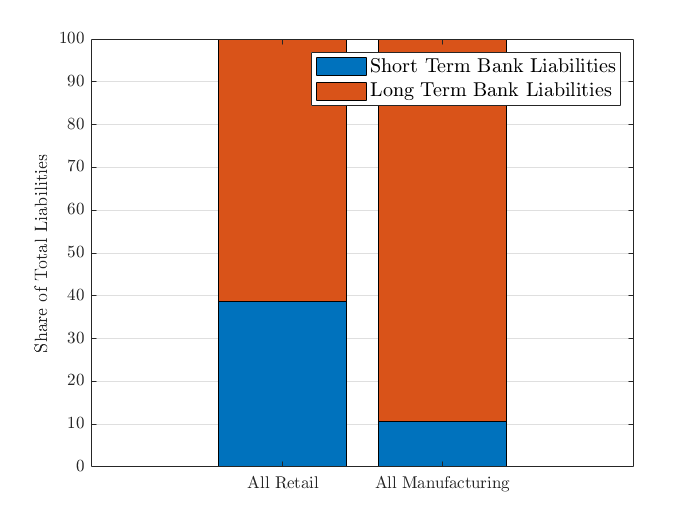}
         \caption{Short vs Long Bank Debt \textit{Normalized}}
         \label{fig:bank_short_vs_long_normalized}
     \end{subfigure}
     \floatfoot{\textbf{Note:} Data comes from the Quarterly Finance Report and is sourced from the St. Louis FRED dataset. }
\end{figure}

First, we carry out a decomposition of US firms' debt, presented in Figure \ref{fig:us_comparison}
On the top left panel, Figure \ref{fig:banks_vs_not} shows that US firms only source between 20\% and 30\% of their financing needs through bank debt for ``Manufacturing'' and ``Retail'' firms respectively. This is, the vast majority of financing needs are sourced through firms issuing commercial papers and/or corporate bonds. This in sharp contrast with the results presented for Argentinean firms, as shown in Figure \ref{fig:Role_Banks_Credit} in Section \ref{sec:appendix_international_comparison}. 

On the top right panel, Figure \ref{fig:short_vs_long} presents the decomposition of firms' total debt into short term and long term debt, defined as maturity lower or greater than one year. Once again, in contrast with the results for Argentina presented in Table \ref{tab:composition_firm_debt_type} in Section \ref{sec:cash_flow_lending}, the vast majority of firms' debt is long term, 16.81\% and 8.70\% for ``Retail'' and ``Manufacturing'' firms respectively. 

On the bottom panels of Figure \ref{fig:us_comparison}, we show that even when focusing only on firms' bank debt, the vast majority of the debt is long term. Figure \ref{fig:bank_short_vs_long} shows the relative importance of short and long term bank debt over firms' total debt. Figure \ref{fig:bank_short_vs_long_normalized} shows the relative importance of short and long term debt over total firms' debt. Overall, the main message these figures convey is that firms in the US borrow from banks at a relatively long-term debt, in sharp contrast with the results presented for Argentinean firms.

\begin{figure}[ht]
    \centering
    \caption{Importance of Net Working Capital over Total Liabilities}
    \label{fig:net_working_capital}
    \includegraphics[width=12cm,height=8cm]{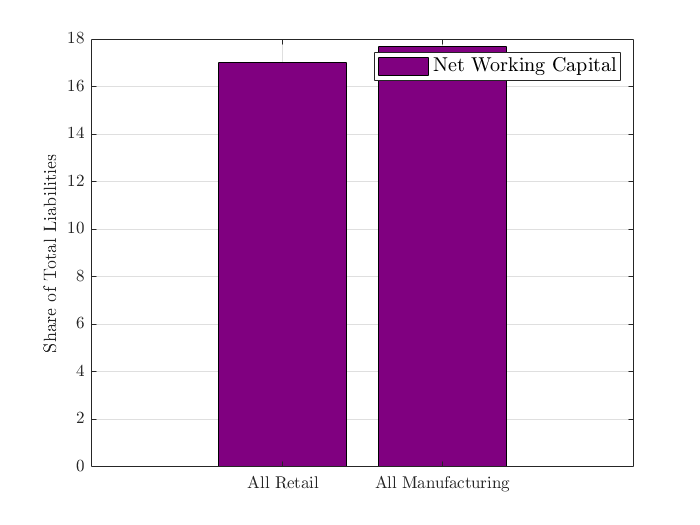}
\end{figure}

Finally, Figure \ref{fig:net_working_capital} presents the share of US firms' net working capital on their total liabilities. For both ``All Retail'' and ``All Manufacturing'' firms, the financing of working capital only explains 17\% of total liabilities.
Once again, this is in stark contrast with the results for Argentinean firms, presented in Table \ref{tab:composition_firm_debt_type}. For the case of Argentinean firms, close to 70\% of firms' liabilities are associated with working capital needs.

\subsection{International Comparison} \label{subsec:appendix_international_comparison_inter}

\noindent
\textbf{Role of Short Term Debt.} In Section \ref{subsec:lending_comparison_international} we argued that firms in Emerging Markets borrow at relatively shorter terms than firms in Advanced Economies. In particular, in Figure \ref{fig:Lpoly_Current_Liabilities} we fitted a local smooth polynomial which related firms' ratio of short term or current liabilities to total liabilities and country's income per capital. Table \ref{tab:ratio_current_gdp} shows that the negative relationship between firms' ratio of short term liabilities to total liabilities and their country's income per capita also emerges in an econometric exercise.
\begin{table}[ht]
    \centering
    \caption{Role of Short Term Debt across Income Levels}
    \label{tab:ratio_current_gdp}
    \scriptsize
\begin{tabular}{lcccccc}
& \multicolumn{4}{c}{Ratio of Current Liabilities to Total Liaibilities} \\
 & (1) & (2) & (3) & (4) & (5) & (6) \\ \hline \hline
 &  &  &  &  &  & \\
$\ln \text{Income per Capita}_{c,t}$ & -0.0709*** & -0.0798*** & -0.0834*** & -0.0863*** & -0.0862*** & -0.0605*** \\
 & (0.000919) & (0.000838) & (0.0101) & (0.00565) & (0.00566) & (0.000912) \\
$\text{Leverage}_{i,t}$ &  &  &  &  & -0.000154 & -0.00207*** \\
 &  &  &  &  & (0.000193) & (0.000350) \\
$\mathbbm{1}\left[\text{Country}_d == US\right]$ &  &  &  &  &  & -0.115*** \\
 &  &  &  &  &  & (0.00230) \\
 &  &  &  &  &  \\
Sector FE  & no & yes & yes & yes & yes & yes \\
Country FE & no & no  & yes & yes & yes & no \\
Firm    FE & no & no  & no  & yes & yes & no \\
 &  &  &  &  &  \\ 
Observations & 94,420 & 94,347 & 94,347 & 94,237 & 94,073 & 94,185 \\
$R^2$ & 0.059 & 0.358 & 0.461 & 0.858 & 0.858 & 0.375 \\ \hline \hline
\multicolumn{6}{c}{ Standard errors in parentheses} \\
\multicolumn{6}{c}{ *** p$<$0.01, ** p$<$0.05, * p$<$0.1} \\
\end{tabular}
\end{table}
Note that the result is present even as we control for firms' leverage and we introduce different fixed effects. 

\noindent
\textbf{Working Capital vs Capital Expenditures.} In Section \ref{subsec:lending_comparison_international} we argue that firms in Emerging Markets primarily borrow to finance working capital needs while firms in Advanced Economies borrow primarily to finance capital expenditures. Figure \ref{fig:wk_vs_ke_main} showed this pattern for the case of firms in Argentina and the US. 

Next, we use data coming from the World Bank's Enterprise survey to show that these pattern that firms' debt in Emerging Markets is primarily destined to cover working capital needs instead of capital expenditures. In particular, we exploit two questions from the survey
\begin{itemize}
    \item Firms using banks to finance working capital (\% of firms)
    
    \item Firms using banks to finance investment (\% of firms)
\end{itemize}
First, note that the questions ask in particular whether firms are relying on banks. Thus, the questions are ignoring whether firms rely in financial markets to finance working capital or investment, i.e. capital expenditures. Second, note that these questions are discrete or dichotomic questions and do not provide us with information on the share of firms' bank debt which is explained by working capital and/or capital expenditures.

\begin{figure}[ht]
    \centering
    \caption{Comparison in Financing Needs \\ \footnotesize International Comparison }
    \label{fig:international_comparison_financing_needs}
     \centering
     \begin{subfigure}[b]{0.3\textwidth}
         \centering
         \includegraphics[width=\textwidth]{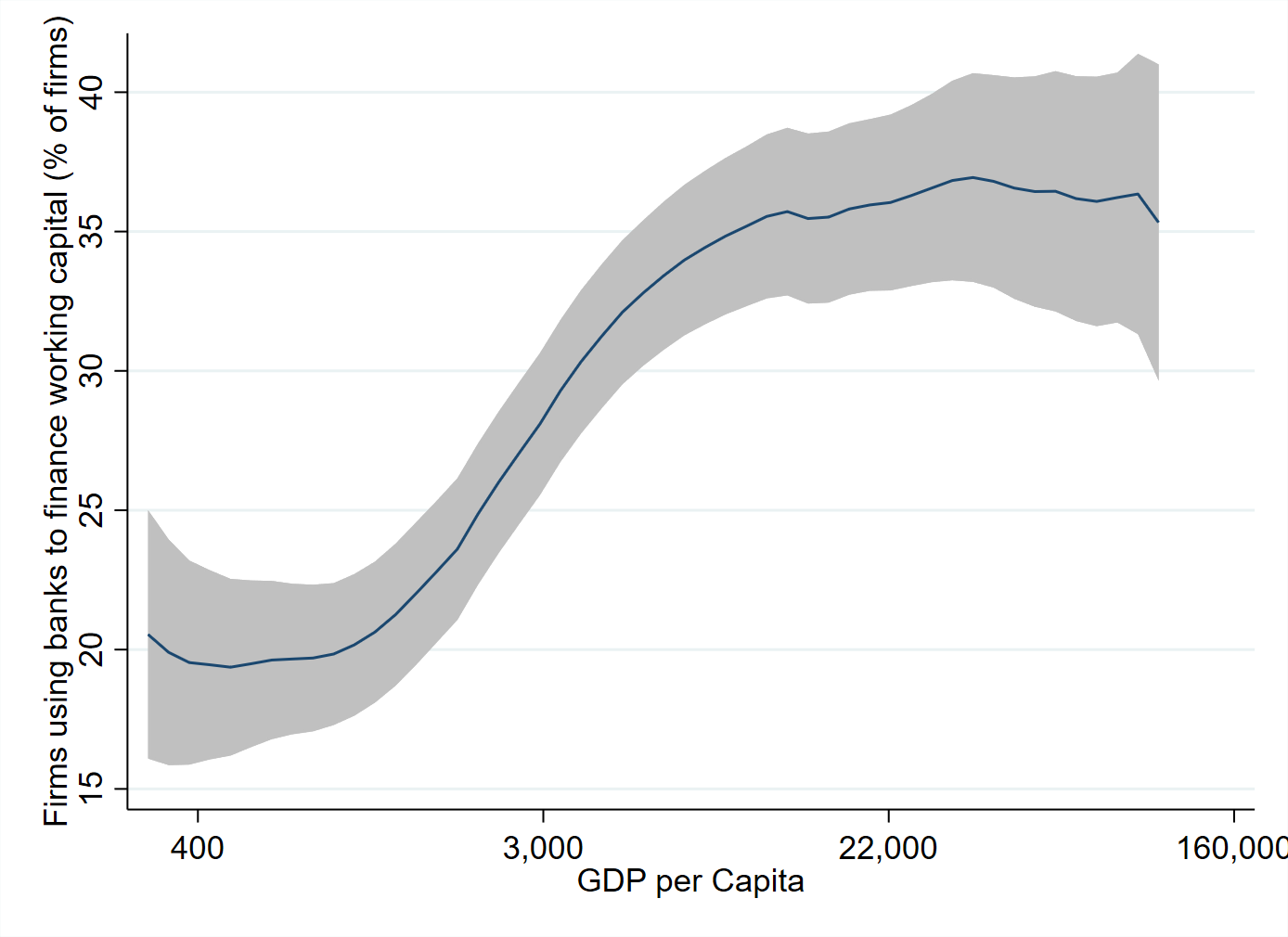}
         \caption{Working Capital}
         \label{fig:Lpoly_Firms_WK_GDP}
     \end{subfigure}
     \hfill
     \begin{subfigure}[b]{0.3\textwidth}
         \centering
         \includegraphics[width=\textwidth]{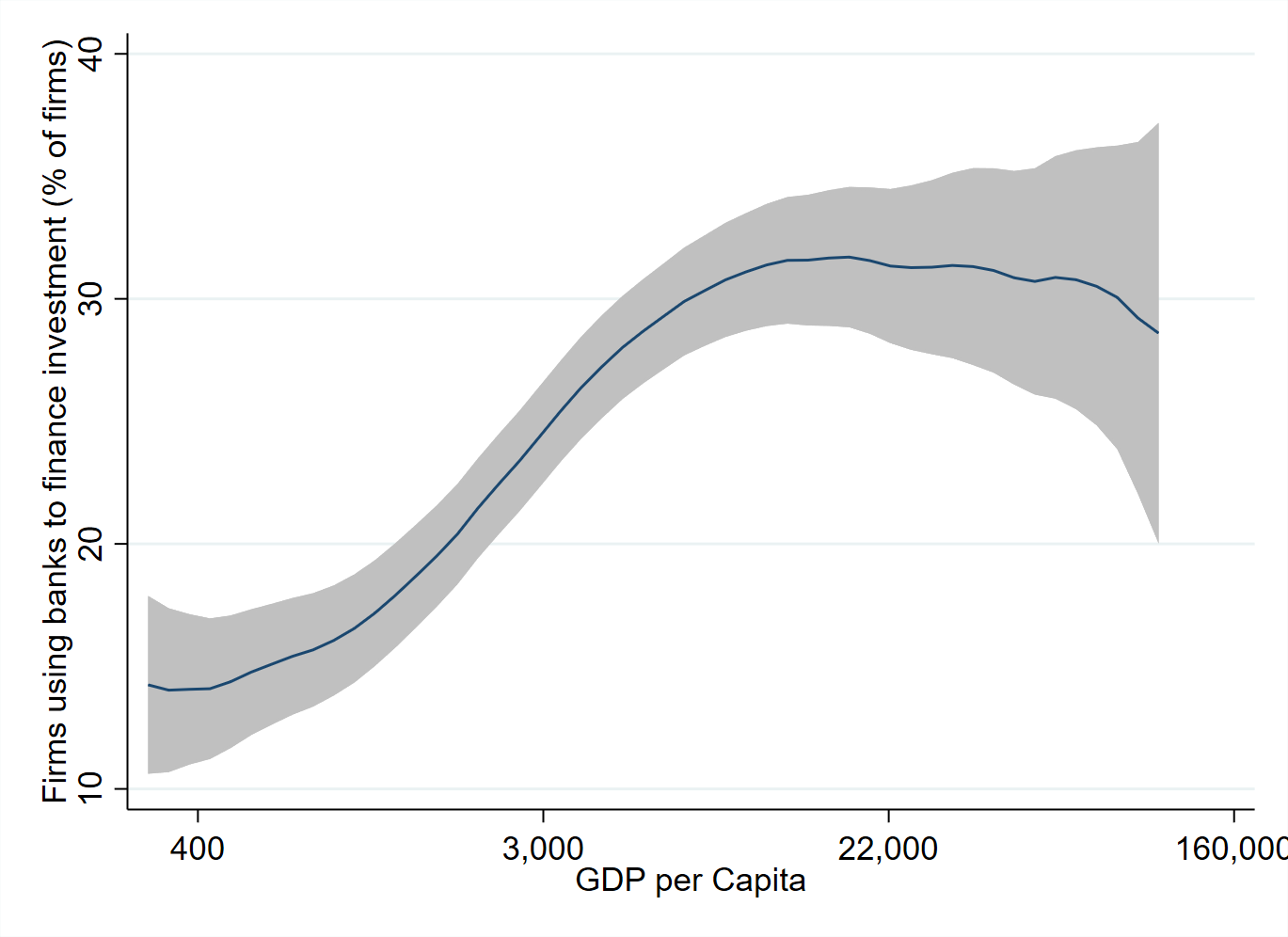}
         \caption{Capital Expenditures}
         \label{fig:Lpoly_Firms_KE_GDP}
     \end{subfigure}
     \hfill
     \begin{subfigure}[b]{0.3\textwidth}
         \centering
         \includegraphics[width=\textwidth]{International_Comparison/Lpoly_Firms_Ratio.png}
         \caption{Ratio}
         \label{fig:Lpoly_Firms_Ratio}
     \end{subfigure}      
     \floatfoot{\textbf{Note:} Figure \ref{fig:Lpoly_Firms_WK_GDP} presents data on the percentage of firms in each country which rely in banks for working capital needs. Figure \ref{fig:Lpoly_Firms_KE_GDP} presents data on the percentage of firms in each country which rely in banks for investment expenditures. Each figure fits a local smooth polynomial to country's real GDP per capital. For each survey question we compute the average response across all ways of the survey. Data on GDP per capita is sourced from the World Bank at constant 2015 USD. Given that waves of the WBES occurred from 2002 to 2020, we compute the average GDP per capita per country for said period.}
\end{figure}

Figure \ref{fig:international_comparison_financing_needs} presents the relationship between the share of firms which rely in banks for working capital and capital expenditure and country's GDP per capita, Figures \ref{fig:Lpoly_Firms_WK_GDP} and \ref{fig:Lpoly_Firms_KE_GDP} respectively.
Across panels, both graphs present a strong and positive relationship between the share of firms which rely in banks and a country's GDP per capita levels. This is in line with the results presented in Figure \ref{fig:lpoly_international}. However, there are some subtle differences across figures. First, across all levels of GDP per capita, the smooth local polynomial for the share of firms for which rely in banks for working capital needs is greater than the one for capital expenditures. This is particular the case for relatively poorer countries. Furthermore, the share of firms which rely in banks for capital expenditures seems to peek somewhere in between levels of GDP per capita between USD 10,000 and USD 15,000, and even seems to decrease for the countries with the highest levels of GDP per capita. This is in with the result presented in Figure \ref{fig:Lpoly_Ratio_GDP} which showed that relatively richer countries rely relatively more in financial markets than banks.

We show that firms in relatively poorer countries rely relatively more in banks for working capital needs than for capital expenditure financing by computing the ratio between the percentage of firms using banks to finance working capital and the percentage of firms using banks to finance investment. Figure \ref{fig:Lpoly_Firms_Ratio} shows that there is a decreasing relationship between this ratio and country's GDP per capita. While it is impossible to control for the fact that firms in richer countries borrow relatively more from financial markets than relatively poorer ones, this result provide supporting evidence that firms in Emerging Markets tend to rely in banks relatively more for working capital needs than for capital expenditure needs. 

\newpage
\section{Additional Stylized Facts} \label{sec:appendix_stylized_facts}

In this Appendix of the paper we present additional evidence which provide further evidence on the role of earning-based debt contracts presented in Section \ref{sec:cash_flow_lending}. 

First, we show that firms, i.e. independent organizations, borrow relatively than independent contractors. To do so, we run a regression of the form
\begin{equation} \label{eq:pfin_gar_type_firm}
    \text{\% of Collateral-Based Debt}_{i} = \gamma \mathbbm{1} \left[Firm\right]_i + \Gamma \text{Sector}_{i} + \epsilon_{i}
\end{equation}
where $\text{\% of Collateral-Based Debt}_{i}$ represents the share of a private agent's total debt which is collateral-based,  $\mathbbm{1} \left[Firm\right]_i$ is an indicator function which takes the value of 1 if the  private agent is a firm and 0 otherwise, and $\text{Sector}_{i}$ is a dummy variable for each sector of activity, i.e., sector fixed effects.
\begin{table}[ht]
    \centering
    \caption{Collateral-Based Lending \& Type of Firm}
    \label{tab:pfin_gar_type_firm}
    \footnotesize
\begin{tabular}{lccc}
 & \multicolumn{3}{c}{Share of Debt which is Collateral-Based} \\ \hline \hline
 &  &  &  \\
$\mathbbm{1} \left[Firm\right]_i$ & -0.350*** & -0.306*** & -1.654*** \\
 & (0.0786) & (0.0791) & (0.106) \\
$\ln \text{Employment}_i$ &  &  & 2.304*** \\
 &  &  & (0.0413) \\
Constant & 10.82*** & 10.80*** & 9.227*** \\
 & (0.0529) & (0.0528) & (0.0609) \\
 &  &  &  \\
Sector FE & No & Yes & Yes \\ 
 &  &  &  \\
Observations & 455,675 & 455,675 & 384,139 \\
R-squared & 0.000 & 0.018 & 0.027 \\ \hline \hline
\multicolumn{4}{c}{ Standard errors in parentheses} \\
\multicolumn{4}{c}{ *** p$<$0.01, ** p$<$0.05, * p$<$0.1} \\
\end{tabular}
\end{table}
The results of estimating Equation \ref{eq:pfin_gar_type_firm} are presented in Table \ref{tab:pfin_gar_type_firm}. Across specifications, firms, identified as independent organizations (excluding households and independent contractors) exhibit lower shares of collateral-based debt. 

Second, we study the relationship between collateral-based debt and firm size. To do so, we run a regression of the form
\begin{equation} \label{eq:reg_pfin_gar_empleo}
    \text{\% of Collateral-Based Debt}_{i} = \gamma \ln \text{Employment}_{i} + \Gamma \text{Sector}_{i} + \epsilon_{i}
\end{equation}
where $\text{\% of Collateral-Based Debt}_{i}$ represents the share of a private agent's total debt which is collateral-based,  $\ln \text{Employment}_i$ is represents firm $i$'s log employment, and $\text{Sector}_{i}$ is a dummy variable for each sector of activity, i.e., sector fixed effects.
\begin{table}[ht]
    \centering
    \caption{Collateral-Based Lending \& Firm Size}
    \footnotesize
    \label{tab:pfin_gar_empleo}
\begin{tabular}{lcccccc}
 & \multicolumn{6}{c}{Share of Debt which is Collateral-Based} \\
 & \multicolumn{2}{c}{Total Firms} & \multicolumn{2}{c}{$L\geq 100$} & \multicolumn{2}{c}{$L\geq 500$} \\ \hline \hline
 &  &  &  &  &  &  \\
$\ln \text{Employment}_{i} $ & 1.892*** & 2.489*** & -1.896*** & -2.469*** & -4.338*** & -5.627*** \\
 & (0.0346) & (0.0523) & (0.385) & (0.396) & (0.943) & (0.970) \\
 &  &  &  &  &  &  \\
Constant & 9.113*** & 7.111*** & 28.08*** & 31.24*** & 45.79*** & 54.91*** \\
 & (0.0583) & (0.127) & (2.155) & (2.210) & (6.723) & (6.909) \\
 &  &  &  &  &  &  \\
Sector FE & No & Yes & No & Yes & No & Yes \\ 
 &  &  &  &  &  &  \\
Observations & 384,139 & 134,790 & 7,237 & 7,237 & 1,259 & 1,259 \\
 R-squared & 0.008 & 0.048 & 0.003 & 0.070 & 0.017 & 0.104 \\ \hline  \hline
\multicolumn{7}{c}{ Standard errors in parentheses} \\
\multicolumn{7}{c}{ *** p$<$0.01, ** p$<$0.05, * p$<$0.1} \\
\end{tabular}
\end{table}
The results of estimating Equation \ref{eq:reg_pfin_gar_empleo} are presented in Table \ref{tab:pfin_gar_empleo}. The first two columns show that for the full sample of firms, there is a positive relationship between firm size and collateral-based lending. However, for a sample of relatively large firms, which focus the vast majority of both total employment and total bank-debt, there is a significant negative relationship between firm size and collateral-based debt. These results present a robustness check to the results presented in Figure \ref{fig:lpoly_cash_flow_size} in Section \ref{sec:cash_flow_lending} which showed an inverted U-shaped relationship between collateral-based lending and firm size. 

Third, in Section \ref{sec:cash_flow_lending} we argued that other papers in the literature which study the composition of firm debt for US firms emphasize the role of age in shaping debt-patterns. While our datasets do not allow us to identify a firms' age, we can construct indirect measures of a firm's age. As described in Appendix \ref{sec:appendix_data_description}, our datasets allow us to identify firms' unique tax identification number. Consequently, across datasets we can observe firms from 1994-2019. Across samples we can identify the first year we \textit{observe} a firm in our samples. This approach is not perfect as it could be the case that the first year we observe a firm coincides with the first year of our dataset.
\begin{table}[ht]
    \centering
    \caption{Share of Credit by Firms' Age}
    \label{tab:share_credit_age}
    \begin{tabular}{l  c }
    Age Category	    &	Share of Total Debt	\\ \hline \hline
    Under 5 years    	&	6.14\%	\\
    Between 5 \& 10	&	14.02\%	\\
    Between 11 \& 20	&	22.13\%	\\
    Higher than 20	    &	57.71\%	\\ \hline \hline
    \end{tabular}
\end{table}
Table \ref{tab:share_credit_age} shows that more than 50\% of total bank-debt is focused on firms which are older than 20 years old, with 46.27\% focused on firms which are \textit{at least} 24 years old (the highest age a firm can be in our sample given the method described above).

\begin{figure}
    \centering
    \caption{Relationship between Collateral-Based Lending \& Firms' Age}
    \label{fig:lpoly_age_figures}
     \centering
     \begin{subfigure}[b]{0.495\textwidth}
         \centering
         \includegraphics[width=\textwidth]{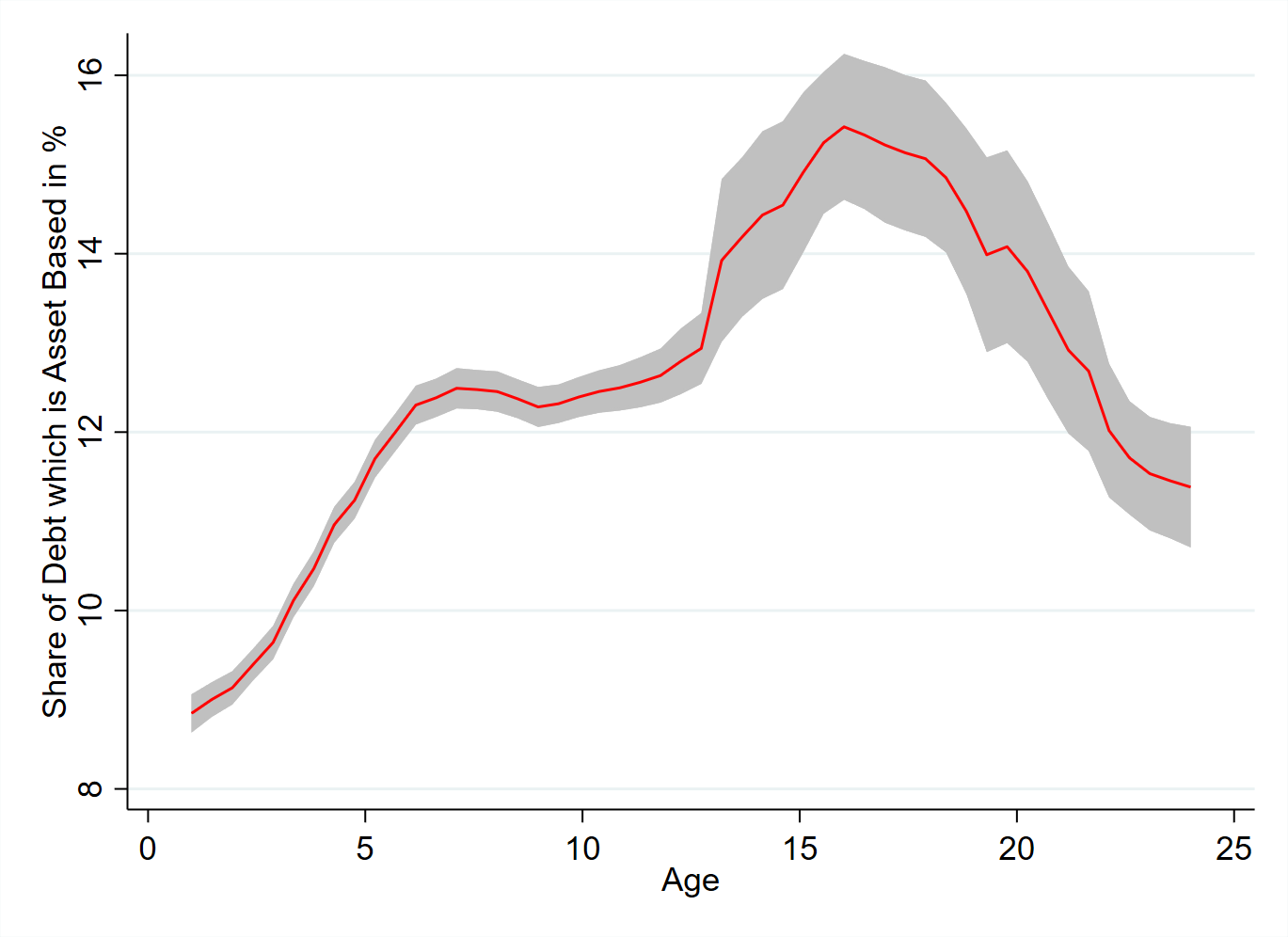}
         \caption{Total Sample}
         \label{fig:lpoly_age}
     \end{subfigure}
     \hfill
     \begin{subfigure}[b]{0.495\textwidth}
         \centering
         \includegraphics[width=\textwidth]{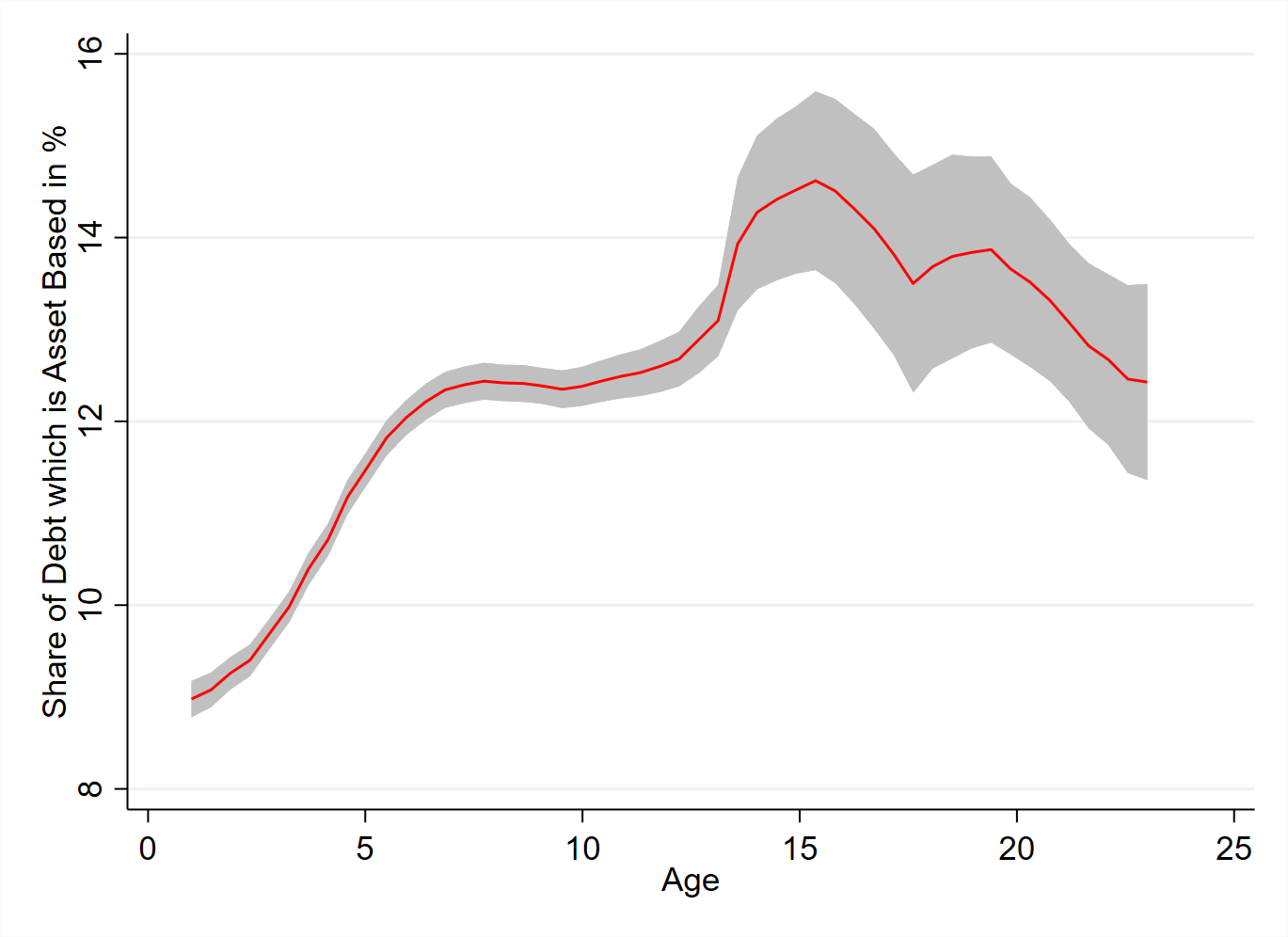}
         \caption{Robustness Check}
         \label{fig:lpoly_age_excluding}
     \end{subfigure}
     \floatfoot{\textbf{Note:} The red full line represents the estimated local polynomial. The grey area represents a 95\% confidence interval. The left panel fits the estimated local polynomial for the whole sample. The right panel fits the estimated local polynomial dropping the firms with age 17 and age 24, which are firms whose first observation coincides with the beginning of our employment and international trade datasets, respectively. The kernel-weighted local polynomial smoothing function is estimated using a Epanechnikov kernel. For the left panel we fit a polynomial of degree 0, a bandwidth of 0.95 and bandwidth for standard error calculation of 1.42. For the right panel we fit a polynomial of degree 0, a bandwidth of 1.17 and bandwidth for standard error calculation of 1.75.}
\end{figure}

In order to test the relationship between firms' age and the share of collateral-based debt we fit a kernel-weighted local polynomial smoothing function, presented in Figure \ref{fig:lpoly_age_figures}.
On left panel, Figure \ref{fig:lpoly_age} presents the fitted polynomial for the whole sample. On the right panel, Figure \ref{fig:lpoly_age_excluding} presents the fitted polynomial for the sample of firms for which age can be computed outside the start of our datasets. In line with the results presented for the relationship between collateral-based debt and firm size, there is an inverted U relationship between collateral-based debt and firm age. The peak of collateral-based lending is reached around the age of 15 years old. Similarly, \cite{lian2021anatomy} find for US firms that the share of collateral-based lending peaks close to 15 years old.

For robustness sake, we estimate the following empirical regression
\begin{equation} \label{eq:pfin_gar_age}
    Y_{i} = \gamma \text{Age}_{i} +  \delta \text{Age}^2_{i} + \Gamma \text{Sector}_{i} + \epsilon_{i}
\end{equation}
where $Y_i$ represents either firm $i$'s share of collateral-based lending or an indicator function which takes the value of 1 if the firm has any collateral-based debt and 0 otherwise. As it was the case for the empirical specifications in Equations \ref{eq:pfin_gar_type_firm} and \ref{eq:reg_pfin_gar_empleo}, $\text{Sector}_{i}$ represent sector fixed effects.
\begin{table}[ht]
    \centering
    \caption{Collateral-Based Lending \& Firm Age}
    \footnotesize
    \label{tab:pfin_gar_age}
\begin{tabular}{lcccc}
 & \multicolumn{2}{c}{Share of Collateral-Based Debt} & \multicolumn{2}{c}{$\mathbbm{1}\left[\text{Collateral-Based Debt}>0 \right]$} \\ 
  & (1) & (2) & (3) & (4) \\  \hline \hline
 &  &  &  &  \\
$\text{Age}$ & 0.755*** & 0.636*** & 0.0134*** & 0.0114*** \\
 & (0.0322) & (0.0321) & (0.000465) & (0.000462) \\
$\text{Age}^2$ & -0.0252*** & -0.0230*** & -7.80e-05*** & -7.65e-05*** \\
 & (0.00153) & (0.00154) & (2.21e-05) & (2.21e-05) \\
 &  &  &  &  \\
Constant & 7.976*** & 8.613*** & 0.114*** & 0.126*** \\
 & (0.136) & (0.136) & (0.00197) & (0.00196) \\
 &  &  &  &  \\
 & No & Yes & No & Yes \\
 &  &  &  &  \\
Observations & 234,708 & 234,708 & 234,708 & 234,708 \\
 R-squared & 0.003 & 0.020 & 0.019 & 0.042 \\ \hline \hline
\multicolumn{5}{c}{ Standard errors in parentheses} \\
\multicolumn{5}{c}{ *** p$<$0.01, ** p$<$0.05, * p$<$0.1} \\
\end{tabular}
\end{table}
Table \ref{tab:pfin_gar_age} presents the results of estimating Equation \ref{eq:pfin_gar_age}. Across specifications and variables $Y_i$, the coefficient on the term $\text{Age}$ is positive and the coefficient on the term $\text{Age}^2$ is negative and statistically significant. 

Next, we turn to estimating how the share of collateral-based debt varies across sectors of activity. To do so, we estimate a regression of the form presented again below
\begin{equation} \label{eq:regression_pfin_gar_sector_appendix}
    \text{\% of Collateral-Based Debt}_{i} = \sum^S_{s} \gamma_s \mathbbm{1}\left[\text{Sector}={s}\right] + \Gamma_i +  \epsilon_{i}
\end{equation}
where $\text{\% of Collateral-Based Debt}_{i}$ represents firm $i$'s share of collateral-based debt, where $\mathbbm{1}\left[\text{Sector}={s}\right]$ is an indicator function if firm $i$ belongs to sector $s$ and zero otherwise, $\Gamma_i$ is a vector of firm level controls (such as log employment, log total bank debt, indicator functions for exporter, importer, age and age squared). 

\begin{table}[ht]
    \centering
    \caption{Share of Collateral-Based Debt across Sectors}
    \tiny
    \label{tab:pfin_gar_sector}
\begin{tabular}{lcccccccc} \hline
 & (1) & (2) & (3) & (4) & (5) & (6) & (7) & (8) \\
 & \multicolumn{8}{c}{Share of Collateral-Based Debt} \\ \hline
 &  &  &  &  &  &  &  &  \\
Fishing & -0.986 & 6.111*** & -2.952 & 4.041* & -4.261 & 1.433 & -6.376 & -10.89 \\
 & (1.790) & (1.877) & (1.988) & (2.217) & (3.399) & (6.449) & (3.894) & (11.83) \\
Mining  & 7.942*** & 9.600*** & 7.283*** & 9.705*** & 3.555** & 6.088** & -4.258** & -3.015 \\
 & (0.874) & (0.894) & (0.989) & (1.071) & (1.686) & (3.042) & (2.132) & (6.292) \\
Manufacturing & -2.486*** & 2.621*** & -2.354*** & 2.152*** & -3.978*** & -2.254 & -6.139*** & -8.938* \\
 & (0.174) & (0.187) & (0.235) & (0.288) & (0.441) & (1.964) & (0.465) & (5.113) \\
Electricity, gas \& water & 4.655*** & 7.527*** & 4.805*** & 8.066*** & 0.0494 & 1.002 & -4.668** & -3.232 \\
 & (0.842) & (0.858) & (0.907) & (0.975) & (1.556) & (2.803) & (2.029) & (5.873) \\
Construction & -2.858*** & 3.307*** & -2.925*** & 3.013*** & -2.381*** & 6.313*** & -4.669*** & 3.591 \\
 & (0.221) & (0.240) & (0.292) & (0.360) & (0.482) & (2.100) & (0.492) & (5.549) \\
Wholesale \& retail & -4.221*** & 1.980*** & -4.695*** & 1.170*** & -5.678*** & -4.604** & -5.836*** & -6.184 \\
 & (0.139) & (0.148) & (0.207) & (0.259) & (0.358) & (2.020) & (0.354) & (5.366) \\
Hotels \& restaurants & -7.742*** & 0.393* & -9.047*** & -0.702 & -7.675*** & -3.258 & -7.835*** & -3.955 \\
 & (0.212) & (0.228) & (0.327) & (0.442) & (0.495) & (2.770) & (0.491) & (7.390) \\
Transportation \& communications & 2.955*** & 8.837*** & 3.395*** & 12.72*** & -0.841** & 17.67*** & -2.120*** & 11.51** \\
 & (0.176) & (0.186) & (0.271) & (0.354) & (0.420) & (2.177) & (0.417) & (5.286) \\
Financial intermediation & -7.843*** & -3.387*** & -9.273*** & -4.553*** & -7.253*** & -14.09*** & -6.530*** & -15.02*** \\
 & (0.384) & (0.396) & (0.616) & (0.751) & (0.938) & (2.838) & (0.977) & (5.770) \\
Real estate & -7.938*** & -1.565*** & -8.494*** & -0.627** & -7.307*** & -3.548* & -7.528*** & -8.158 \\
 & (0.154) & (0.165) & (0.224) & (0.290) & (0.372) & (2.023) & (0.369) & (5.065) \\
Public administration & -4.898*** & -0.351 & -5.020*** & -4.709 & 2.619** & -2.501 & 1.283 & -3.189 \\
 & (0.707) & (1.906) & (0.747) & (5.946) & (1.190) & (7.714) & (1.302) & (8.695) \\
Education services & -8.718*** & -0.843* & -9.416*** & -1.327* & -7.044*** & -6.741*** & -7.726*** & -0.614 \\
 & (0.417) & (0.458) & (0.587) & (0.793) & (0.770) & (2.235) & (0.955) & (7.755) \\
Health \& social services & -6.340*** & -0.607** & -5.880*** & 4.252*** & -5.804*** & -0.203 & -6.784*** & -14.15** \\
 & (0.212) & (0.254) & (0.295) & (0.553) & (0.522) & (2.542) & (0.528) & (6.314) \\
Other services & -8.825*** & 0.248 & -10.37*** & 1.255** & -8.012*** & -3.805 & -8.113*** & -9.208 \\
 & (0.198) & (0.235) & (0.277) & (0.505) & (0.452) & (2.606) & (0.449) & (6.312) \\
exporter &  & -5.252*** &  & -3.954*** &  & -6.197*** &  & -7.212*** \\
 &  & (0.337) &  & (0.366) &  & (1.097) &  & (2.569) \\
importer &  & -5.941*** &  & -4.772*** &  & -2.830*** &  & -3.580 \\
 &  & (0.228) &  & (0.261) &  & (0.988) &  & (2.241) \\
ln\_empleo &  & -0.396*** &  & -0.337*** &  & -3.822*** &  & -5.435*** \\
 &  & (0.0393) &  & (0.0633) &  & (0.443) &  & (1.161) \\
ln\_credit &  & 4.094*** &  & 3.288*** &  & 2.169*** &  & 1.518*** \\
 &  & (0.0203) &  & (0.0297) &  & (0.112) &  & (0.228) \\
Constant & 14.78*** & -34.40*** & 14.95*** & -25.42*** & 13.46*** & 10.07*** & 13.27*** & 37.10*** \\
 & (0.121) & (0.268) & (0.175) & (0.401) & (0.320) & (3.112) & (0.317) & (9.239) \\
 &  &  &  &  &  &  &  &  \\
Observations & 454,640 & 369,924 & 205,845 & 121,129 & 78,343 & 6,637 & 72,365 & 1,014 \\
 R-squared & 0.018 & 0.123 & 0.022 & 0.132 & 0.012 & 0.130 & 0.010 & 0.167 \\ \hline
\multicolumn{9}{c}{ Standard errors in parentheses} \\
\multicolumn{9}{c}{ *** p$<$0.01, ** p$<$0.05, * p$<$0.1} \\
\end{tabular}
\end{table}

Table \ref{tab:pfin_gar_sector_short} presents the results of estimating Equation \ref{eq:regression_pfin_gar_sector_appendix}.
\begin{table}[ht]
    \centering
        \caption{Collateral-Based Lending across Sectors \\ \footnotesize Regression Analysis}
        \scriptsize
    \label{tab:pfin_gar_sector_short}
\begin{tabular}{lcc}
& \multicolumn{2}{c}{Share of Collateral-Based Debt} \\ \hline \hline
Fishing & -0.986 & 6.111*** \\
 & (1.790) & (1.877) \\
Mining & 7.942*** & 9.600*** \\
 & (0.874) & (0.894) \\
Manufacturing  & -2.486*** & 2.621*** \\
 & (0.174) & (0.187) \\
Electricity, gas \& water & 4.655*** & 7.527*** \\
 & (0.842) & (0.858) \\
Construction  & -2.858*** & 3.307*** \\
 & (0.221) & (0.240) \\
Wholesale \& retail  & -4.221*** & 1.980*** \\
 & (0.139) & (0.148) \\
Hotels \& restaurants & -7.742*** & 0.393* \\
 & (0.212) & (0.228) \\
Transportation \& communications & 2.955*** & 8.837*** \\
 & (0.176) & (0.186) \\
Financial intermediation  & -7.843*** & -3.387*** \\
 & (0.384) & (0.396) \\
Real estate, business and rental activities  & -7.938*** & -1.565*** \\
 & (0.154) & (0.165) \\
Public administration & -4.898*** & -0.351 \\
 & (0.707) & (1.906) \\
Education services & -8.718*** & -0.843* \\
 & (0.417) & (0.458) \\
Health \& social services  & -6.340*** & -0.607** \\
 & (0.212) & (0.254) \\
Other Services & -8.825*** & 0.248 \\
 & (0.198) & (0.235) \\
 &  &  \\
Firm Controls & No & Yes \\ 
 &  &  \\
Observations & 454,640 & 369,924 \\ \hline \hline
\multicolumn{3}{c}{ Standard errors in parentheses} \\ 
\multicolumn{3}{c}{ *** p$<$0.01, ** p$<$0.05, * p$<$0.1} \\
\end{tabular}
\end{table}
Results are broadly in line with the results presented in the main body of the paper. 

Finally, we present supporting evidence of the positive firm level correlation between interest payments and cash flows presented in Section \ref{subsec:interest_sensitive_comparison}.
\begin{figure}[ht]
\centering
    \caption{Correlation between Interest Payments \& Cash Flows}
    \label{fig:Correlation_Interest_EBITDA_Firm}
\includegraphics[width=12cm,height=8cm]{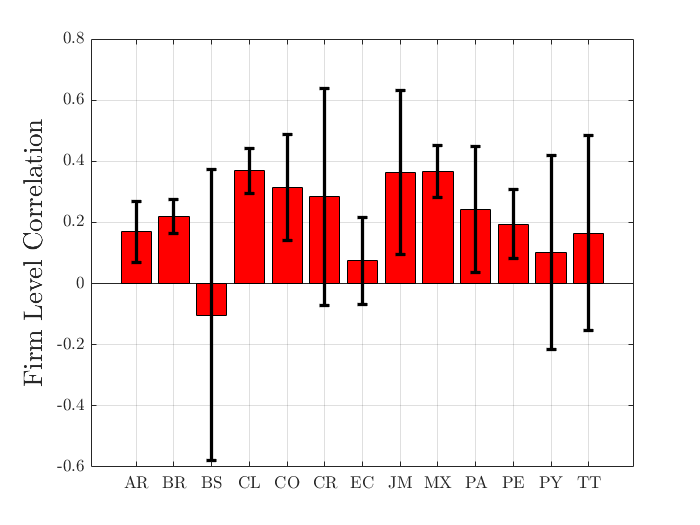}
\end{figure}

\noindent
\textbf{Additional Evidence on the Fall in the Share of Collateral-Based Lending.} In Table \ref{subsec:lending_heterogeneity} we showed: (i) credit lines associated with capital expenditures exhibit a greater share of collateral-based lending while credit lines associated working capital expenditures exhibit a greater share of cash flow-based lending; (ii) firms primarily borrow from banks to finance working capital expenditures. Table \ref{tab:composition_firm_debt_type_appendix} shows the evolution of cash flow-based lending across credit lines for both the year 2000 and 2017.
\begin{table}[ht]
    \centering
    \caption{Cash Flow-Based Debt across Credit Lines \\ \footnotesize Across Time}
    \small
    \label{tab:composition_firm_debt_type_appendix}
    \begin{tabular}{l l c c}
 & & \multicolumn{2}{c}{Share of Cash Flow-Based Debt in \%} \\ 
\multicolumn{2}{l}{Category} & 2000	&	2017	\\ \hline \hline
\multicolumn{2}{l}{Automotive loans} & 61.1	&	4.80	\\
\multicolumn{2}{l}{Machinery \& Equipment loans} & 23.2	&	38.43	\\
\multicolumn{2}{l}{Real estate loans} & 8.3	&	43.05	\\
 &	Dwelling & 2.3	&	48.40	\\
 &	Commercial & 8.4	&	42.82	\\
\multicolumn{2}{l}{Credits for financial leasing} & 81.8	&	47.9	\\
\multicolumn{2}{l}{Un-collateralized or discounted documents} & 	77.8	&	89.3	\\
\multicolumn{2}{l}{Short term credit lines ($<$30 days)} &		82.8	&	92.6	\\
\multicolumn{2}{l}{Financing of working capital for exporting}		& no data	&	96.5	\\
\multicolumn{2}{l}{Credit card debt} & 61.3	& 99.5	\\ \hline \hline
    \end{tabular}
   \floatfoot{\textbf{Note:} This table presents the share of firm-bank debt which is cash flow-based according to different type of debt contracts reported by banks to the Central Bank to meet with current regulation. Data is presented for the years 2000 and 2017. The reasoning behind this choice of years is to not taint our sample with the effects of financial crises, default or severe financial repression episodes. The types of credit lines are in descending order according to the share of cash flow-based contracts in the year 2017. Financial leasing is a debt contract through which a bank acquires a productive asset, which has previously been selected by the tenant, and delivers it to the latter for use in exchange for a fee. There is no data for ``Financing of working capital for exporting'' for the year 2000 as this type of credit line was introduced in the year 2014. We drop debt contract types which are defined as ``\textit{other loans of ...}'' of different nature.}
\end{table}
The takeaway from Table \ref{tab:composition_firm_debt_type_appendix} is that the share of cash flow-based lending has increased significantly during the years 2000 and 2017 across almost all types of credit lines.\footnote{The only exemption is the credit line for ``Automotive loans'' which saw a sharp reduction in the share of cash flow-based lending from 61.1\% in the year 2000 to only 4.80\% in the year 2017. This increase in the prevalence of collateral-based lending may be attributed to changes in the main suppliers of ``Automotive loans'' credit lines in the last two decades. While banks explained the vast majority of ``Automotive loans'' in the year 2000, after the 2001-2002 sovereign default crisis actual car manufacturers started financing and insurance companies which focused on providing financing of automotive loans. A brief description of this type of loans can be found in the following article  \url{https://www.ambito.com/economia/autos/como-comprar-un-auto-credito-prendario-la-modalidad-que-es-furor-el-mercado-automotriz-n5374828}. As the article states ``This type of loan is granted for the purchase of new or used vehicles, and works as follows: the good acquired from the financing, in this case a car, is "pledged" in favor of the entity that grants the loan as collateral until the loans total cancellation. During this period, the borrower is prevented from selling the vehicle until all installments are paid.} Furthermore, the share of cash flow-based debt in credit lines associated to the purchase of real estate and or the purchase of physical capital has increased significantly, explaining close to 50\% in the year 2017. This is in line with evidence presented by \cite{greenwald2018mortgage}, which show for the US that while most of mortgage loans are collateral-based, a significant share of mortgage loans are cash flow-based, i.e. related to a households and/or firms' cash-flows. Overall, this trend suggests the increasing importance of cash flow-based lending in Argentina.

\newpage
\section{Complimentary Stylized Facts} \label{sec:appendix_complimentary_facts}

In this Appendix, we present additional and complementary stylized facts to those presented in the main-body of the paper. 

\subsection{Debt-Patterns across Firms' International Trade Participation } \label{subsec:appendix_debt_patterns_international_trade}

In this Appendix we complement the results presented on firms' debt-patterns across their participation on international trade in Section \ref{subsec:lending_heterogeneity}. While in Section \ref{subsec:lending_heterogeneity} we showed that exporting firms exhibit a greater share of cash flow-based debt, we now turn to analyzing how firms' international trade participation is related to their total bank-debt. To do so, we estimate variations of the following regression
\begin{align} \label{eq:credit_trade}
    \ln \text{Total-Bank-Debt}_{i} = \beta \mathbbm{1}\left[\text{Exporting-Performance}\right]_i + \alpha \mathbbm{1}\left[\text{Importing-Performance}\right]_i + \Gamma_i + \epsilon_i
\end{align}
where $\mathbbm{1}\left[\text{Exporting-Performance}\right]_i$ and  $\mathbbm{1}\left[\text{Importing-Performance}\right]_i$ are indicator functions of a firm's exporting and importing performance, and $\Gamma_i$ is a vector of firm controls.    

Table \ref{tab:credit_trade} presents the results of estimating different variations of Equation \ref{eq:credit_trade}. Column (1) show that exporting firms exhibit a greater amount of total bank-debt than non-exporting firms.
\begin{table}[ht]
    \centering
    \caption{Firm Bank Debt \& International Trade Participation}
    \label{tab:credit_trade}
    \footnotesize
\begin{tabular}{lccccc}
 & \multicolumn{5}{c}{$ \ln \text{Total-Bank-Debt}_{i} $} \\
 & (1) & (2) & (3) & (4) & (5) \\ \hline \hline
 &  &  &  &  &  \\
$\mathbbm{1}\left[\text{Exporter}\right]_i$ & 2.528*** & 1.384*** & 0.825*** &  &  \\
 & (0.0256) & (0.0253) & (0.0273) &  &  \\
$\mathbbm{1}\left[\text{Importer}\right]_i$ &  &  & 0.988*** &  & 0.334*** \\
 &  &  & (0.0184) &  & (0.0704) \\
$\mathbbm{1}\left[\text{Above Mean Exports}\right]_i$ &  &  &  & 1.128*** & 1.150*** \\
 &  &  &  & (0.106) & (0.106) \\
 &  &  &  &  &  \\
Sector Fixed Effects & Yes & Yes & Yes & Yes & Yes \\
Firm Employment      & No  & Yes & Yes & Yes & Yes \\
 &  &  &  &  &  \\ 
Observations & 370,851 & 370,517 & 370,517 & 7,708 & 7,708 \\
$R^{2}$ & 0.079 & 0.188 & 0.194 & 0.357 & 0.359 \\ \hline \hline
\multicolumn{6}{c}{ Standard errors in parentheses} \\
\multicolumn{6}{c}{ *** p$<$0.01, ** p$<$0.05, * p$<$0.1} \\
\end{tabular}
\floatfoot{\textbf{Note:} The table presents results from estimating Equation \ref{eq:credit_trade} using data for the year 2017. Columns (1) through (3) use indicator variables $\mathbbm{1}\left[\text{Exporter}\right]_i$ and $\mathbbm{1}\left[\text{Importer}\right]_i$ which are indicator functions which takes the value of one if the firm has exported or imported in the calendar year and zero otherwise, respectively. Columns (4) and (5) use explanatory variable $\mathbbm{1}\left[\text{Above Mean Exports}\right]_i$ which is an indicator function which takes the value of one if the firm's exported value is greater than the sector's mean. }
\end{table}
Column (2) shows that this result still holds when controlling for firms' employment level (as firm employment and other measures of firm size are positively correlated with firms' exporter status). Column (3) shows that both exporting and importing firms exhibit a greater amount of bank debt. Columns (4) shows that within exporting firms, those with exported value above the sector's mean exhibit a greater amount of bank-debt than firms below the sector's mean. Column (5) shows that this result still holds even when controlling for firms' importing status. This result is not surprising as exporting activities are intensive in credit (see \cite{manova2013credit} and \cite{camara2021crisis}).

\newpage
\section{Additional Details on BCRA Regulations} \label{sec:appendix_details_regulations}

\subsection{Banks' Monitoring Role over Firms} \label{subsec:appendix_BCRA_Regulations_monitoring}

Next, we describe banks' monitoring role over firms in Argentina. As argued in Section \ref{subsec:lending_comparison_international}, the banking and finance literature has found clear evidence of banks' monitoring role and informational advantage. We argue that this role is also played by banks in Argentina.

The monitoring role of banks is clearly stipulated by the Argentinean Central Bank regulations. Two distinct set of regulations address this role. First, bank must update firms' status at the Central Bank of Argentina's ``\textit{Central de Deudores}'' at a monthly frequency. As we described in Section \ref{sec:appendix_data_description}, this implies that banks must report firms' borrowed amounts by type of credit line, whether the debt-contract is in normal situation, if any stipulation in the debt contract has been violated and if so the status of the debt contract.\footnote{This information is easily accessible to the public and can be accessed through the following website
\url{http://www.bcra.gob.ar/BCRAyVos/Situacion_Crediticia.asp}. Furthermore, this dataset provides information on whether firms have recently issued any non-sufficient-fund checks.}  Banks compliance with these informational requirements is part of the BCRA's macro-prudential and credit regulations which banks must report to at a monthly frequency. In particular, banks or any other financial entity under the BCRA, faces capital requirements depending on the riskiness of the assets in their balance sheet.\footnote{See \url{http://www.bcra.gov.ar/Pdfs/Texord/t-capmin.pdf} } If a firm breaches its debt-contract and is categorized under ``Risk status \# 3 or at medium risk of default'', a bank's capital requirements increase in an amount proportional in an amount equal to all of its exposure to said firm (i.e., not only the exposure to the specific debt-contract breached).\footnote{See  BCRA's ``Policy on minimum capital requirements of financial institutions'', Section 2 on ``Capital Requirements due to Credit Risk'', bullet point 2.5.7.} 

Banks monitoring role over firms is also embedded on the BCRA's regulations on banks' credit-policy and debtor classification (or ``\textit{Clasificaci\'on de Deudores}'').\footnote{See \url{https://www.bcra.gob.ar/Pdfs/Texord/t-cladeu.pdf} for the full text.} The regulations stipulate that banks must re-evaluate and re-asses firms' credit status and repayment capacity periodically according to its total amount borrowed.\footnote{See Section 3.2 of BCRA's ``\textit{Clasificaci\'on de Deudores}'': The classification of debtors must be carried out with a frequency that takes into account their importance –considering all financing included–, and in all cases the analysis carried out must be documented.} As we explain in detail in Section \ref{subsec:lending_comparison_international}, these bank assessments focus on firms' cash-flow projections and keeping adequate debt-payment to cash-flow ratios. Under normal circumstances, assessments are carried out quarterly or semi-annually depending on firms borrowing as a share of banks total lending.\footnote{A firm's status and re-assessment must be carried out quarterly if its total borrowing is equal to or higher than 5\% of the bank's net-worth or equal to or higher than 5\% of the bank's total assets. If a firm's borrowing is lower than 5\% of the bank's net-worth or total assets then its credit-status and re-assessment must be carried out semi-annually or twice a year.} However, banks must carry out mandatory and immediate firm re-assessments if: (i) the firm violates any condition of the original debt contract or of any debt renegotiation (such as being late on any payment for a period of time greater than 30 days), (ii) the firm's status with another bank at the BCRA's ``\textit{Central de deudores}'' deteriorates to a status below the current one at the bank.\footnote{See Section 6.4 ``Mandatory reconsideration of debtor classification'' or ``\textit{Reconsideraci\'on obligatoria de clasificaci\'on de deudores}''.}  Consequently, firms are subject to constant monitoring and scrutiny by banks which amplify the impact of any change in firms' financing needs.

Next, the document presents with detailed guidelines on how to classify a debtor according to its risk of default. We produce below the sections of the document which provide key insights on the role of firms' cash flows on determining a debtor's riskiness and or financial situation. 
\begin{itemize}
    \item \underline{Section 6.2} ``Debtors' classification criterion''. In this section of the document, the regulation states that the basic evaluation criterion of a debtor's repayment capacity must be based on the estimated financial flow and, only secondly, based on the liquidation value of the client's assets, given that the debt-financing must respond to the client's true needs. Furthermore, debt-financing should be carried out under amortization conditions in accordance with the real possibilities of repayment that its activity and its cash flow may allow. For the case of debt contracts in foreign currency, the regulation states that debtors should be able to reflect a regular and predictable cash-flow in foreign currency.
    
    \item \underline{Section 6.5.1} ``Classification of a debtor as \textit{normal} or low-risk.'' In this section of the document, regulation state that a debtor's financial situation should be classified as \textit{normal} or low risk if debt structure to cash flow relationship is \textit{adequate}, with cash-flows being able to repay the capital and interests of its obligations with the bank. Even more, the regulation stipulate that a debtor's situation could be classified as \textit{normal} or low-risk if their main sector of activity is projected to report increases in cash-flows.
    
    \item \underline{Section 6.5.3} ``Classification of a debtor as \textit{reporting problems}.'' In this section of the document the regulations state that a debtor's financial situation should be classified as ``reporting problems'' if the analysis of the debtor's cash flow shows that it has problems meeting all of its financial commitments and that, if not corrected, these problems could result in a loss for the financial institution and/or bank. The debtor presents an illiquid financial situation and a level of cash flow that does not allow it to meet the payment of the entire capital and interest on debts, being able to cover only the latter. The debtor shows low cash flows and the projection of its cash flows show a progressive deterioration and a high sensitivity to minor and foreseeable modifications of its own variables and/or of the economic environment, further weakening its payment possibilities.
    
    \item \underline{Section 6.5.4} ``Classification of a debtor as \textit{high-risk} of insolvency and/or default.'' In this section of the document the regulations state that a debtor's financial situation should be classified as ``high-risk of insolvency and/or default''. The analysis of the debtor's cash flow shows that it is highly unlikely that it will be able to meet all of his financial commitments. The debtor's cash of funds is manifestly insufficient, not enough to cover the payment of interest, and it is feasible to presume that it will also have difficulties in complying with possible refinancing agreements.
    
    \item \underline{Section 6.5.5} ``Classification of a debtor as irrecoverable.'' In this section of the document the regulations state that a debtor's financial situation should be classified as ``irrecoverable''. Debts included in this category are considered noncollectable. While these assets might have some recovery value under a certain set of future circumstances, their \textit{uncollectibility} is evident at the time of analysis. Particularly, the debtor's cash flow is not enough to cover production costs. 
\end{itemize}

\subsection{Identification of Cash Flow and Collateral-Based Contracts} \label{subsec:appendix_BCRA_Regulations_ABC}

The Argentinean Central Bank's ``\textit{Central de Deudores}'' datasets described in Section \ref{sec:appendix_data_description} reports whether firms' debt contracts are backed by ``Preferred assets or collaterals'' (``\textit{Garant\'ias Preferidas}'') or not.\footnote{Additional details on Argentinean Central Banks' regulation of collateral-based debt contracts can be found at \url{https://www.bcra.gob.ar/Pdfs/Texord/t-garant.pdf}.} BCRA regulations contemplate two distinct classes of assets which can be considered as collateral: ``Type A'' assets and ``Type B'' assets. On the one hand, ``Type A'' assets are comprised of the transfer or surety of rights with respect to financial assets, equity or other financial documents of any nature that, reliably instrumented, ensure that the bank or lending entity will be able to dispose of the funds by way of cancellation of the obligation contracted by the client, without the need to previously require the debtor to sell the assets or repay the debt. This implies that these assets liquidation depends on solvent third parties or on the existence of sufficiently liquid markets in which the aforementioned assets, securities or documents, or the effects they represent, can be directly liquidated and/or settled.\footnote{Note that this does not imply that the assets or securities' expiration dates coincides with the expiration of the loan or with the committed periodic payments or that the proceeds be applied to the cancellation of the debt or transferred directly to the entity for that purpose.} This asset-classification comprises a diverse type of assets such as cash or highly liquid assets in \textit{pesos} (domestic currency) or several other foreign currencies (taking into account the evolution of their spot market price), gold (taking into account the evolution of its market price), financial securities such as Argentina's sovereign bonds, Central Bank's interest bearing liabilities and private equity securities. 

On the other hand, type B ``preferred assets'' are comprised of assets or rights over assets or commitments of third parties that, reliably instrumented, ensure that the bank or lending entity will be able to dispose of the funds by way of cancellation of the obligation contracted by the client, previously complying with the procedures established for the execution of these assets. Once again, the regulation implies that the comprised assets should efficiently liquidated in transparent and liquid markets. This type of assets comprises claims over real estate (built property land lots and construction trusts), claims over automotive vehicles and agricultural, road and industrial machinery (to the extent that they are registered in the pertinent national registry of automotive property and have a market that allows obtaining a reference value), or a fixed pledge with registration on bovine cattle; financial leasing contracts with respect to the purchase of capital goods and/or other machinery.\footnote{Under financial leasing contracts for firms' purchase of capital goods and/or other machinery, the Argentinean firm finances the purchase of a productive good and keeps claims over the good while the borrowing firms makes use of it for production.} A private agent which borrows backed by these assets must present documentation which prove ownership and provide significant detail on assets. A subset of this documentation is the registration of assets, such as real estate, automotive vehicles, agricultural machinery and cattle stock under public registration entities.\footnote{For instance, real estate ownership and use must be registered in province specific registration entities. For the case of the capital city see \url{https://www.argentina.gob.ar/justicia/propiedadinmueble}. For automotive vehicles (including individual and company owned vehicles) there is a national registry which could be accessed at \url{https://www.dnrpa.gov.ar/portal_dnrpa/}. This registry also incldues agricultural machinery, see \url{http://servicios.infoleg.gob.ar/infolegInternet/anexos/45000-49999/46096/norma.htm}. Cattle stock and other agricultural assets must be declared for both registration purposes and for phytosanitary conditions/regulations. For more information on the registry of cattle stocks see \url{https://www.argentina.gob.ar/senasa/programas-sanitarios/cadenaanimal/bovinos-y-bubalinos/bovinos-y-bubalinos-produccion-primaria/registros-y-habilitaciones/bovinos-y-bubalinos-produccion-primaria/identificacion-animal} and \url{https://www.argentina.gob.ar/senasa/programas-sanitarios/cadenaanimal/bovinos-y-bubalinos/bovinos-y-bubalinos-produccion-primaria/registros-y-habilitaciones/bovinos-y-bubalinos-produccion-primaria/identificacion-animal}. For the registry of fishery machinery and equipment see \url{https://www.argentina.gob.ar/palabras-clave/registro-de-la-pesca}.}

The Central Bank's regulation stipulates constraints on how much firms can borrow given the ``preferred assets'' that back debt contracts.\footnote{These constraints on the amount firms can borrow as a function of the value of ``preferred assets'' are usually referred within the Central Bank's regulation as ``m\'argenes de cobertura''.}  Table \ref{tab:bounds_preffered_assets_appendix} shows bounds as a percentage of asset values for several types of both Type ``A'' and Type ``B'' preferred assets debt contracts.
\begin{table}[ht]
    \centering
    \caption{Collateral-Based Bank Debt \\ \footnotesize Bounds and Borrowing Limits}
    \label{tab:bounds_preffered_assets_appendix}
    \footnotesize
    \begin{tabular}{l l c}
                         &  &  Bound as \% of Asset Value \\ \hline \hline
    \multicolumn{3}{c}{ }  \\
    \multicolumn{2}{l}{\underline{Preferred Type ``A'' assets}} &  \\
     & Cash or Highly-Liquid Domestic Currency assets & 100\% \\
     & Cash or Highly-Liquid Foreign Currency assets & 80\% \\
     & Gold & 80\% \\
     & Sovereign Bonds & 75\% \\
     & Central Bank Liabilities & 100\% \\
     & Private Equity Claims    & 70\% \\     
    \multicolumn{3}{c}{ }  \\
    \multicolumn{2}{l}{\underline{Preferred Type ``B'' assets}} &  \\     & Real Estate  & 50\%-100\% \\   
       & Automotive vehicles    \& agricultural machinery & 60\%-75\% \\
       & Road \& industrial machinery        & 60\% \\
       & Cattle stock                        & 60\% \\
    \multicolumn{3}{c}{ }  \\
     \hline \hline 
    \end{tabular}
    \floatfoot{\textbf{Note:} The table presents bounds on a subset of assets which are usually considered as collateral of private agents'. Consequently, the table is not exhaustive. All bounds are computed as a fraction of current market value of the assets. Asset types with bounds on the value which can back debt contracts represent heterogeneity in bounds within asset types. For instance, how much an agent can borrow backed by real estate depends on whether the property is used as living place or not, whether the property is an empty lot, a lot with construction built on it and/or an agricultural lot.}
\end{table}
The top panel shows the bounds on Type ``A'' assets. How much a private agent can borrow is higher for financial assets in domestic currency than for financial assets in foreign currency. Additionally, private agents can pledge a higher fraction of more liquid assets, such as cash or central bank liabilities, than of relatively less liquid assets such as sovereign bonds and/or private equity. The bottom panel shows private agents' bounds for several Type ``B'' assets. Depending on the type of asset and the debtor's credit-worthiness (more on this in the next section) these debt contracts exhibit bounds between 60\% and 100\% of the market value of the Type ``B'' of preferred assets.

\section{Evidence of Binding Borrowing Constraints \& its Implications} \label{sec:interest_sensitive_binding_borrowing}

This previous analysis in this section characterizes firms' borrowing constraints according to BCRA regulations and evidence from corporate bonds. A natural follow up question is whether these borrowing constraints bind, and if so, what are the implications of binding borrowing constraints. In this section, we present evidence that violations of these borrowing constraints around the the Argentinean 2001-2002 sovereign default crisis have a significant impact on firms access to banking debt. Furthermore, we argue that this drops is quantitatively larger for firms which borrowed from banks that are relatively more exposed to the crisis.

\noindent
\textbf{Macroeconomic background.} The 1990s were a period of drastic reforms in Argentina. First, in 1991 a currency board was implemented which pegged the peso to the US dollar at \$1 peso = US\$ 1 dollar. Second, the government adopted a set of market-oriented reforms, privatizations of public enterprises, and a re-structuring of their sovereign debt as part of the Brady plan. The banking industry experienced the implementation of international regulations such as deposit insurance and Basel guidelines, and the entry of foreign institutions through mergers and privatizations. These reforms were able to tame inflation and lead to an expansion of output and credit intermediation. 

\begin{figure}[ht]
    \centering
    \includegraphics[width=16cm,height=12cm]{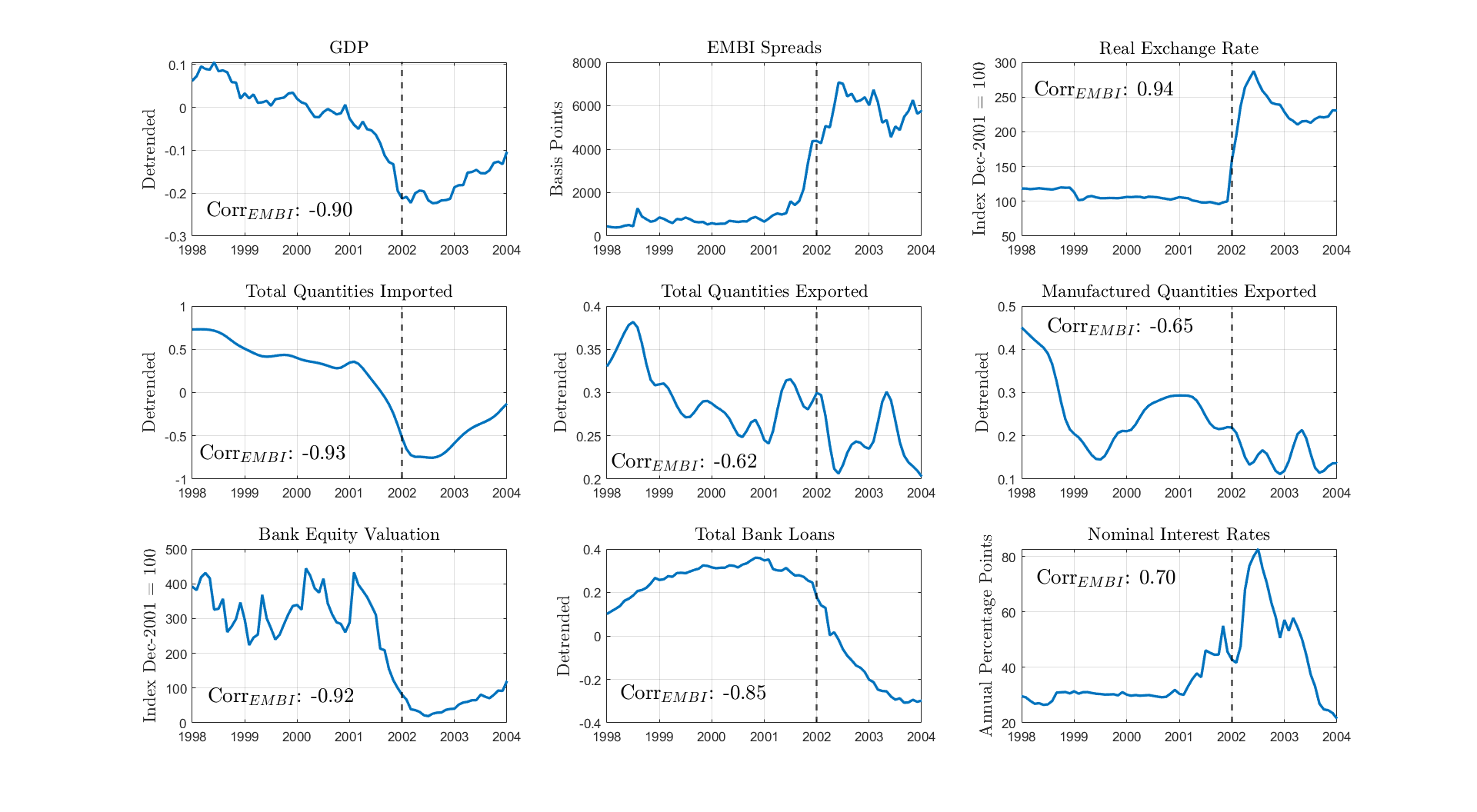}
    \caption{Macroeconomic \& Financial Dynamics around the Sovereign Crisis}
    \label{fig:Macro_Crisis_2001}
    \floatfoot{\textbf{Note:} There are 9 sub-figures arranged in three rows and three columns. Ordered from top to bottom and from left to right the variables are: (i) GDP, (ii) EMBI spreads, (iii) Real exchange rate with the US dollar, (iv) total quantities imported, (v) total quantities exported, (vi) manufacturing quantities exported, (vii) the mean bank equity valuation, (vii) total real bank loans, (ix) the nominal interest rate.}
\end{figure}

Figure \ref{fig:Macro_Crisis_2001} shows the dynamics of key macroeconomic variables around the crisis.
The economy as well as the growth of private credit slowed down and the peso appreciated noticeably late in 1998 in response to the East Asian and Brazilian crisis. Sovereign spreads increased significantly which lead to a drop in GDP close to 5\% between 1999 and 2000. In March 2001, a crisis within the president’s cabinet induced the first run on bank deposits, debilitating banks' health and curtailing private credit. Figure shows that the domestic banking industry exhibited a significant exposure to both sovereign risk (measured as the fraction of sovereign bonds to total assets) and to the risk of a devaluation (measured as the fraction of foreign currency liabilities to total assets).
\begin{figure}[ht]
    \centering
    \caption{Banking Industry Pre Crisis Exposure }
    \label{fig:bank_exposure_crisis}
     \centering
     \begin{subfigure}[b]{0.495\textwidth}
         \centering
         \includegraphics[height=6cm,width=8cm]{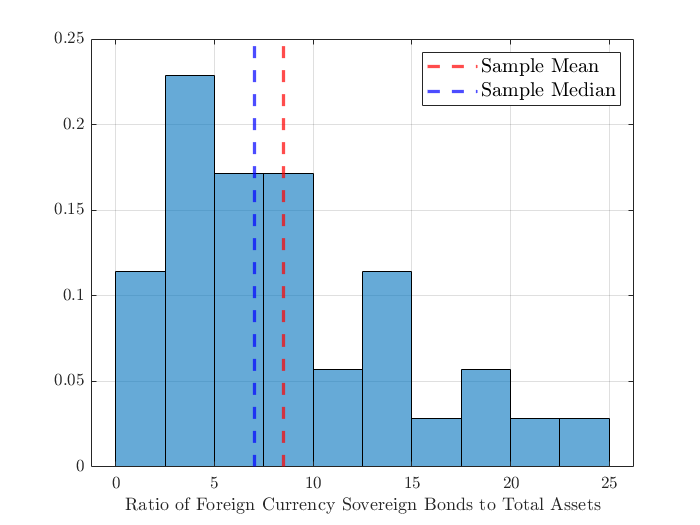}
         \caption{Exposure to Sovereign Risk}
         \label{fig:Histogram_titulos_publ_FC_a_2001}
     \end{subfigure}
     \hfill
     \begin{subfigure}[b]{0.495\textwidth}
         \centering
         \includegraphics[height=6cm,width=8cm]{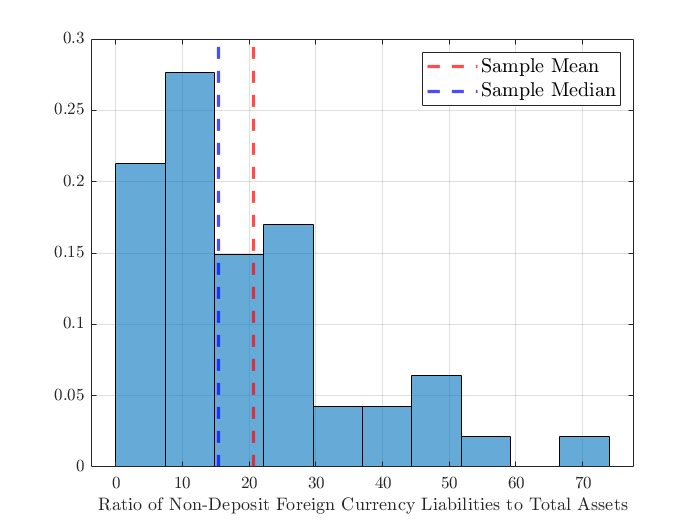}
         \caption{Exposure to Devaluation Risk}
         \label{fig:Histogram_non_dep_liab_FC_a_2001}
     \end{subfigure} 
     \floatfoot{\textbf{Note:} Figures are constructed using the simple average for the year 2001. The left panel presents the results for banks' exposure to sovereign risk proxied as the ratio of foreign currency sovereign bonds to total assets. The right panel presents the results for banks' exposure to a devaluation risk proxied as the ratio of non-deposit foreign currency liabilities to total assets. Each plot shows the correlation between each of the variables and the EMBI spreads.}
\end{figure}
By mid 2001, spreads on government debt increased to over 1300 basis points over U.S. Treasuries and bank deposits decline was greater than 10\%. During the year 2001 alone, banks valuation decreased by 50\%, the nominal lending interest rate to firms doubled from 30\% to 60\%, and total bank loans decreased by close to 15\%. In order to stop the run on bank deposits the government announced a ``\textit{Corralito}'' through which it froze all bank accounts on December 2001.\footnote{The initial measure stated that the freeze on all bank accounts would only last 90 days.} Only a small amount of cash was allowed for withdrawal on a weekly basis, and only from deposit accounts denominated in pesos.

The Argentinean economy collapsed due to the drastic reduction in bank and overall financial intermediation. The government announced the default on sovereign debt on December 26 2001 and formally abandoned the currency board on January 7, 2002. By February 2002, a dirty-float exchange rate regime was implemented which resulted in a that led to an increase in the nominal exchange rate close 250\% in 2 months. Together with the new exchange rate regime, the government announced the ``\textit{Corral\'on}''. The first part of this economic program was the announcement of a “\textit{pesification}”, i.e., the conversion of financial assets and liabilities denominated in US dollars to pesos at different rates.\footnote{All dollar denominated deposits were converted to pesos at the exchange rate of \$1.4 = US\$1. All loans to the private sector were converted at the rate of \$1 = US\$1. Sovereign bonds and public credit were converted at \$1.4 = US\$1.} Second, this program maintained the freeze on deposits which lasted until December 2002.  

In summary, in detrended terms GDP fell close to 30\% between its peak in 1998 and the 2002-2003. EMBI spreads would remain at all time high during 2003 and 2004, up until the sovereign debt re-structuring in the year 2005. Domestic financial conditions continued significantly tight post-default and devaluation as banks valuations and total loans remained significantly below pre-crisis levels.

\noindent
\textbf{Evidence on the violation of borrowing constraints.} Our credit registry data set allows us to observe the status of firms' credit lines. The status is defined as one of the six categories of riskiness presented in Section \ref{sec:interest_sensitive_borrowing_constraints} and described in greater detail in Appendix \ref{sec:appendix_details_regulations}. In particular, we group firms' debt into three financial situations or status: (i) ``non-violation'' of borrowing constraint or ``normal situation'', (ii) ``violation'' of borrowing constraint situation'' or firms in ``medium risk of default'' as defined in Section \ref{sec:interest_sensitive_borrowing_constraints}, (iii) ``written off' or ``irrecoverable'' as defined in Section \ref{sec:interest_sensitive_borrowing_constraints}.

\begin{figure}[ht]
    \centering
    \caption{GDP \& Borrowing Constraint Violations}
    \label{fig:GDP_violations}
     \begin{subfigure}[b]{0.495\textwidth}
         \centering
         \includegraphics[width=\textwidth]{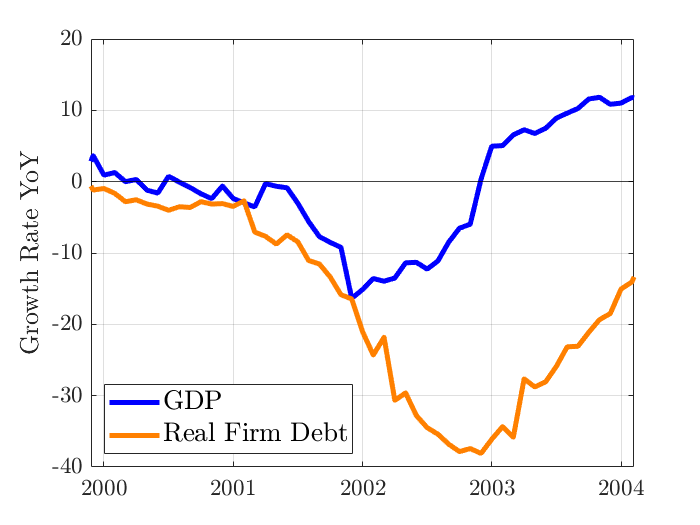}
         \caption{GDP \& Firm Debt Dynamics}
         \label{fig:GDP_violations_GDP}
     \end{subfigure}
     \hfill
     \begin{subfigure}[b]{0.495\textwidth}
         \centering
         \includegraphics[width=\textwidth]{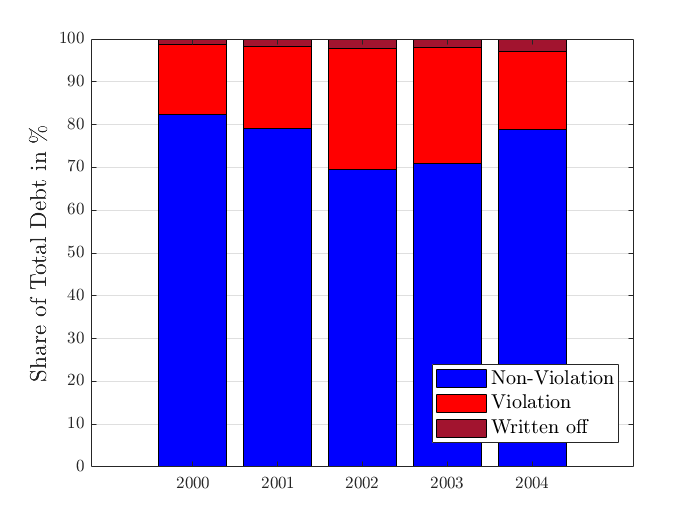}
         \caption{Borrowing Constraint Violations}
         \label{fig:GDP_violations_violations}
     \end{subfigure} 
\end{figure}

We now document a novel fact of firms' bank debt around sovereign default crisis: a significant share of firm debt breach their debt contracts and/or borrowing constraints, particularly around sovereign default. Figure \ref{fig:GDP_violations} presents evidence of this fact around the 2001-2002 crisis. On the left panel, Figure \ref{fig:GDP_violations_GDP} shows the dynamics of GDP and firm debt from the year 2002 to 2004.\footnote{In particular, Figure \ref{fig:GDP_violations_GDP} presents a close up of the dynamics presented in Figure \ref{fig:Growth_Rates} around the sovereign default crisis.} On the right panel, Figure \ref{fig:GDP_violations_violations} presents the share of total debt, measured in volume, according to the three financial situations defined above.
The first conclusion from this figure is that a significant share of firms' debt contracts is in violation. For instance, close to 18\% of firms' debt contracts, measured in volume, are in violation during the year 2000. The second conclusion, is that debt contract violations are pro-cyclical, i.e., as the sovereign crises lead to an economic depression, the share of total debt in violation almost doubled to more than 30\% by the year 2002. 

The slow recovery of firm credit after the sovereign crisis is in line with the persistent increase in the share of firm debt under contract violation after the crisis. On the left panel, Figure \ref{fig:GDP_violations_GDP} shows that GDP was already growing, year-to-year, by the end of 2002. Even as GDP reached double digits growth rate during the years 2003 and 2004, firm debt continued exhibiting a negative growth rate during this period. On the right panel, Figure \ref{fig:GDP_violations_violations} shows that after peaking during the year 2002, the share of firm debt remained high during the years 2003 and 2004, compared to the pre-crisis year 2000. 

\noindent
\textbf{Research designs.} Our research design builds on banks' heterogeneous exposure to sovereign and devaluation risk before the crisis. The financial crisis brought a complete collapse of banks and overall financial intermediation. However, banks exposure to sovereign and foreign currency liabilities differed significantly across banks, as Figure \ref{fig:bank_exposure_crisis} shows. For instance, Figure \ref{fig:Histogram_titulos_publ_FC_a_2001} shows that the mean ratio of sovereign bonds to total assets in the year 2001 was 8\% some banks exhibited ratios above 20\%. Figure \ref{fig:Histogram_non_dep_liab_FC_a_2001} shows a similar heterogeneity on banks' exposure to foreign currency liabilities. 

Our research designs exploits the heterogeneity in banks' exposure to the financial crisis before the sovereign default as a source exogenous variation in firms' availability of credit. Furthermore, we assess whether banks' exposure to the financial crisis affects in particular the access to credit of firms which have breached their debt contract. We test this hypothesis through regressions similar to those proposed by \cite{chodorow2022loan}
\begin{align} \label{eq:regression_covenant}
    L_{i,b,t} &= \beta_0 + \beta_1 \left[\text{Exposed Bank}_b\right] + \beta_2 \left[\text{Breached Contract}_{i,t-1,t}\right] + \nonumber \\
    & \quad + \beta_3 \left[\text{Exposed Bank}_b\right] \times + \left[\text{Breached Contract}_{i,t-1,t}\right] \\
    & \quad + \gamma'X_{i,b,t} + \epsilon_{i,b,t} \nonumber
\end{align}
where $L_{i,b,t}$ denotes the amount of bank debt of firm $i$ from bank $b$ in period $t$, $\left[\text{Exposed Bank}_b\right]$ measures the health of bank $b$, $\left[\text{Breached Contract}_{i,t-1,t}\right]$ is a measure of the health of the debt contract between firm $i$ and bank $b$, and $X_{i,b,t}$ includes any covariates. 

\begin{figure}[ht]
    \centering
    \caption{Exposure to Financial Crisis \& Post Default Lending \\ \footnotesize Bank Level Analysis}
    \label{fig:scatter_growth_f1}
     \centering
     \begin{subfigure}[b]{0.495\textwidth}
         \centering
         \includegraphics[height=6cm,width=8cm]{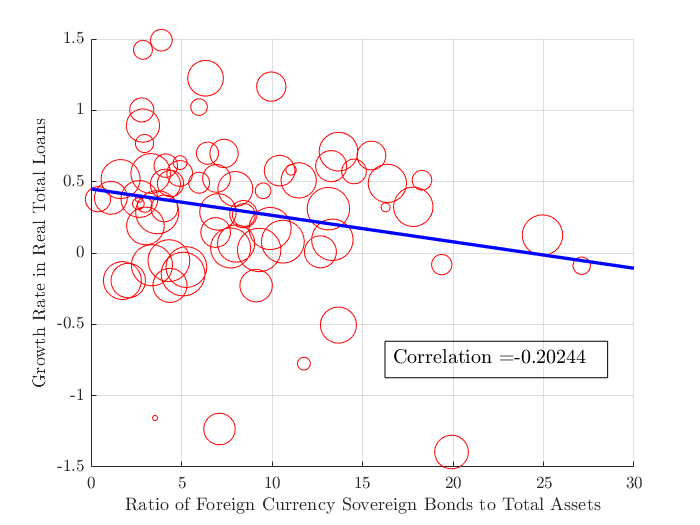}
         \caption{Exposure to Sovereign Risk}
         \label{fig:scatter_titulos_publ_FC_a_2001}
     \end{subfigure}
     \hfill
     \begin{subfigure}[b]{0.495\textwidth}
         \centering
         \includegraphics[height=6cm,width=8cm]{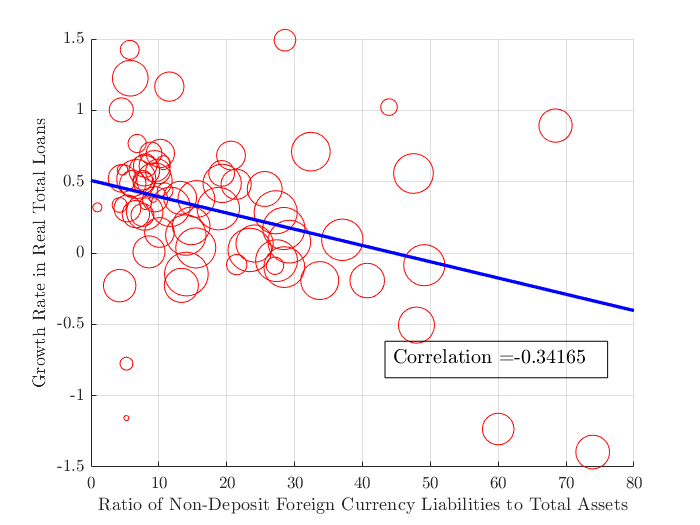}
         \caption{Exposure to Devaluation Risk}
         \label{fig:scatter_non_dep_FC_a_2001}
     \end{subfigure} 
     \floatfoot{\textbf{Note:} On the $x$-axis, the indicators of exposure to sovereign risk and devaluation risk are constructed as the simple average for the year 2001. On the $y$-axis the growth rate of total bank level loans is constructed as $\left(L_{b,2004}-L_{b,2003}\right)/ \left[0.5\times\left(L_{b,2004}-L_{b,2003}\right)\right]$.The size of the markers reflects banks' size measured as the value of their assets in the year 2001.}
\end{figure}

First, we show how different dimensions of banks' exposure to the financial crisis affected their aggregate loan performance and the construction of variable $\left[\text{Exposed Bank}_b\right]$. We begin by showing how banks' exposure to sovereign and devaluation risk \textit{before} the crisis impacted their lending performance \textit{after} the government announced the sovereign default and the end of the currency board. Figure \ref{fig:scatter_growth_f1} shows banks' growth rate in real lending after the \textit{de-freezing} of deposits in late 2002 plotted against their \textit{pre-crisis} exposure to sovereign and devaluation risk.
Figures \ref{fig:scatter_titulos_publ_FC_a_2001} and \ref{fig:scatter_non_dep_FC_a_2001} show evidence that banks which exhibited a greater exposure to sovereign and devaluation risk before the crisis experienced a larger drop in their lending once banks re-opened. 

We construct our indicator of banks' exposure to the financial crisis following \cite{chodorow2022loan}. We combine the two indicators of banks' exposure to the financial crisis presented above, extract the first principal component and construct a rank-normalized variable as the rank of the first principal component divided by the number of banks in the sample.\footnote{This implies that once the principal component has been recovered and predicted for each bank $b$, the banks are ranked according to the value of their predicted principal component with the highest value awarded the top ranking and the lowest value awarded the bottom ranking. Finally, the normalized ranking is computed as $\left(r_b-1\right)/(R-1)$ where $r_b$ is bank $b$'s ranking and $R-1$ is the total amount of observations. }
\begin{figure}[ht]
    \centering
    \caption{Change in Share of Total Loans \& Exposure to the Crisis \\ \footnotesize According to $\text{Exposed Bank}_b$}
    \includegraphics[width=10cm,height=7cm]{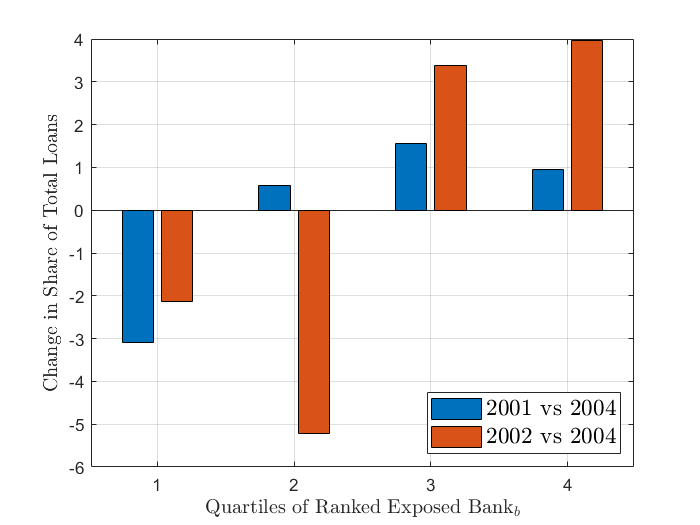}
    \floatfoot{\textbf{Note:} Figure reports the change in the share of total loan supply from banks in the four quartiles of the constructed variable $\text{Exposed Bank}_b$. The first bar or blue bar for each quartile represents the change between 2001 and 2004, while the second or red bar presents the change between between 2002 and 2004.}
    \label{fig:Change_Shares_Banks}
\end{figure}
Figure \ref{fig:Change_Shares_Banks} shows the change in the share of aggregate bank loans across the fourth quartiles of variable $\text{Exposed Bank}_b$ between 2001 and 2004. This figure shows that relatively more exposed banks exhibited a large drop in their share of total loans while those less exposed exhibited a significant increase. 

Estimating Equation \ref{eq:regression_covenant} with ordinary least squares (OLS) raises two main concerns. First, firms that breach their debt contracts and borrow from relatively more exposed banks may differ along other dimensions from firms that breach their debt contracts and have healthier lenders. The second concern is that a firm breaching its debt contract may also correlate with other firm characteristics and bank debt may also depend on the interplay between a banks exposure to the crisis and these other characteristics.

\newpage
\section{Additional Details on Section \ref{sec:model_simple} Model} \label{sec:appendix_details_simple_model}

\subsection{Model with Cash Flow \& Collateral Constraints} \label{subsec:appendix_real_model_collatera}

\subsubsection{Households} \label{subsec:trend_quant_households_wk}

The household seeks to maximize the utility function
\begin{align*}
    \mathbb{E}_0 \sum^{\infty}_{t=0} \beta^{t} \frac{\left(C_t - \zeta X_{t-1} n^{\theta}_{t}\right)^{1-\sigma}-1}{1-\sigma} 
\end{align*}
where $X_t$ is a trend productivity component, $C_t$ is consumption, $n_t$ is labor effort, $\beta$ is a discount factor, $<0\beta<1$. Every period, households face a budget constraint
\begin{align*}
    W_t n_t + B_{t-1}+s_{t-1}\left(D_t + P_t\right) = \frac{B_t}{1+r_t} + s_t P_t + C_t + T_t
\end{align*}
where $W_t$ is the wage in terms of the numeraire goods, $B_{t-1}$ is the stock of freely traded bonds, $s_t$ is the household's shares of domestic firms, $D_t$ is the paid dividend and $P_t$ is the price of the shares. Taxation $T_t$, which is the household takes as given, is used to finance the firms' tax benefit enjoyed by firms when borrowing
\begin{align*}
    T_t = \frac{B^{f}}{R_t}-\frac{B^{f}}{1+r_t}
\end{align*}
where $B^{f}_t$ represents the firm's stock of debt and $R_t = 1 + r_t \left(1-\tau\right)$.

The household's Lagrangean can be written down as
\begin{align*}
    \mathcal{L}:& \quad \beta^{t} \frac{\left(C_t - \zeta X_{t-1} n^{\theta}_{t}\right)^{1-\sigma}-1}{1-\sigma} + \beta^{t} X^{-\sigma}_{t-1} \lambda_t \left[W_t n_t + B_{t-1} + s_{t-1} \left(D_t + P_t\right) - \frac{B_t}{1+r_t}-s_t P_t - C_t - T_t \right]
\end{align*}
The first order conditions with respect to $C_t$, $n_t$, $B_t$ and $s_t$ are
\begin{align*}
    \frac{\partial \mathcal{L}}{\partial C_t}:& \quad \left(\frac{C_t}{X_{t-1}} - \zeta n^{\theta}_t \right)^{-\sigma} = \lambda_t \\
    \frac{\partial \mathcal{L}}{\partial n_t}:& \quad \left(\frac{C_t}{X_{t-1}} - \zeta n^{\theta}_t \right)^{-\sigma} \zeta \theta n^{\theta-1}_t = \lambda_t \frac{W_t}{X_{t-1}} \\
    \frac{\partial \mathcal{L}}{\partial B_t}:& \quad 1 = g^{-\sigma}_t \left(1+r_t\right) \beta \mathbb{E}_t \frac{\lambda_{t+1}}{\lambda_{t}} \\
    \frac{\partial \mathcal{L}}{\partial s_t}:& \quad P_t = g^{-\sigma}_t \beta \mathbb{E}_t \frac{\lambda_{t+1}}{\lambda_{t}} \left(D_{t+1} + P_{t+1} \right)
\end{align*}
Note that $C_t$, $W_t$, $P_t$, and $D_t$ exhibit trend growth. We define $c_t \equiv C_t / X_{t-1}$, $w_t \equiv W_t / X_{t-1}$, $p_t \equiv P_t / X_{t-1}$ and $d_t \equiv D_t / X_{t-1}$ and re-write the first order conditions in terms of detrended variables as
\begin{align}
    \frac{\partial \mathcal{L}}{\partial C_t}:& \quad \left(c_t - \zeta n^{\theta}_t \right)^{-\sigma} = \lambda_t \\
    \frac{\partial \mathcal{L}}{\partial n_t}:& \quad \left(c_t - \zeta n^{\theta}_t \right)^{-\sigma} \zeta \theta n^{\theta-1}_t = \lambda_t w_t  \\
    \frac{\partial \mathcal{L}}{\partial B_t}:& \quad 1 = g^{-\sigma}_t \left(1+r_t\right) \beta \mathbb{E}_t \frac{\lambda_{t+1}}{\lambda_{t}} \\
    \frac{\partial \mathcal{L}}{\partial s_t}:& \quad p_t = g^{1-\sigma}_t \beta \mathbb{E}_t \frac{\lambda_{t+1}}{\lambda_{t}} \left(d_{t+1} + p_{t+1} \right)
\end{align}
Note that if we set the growth rate of trend productivity equal to 1, i.e. $g_t = 1 \Rightarrow X_t = X_{t-1} \forall t$, we obtain the same first order conditions than in the model without trend growth. 

\subsubsection{Firms} \label{subsec:trend_quant_firms_wk}

Firms face the same frictions as in \cite{jermann2012macroeconomic} but in a small open economy with productivity trend growth as in \cite{garcia2010real}. At any time $t$, a representative firm produces by hiring labor $n_t$, buying imported intermediate inputs $M_t$, and using accumulated capital $K_{t-1}$, using a Cobb-Douglas production function
\begin{align*}
    Y_t = a_t K^{\alpha}_{t-1} M^{\beta}_t \left(X_t n_t\right)^{1-\alpha-\beta}
\end{align*}
where $Y_t$ denotes output in period $t$, $K_{t-1}$ denotes the stock of capital installed in period $t$ which is used for production in period $t$, $n_t$ denotes hired labor, and $a_t$ and $X_t$ represent productivity shocks.

The productivity shock $a_t$ is assumed to follow a first-order auto-regressive process in logs. That is,
\begin{align}
    \ln a_t = \rho_a \ln a_{t-1} + \epsilon^{a}_t, \quad \epsilon^{a}_t \sim \mathcal{N} \left(0,\sigma^{a}\right)
\end{align}
The productivity process shock $X_t$ is non-stationary. Let
\begin{align*}
    g_t \equiv \frac{X_t}{X_{t-1}}
\end{align*}
denotes the gross growth rate of $X_t$. We assume that the logarithm of $g_t$ follows a first-order auto-regressive process of the form
\begin{align*}
    \ln \left(\frac{g_t}{\bar{g}}\right) = \rho_{g} \ln \left(\frac{g_{t-1}}{\bar{g}}\right) + \epsilon^{g}_{t}, \quad \sim \mathcal{N} \left(0,\sigma^{g}\right)
\end{align*}
Capital accumulation is subject to investment adjustment costs
\begin{align*}
    K_t = \left(1-\delta \right) K_{t-1} + I_t \left[1-\frac{\phi}{2} \left(\frac{I_{t}}{I_{t-1}}- \bar{g}\right)^{2} \right]
\end{align*}
where $I_t$ is investment, $\delta$ is the depreciation rate of capital, $\phi$  is a parameter which governs the curvature of the investment adjustment cost. Note that as the economy exhibits trend growth rate, the adjustment costs of investments exhibit the deterministic component of trend productivity growth rate $\bar{g}$ inside the parenthesis. 

As in \cite{jermann2012macroeconomic}, firms can finance investment by issuing equity $D_t$ or debt $B^{f}_t$. Reducing equity payouts to finance investment projects does not affect a firm’s tax liabilities in the same way as issuing new debt. As a result, firms prefer debt to equity finance in this model. This preference for debt finance is captured by a constant tax benefit, or subsidy. The effective interest rate faced by firms is $R_t = 1 + r_t \left(1-\tau\right)$. We introduce the presence of function $\Psi \left(d_t\right)$ which captures costs related to equity payouts and issuance compared to some deflated steady state value $d_t = D_t/X_{t-1}$:
\begin{align*}
    \Psi \left(d_t\right) = d_t + \psi \left(d_t - \bar{d}\right)^2
\end{align*}
where $\bar{d}$ is the steady state value of dividend payout deflated by technology. Analogously, we can think about an un-detrended cost function such as
\begin{align*}
    \Upsilon\left(D_t\right) = D_t + X_{t-1} \psi \left(\frac{D_t}{X_{t-1}} - \bar{d} \right)^2
\end{align*}
where the $X_{t-1}$ component in the second term allows the costs to grow at the same rate at $D_t$. This implies the following derivative
\begin{align*}
    \frac{\partial \Upsilon \left(D_t\right)}{\partial D_t}&= 1 + 2 \psi \left(d_t - \bar{d} \right) 
\end{align*}

\noindent
We assume that firms face two types of financial frictions. First, firms face a working capital constraint and must finance a fraction $\lambda^{wk}$ of their labor and imported inputs expenses. Second, firms face a borrowing constraint, expressed in terms of the numeraire good, as a function of the value of their capital and their earnings, following \cite{drechsel2019earnings}
\begin{align*}
    \frac{B^{f}}{1+r_t} + R_t \lambda^{wk} \left(P^M m_t + W_t n_t \right) \leq& \omega \theta_{\pi} \left[a_tK^{\alpha}_{t-1} M^{\beta}_{t}\left(X_t n_t\right)^{1-\alpha-\beta} - \left(1-\lambda^{wk} + \lambda^{wk} R_t \right) \left( P^{M}_t M_t +  W_t n_t \right)\right] \\ 
    & \quad + \left(1-\omega\right) \theta_{k} \mathbb{E}_t p^{k}_{t+1} \left(1-\delta\right) K_t 
\end{align*}
The objective of the firm is to maximize the expected discounted stream of the dividends paid to the firms' owners (the representative household). We can write the firm's constrained optimization problem as

\begin{equation}
    \begin{aligned}
    \max \quad &  \mathbb{E}_0 \sum^{\infty}_{t=0} \Lambda_t D_t \\
    \textrm{s.t.} \quad & Y_t - \left(1-\lambda^{wk}+\lambda^{wk}R_t\right) \left(P^{M}_t M_t + W_t n_t\right) - I_t  - B^{f}_{t-1} + \frac{B^{f}_t}{R_t} - D_t - X_{t-1}\left(\frac{D_t}{X_{t-1}} - \bar{d}\right)^2 = 0 \\
      &  K_t = \left(1-\delta\right) K_{t-1} + I_t \left[1-\frac{\phi}{2} \left(\frac{I_t}{I_{t-1}}-\bar{g}\right)^2\right]   \\
      & \frac{B^{f}}{1+r_t} + R_t \lambda^{wk} \left(P^M M_t + W_t n_t \right) = \omega \theta_{\pi} \bigg\{a_tK^{\alpha}_{t-1} M^{\beta}_{t}\left(X_t n_t\right)^{1-\alpha-\beta} - \\
      & \qquad \left(1-\lambda^{wk} + \lambda^{wk} R_t \right) \left( P^{M}_t M_t -  W_t n_t \right)\bigg\} + \left(1-\omega\right) \theta_{k} \mathbb{E}_t p^{k}_{t+1} \left(1-\delta\right) K_t 
    \end{aligned}
\end{equation}
The firm problem's Lagrangean can be written as
\begin{align*}
    \mathcal{L}^f:& \Lambda_t D_t \quad + \\
    & + \Lambda_t \gamma_t \bigg\{ a_tK^{\alpha}_{t-1} M^{\beta}_t \left(X_t n_t\right)^{1-\alpha-\beta} -\left(1-\lambda^{wk}+\lambda^{wk}R_t\right)\left(P^{M}_t M_t + W_t n_t\right) + \frac{B^{f}_t}{R_t} - I_t \\
    & \qquad - B^f_{t-1} - D_t - X_{t-1}\left(\frac{D_t}{X_{t-1}} - \bar{d}\right)^2  \bigg\} +   \\
    & + \Lambda_t Q_t \left[ \left(1-\delta\right) K_{t-1} + I_t \left[1-\frac{\phi}{2} \left(\frac{I_t}{I_{t-1}}-\bar{g}\right)^2\right]-  K_t\right] \\
    & + \Lambda_t \mu_t \bigg\{ \omega \theta_{\pi} \left[a_tK^{\alpha}_{t-1}M^{\beta}_t\left(X_t n_t\right)^{1-\alpha-\beta} - \left(1-\lambda^{wk}+\lambda^{wk}R_t \right)\left(P^{M}_t M_t + W_t n_t \right) \right] \\
    & \qquad + \left(1-\omega\right) \theta_{k} \mathbb{E}_t p^{k}_{t+1} \left(1-\delta\right) K_t  -  \frac{B^{f}_t}{1+r_t} - \lambda^{wk}R_t\left(P^{M}_t M_t + W_t n_t\right) \bigg\}
\end{align*}

\noindent
The first order conditions with respect to $D_t$, $B_t$, $K_t$ and $I_t$ are
\begin{align*}
    \frac{\partial \mathcal{L}^f}{\partial D_t}:& \quad \gamma_t = \frac{1}{\Upsilon'\left(D_t\right)} = \frac{1}{1+2\psi\left(\frac{D_t}{X_{t-1}}-\bar{d}\right)} \\
    \frac{\partial \mathcal{L}^f}{\partial K_t}:& \quad Q_t = \frac{\Lambda_{t+1}}{\Lambda_t}\bigg\{ a_{t+1} \alpha \left( \frac{Y_{t+1}}{K_t}\right) \left[ \frac{1}{1+2\psi \left(\frac{D_t}{X_{t-1}} -\bar{d} \right)} + \mu_{t+1}\omega \theta_{\pi} \bigg\}\right]   + \mu_t \theta_{k} \left(1-\omega\right) \mathbb{E}_t p^{k}_{t+1}\left(1-\delta\right) \\
    \frac{\partial \mathcal{L}^f}{\partial B_t}:& \quad 1 = \frac{\Upsilon'\left(D_{t+1}\right)}{\Upsilon'\left(D_{t}\right)} \frac{\Lambda_{t+1}}{\Lambda_{t}} R_t + \frac{\mu_t}{\Upsilon'\left(D_{t}\right)} \frac{R_t}{1+r_t} \\
    \frac{\partial \mathcal{L}^f}{\partial I_t}:& \quad \frac{1}{\Upsilon'\left(D_{t}\right)} = Q_t \frac{\partial \Phi \left(I_t,I_{t-1}\right) }{\partial I_t} + \frac{\Lambda_{t+1}}{\Lambda_{t}} Q_{t+1} \frac{\partial \Phi \left(I_{t+1},I_{t}\right) }{\partial I_t}
\end{align*}
Given that the model has trend growth, we need to deflate the equilibrium conditions 
\begin{align*}
    \frac{\partial \mathcal{L}^f}{\partial D_t}:& \quad \gamma_t = \frac{1}{\Upsilon'\left(D_t\right)} = \frac{1}{1+2\psi\left(d_t-\bar{d}\right)} \\
    \frac{\partial \mathcal{L}^f}{\partial K_t}:& \quad Q_t = \mathbb{E}_t \beta \frac{\lambda_{t+1}}{\lambda_{t}} g^{-\sigma}_t \bigg\{ a_{t+1} \alpha \left(\frac{y_{t+1}}{k_t}\right) \left[ \frac{1}{1+2\psi \left(d_t-\bar{d}\right)} + \mu_{t+1} \omega \theta_{\pi}   \right] \bigg\}  + \mu_t \theta_{k} \left(1-\omega\right) \mathbb{E}_t p^{k}_{t+1}\left(1-\delta\right) \\
    \frac{\partial \mathcal{L}^f}{\partial B_t}:& \quad 1 = \beta \mathbb{E}_t g^{-\sigma}_t \frac{\lambda_{t+1}}{\lambda_{t}} \frac{1+2\psi\left(d_t-\bar{d}\right)}{1+2\psi\left(d_{t+1}-\bar{d}\right)} R_t + \mu_t \left[1+2\psi\left(d_t-\bar{d}\right) \right] \frac{R_t}{1+r_t} \\
    \frac{\partial \mathcal{L}^f}{\partial I_t}:& \quad \frac{1}{1+2\psi\left(d_t-\bar{d}\right)} = Q_t \left[ 1-\frac{\phi}{2}\left(\frac{i_{t}}{i_{t-1}}g_{t-1}-\bar{g}\right)- \phi \left( \frac{i_{t}}{i_{t-1}}g_{t-1} - \bar{g} \right)  \frac{i_{t}}{i_{t-1}}g_{t-1} \right] \\
    & \quad \quad \quad - \mathbb{E}_t \beta \frac{\lambda_{t+1}}{\lambda_{t}} g^{-\sigma}_t Q_{t+1} \left[ \phi \left(\frac{i_{t+1}}{i_{t}}g_t - \bar{g} \right) \right] \left(\frac{i_{t+1}}{i_{t}}g_t\right)^2
\end{align*}

\noindent
Next, we compute the first order conditions of $n_t$ and $M_t$. We start by computing the first order condition with respect to labor. Note that $n_t$ enters the firms' flow of funds constraint (or budget constraint) in the production function and as a cost in the wage-bill, and in the borrowing constraint by affecting revenue and the amount borrowed.
\begin{align*}
    \frac{\partial \mathcal{L}^f}{\partial n_t}:& \quad \Lambda_t \gamma_t \bigg\{a_t K^{\alpha}_{t-1} M^{\beta}_t \left(1-\alpha-\beta\right) X^{1-\alpha-\beta}_t n^{-\alpha-\beta}_t - \left(1-\lambda^{wk} + \lambda^{wk}R_t\right) W_t \bigg\} +\\
    & \quad + \Lambda_t \mu_t \bigg\{\omega \theta_{\pi} \left[a_t K^{\alpha}_{t-1} M^{\beta}_t \left(1-\alpha-\beta\right) X^{1-\alpha-\beta}_t n^{-\alpha-\beta}_t - \left(1-\lambda^{wk} + \lambda^{wk}R_t\right) W_t \right] - \\
    & \qquad \lambda^{wk} R_t W_t \bigg\} = 0
\end{align*}
We define the marginal product of labor as
\begin{align*}
    MPL_t = a_t K^{\alpha}_{t-1} M^{\beta}_t \left(1-\alpha-\beta\right) X^{1-\alpha-\beta}_t n^{-\alpha-\beta}_t
\end{align*}
Re-arranging the terms in the first order condition for labor, we find
\begin{align*}
    \Lambda_t MPL_t \left(\gamma_t+\omega\theta_{\pi}\mu_t\right) =  \Lambda_t \left(1-\lambda^{wr}+\lambda^{wr}R_t\right) W_t \left(\gamma_t+\omega\theta_{\pi}\mu_t\right) + \Lambda_t \mu_t \lambda^{wr} R_t W_t
\end{align*}
Dividing both sides of the equation by $\Lambda_t$
\begin{align*}
    MPL_t \left(\gamma_t+\omega\theta_{\pi}\mu_t\right) =  \left(1-\lambda^{wr}+\lambda^{wr}R_t\right) W_t \left(\gamma_t+\omega\theta_{\pi}\mu_t\right) + \mu_t \lambda^{wr} R_t W_t
\end{align*}
Next, we divide both sides of the equation by $\left(\gamma_t+\omega\theta_{\pi}\mu_t\right)$
\begin{align*}
    MPL_t =  \left(1-\lambda^{wr}+\lambda^{wr}R_t\right) W_t + \frac{\mu_t}{\left(\gamma_t+\omega\theta_{\pi}\mu_t\right)} \lambda^{wr} R_t W_t
\end{align*}
Factoring out $W_t$ from the right hand side, the equilibrium condition for labor 
\begin{align*}
    MPL_t =  W_t \left[ \underbrace{\left(1-\lambda^{wr}+\lambda^{wr}R_t\right)}_{\text{Working Capital Constraint}} + \underbrace{\frac{\mu_t}{\left(\gamma_t+\omega\theta_{\pi}\mu_t\right)} \lambda^{wr} R_t}_{\text{Borrowing Constraint - Wedge}} \right]
\end{align*}
sets the marginal product of labor equal to the cost adjusted by both the working capital constraint and a wedge generated by working capital being part of the borrowing constraint. 

Note that if $\lambda^{wr}=0$, i.e., there are no working capital constraints, then we obtain the same first order constraint we had in the previous model
\begin{align*}
    MPL_t =  W_t
\end{align*}
If the intra-temporal working capital borrowing is not part of the borrowing constraint then $\mu_t=0$ and the equilibrium condition for capital becomes
\begin{align*}
    MPL_t =  W_t \left(1-\lambda^{wk} + \lambda^{wk}R_t\right)
\end{align*}
which is in line with \cite{neumeyer2005business}. If we set $\omega$ equal to zero, making the borrowing constraint completely asset-based, we obtain
\begin{align*}
    MPL_t =  W_t \left[ \underbrace{\left(1-\lambda^{wr}+\lambda^{wr}R_t\right)}_{\text{Working Capital Constraint}} + \underbrace{\frac{\mu_t}{\gamma_t} \lambda^{wr} R_t}_{\text{Borrowing Constraint - Wedge}} \right]
\end{align*}
which is in line with the specification of \cite{mendoza2010sudden}.

We can obtain a similar first order condition for firm's demand of intermediate inputs
\begin{align*}
    MPM_t =  P^{M}_t \left[ \underbrace{\left(1-\lambda^{wr}+\lambda^{wr}R_t\right)}_{\text{Working Capital Constraint}} + \underbrace{\frac{\mu_t}{\left(\gamma_t+\omega\theta_{\pi}\mu_t\right)} \lambda^{wr} R_t}_{\text{Borrowing Constraint - Wedge}} \right]
\end{align*}

\subsubsection{Aggregation \& Closing the SOE} \label{subsec:trend_quant_aggregation_wk}

First, only domestic households can own shares on domestic firms, which implies that
\begin{align}
    s_t = 1 \quad \forall t
\end{align}
Next, we combine the firm's and household's budget constraint 
\begin{align*}
    Y_t - \left(1-\lambda^{wk}+\lambda^{wk}R_t\right)\left(P^{M}_t M_t + W_t n_t\right) + \frac{B^{f}_t}{R_t} - I_t  - B^f_{t-1} - D_t - X_{t-1}\psi \left(\frac{D_t}{X_{t-1}}-\bar{d}\right)^2 &= 0 \\
    W_t n_t + B_{t-1}+ s_{t-1} \left(D_t + P_t\right) - \frac{B_{t}}{1+r_t} - s_t P_t - C_t - T_t &=0
\end{align*}
First, we need to de-trend these budget constraints by dividing by $X_{t-1}$
\begin{align*}
    y_t -\left(1-\lambda^{wk}+\lambda^{wk}R_t\right)\left(P^{M}_t m_t + w_t n_t\right) + \frac{b^{f}_t}{R_t}g_t - i_t  - b^f_{t-1} - d_t - \psi \left(d_t-\bar{d}\right)^2 &= 0 \\
    w_t n_t + b_{t-1}+ s_{t-1} \left(d_t + p_t\right) - \frac{b_{t}}{1+r_t}g_t - s_t p_t - c_t - t_t &=0
\end{align*}
where 
\begin{align*}
    t_t &= \frac{T_t}{X_{t-1}} \\
        &= \frac{B^{f}}{R_t} \frac{1}{X_{t-1}} - \frac{B^{f}}{1+r_t} \frac{1}{X_{t-1}} \\
        &= \frac{B^{f}}{R_t} \frac{1}{X_{t-1}} \frac{X_t}{X_t} - \frac{B^{f}}{1+r_t} \frac{1}{X_{t-1}} \frac{X_t}{X_t} \\
        &= \frac{b^{f}}{R_t} g_t - \frac{b^{f}}{1+r_t} g_t 
\end{align*}
Next, consolidating the budget constraints we obtain
\begin{align*}
    Y_t - \lambda^{wk} W_t n_t \left(R_t-1\right)-&\left(1-\lambda^{wk}+\lambda^{wk}R_t\right)P^{M} M_t - I_t \\
    &\quad - C_t - T_t - X_{t-1} \psi \left(\frac{D_t}{X_{t-1}} -\bar{d}\right)^2 - B^{f}_{t-1} + \frac{B^f_t}{R_t} + B_{t-1} - \frac{B_t}{1+r_t} &= 0
\end{align*}
which after deflating becomes
\begin{align*}
    y_t -\lambda^{wk} w_t n_t \left(R_t-1\right) - &\left(1-\lambda^{wk}+\lambda^{wk}R_t\right)P^{M} m_t - i_t \\
    & \quad - c_t - t_t - \psi\left(d_t-\bar{d}\right)^2 - b^{f}_{t-1} + \frac{b^f_t}{R_t}g_t + b_{t-1} - \frac{b_t}{1+r_t}g_t = 0
\end{align*}
Next, we need to define the trade balance. We can define the trade balance as
\begin{align}
    TB_t = g_t\left(\frac{b_t-b^{f}_t}{1+r_t}\right) -  \left(b_{t-1}-b^{f}_{t-1}\right) 
\end{align}
or possible better as
\begin{align*}
    TB_t = g_t\left(\frac{b_t-b^{f}_t}{1+r_t}\right) -  \left(b_{t-1}-b^{f}_{t-1}\right) + \lambda^{wk} w_t n_t \left(R_t-1\right) + \lambda^{wk} P^{M}_t m_t \left(R_t-1\right) 
\end{align*}
and the holding country's net foreign asset position as
\begin{align}
    NFA_t = g_t\left(b_t - b^{f}_t\right) 
\end{align}
Note, we can re-write the trade balance as
\begin{align*}
    TB_t = y_t - c_t - i_t - t_t - \psi\left(d_t - \bar{d}\right)^2 - P^M_t m_t 
\end{align*}

\subsubsection{Endogenous Variables \& Equilibrium Conditions}

This economy is comprised of 24 endogenous variables
\begin{itemize}
    \item Two (2) Interest rates: $r_t$, $R_t$
    
    \item Three (3) Financial assets: $b_t$, $b^{f}_t$, $s_t$
    
    \item Three (3) Household variables: $c_t$, $n_t$, $\lambda_t$
    
    \item Three (4) Prices: $w_t$, $Q_t$, $p_t$, $p^{k}_t$
    
    \item Seven (8) Firm's quantities: $d_t$, $g_t$, $a_t$, $k_t$, $i_t$, $\gamma_t$, $\mu_t$, $m_t$
    
    \item Four (4) Aggregate variables: $y_t$, $TB_t$, $T_t$, $NFA_t$ 
\end{itemize}
Thus, this economy is characterized by 24 equilibrium conditions, starting with the (4) three conditions which characterize the law of motion of exogenous variables and the relationship between $R_t$ and $1+r_t$, (6) the definition of six aggregate variables, (5) four household's first order conditions and the household's budget constraint, (8) six of firms' first order conditions, the law of motion of capital and the borrowing constraint, (1) and finally an expression for the price of capital (using market valuation)

\begin{align}
    \ln a_t &= \rho_{a} \ln a_{t-1} + \epsilon^{a}_{t} \\
    \ln \frac{g_t}{\bar{g}} &= \rho_{g} \ln \frac{g_{t-1}}{\bar{g}} + \epsilon^{g}_{t} \\
    \left(1+r_t\right) &= \left(1+r^{*}\right) \exp\left[-\xi\left(NFA_t-\bar{NFA}\right) \right] \\
    R_t &= 1+r_t\left(1-\tau\right) \\
    NFA_t &= g_t \left(b_t - b^{f}_t\right) \\
    TB_t  &= g_t\left(\frac{b_t-b^{f}_t}{1+r_t}\right) -  \left(b_{t-1}-b^{f}_{t-1}\right) + \lambda^{wk} w_t n_t \left(R_t-1\right) + \lambda^{wk} P^{M}_t m_t \left(R_t-1\right) \\
    TB_t &= y_t -c_t - i_t - P^M_t m_t - \psi \left(d_t - \bar{d}\right)^2 \\
    s_t  &= 1 \\
    t_t  &= g_t \left(\frac{b^{f}_t}{R_t}-\frac{b^{f}_t}{1+r_t}\right) \\
    y_t &= a_t k^{\alpha}_{t-1} m^{\beta}_t g^{1-\alpha-\beta}_t n^{1-\alpha-\beta}_t \\
    \lambda_t &= \left(c_t - \zeta n^{\theta}_t\right)^{-\sigma} \\
    \lambda_t w_t &= \zeta \theta n^{\theta-1}_t\left(c_t - \zeta n^{\theta}_t\right)^{-\sigma} \\
    1 &= g^{-\sigma}_t \left(1+r_t\right) \beta \mathbb{E}_t \frac{\lambda_{t+1}}{\lambda_t} \\
    p_t &= g^{1-\sigma}_t \beta \mathbb{E}_t \frac{\lambda_{t+1}}{\lambda_t} \left(d_{t+1}+p_{t+1}\right) \\
    0 &= g_t \frac{b_t}{1+r_t} + s_t p_t + c_t +t_t - \left[w_t n_t + b_{t-1} + s_{t-1}\left(d_t+p_t\right) \right] \\ 
    w_t &= a_t \left(1-\alpha-\beta\right) g^{1-\alpha-\beta}_t k^{\alpha}_{t-1} m^{\beta}_t n^{-\alpha-\beta}_t \times \left[\left(1-\lambda^{wr}+\lambda^{wr}R_t\right) + \frac{\mu_t}{\left(\gamma_t+\omega\theta_{\pi}\mu_t\right)} \lambda^{wr} R_t \right]^{-1} \\
    P^{M}_t &= a_t \beta g^{1-\alpha-\beta}_t k^{\alpha}_{t-1} m^{\beta-1}_t n^{1-\alpha-\beta}_t \times \left[\left(1-\lambda^{wr}+\lambda^{wr}R_t\right) + \frac{\mu_t}{\left(\gamma_t+\omega\theta_{\pi}\mu_t\right)} \lambda^{wr} R_t \right]^{-1} \\
    \gamma_t &= \frac{1}{1+2\psi \left(d_t-\bar{d}\right)} \\
    1 &= \beta \mathbb{E}_t g^{-\sigma}_t \frac{\lambda_{t+1}}{\lambda_{t}} \frac{1+2\psi\left(d_t-\bar{d}\right)}{1+2\psi\left(d_{t+1}-\bar{d}\right)} R_t + \mu_t \left[1+2\psi\left(d_t-\bar{d}\right) \right] \frac{R_t}{1+r_t} \\
    Q_t &= \beta \mathbb{E}_t g^{-\sigma}_t \frac{\lambda_{t+1}}{\lambda_t} \left[\alpha \frac{y_{t+1}}{k_t} \left(\frac{1}{1+2\psi\left(d_t - \bar{d}\right)} + \mu_{t+1} \omega \theta_{\pi} \right) + Q_{t+1} \left(1-\delta\right)\right] + \\
    & \qquad + \mathbb{E}_t \mu_t \left(1-\omega\right) \theta_{k} p^{k}_{t+1} \left(1-\delta\right) \nonumber \\
    \frac{1}{1+2\psi\left(d_t-\bar{d}\right)} &= Q_t \left[ 1-\frac{\phi}{2}\left(\frac{i_{t}}{i_{t-1}}g_{t-1}-\bar{g}\right)- \phi \left( \frac{i_{t}}{i_{t-1}}g_{t-1} - \bar{g} \right)  \frac{i_{t}}{i_{t-1}}g_{t-1} \right] \\
    & \quad - \mathbb{E}_t \beta \frac{\lambda_{t+1}}{\lambda_{t}} g^{-\sigma}_t Q_{t+1} \left[ \phi \left(\frac{i_{t+1}}{i_{t}}g_t - \bar{g} \right) \right] \left(\frac{i_{t+1}}{i_{t}}g_t\right)^2 \nonumber \\
    k_{t} g_t &= \left(1-\delta\right) k_{t-1} + i_t \left[1-\frac{\phi}{2}\left(\frac{i_t}{i_{t-1}}g_{t-1} - \bar{g} \right)^2 \right]    
\end{align}
\begin{align} 
    \frac{b^{f}_t}{1+r_t} g_t &+ \lambda^{wk} R_t \left(P^{M_t} m_t + w_t n_t\right) = \omega \theta_{\pi} \left[a_{t} \left(k_{t-1}\right)^{\alpha} m^{\beta}_t g^{1-\alpha-\beta}_t n^{1-\alpha-\beta}_t -\left(1-\lambda^{wk}+\lambda^{wk}\right)\left(P^{M}m_t+w_t n_t\right)  \right] \nonumber \\
    & \qquad + \theta_{k} \left(1-\omega\right)\mathbb{E}_t p^{k}_{t+1} k_t \left(1-\delta\right)g_t  \\
    p^{k}_t &= Q_t
\end{align}

\noindent
\textbf{Analytically computing the steady state.} From the steady state equilibrium conditions we know that transitory productivity process is equal to
\begin{align*}
    a = 1
\end{align*}
and the growth rate of the trend productivity process is equal to
\begin{align*}
    g = \bar{g}
\end{align*}
For a given discount factor $\beta$, we know that 
\begin{align*}
    \left(1+r^{*}\right) &= \frac{\bar{g}^{\sigma}}{\beta}  \\
    r^{*} &= \frac{\bar{g}^{\sigma}-\beta}{\beta}
\end{align*}
which leads to a straightforward expression for the interest rate which firms face, given a tax advantage $\tau$,
\begin{align*}
    R = 1+r^{*}\left(1-\tau\right)
\end{align*}
We also know that 
\begin{align*}
    \gamma &= 1 \\
    Q  &= 1 \\
    p^{k} &= 1
\end{align*}
We can compute the Lagrange multiplier on firms' borrowing constraint
\begin{align*}
    \mu = \left(1+r^{*}\right) \left[\frac{1}{R} - \beta \bar{g}^{-\sigma} \right] 
\end{align*}

\noindent
Next, we turn to computing the steady state value of the following four variables
\begin{align*}
    \{k,n,m,w\}
\end{align*}
To do so, we exploit the following four equilibrium conditions
\begin{align*}
    w \left(1-\lambda^{wk}+\lambda^{wk} R\right)&= a \left(1-\alpha - \beta\right) \bar{g}^{1-\alpha - \beta} k^{\alpha} m^{\beta} n^{-\alpha -\beta} \\
    P^{M} \left(1-\lambda^{wk}+\lambda^{wk} R\right) &= a \beta \bar{g}^{1-\alpha -\beta} k^{\alpha} m^{\beta-1} n^{1-\alpha -\beta} \\
    1 &= \beta  \bar{g}^{-\sigma} \left[\alpha \frac{y}{k} \left(1 + \mu \omega \theta_{\pi} \right) + \left(1-\delta\right)\right]
    + \mu \left(1-\omega\right) \theta_{k} p^{k} \left(1-\delta\right) \\
    w &= \zeta \theta n^{\theta -1} 
\end{align*}
First, we re-organize
\begin{align*}
    1 &= \beta  \bar{g}^{-\sigma} \left[\alpha \frac{y}{k} \left(1 + \mu \omega \theta_{\pi} \right) + \left(1-\delta\right)\right] +
    + \mu \left(1-\omega\right) \theta_{k} p^{k} \left(1-\delta\right)
\end{align*}
to obtain an expression for $y/k$
\begin{align*}
    \alpha \frac{y}{k} &= \frac{1}{1+\mu \omega \theta_{\pi}} \bigg\{ \left(\frac{1}{\beta \bar{g}^{-\sigma}}\right) \left[1-\mu\left(1-\omega\right)\theta_{k}p^{k}\left(1-\delta\right) \right]   - \left(1-\delta\right)\bigg\} \\
    \alpha k^{\alpha-1}m^{\beta}n^{1-\alpha-\beta} &= \underbrace{\frac{1}{1+\mu \omega \theta_{\pi}} \bigg\{ \left(\frac{1}{\beta \bar{g}^{-\sigma}}\right) \left[1-\mu\left(1-\omega\right)\theta_{k}p^{k}\left(1-\delta\right) \right]   - \left(1-\delta\right)\bigg\}}_{P^{K}}
\end{align*}
Hence, our system of four equations and four unknowns becomes
\begin{align*}
    w \left[\left(1-\lambda^{wr}+\lambda^{wr}R_t\right) + \frac{\mu_t \lambda^{wr} R_t}{\left(\gamma_t+\omega\theta_{\pi}\mu_t\right)}  \right] &= a \left(1-\alpha - \beta\right) \bar{g}^{1-\alpha - \beta} k^{\alpha} m^{\beta} n^{-\alpha -\beta} \\
    P^{M} \left[\left(1-\lambda^{wr}+\lambda^{wr}R_t\right) + \frac{\mu_t \lambda^{wr} R_t}{\left(\gamma_t+\omega\theta_{\pi}\mu_t\right)}  \right]  &= a \beta \bar{g}^{1-\alpha -\beta} k^{\alpha} m^{\beta-1} n^{1-\alpha -\beta} \\
    P^K &= \alpha k^{\alpha-1}m^{\beta}\bar{g}^{1-\alpha-\beta}n^{1-\alpha-\beta}\\
    w &= \zeta \theta n^{\theta -1} 
\end{align*}
Note that in steady state $\left[\left(1-\lambda^{wr}+\lambda^{wr}R_t\right) + \frac{\mu_t}{\left(\gamma_t+\omega\theta_{\pi}\mu_t\right)} \lambda^{wr} R_t \right]$ becomes
\begin{align*}
    \text{Wedge}^{SS} = \left[\left(1-\lambda^{wr}+\lambda^{wr}R\right) + \frac{\mu \lambda^{wr} R}{\left(1+\omega\theta_{\pi}\mu\right)}  \right]
\end{align*}
as $\gamma = 1$ in steady state. We can re-write our system of equations as
\begin{align*}
    w \text{Wedge}^{SS} &= a \left(1-\alpha - \beta\right) \bar{g}^{1-\alpha - \beta} k^{\alpha} m^{\beta} n^{-\alpha -\beta} \\
    P^{M} \text{Wedge}^{SS}  &= a \beta \bar{g}^{1-\alpha -\beta} k^{\alpha} m^{\beta-1} n^{1-\alpha -\beta} \\
    P^K &= \alpha k^{\alpha-1}m^{\beta}\bar{g}^{1-\alpha-\beta}n^{1-\alpha-\beta}\\
    w &= \zeta \theta n^{\theta -1} 
\end{align*}

\noindent
First, we combine the second and third equations to obtain an expression for $m$ as a function of $k$ and parameters
\begin{align*}
    \frac{P^{K}}{P^{M}\text{Wedge}^{SS}} &= \frac{\alpha}{\beta} \frac{m}{k} \\
    \Rightarrow m &= \frac{P^{K}}{P^{M}\text{Wedge}^{SS}} \frac{\beta}{\alpha} k  
\end{align*}
Next, we plug in the expression for $m$ in the third equation
\begin{align*}
    P^K &= \alpha k^{\alpha-1}m^{\beta}\bar{g}^{1-\alpha-\beta}n^{1-\alpha-\beta} \\
    P^K &= \alpha k^{\alpha-1}\left[\frac{P^{K}}{P^{M} \text{Wedge}^{SS}} \frac{\beta}{\alpha} k \right]^{\beta}\bar{g}^{1-\alpha-\beta}n^{1-\alpha-\beta} \\
    P^K &= \alpha \left[\frac{P^{K}}{P^{M} \text{Wedge}^{SS}} \frac{\beta}{\alpha} \right]^{\beta} k^{\alpha + \beta-1} \bar{g}^{1-\alpha-\beta}n^{1-\alpha-\beta} \\
    P^K &= \alpha \left[\frac{P^{K}}{P^{M} \text{Wedge}^{SS}} \frac{\beta}{\alpha} \right]^{\beta} \left(\frac{k}{\bar{g}n}\right)^{\alpha+\beta-1}
\end{align*}
which allows us to construct an expression for the ratio $k/\bar{g}n$
\begin{align*}
    \left(\frac{k}{\bar{g}n}\right)^{\alpha+\beta-1} &= P^{K} \frac{1}{\alpha} \left[\frac{P^{K}}{P^{M} \text{Wedge}^{SS}} \frac{\beta}{\alpha} \right]^{-\beta} \\
    \frac{k}{\bar{g}n} &= \left[P^{K} \frac{1}{\alpha} \left[\frac{P^{K}}{P^{M} \text{Wedge}^{SS}} \frac{\beta}{\alpha} \right]^{-\beta} \right]^{\frac{1}{\alpha+\beta-1}}
\end{align*}
Following, we use the first equation in our system, and start by introducing our expression for $m$ depending on $k$
\begin{align*}
    w &= a \left(1-\alpha - \beta\right) \bar{g}^{1-\alpha - \beta} k^{\alpha} m^{\beta} n^{-\alpha -\beta} \\
    w &= a \left(1-\alpha - \beta\right) \bar{g}^{1-\alpha - \beta} k^{\alpha} \left[\frac{P^{K}}{P^{M} \text{Wedge}^{SS}} \frac{\beta}{\alpha} k \right]^{\beta} n^{-\alpha -\beta} \\
    w &= a \left(1-\alpha-\beta\right) \bar{g}^{1-\alpha - \beta} \left[\frac{P^{K}}{P^{M} \text{Wedge}^{SS}} \frac{\beta}{\alpha} \right]^{\beta} k^{\alpha+\beta} n^{-\alpha-\beta}
\end{align*}
Re-arranging the terms, we can express wages $w$ as a function of known parameters
\begin{align*}
    w &= a \left(1-\alpha-\beta\right)  \left[\frac{P^{K}}{P^{M} \text{Wedge}^{SS}} \frac{\beta}{\alpha} \right]^{\beta} \left(\frac{k}{\bar{g}n}\right)^{\alpha+\beta} \bar{g}
\end{align*}
Next, we use the last equation, to compute the steady state labor
\begin{align*}
    w &= \zeta \theta n^{\theta -1} \\
    n &= \left(\frac{w}{\zeta \theta}\right)^{\theta-1}
\end{align*}
Next, we can back out the steady state stock of capital as
\begin{align*}
    k = \frac{k}{\bar{g}n} \times \bar{g} n
\end{align*}
Finally, we use the expression for capital to compute the steady state level of intermediate inputs
\begin{align*}
    m &= \frac{P^{K}}{P^{M} \text{Wedge}^{SS}} \frac{\beta}{\alpha} k  
\end{align*}

Next, we can compute the steady state investment exploiting the law of motion for capital
\begin{align*}
    k \bar{g} &= \left(1-\delta\right) k + i \\
    i  &= \left[\bar{g} - \left(1-\delta\right) \right] k  
\end{align*}
and the level of production using the production function
\begin{align*}
    y = k^{\alpha} m^{\beta} \left(\bar{g} n \right)^{1-\alpha-\beta}
\end{align*}
We can compute the steady state level of debt
\begin{align*}
    \frac{b^{f}}{1+r^{*}} \bar{g} +\lambda^{wk} R \left(P^{M} m + wn\right) &= \omega \theta_{\pi} \left[k^{\alpha} m^{\beta} \bar{g}^{1-\alpha-\beta} n^{1-\alpha-\beta} - \left(1-\lambda^{wk}+\lambda^{wk}R\right)\left( P^{M}m + w n\right)\right] + \\
    & + \quad \quad \theta_{k} \left(1-\omega\right) p^{k} k \left(1-\delta\right) \bar{g} \\
    b^{f} &= \left(\frac{1+r^{*}}{\bar{g}} \right) \bigg\}[\omega \theta_{\pi} \left[k^{\alpha} m^{\beta} \bar{g}^{1-\alpha-\beta} n^{1-\alpha-\beta}  \left(1-\lambda^{wk}+\lambda^{wk}R\right)\left( P^{M}m + w n\right)\right] \\
    & \quad \quad +\theta_{k} \left(1-\omega\right) p^{k} k \left(1-\delta\right) \bar{g}  - \lambda^{wk} R \left(P^{M} m + wn\right)\bigg\}
\end{align*}
For any given $\bar{NFA}$, we can compute the steady state
\begin{align*}
    NFA_t     &= g_t \left(b_t - b^{f}_t\right) \\
    \bar{NFA} &= \bar{g} \left(b - b^{f}\right) \\
    \Rightarrow b &= \frac{\bar{NFA}}{\bar{g}} + b^{f}
\end{align*}
and the trade balance in steady state as
\begin{align*}
    TB_t &= g_t \left(\frac{b_t-b^{f}_t}{1+r_t}\right) - \left(b_{t-1} - b^{f}_{t-1}\right) \\
    TB   &= \bar{g} \left(\frac{b-b^{f}}{1+r^{*}}\right) - \left(b - b^{f} \right)    \\
    TB   &= \left(b-b^{f}\right) \left( \frac{\bar{g}}{1+r^{*}}-1\right) 
\end{align*}
Knowing the steady state level of firm debt we can compute the taxation level of households
\begin{align*}
    t  &= \bar{g} b^{f} \left(\frac{1}{1 +r^{*}\left(1-\tau\right)} - \frac{1}{1 +r^{*}}\right)
\end{align*}
With the aggregate variables $y$, $TB$ and $i$ we can back-out
\begin{align*}
    c = y - i - TB - P^{M} m
\end{align*}
Once we have computed the steady state level of consumption we can compute the steady state level of the household's budget constraint Lagrange multiplier
\begin{align*}
    \lambda = \left(c- \zeta n^{\theta} \right)^{-\sigma} 
\end{align*}
The last two variables that we have not computed steady state values so far are the price of shares $p_t$ and the dividend $d_t$, i.e., $\bar{d}$. Note that dividends enter the household's budget constraint, which is an equilibrium condition. We can use this equilibrium condition to back out an expression for the steady-state value of dividends
\begin{align*}
    d = \bar{d} = c + t - wn + \bar{g} \frac{b}{1+r} - b
\end{align*}
Finally, the expression for the price of shares
\begin{align*}
    p = \frac{\beta \bar{g}^{1-\sigma} }{1-\bar{g}^{1-\sigma}\beta} d
\end{align*}
This concludes the analytical computation of the steady state.

\subsection{Model with Interest Coverage Constraints} \label{subsec:appendix_real_model_IC}

\subsubsection{Households} \label{subsec:trend_quant_households_wk_IC}

The household seeks to maximize the utility function
\begin{align*}
    \mathbb{E}_0 \sum^{\infty}_{t=0} \beta^{t} \frac{\left(C_t - \zeta X_{t-1} n^{\theta}_{t}\right)^{1-\sigma}-1}{1-\sigma} 
\end{align*}
where $X_t$ is a trend productivity component, $C_t$ is consumption, $n_t$ is labor effort, $\beta$ is a discount factor, $<0\beta<1$. Every period, households face a budget constraint
\begin{align*}
    W_t n_t + B_{t-1}+s_{t-1}\left(D_t + P_t\right) = \frac{B_t}{1+r_t} + s_t P_t + C_t + T_t
\end{align*}
where $W_t$ is the wage in terms of the numeraire goods, $B_{t-1}$ is the stock of freely traded bonds, $s_t$ is the household's shares of domestic firms, $D_t$ is the paid dividend and $P_t$ is the price of the shares. Taxation $T_t$, which is the household takes as given, is used to finance the firms' tax benefit enjoyed by firms when borrowing
\begin{align*}
    T_t = \frac{B^{f}}{R_t}-\frac{B^{f}}{1+r_t}
\end{align*}
where $B^{f}_t$ represents the firm's stock of debt and $R_t = 1 + r_t \left(1-\tau\right)$.

The household's Lagrangean can be written down as
\begin{align*}
    \mathcal{L}:& \quad \beta^{t} \frac{\left(C_t - \zeta X_{t-1} n^{\theta}_{t}\right)^{1-\sigma}-1}{1-\sigma} + \beta^{t} X^{-\sigma}_{t-1} \lambda_t \left[W_t n_t + B_{t-1} + s_{t-1} \left(D_t + P_t\right) - \frac{B_t}{1+r_t}-s_t P_t - C_t - T_t \right]
\end{align*}
The first order conditions with respect to $C_t$, $n_t$, $B_t$ and $s_t$ are
\begin{align*}
    \frac{\partial \mathcal{L}}{\partial C_t}:& \quad \left(\frac{C_t}{X_{t-1}} - \zeta n^{\theta}_t \right)^{-\sigma} = \lambda_t \\
    \frac{\partial \mathcal{L}}{\partial n_t}:& \quad \left(\frac{C_t}{X_{t-1}} - \zeta n^{\theta}_t \right)^{-\sigma} \zeta \theta n^{\theta-1}_t = \lambda_t \frac{W_t}{X_{t-1}} \\
    \frac{\partial \mathcal{L}}{\partial B_t}:& \quad 1 = g^{-\sigma}_t \left(1+r_t\right) \beta \mathbb{E}_t \frac{\lambda_{t+1}}{\lambda_{t}} \\
    \frac{\partial \mathcal{L}}{\partial s_t}:& \quad P_t = g^{-\sigma}_t \beta \mathbb{E}_t \frac{\lambda_{t+1}}{\lambda_{t}} \left(D_{t+1} + P_{t+1} \right)
\end{align*}
Note that $C_t$, $W_t$, $P_t$, and $D_t$ exhibit trend growth. We define $c_t \equiv C_t / X_{t-1}$, $w_t \equiv W_t / X_{t-1}$, $p_t \equiv P_t / X_{t-1}$ and $d_t \equiv D_t / X_{t-1}$ and re-write the first order conditions in terms of detrended variables as
\begin{align}
    \frac{\partial \mathcal{L}}{\partial C_t}:& \quad \left(c_t - \zeta n^{\theta}_t \right)^{-\sigma} = \lambda_t \\
    \frac{\partial \mathcal{L}}{\partial n_t}:& \quad \left(c_t - \zeta n^{\theta}_t \right)^{-\sigma} \zeta \theta n^{\theta-1}_t = \lambda_t w_t  \\
    \frac{\partial \mathcal{L}}{\partial B_t}:& \quad 1 = g^{-\sigma}_t \left(1+r_t\right) \beta \mathbb{E}_t \frac{\lambda_{t+1}}{\lambda_{t}} \\
    \frac{\partial \mathcal{L}}{\partial s_t}:& \quad p_t = g^{1-\sigma}_t \beta \mathbb{E}_t \frac{\lambda_{t+1}}{\lambda_{t}} \left(d_{t+1} + p_{t+1} \right)
\end{align}
Note that if we set the growth rate of trend productivity equal to 1, i.e. $g_t = 1 \Rightarrow X_t = X_{t-1} \forall t$, we obtain the same first order conditions than in the model without trend growth. 

\subsubsection{Firms} \label{subsec:trend_quant_firms_wk_IC}

Firms face the same frictions as in \cite{jermann2012macroeconomic} but in a small open economy with productivity trend growth as in \cite{garcia2010real}. At any time $t$, a representative firm produces by hiring labor $n_t$, buying imported intermediate inputs $M_t$, and using accumulated capital $K_{t-1}$, using a Cobb-Douglas production function
\begin{align*}
    Y_t = a_t K^{\alpha}_{t-1} M^{\beta}_t \left(X_t n_t\right)^{1-\alpha-\beta}
\end{align*}
where $Y_t$ denotes output in period $t$, $K_{t-1}$ denotes the stock of capital installed in period $t$ which is used for production in period $t$, $n_t$ denotes hired labor, and $a_t$ and $X_t$ represent productivity shocks.

The productivity shock $a_t$ is assumed to follow a first-order auto-regressive process in logs. That is,
\begin{align}
    \ln a_t = \rho_a \ln a_{t-1} + \epsilon^{a}_t, \quad \epsilon^{a}_t \sim \mathcal{N} \left(0,\sigma^{a}\right)
\end{align}
The productivity process shock $X_t$ is non-stationary. Let
\begin{align*}
    g_t \equiv \frac{X_t}{X_{t-1}}
\end{align*}
denotes the gross growth rate of $X_t$. We assume that the logarithm of $g_t$ follows a first-order auto-regressive process of the form
\begin{align*}
    \ln \left(\frac{g_t}{\bar{g}}\right) = \rho_{g} \ln \left(\frac{g_{t-1}}{\bar{g}}\right) + \epsilon^{g}_{t}, \quad \sim \mathcal{N} \left(0,\sigma^{g}\right)
\end{align*}
Capital accumulation is subject to investment adjustment costs
\begin{align*}
    K_t = \left(1-\delta \right) K_{t-1} + I_t \left[1-\frac{\phi}{2} \left(\frac{I_{t}}{I_{t-1}}- \bar{g}\right)^{2} \right]
\end{align*}
where $I_t$ is investment, $\delta$ is the depreciation rate of capital, $\phi$  is a parameter which governs the curvature of the investment adjustment cost. Note that as the economy exhibits trend growth rate, the adjustment costs of investments exhibit the deterministic component of trend productivity growth rate $\bar{g}$ inside the parenthesis. 

As in \cite{jermann2012macroeconomic}, firms can finance investment by issuing equity $D_t$ or debt $B^{f}_t$. Reducing equity payouts to finance investment projects does not affect a firm’s tax liabilities in the same way as issuing new debt. As a result, firms prefer debt to equity finance in this model. This preference for debt finance is captured by a constant tax benefit, or subsidy. The effective interest rate faced by firms is $R_t = 1 + r_t \left(1-\tau\right)$. We introduce the presence of function $\Psi \left(d_t\right)$ which captures costs related to equity payouts and issuance compared to some deflated steady state value $d_t = D_t/X_{t-1}$:
\begin{align*}
    \Psi \left(d_t\right) = d_t + \psi \left(d_t - \bar{d}\right)^2
\end{align*}
where $\bar{d}$ is the steady state value of dividend payout deflated by technology. Analogously, we can think about an un-detrended cost function such as
\begin{align*}
    \Upsilon\left(D_t\right) = D_t + X_{t-1} \psi \left(\frac{D_t}{X_{t-1}} - \bar{d} \right)^2
\end{align*}
where the $X_{t-1}$ component in the second term allows the costs to grow at the same rate at $D_t$. This implies the following derivative
\begin{align*}
    \frac{\partial \Upsilon \left(D_t\right)}{\partial D_t}&= 1 + 2 \psi \left(d_t - \bar{d} \right) 
\end{align*}

\noindent
We assume that firms face two types of financial frictions. First, firms face a working capital constraint and must finance a fraction $\lambda^{wk}$ of their labor and imported inputs expenses. Second, firms face a borrowing constraint, expressed in terms of the numeraire good, as a function of the value of their capital and their earnings, following \cite{drechsel2019earnings}
\begin{align*}
    \frac{r_t B^f_t}{1+r_t} + R_t \lambda^{wk} \left(P^{M}m_t + W_t n_t\right) \leq \theta^{IC}  \left[a_tK^{\alpha}_{t-1} M^{\beta}_{t}\left(X_t n_t\right)^{1-\alpha-\beta} - \left(1-\lambda^{wk} + \lambda^{wk} R_t \right) \left( P^{M}_t M_t +  W_t n_t \right)\right]
\end{align*}
The objective of the firm is to maximize the expected discounted stream of the dividends paid to the firms' owners (the representative household). We can write the firm's constrained optimization problem as

\begin{equation}
    \begin{aligned}
    \max \quad &  \mathbb{E}_0 \sum^{\infty}_{t=0} \Lambda_t D_t \\
    \textrm{s.t.} \quad & Y_t - \left(1-\lambda^{wk}+\lambda^{wk}R_t\right) \left(P^{M}_t M_t + W_t n_t\right) - I_t  - B^{f}_{t-1} + \frac{B^{f}_t}{R_t} - D_t - X_{t-1}\left(\frac{D_t}{X_{t-1}} - \bar{d}\right)^2 = 0 \\
      &  K_t = \left(1-\delta\right) K_{t-1} + I_t \left[1-\frac{\phi}{2} \left(\frac{I_t}{I_{t-1}}-\bar{g}\right)^2\right]   \\
      & \frac{r_t B^f_t}{1+r_t} + R_t \lambda^{wk} \left(P^{M}m_t + W_t n_t\right) \leq \theta^{IC}  \bigg\{a_tK^{\alpha}_{t-1} M^{\beta}_{t}\left(X_t n_t\right)^{1-\alpha-\beta} \\
      & \qquad \qquad \qquad \qquad \qquad \qquad \qquad \qquad \qquad \qquad - \left(1-\lambda^{wk} + \lambda^{wk} R_t \right) \left( P^{M}_t M_t +  W_t n_t \right)\bigg\} 
    \end{aligned}
\end{equation}
The firm problem's Lagrangean can be written as
\begin{align*}
    \mathcal{L}^f:& \Lambda_t D_t \quad + \\
    & + \Lambda_t \gamma_t \bigg\{ a_tK^{\alpha}_{t-1} M^{\beta}_t \left(X_t n_t\right)^{1-\alpha-\beta} -\left(1-\lambda^{wk}+\lambda^{wk}R_t\right)\left(P^{M}_t M_t + W_t n_t\right) + \frac{B^{f}_t}{R_t} - I_t \\
    & \qquad - B^f_{t-1} - D_t - X_{t-1}\left(\frac{D_t}{X_{t-1}} - \bar{d}\right)^2  \bigg\} +   \\
    & + \Lambda_t Q_t \left[ \left(1-\delta\right) K_{t-1} + I_t \left[1-\frac{\phi}{2} \left(\frac{I_t}{I_{t-1}}-\bar{g}\right)^2\right]-  K_t\right] \\
    & + \Lambda_t \mu_t \bigg\{ \theta^{IC}  \bigg\{a_tK^{\alpha}_{t-1} M^{\beta}_{t}\left(X_t n_t\right)^{1-\alpha-\beta} \\
      &  \qquad \qquad \qquad \qquad  - \left(1-\lambda^{wk} + \lambda^{wk} R_t \right) \left( P^{M}_t M_t +  W_t n_t \right)\bigg\}   -  \frac{r_t B^{f}_t}{1+r_t} - \lambda^{wk}R_t\left(P^{M}_t M_t + W_t n_t\right) \bigg\}
\end{align*}

\noindent
The first order conditions with respect to $D_t$, $B_t$, $K_t$ and $I_t$ are
\begin{align*}
    \frac{\partial \mathcal{L}^f}{\partial D_t}:& \quad \gamma_t = \frac{1}{\Upsilon'\left(D_t\right)} = \frac{1}{1+2\psi\left(\frac{D_t}{X_{t-1}}-\bar{d}\right)} \\
    \frac{\partial \mathcal{L}^f}{\partial K_t}:& \quad Q_t = \frac{\Lambda_{t+1}}{\Lambda_t}\bigg\{ a_{t+1} \alpha \left( \frac{Y_{t+1}}{K_t}\right) \left[ \frac{1}{1+2\psi \left(\frac{D_t}{X_{t-1}} -\bar{d} \right)} + \mu_{t+1} \theta^{IC} \bigg\}\right]  \\
    \frac{\partial \mathcal{L}^f}{\partial B_t}:& \quad 1 = \frac{\Upsilon'\left(D_{t+1}\right)}{\Upsilon'\left(D_{t}\right)} \frac{\Lambda_{t+1}}{\Lambda_{t}} R_t + R_t \frac{\mu_t}{\Upsilon'\left(D_{t}\right)} \frac{r_t}{1+r_t} \\
    \frac{\partial \mathcal{L}^f}{\partial I_t}:& \quad \frac{1}{\Upsilon'\left(D_{t}\right)} = Q_t \frac{\partial \Phi \left(I_t,I_{t-1}\right) }{\partial I_t} + \frac{\Lambda_{t+1}}{\Lambda_{t}} Q_{t+1} \frac{\partial \Phi \left(I_{t+1},I_{t}\right) }{\partial I_t}
\end{align*}
Given that the model has trend growth, we need to deflate the equilibrium conditions 
\begin{align*}
    \frac{\partial \mathcal{L}^f}{\partial D_t}:& \quad \gamma_t = \frac{1}{\Upsilon'\left(D_t\right)} = \frac{1}{1+2\psi\left(d_t-\bar{d}\right)} \\
    \frac{\partial \mathcal{L}^f}{\partial K_t}:& \quad Q_t = \mathbb{E}_t \beta \frac{\lambda_{t+1}}{\lambda_{t}} g^{-\sigma}_t \bigg\{ a_{t+1} \alpha \left(\frac{y_{t+1}}{k_t}\right) \left[ \frac{1}{1+2\psi \left(d_t-\bar{d}\right)} + \mu_{t+1} \omega \theta_{\pi}   \right] \bigg\} \\
    \frac{\partial \mathcal{L}^f}{\partial B_t}:& \quad 1 = \beta \mathbb{E}_t g^{-\sigma}_t \frac{\lambda_{t+1}}{\lambda_{t}} \frac{1+2\psi\left(d_t-\bar{d}\right)}{1+2\psi\left(d_{t+1}-\bar{d}\right)} R_t + \mu_t \left[1+2\psi\left(d_t-\bar{d}\right) \right] \frac{r_t}{1+r_t} R_t \\
    \frac{\partial \mathcal{L}^f}{\partial I_t}:& \quad \frac{1}{1+2\psi\left(d_t-\bar{d}\right)} = Q_t \left[ 1-\frac{\phi}{2}\left(\frac{i_{t}}{i_{t-1}}g_{t-1}-\bar{g}\right)- \phi \left( \frac{i_{t}}{i_{t-1}}g_{t-1} - \bar{g} \right)  \frac{i_{t}}{i_{t-1}}g_{t-1} \right] \\
    & \quad \quad \quad - \mathbb{E}_t \beta \frac{\lambda_{t+1}}{\lambda_{t}} g^{-\sigma}_t Q_{t+1} \left[ \phi \left(\frac{i_{t+1}}{i_{t}}g_t - \bar{g} \right) \right] \left(\frac{i_{t+1}}{i_{t}}g_t\right)^2
\end{align*}
Next, we compute the first order conditions of $n_t$ and $M_t$. We start by computing the first order condition with respect to labor. Note that $n_t$ enters the firms' flow of funds constraint (or budget constraint) in the production function and as a cost in the wage-bill, and in the borrowing constraint by affecting revenue and the amount borrowed.
\begin{align*}
    \frac{\partial \mathcal{L}^f}{\partial n_t}:& \quad \Lambda_t \gamma_t \bigg\{a_t K^{\alpha}_{t-1} M^{\beta}_t \left(1-\alpha-\beta\right) X^{1-\alpha-\beta}_t n^{-\alpha-\beta}_t - \left(1-\lambda^{wk} + \lambda^{wk}R_t\right) W_t \bigg\} +\\
    & \quad + \Lambda_t \mu_t \bigg\{\omega \theta_{\pi} \left[a_t K^{\alpha}_{t-1} M^{\beta}_t \left(1-\alpha-\beta\right) X^{1-\alpha-\beta}_t n^{-\alpha-\beta}_t - \left(1-\lambda^{wk} + \lambda^{wk}R_t\right) W_t \right] - \\
    & \qquad \lambda^{wk} R_t W_t \bigg\} = 0
\end{align*}
We define the marginal product of labor as
\begin{align*}
    MPL_t = a_t K^{\alpha}_{t-1} M^{\beta}_t \left(1-\alpha-\beta\right) X^{1-\alpha-\beta}_t n^{-\alpha-\beta}_t
\end{align*}
Re-arranging the terms in the first order condition for labor, we find
\begin{align*}
    \Lambda_t MPL_t \left(\gamma_t+\omega\theta_{\pi}\mu_t\right) =  \Lambda_t \left(1-\lambda^{wr}+\lambda^{wr}R_t\right) W_t \left(\gamma_t+\omega\theta_{\pi}\mu_t\right) + \Lambda_t \mu_t \lambda^{wr} R_t W_t
\end{align*}
Dividing both sides of the equation by $\Lambda_t$
\begin{align*}
    MPL_t \left(\gamma_t+\omega\theta_{\pi}\mu_t\right) =  \left(1-\lambda^{wr}+\lambda^{wr}R_t\right) W_t \left(\gamma_t+\omega\theta_{\pi}\mu_t\right) + \mu_t \lambda^{wr} R_t W_t
\end{align*}
Next, we divide both sides of the equation by $\left(\gamma_t+\omega\theta_{\pi}\mu_t\right)$
\begin{align*}
    MPL_t =  \left(1-\lambda^{wr}+\lambda^{wr}R_t\right) W_t + \frac{\mu_t}{\left(\gamma_t+\omega\theta_{\pi}\mu_t\right)} \lambda^{wr} R_t W_t
\end{align*}
Factoring out $W_t$ from the right hand side, the equilibrium condition for labor 
\begin{align*}
    MPL_t =  W_t \left[ \underbrace{\left(1-\lambda^{wr}+\lambda^{wr}R_t\right)}_{\text{Working Capital Constraint}} + \underbrace{\frac{\mu_t}{\left(\gamma_t+\omega\theta_{\pi}\mu_t\right)} \lambda^{wr} R_t}_{\text{Borrowing Constraint - Wedge}} \right]
\end{align*}
sets the marginal product of labor equal to the cost adjusted by both the working capital constraint and a wedge generated by working capital being part of the borrowing constraint. 

Note that if $\lambda^{wr}=0$, i.e., there are no working capital constraints, then we obtain the same first order constraint we had in the previous model
\begin{align*}
    MPL_t =  W_t
\end{align*}
If the intra-temporal working capital borrowing is not part of the borrowing constraint then $\mu_t=0$ and the equilibrium condition for capital becomes
\begin{align*}
    MPL_t =  W_t \left(1-\lambda^{wk} + \lambda^{wk}R_t\right)
\end{align*}
which is in line with \cite{neumeyer2005business}. If we set $\omega$ equal to zero, making the borrowing constraint completely asset-based, we obtain
\begin{align*}
    MPL_t =  W_t \left[ \underbrace{\left(1-\lambda^{wr}+\lambda^{wr}R_t\right)}_{\text{Working Capital Constraint}} + \underbrace{\frac{\mu_t}{\gamma_t} \lambda^{wr} R_t}_{\text{Borrowing Constraint - Wedge}} \right]
\end{align*}
which is in line with the specification of \cite{mendoza2010sudden}.

We can obtain a similar first order condition for firm's demand of intermediate inputs
\begin{align*}
    MPM_t =  P^{M}_t \left[ \underbrace{\left(1-\lambda^{wr}+\lambda^{wr}R_t\right)}_{\text{Working Capital Constraint}} + \underbrace{\frac{\mu_t}{\left(\gamma_t+\omega\theta_{\pi}\mu_t\right)} \lambda^{wr} R_t}_{\text{Borrowing Constraint - Wedge}} \right]
\end{align*}

\subsubsection{Aggregation \& Closing the SOE} \label{subsec:trend_quant_aggregation_wk_IC}

First, only domestic households can own shares on domestic firms, which implies that
\begin{align}
    s_t = 1 \quad \forall t
\end{align}
Next, we combine the firm's and household's budget constraint 
\begin{align*}
    Y_t - \left(1-\lambda^{wk}+\lambda^{wk}R_t\right)\left(P^{M}_t M_t + W_t n_t\right) + \frac{B^{f}_t}{R_t} - I_t  - B^f_{t-1} - D_t - X_{t-1}\psi \left(\frac{D_t}{X_{t-1}}-\bar{d}\right)^2 &= 0 \\
    W_t n_t + B_{t-1}+ s_{t-1} \left(D_t + P_t\right) - \frac{B_{t}}{1+r_t} - s_t P_t - C_t - T_t &=0
\end{align*}
First, we need to de-trend these budget constraints by dividing by $X_{t-1}$
\begin{align*}
    y_t -\left(1-\lambda^{wk}+\lambda^{wk}R_t\right)\left(P^{M}_t m_t + w_t n_t\right) + \frac{b^{f}_t}{R_t}g_t - i_t  - b^f_{t-1} - d_t - \psi \left(d_t-\bar{d}\right)^2 &= 0 \\
    w_t n_t + b_{t-1}+ s_{t-1} \left(d_t + p_t\right) - \frac{b_{t}}{1+r_t}g_t - s_t p_t - c_t - t_t &=0
\end{align*}
where 
\begin{align*}
    t_t &= \frac{T_t}{X_{t-1}} \\
        &= \frac{B^{f}}{R_t} \frac{1}{X_{t-1}} - \frac{B^{f}}{1+r_t} \frac{1}{X_{t-1}} \\
        &= \frac{B^{f}}{R_t} \frac{1}{X_{t-1}} \frac{X_t}{X_t} - \frac{B^{f}}{1+r_t} \frac{1}{X_{t-1}} \frac{X_t}{X_t} \\
        &= \frac{b^{f}}{R_t} g_t - \frac{b^{f}}{1+r_t} g_t 
\end{align*}
Next, consolidating the budget constraints we obtain
\begin{align*}
    Y_t - \lambda^{wk} W_t n_t \left(R_t-1\right)-&\left(1-\lambda^{wk}+\lambda^{wk}R_t\right)P^{M} M_t - I_t \\
    &\quad - C_t - T_t - X_{t-1} \psi \left(\frac{D_t}{X_{t-1}} -\bar{d}\right)^2 - B^{f}_{t-1} + \frac{B^f_t}{R_t} + B_{t-1} - \frac{B_t}{1+r_t} &= 0
\end{align*}
which after deflating becomes
\begin{align*}
    y_t -\lambda^{wk} w_t n_t \left(R_t-1\right) - &\left(1-\lambda^{wk}+\lambda^{wk}R_t\right)P^{M} m_t - i_t \\
    & \quad - c_t - t_t - \psi\left(d_t-\bar{d}\right)^2 - b^{f}_{t-1} + \frac{b^f_t}{R_t}g_t + b_{t-1} - \frac{b_t}{1+r_t}g_t = 0
\end{align*}
Next, we need to define the trade balance. We can define the trade balance as
\begin{align}
    TB_t = g_t\left(\frac{b_t-b^{f}_t}{1+r_t}\right) -  \left(b_{t-1}-b^{f}_{t-1}\right) 
\end{align}
or possible better as
\begin{align*}
    TB_t = g_t\left(\frac{b_t-b^{f}_t}{1+r_t}\right) -  \left(b_{t-1}-b^{f}_{t-1}\right) + \lambda^{wk} w_t n_t \left(R_t-1\right) + \lambda^{wk} P^{M}_t m_t \left(R_t-1\right) 
\end{align*}
and the holding country's net foreign asset position as
\begin{align}
    NFA_t = g_t\left(b_t - b^{f}_t\right) 
\end{align}
Note, we can re-write the trade balance as
\begin{align*}
    TB_t = y_t - c_t - i_t - t_t - \psi\left(d_t - \bar{d}\right)^2 - P^M_t m_t 
\end{align*}

\subsubsection{Endogenous Variables \& Equilibrium Conditions}

This economy is comprised of 24 endogenous variables
\begin{itemize}
    \item Two (2) Interest rates: $r_t$, $R_t$
    
    \item Three (3) Financial assets: $b_t$, $b^{f}_t$, $s_t$
    
    \item Three (3) Household variables: $c_t$, $n_t$, $\lambda_t$
    
    \item Three (4) Prices: $w_t$, $Q_t$, $p_t$, $p^{k}_t$
    
    \item Seven (8) Firm's quantities: $d_t$, $g_t$, $a_t$, $k_t$, $i_t$, $\gamma_t$, $\mu_t$, $m_t$
    
    \item Four (4) Aggregate variables: $y_t$, $TB_t$, $T_t$, $NFA_t$ 
\end{itemize}
Thus, this economy is characterized by 24 equilibrium conditions, starting with the (4) three conditions which characterize the law of motion of exogenous variables and the relationship between $R_t$ and $1+r_t$, (6) the definition of six aggregate variables, (5) four household's first order conditions and the household's budget constraint, (8) six of firms' first order conditions, the law of motion of capital and the borrowing constraint, (1) and finally an expression for the price of capital (using market valuation)

\begin{align}
    \ln a_t &= \rho_{a} \ln a_{t-1} + \epsilon^{a}_{t} \\
    \ln \frac{g_t}{\bar{g}} &= \rho_{g} \ln \frac{g_{t-1}}{\bar{g}} + \epsilon^{g}_{t} \\
    \left(1+r_t\right) &= \left(1+r^{*}\right) \exp\left[-\xi\left(NFA_t-\bar{NFA}\right) \right] \\
    R_t &= 1+r_t\left(1-\tau\right) \\
    NFA_t &= g_t \left(b_t - b^{f}_t\right) \\
    TB_t  &= g_t\left(\frac{b_t-b^{f}_t}{1+r_t}\right) -  \left(b_{t-1}-b^{f}_{t-1}\right) + \lambda^{wk} w_t n_t \left(R_t-1\right) + \lambda^{wk} P^{M}_t m_t \left(R_t-1\right) \\
    TB_t &= y_t -c_t - i_t - P^M_t m_t - \psi \left(d_t - \bar{d}\right)^2 \\
    s_t  &= 1 \\
    t_t  &= g_t \left(\frac{b^{f}_t}{R_t}-\frac{b^{f}_t}{1+r_t}\right) \\
    y_t &= a_t k^{\alpha}_{t-1} m^{\beta}_t g^{1-\alpha-\beta}_t n^{1-\alpha-\beta}_t \\
    \lambda_t &= \left(c_t - \zeta n^{\theta}_t\right)^{-\sigma} \\
    \lambda_t w_t &= \zeta \theta n^{\theta-1}_t\left(c_t - \zeta n^{\theta}_t\right)^{-\sigma} \\
    1 &= g^{-\sigma}_t \left(1+r_t\right) \beta \mathbb{E}_t \frac{\lambda_{t+1}}{\lambda_t} \\
    p_t &= g^{1-\sigma}_t \beta \mathbb{E}_t \frac{\lambda_{t+1}}{\lambda_t} \left(d_{t+1}+p_{t+1}\right) \\
    0 &= g_t \frac{b_t}{1+r_t} + s_t p_t + c_t +t_t - \left[w_t n_t + b_{t-1} + s_{t-1}\left(d_t+p_t\right) \right] \\ 
    w_t &= a_t \left(1-\alpha-\beta\right) g^{1-\alpha-\beta}_t k^{\alpha}_{t-1} m^{\beta}_t n^{-\alpha-\beta}_t \times \left[\left(1-\lambda^{wr}+\lambda^{wr}R_t\right) + \frac{\mu_t}{\left(\gamma_t+\omega\theta_{\pi}\mu_t\right)} \lambda^{wr} R_t \right]^{-1} \\
    P^{M}_t &= a_t \beta g^{1-\alpha-\beta}_t k^{\alpha}_{t-1} m^{\beta-1}_t n^{1-\alpha-\beta}_t \times \left[\left(1-\lambda^{wr}+\lambda^{wr}R_t\right) + \frac{\mu_t}{\left(\gamma_t+\omega\theta_{\pi}\mu_t\right)} \lambda^{wr} R_t \right]^{-1} \\
    \gamma_t &= \frac{1}{1+2\psi \left(d_t-\bar{d}\right)} \\
    1 &= \beta \mathbb{E}_t g^{-\sigma}_t \frac{\lambda_{t+1}}{\lambda_{t}} \frac{1+2\psi\left(d_t-\bar{d}\right)}{1+2\psi\left(d_{t+1}-\bar{d}\right)} R_t + \mu_t \left[1+2\psi\left(d_t-\bar{d}\right) \right] \frac{r_t}{1+r_t} R_t \\
    Q_t &= \beta \mathbb{E}_t g^{-\sigma}_t \frac{\lambda_{t+1}}{\lambda_t} \left[\alpha \frac{y_{t+1}}{k_t} \left(\frac{1}{1+2\psi\left(d_t - \bar{d}\right)} + \mu_{t+1} \omega \theta_{\pi} \right) + Q_{t+1} \left(1-\delta\right)\right] + \\
    & \qquad + \mathbb{E}_t \mu_t \left(1-\omega\right) \theta_{k} p^{k}_{t+1} \left(1-\delta\right) \nonumber \\
    \frac{1}{1+2\psi\left(d_t-\bar{d}\right)} &= Q_t \left[ 1-\frac{\phi}{2}\left(\frac{i_{t}}{i_{t-1}}g_{t-1}-\bar{g}\right)- \phi \left( \frac{i_{t}}{i_{t-1}}g_{t-1} - \bar{g} \right)  \frac{i_{t}}{i_{t-1}}g_{t-1} \right] \\
    & \quad - \mathbb{E}_t \beta \frac{\lambda_{t+1}}{\lambda_{t}} g^{-\sigma}_t Q_{t+1} \left[ \phi \left(\frac{i_{t+1}}{i_{t}}g_t - \bar{g} \right) \right] \left(\frac{i_{t+1}}{i_{t}}g_t\right)^2 \nonumber \\
    k_{t} g_t &= \left(1-\delta\right) k_{t-1} + i_t \left[1-\frac{\phi}{2}\left(\frac{i_t}{i_{t-1}}g_{t-1} - \bar{g} \right)^2 \right]    
\end{align}
\begin{align} 
    \frac{b^{f}_t}{1+r_t} g_t &+ \lambda^{wk} R_t \left(P^{M_t} m_t + w_t n_t\right) = \omega \theta_{\pi} \left[a_{t} \left(k_{t-1}\right)^{\alpha} m^{\beta}_t g^{1-\alpha-\beta}_t n^{1-\alpha-\beta}_t -\left(1-\lambda^{wk}+\lambda^{wk}\right)\left(P^{M}m_t+w_t n_t\right)  \right] \nonumber \\
    & \qquad + \theta_{k} \left(1-\omega\right)\mathbb{E}_t p^{k}_{t+1} k_t \left(1-\delta\right)g_t  \\
    p^{k}_t &= Q_t
\end{align}

\noindent
\textbf{Analytically computing the steady state.} From the steady state equilibrium conditions we know that transitory productivity process is equal to
\begin{align*}
    a = 1
\end{align*}
and the growth rate of the trend productivity process is equal to
\begin{align*}
    g = \bar{g}
\end{align*}
For a given discount factor $\beta$, we know that 
\begin{align*}
    \left(1+r^{*}\right) &= \frac{\bar{g}^{\sigma}}{\beta}  \\
    r^{*} &= \frac{\bar{g}^{\sigma}-\beta}{\beta}
\end{align*}
which leads to a straightforward expression for the interest rate which firms face, given a tax advantage $\tau$,
\begin{align*}
    R = 1+r^{*}\left(1-\tau\right)
\end{align*}
We also know that 
\begin{align*}
    \gamma &= 1 \\
    Q  &= 1 \\
    p^{k} &= 1
\end{align*}
We can compute the Lagrange multiplier on firms' borrowing constraint
\begin{align*}
    \mu = \left(1+r^{*}\right) \left[\frac{1}{R} - \beta \bar{g}^{-\sigma} \right] 
\end{align*}

\noindent
Next, we turn to computing the steady state value of the following four variables
\begin{align*}
    \{k,n,m,w\}
\end{align*}
To do so, we exploit the following four equilibrium conditions
\begin{align*}
    w \left(1-\lambda^{wk}+\lambda^{wk} R\right)&= a \left(1-\alpha - \beta\right) \bar{g}^{1-\alpha - \beta} k^{\alpha} m^{\beta} n^{-\alpha -\beta} \\
    P^{M} \left(1-\lambda^{wk}+\lambda^{wk} R\right) &= a \beta \bar{g}^{1-\alpha -\beta} k^{\alpha} m^{\beta-1} n^{1-\alpha -\beta} \\
    1 &= \beta  \bar{g}^{-\sigma} \left[\alpha \frac{y}{k} \left(1 + \mu \theta^{IC} \right) + \left(1-\delta\right)\right]\\
    w &= \zeta \theta n^{\theta -1} 
\end{align*}
First, we re-organize
\begin{align*}
    1 &= \beta  \bar{g}^{-\sigma} \left[\alpha \frac{y}{k} \left(1 + \mu \theta^{IC} \right) + \left(1-\delta\right)\right]
\end{align*}
to obtain an expression for $y/k$
\begin{align*}
    \alpha \frac{y}{k} &= \frac{1}{1+\mu \theta^{IC}} \bigg\{ \left(\frac{1}{\beta \bar{g}^{-\sigma}}\right) - \left(1-\delta\right)\bigg\} \\
    \alpha k^{\alpha-1}m^{\beta}n^{1-\alpha-\beta} &= \underbrace{\frac{1}{1+\mu \theta^{IC}} \bigg\{ \left(\frac{1}{\beta \bar{g}^{-\sigma}}\right) - \left(1-\delta\right)\bigg\}}_{P^{K}}
\end{align*}
Hence, our system of four equations and four unknowns becomes
\begin{align*}
    w \left[\left(1-\lambda^{wr}+\lambda^{wr}R_t\right) + \frac{\mu_t \lambda^{wr} R_t}{\left(\gamma_t+\theta^{IC}\mu_t\right)}  \right] &= a \left(1-\alpha - \beta\right) \bar{g}^{1-\alpha - \beta} k^{\alpha} m^{\beta} n^{-\alpha -\beta} \\
    P^{M} \left[\left(1-\lambda^{wr}+\lambda^{wr}R_t\right) + \frac{\mu_t \lambda^{wr} R_t}{\left(\gamma_t+\theta^{IC}\mu_t\right)}  \right]  &= a \beta \bar{g}^{1-\alpha -\beta} k^{\alpha} m^{\beta-1} n^{1-\alpha -\beta} \\
    P^K &= \alpha k^{\alpha-1}m^{\beta}\bar{g}^{1-\alpha-\beta}n^{1-\alpha-\beta}\\
    w &= \zeta \theta n^{\theta -1} 
\end{align*}
Note that in steady state $\left[\left(1-\lambda^{wr}+\lambda^{wr}R_t\right) + \frac{\mu_t}{\left(\gamma_t+\theta^{IC}\mu_t\right)} \lambda^{wr} R_t \right]$ becomes
\begin{align*}
    \text{Wedge}^{SS} = \left[\left(1-\lambda^{wr}+\lambda^{wr}R\right) + \frac{\mu \lambda^{wr} R}{\left(1+\theta^{IC}\mu\right)}  \right]
\end{align*}
as $\gamma = 1$ in steady state. We can re-write our system of equations as
\begin{align*}
    w \text{Wedge}^{SS} &= a \left(1-\alpha - \beta\right) \bar{g}^{1-\alpha - \beta} k^{\alpha} m^{\beta} n^{-\alpha -\beta} \\
    P^{M} \text{Wedge}^{SS}  &= a \beta \bar{g}^{1-\alpha -\beta} k^{\alpha} m^{\beta-1} n^{1-\alpha -\beta} \\
    P^K &= \alpha k^{\alpha-1}m^{\beta}\bar{g}^{1-\alpha-\beta}n^{1-\alpha-\beta}\\
    w &= \zeta \theta n^{\theta -1} 
\end{align*}

\noindent
First, we combine the second and third equations to obtain an expression for $m$ as a function of $k$ and parameters
\begin{align*}
    \frac{P^{K}}{P^{M}\text{Wedge}^{SS}} &= \frac{\alpha}{\beta} \frac{m}{k} \\
    \Rightarrow m &= \frac{P^{K}}{P^{M}\text{Wedge}^{SS}} \frac{\beta}{\alpha} k  
\end{align*}
Next, we plug in the expression for $m$ in the third equation
\begin{align*}
    P^K &= \alpha k^{\alpha-1}m^{\beta}\bar{g}^{1-\alpha-\beta}n^{1-\alpha-\beta} \\
    P^K &= \alpha k^{\alpha-1}\left[\frac{P^{K}}{P^{M} \text{Wedge}^{SS}} \frac{\beta}{\alpha} k \right]^{\beta}\bar{g}^{1-\alpha-\beta}n^{1-\alpha-\beta} \\
    P^K &= \alpha \left[\frac{P^{K}}{P^{M} \text{Wedge}^{SS}} \frac{\beta}{\alpha} \right]^{\beta} k^{\alpha + \beta-1} \bar{g}^{1-\alpha-\beta}n^{1-\alpha-\beta} \\
    P^K &= \alpha \left[\frac{P^{K}}{P^{M} \text{Wedge}^{SS}} \frac{\beta}{\alpha} \right]^{\beta} \left(\frac{k}{\bar{g}n}\right)^{\alpha+\beta-1}
\end{align*}
which allows us to construct an expression for the ratio $k/\bar{g}n$
\begin{align*}
    \left(\frac{k}{\bar{g}n}\right)^{\alpha+\beta-1} &= P^{K} \frac{1}{\alpha} \left[\frac{P^{K}}{P^{M} \text{Wedge}^{SS}} \frac{\beta}{\alpha} \right]^{-\beta} \\
    \frac{k}{\bar{g}n} &= \left[P^{K} \frac{1}{\alpha} \left[\frac{P^{K}}{P^{M} \text{Wedge}^{SS}} \frac{\beta}{\alpha} \right]^{-\beta} \right]^{\frac{1}{\alpha+\beta-1}}
\end{align*}
Following, we use the first equation in our system, and start by introducing our expression for $m$ depending on $k$
\begin{align*}
    w &= a \left(1-\alpha - \beta\right) \bar{g}^{1-\alpha - \beta} k^{\alpha} m^{\beta} n^{-\alpha -\beta} \\
    w &= a \left(1-\alpha - \beta\right) \bar{g}^{1-\alpha - \beta} k^{\alpha} \left[\frac{P^{K}}{P^{M} \text{Wedge}^{SS}} \frac{\beta}{\alpha} k \right]^{\beta} n^{-\alpha -\beta} \\
    w &= a \left(1-\alpha-\beta\right) \bar{g}^{1-\alpha - \beta} \left[\frac{P^{K}}{P^{M} \text{Wedge}^{SS}} \frac{\beta}{\alpha} \right]^{\beta} k^{\alpha+\beta} n^{-\alpha-\beta}
\end{align*}
Re-arranging the terms, we can express wages $w$ as a function of known parameters
\begin{align*}
    w &= a \left(1-\alpha-\beta\right)  \left[\frac{P^{K}}{P^{M} \text{Wedge}^{SS}} \frac{\beta}{\alpha} \right]^{\beta} \left(\frac{k}{\bar{g}n}\right)^{\alpha+\beta} \bar{g}
\end{align*}
Next, we use the last equation, to compute the steady state labor
\begin{align*}
    w &= \zeta \theta n^{\theta -1} \\
    n &= \left(\frac{w}{\zeta \theta}\right)^{\theta-1}
\end{align*}
Next, we can back out the steady state stock of capital as
\begin{align*}
    k = \frac{k}{\bar{g}n} \times \bar{g} n
\end{align*}
Finally, we use the expression for capital to compute the steady state level of intermediate inputs
\begin{align*}
    m &= \frac{P^{K}}{P^{M} \text{Wedge}^{SS}} \frac{\beta}{\alpha} k  
\end{align*}

Next, we can compute the steady state investment exploiting the law of motion for capital
\begin{align*}
    k \bar{g} &= \left(1-\delta\right) k + i \\
    i  &= \left[\bar{g} - \left(1-\delta\right) \right] k  
\end{align*}
and the level of production using the production function
\begin{align*}
    y = k^{\alpha} m^{\beta} \left(\bar{g} n \right)^{1-\alpha-\beta}
\end{align*}

\newpage
\section{Additional Details on Section \ref{sec:nominal_frictions_policy} New Keynesian Model} \label{sec:appendix_details_simple_model_nk}

\subsection{Households} \label{subsec:households}

The utility function is
\begin{align*}
    &\mathbb{E}_0 \sum^{\infty}_{t=0} \beta^{t} \bigg\{\frac{\left[c_t - \theta \omega^{-1} L^{\omega}_t \right]^{1-\sigma}-1}{1-\sigma} + h_t \left(\frac{S_t D_t}{P^{c}_t} \right)  + \\
    & \frac{v_t}{P_t} \left[ S_tR^{*}_{t-1}D^{*}_{t-1} + R_{d,t-1} D_{t-1} + W_t L_t + \left(P^{E}_t + P_tD^{E}_t\right) s^{E}_{t-1} - \left(D_t + S_t D^{*}_t + P^{c}_t c_t + s^{E}_t P^{E}_t + T_t \right)    \right] \bigg\}
\end{align*}
where function $h_t \left(\frac{S_t D_t}{P^{c}_t} \right)$ is given by
\begin{align*}
    h_t \left(\frac{S_t D_t}{P^{c}_t} \right) = -\frac{1}{2} \gamma \left(\frac{S_t D^{*}_t}{P^{c}_t} - \Upsilon^{*}\right)^2
\end{align*}
and $v_t/P_t$ is the Lagrange multiplier. Note that we are assuming that the dividends the household receives from owning shares/claims to the firm are paid in terms of the domestic homogeneous good. 

\noindent
Note that the derivative of function $h_t \left(\frac{S_t D_t}{P^{c}_t} \right)$ with respect to $D^{*}_t$ is
\begin{align*}
    \frac{\partial h_t \left(\frac{S_t D_t}{P^{c}_t} \right)}{\partial D^{*}_t} &= h'_t \left(\frac{S_t D_t}{P^{c}_t} \right) \frac{S_t}{P^{c}_t}  \\
    &= - \gamma \left(\frac{S_t D^{*}_t}{P^{c}_t} - \Upsilon^{*}\right) \frac{S_t}{P^{c}_t} \\
    &= - \gamma \left(\frac{S_t D^{*}_t}{P_t}\frac{P_t}{P^{c}_t} - \Upsilon^{*}\right) \frac{S_t}{P^{c}_t}
\end{align*}
Defining $d^{*}_t = S_t D^{*}_t/P_t$ and $p^{c}_t = P^{c}_t / P_t$, the derivative above becomes
\begin{align*}
    \frac{\partial h_t \left(\frac{S_t D_t}{P^{c}_t} \right)}{\partial D^{*}_t} &= - \gamma \left(\frac{d^{*}_t}{p^{c}_t}- \Upsilon^{*}\right) \frac{S_t}{P^{c}_t}
\end{align*}

Next, we compute the household's first order conditions. We start by computing the first order condition with respect to quantities of consumption, $c_t$.
\begin{align*}
    \left[c_t - \theta \omega^{-1} L^{\omega}_t \right]^{-\gamma} = \frac{v_t P^{c}_t}{P_t} = v_t p^{c}_t
\end{align*}
The first order with respect to hours worked, $L_t$, is given by
\begin{align*}
    \left[c_t - \theta \omega^{-1} L^{\omega}_t \right]^{-\gamma} \theta L^{\omega-1}_t = \frac{v_t W_t}{P_t} = v_t w_t
\end{align*}
Dividing the FOC for $L_t$ by the FOC for $c_t$
\begin{align*}
     \theta L^{\omega-1}_t = \frac{v_t w_t}{v_t p^{c}_t} = \frac{w_t}{p^{c}_t} 
\end{align*}
The first order condition with respect to $D^{*}_t$ is 
\begin{align*}
    - \gamma \left(\frac{S_t D^{*}_t}{P^{c}_t} - \Upsilon^{*}\right) \frac{S_t}{P^{c}_t} - \frac{S_t v_t}{P_t} + \beta \frac{v_{t+1} S_{t+1}}{P_{t+1}} R^{*}_t = 0
\end{align*}
Multiplying by $P^{c}_t/S_t$ we obtain
\begin{align*}
    - \gamma \left(\frac{S_t D^{*}_t}{P^{c}_t} - \Upsilon^{*}\right) - \frac{S_t v_t}{P_t} \frac{P^{c}_t}{S_t} + \beta \frac{v_{t+1} S_{t+1}}{P_{t+1}} \frac{P^{c}_t}{S_t} R^{*}_t &= 0
\end{align*}
Next, we replace $v_t$ and $v_{t+1}$ by
\begin{align*}
    v_t &= \left[c_t - \theta \omega^{-1} L^{\omega}_t \right]^{-\gamma} \frac{P_t}{P^{c}_t} \\
    v_{t+1} &= \left[c_{t+1} - \theta \omega^{-1} L^{\omega}_{t+1} \right]^{-\gamma} \frac{P_{t+1}}{P^{c}_{t+1}}
\end{align*}
to obtain
\begin{align*}
    - \gamma \left(\frac{S_t D^{*}_t}{P^{c}_t} - \Upsilon^{*}\right) - \frac{P^{c}_t}{P_t} \left[c_t - \theta \omega^{-1} L^{\omega}_t \right]^{-\gamma} \frac{P_t}{P^{c}_t} + \beta \left[c_{t+1} - \theta \omega^{-1} L^{\omega}_{t+1} \right]^{-\gamma} \frac{P_{t+1}}{P^{c}_{t+1}} \frac{s_{t+1} P^{c}_t}{P_{t+1}} R^{*}_t &= 0
\end{align*}
which becomes
\begin{align*}
    - \gamma \left(\frac{d^{*}_t}{p^{c}_t} - \Upsilon^{*}\right) - \left[c_t - \theta \omega^{-1} L^{\omega}_t \right]^{-\gamma} + \beta \frac{s_{t+1}}{\pi^{c}_{t+1}} R^{*}_t \left[c_{t+1} - \theta \omega^{-1} L^{\omega}_{t+1} \right]^{-\gamma} =0
\end{align*}
The first order condition with respect to $D_t$ is
\begin{align*}
    \frac{v_t}{P_t} = \beta \frac{v_{t+1}}{P_{t+1}} R_{d,t} 
\end{align*}
replacing $v_t$ and $v_{t+1}$
\begin{align*}
    \frac{1}{P_t} \left[c_t - \theta \omega^{-1} L^{\omega}_t \right]^{-\gamma} \frac{P_t}{P^{c}_t} &= \beta \frac{1}{P_{t+1}} R_{d,t} \left[c_{t+1} - \theta \omega^{-1} L^{\omega}_{t+1} \right]^{-\gamma} \frac{P_{t+1}}{P^{c}_{t+1}} \\
    \left[c_t - \theta \omega^{-1} L^{\omega}_t \right]^{-\gamma} \frac{1}{P^{c}_t} &= \beta R_{d,t} \left[c_{t+1} - \theta \omega^{-1} L^{\omega}_{t+1} \right]^{-\gamma} \frac{1}{P^{c}_{t+1}} \\
    1 &= \beta R_{d,t} \left(\frac{c_t - \theta \omega^{-1} L^{\omega}_t}{c_{t+1} - \theta \omega^{-1} L^{\omega}_{t+1}} \right)^{-\gamma} \frac{P^{c}_t}{P^{c}_{t+1}} \\
    1 &= \beta \frac{R_{d,t}}{\pi^c_{t+1}} \left(\frac{c_t - \theta \omega^{-1} L^{\omega}_t}{c_{t+1} - \theta \omega^{-1} L^{\omega}_{t+1}} \right)^{\gamma}
\end{align*}
Lastly, we need to compute the first order condition with respect to the household's optimal holdings of intermediate firm shares, $s^{E}_t$.
\begin{align*}
    \frac{v_t}{P_t} P^{E}_t &= \beta \frac{v_{t+1}}{P_{t+1}} \left(P^{E}_{t+1} + P_t D^{E}_{t+1} \right) \\
    v_t p^{E}_t &= \beta v_{t+1} \left(p^{E}_{t+1} + D^{E}_{t+1} \right)
\end{align*}
Using the expressions for $v_t$ and $v_{t+1}$ we obtain
\begin{align*}
    1 &= \beta \left(\frac{c_t - \theta \omega^{-1} L^{\omega}_t}{c_{t+1} - \theta \omega^{-1} L^{\omega}_{t+1}} \right)^{\gamma} \frac{\pi_t}{\pi^{c}_{t+1}} \left(\frac{p^{E}_{t+1}+D^{E}_{t+1}}{p^{E}_{t}} \right)
\end{align*}

\noindent
\textbf{Household specific variables:} \textbf{\underline{7}} $c_t$, $L_t$, $d^{*}_t$, $d_t$, $p^{E}_t$, $s^{E}_t$, $v_t$ coming from 5 FOC conditions and share market clearing.

\subsection{Goods Production}

A domestic good, $Y_t$, is produced in the same way that it happens in a closed economy model. That good is used in two separate production functions to produce final output goods. The first production function combines the homogeneous good with an imported good to produce a final consumption good. The second production function combines the homogeneous good with imported goods to produce an investment good, which in turn are an input into the production of capital.

\subsubsection{Domestic Homogeneous Goods} \label{subsub:domestic_homogeneous_goods}

\noindent
\textbf{Final goods.} Production of domestic final goods is carried out aggregating different varieties of intermediate goods
\begin{align*}
    Y_t = \left[\int^{1}_{0} y^{\frac{1}{\eta}}_{i,t} di \right]^{\eta}, \quad \eta>1
\end{align*}
The optimal behavior of the final good producing firm leads to the following demand curve:
\begin{align*}
    p_{i,t} = P_t Y^{\frac{\eta-1}{\eta}}_t y^{\frac{1-\eta}{\eta}}_{i,t}
\end{align*}
The aggregate price index of the final good becomes
\begin{align*}
    P_t = \left(\int^{1}_0 p^{\frac{1}{1-\eta}}_{i,t}\right)^{1-\eta}
\end{align*}
It is useful to define the intermediate firms' real demand and real revenue as a function of aggregate variables $Y_t$, $P_t$ and the quantities $Y_{i,t}$. Real demand, i.e., in terms of the domestic homogeneous good:
\begin{align*}
    D_{i,t} = \frac{p_{i,t}}{P_t} = Y^{\frac{\eta-1}{\eta}}_t y^{\frac{1-\eta}{\eta}}_{i,t}
\end{align*}
Real revenues, i.e., in terms of the domestic homogeneous good:
\begin{align*}
    F_{i,t} = \frac{p_{i,t} y_{i,t}}{P_t} = Y^{\frac{\eta-1}{\eta}}_t y^{\frac{1}{\eta}}_{i,t}
\end{align*}

\noindent
\textbf{Intermediate goods.} A continuum of size 1 of firms produce intermediate goods. They set prices subject to a Rotemberg adjustment cost. All intermediate producers operate the technology 
\begin{align*}
    y_{i,t} = z_t \left(k_{i,t-1}\right)^{\alpha} m^{\gamma}_t n^{1-\alpha-\eta}_{i,t}
\end{align*}
where $\alpha,\eta \in (0,1)$ are the capital and imported inputs share in production, $n_{i,t}$ denotes the amount of labor that firm $i$ hires, $z$ is total factor productivity. The firm's per period cash-flow in terms of the domestic final good is defined as
\begin{align*}
    \pi_{i,t} = \frac{p_{i,t} y_{i,t} - \left(1-\phi + \phi \tilde{R}_t \right) \left(W_t n_{i,t} + P^{m}_t m_{i,t} \right)}{P_t}
\end{align*}

Capital $k_{i,t-1}$ is owned and accumulated by firms. Its law of motion is
\begin{align*}
    k_{i,t} = \left(1-\delta\right) k_{i,t-1} + \nu_{i,t} \left[ 1-\Phi \left(\frac{i_{i,t}}{i_{i,t-1}}\right)\right] i_t
\end{align*}
where $\delta$ is the depreciation rate and the $\Phi (.)$ function introduces investment adjustment costs. The term $\nu_{i,t}$ is a stochastic disturbance, which captures both investment-specific technology as well as the marginal efficiency of investment. Both the presence of investment adjustment costs as well as the stochastic disturbance $\nu_{i,t}$ lead to variation in the market value of capital.

Consider the firm's flow of funds constraint
\begin{align*}
    P_t F_t + B_t = \tilde{R}_{t} B_{t-1}  + \left(1-\phi+\phi \tilde{R}_t\right) \left(W_t n_t + P^{m}_t m_{t} \right) + P_t \Psi \left(D^{E}_t\right) + P^{I}_t i_t + P_t \Upsilon^{P}_t \left(p_{i,t-1},p_{i,t},Y_t \right)
\end{align*}
where $\Psi\left(d_t\right)$ is a dividend adjustment cost function, $\Upsilon \left(P_{i,t-1}, P_{i,t}\right)$ is the Rotemberg price adjustment costs, $P^{i}_t$ is the price of investment goods. The flow of funds constraint, in terms of the domestic homogeneous good, is given by
\begin{align*}
    F_t + \frac{B_t}{P_t} = \frac{\tilde{R}_{t} B_{t-1}}{P_t}  + \left(1-\phi + \phi \tilde{R}_t\right) \left(w_t n_t +p^{m}_t m_t \right) + \Psi \left(D^{E}_t\right) + p^{I}_t i_t + \Upsilon^{P}_t \left(p_{i,t-1},p_{i,t},Y_t \right)
\end{align*}
where $p^{i}_t = P^{i}_t / P_t$ is the relative price of the investment good.

Next, we address the intermediate firm's financing. We assume that firm finances a fraction $\phi$ of total debt in domestic currency and finances a fraction $1-\phi$ of total debt is financed in terms of foreign currency. For intertemporal debt, this implies that
\begin{align*}
    B^{\text{peso}}_t &= \phi B_t \\
    S_t B^{\text{dollar}}_t &= \left(1-\phi\right) B_t \\
\end{align*}
where $B^{\text{peso}}_t$ represents the amount of domestic currency debt which is financed at interest rate
\begin{align*}
    \tilde{R}_{d,t} = 1 + \left(R_{d,t}-1\right)\times\tau
\end{align*}
and $B^{\text{dollar}}_t$ represents the amounts of dollars borrowed at the non-state contingent rate
\begin{align*}
    \tilde{R}^{*}_{t} = 1 + \left(R^{*}_{t}-1\right)\times\tau
\end{align*}
where $\tau$ represents a tax advantage for firms financing through debt. Note that while interest rates paid in period $t+1$ are fixed in period $t$ are non-state contingent, the effective interest rate paid is
\begin{align*}
    \tilde{R}_{t} = s_t \left(1-\phi\right) \tilde{R}^{*}_{t-1} + \phi \tilde{R}_{d,t-1}
\end{align*}
Note, we can define debt in real-terms as
\begin{align*}
    b_t &= \frac{B_t}{P_t} \\
    b^{\text{peso}}_t &= \frac{B^{\text{peso}}_t}{P_t}\\
    b^{\text{dollar}}_t &= \frac{S_t B^{\text{dollar}}_t}{P_t}
\end{align*}
which implies
\begin{align*}
    b_t = b^{\text{peso}}_t + b^{\text{dollar}}_t
\end{align*}

We assume that firms face a working capital requirement as stipulated above. Firms must finance a fraction $\phi$ of their labor and imported input purchases before production which must be financed at rate $\tilde{R}_t$. We denote their intra-temporal working capital borrowing in nominal and real terms as
\begin{align*}
    L_t &= \phi \times \left(W_t n_t + P^{m}_t m_t  \right) \\
    l_t &= \phi \times \left(w_t n_t + p^{m}_t m_t  \right)
\end{align*}

We assume that the intermediate firm faces the following borrowing constraint as a function of its cash-flow
\begin{align*}
    r_t B^{IC}_{t} &\leq \theta^{IC} \pi_{i,t} 
\end{align*}
where $\tilde{r}_t = \tilde{R}_t - 1$ is the net interest rate, and $B^{IC}_t$ is given by
\begin{align*}
    B^{IC}_{t} = B_t + L_t
\end{align*}

The firm's objective function is to maximize the expected stream of real dividends $D^{E}_t$, discounted at the owner's discount factor $\Lambda_t$:
\begin{small}
\begin{align*}
    & \mathcal{L}^{i}: \quad \mathbb{E}_0 \sum^{\infty}_{t=0} \Lambda_t \times \bigg\{ D^{E}_t + \\
    & \qquad + \lambda^{i}_t \left[ P_t F_t + B_t - \tilde{R}_{t} B_{t-1}  - \left(1-\phi+\phi \tilde{R}_t \right)\left(W_t n_t + P^{m}_t m_t\right) - P_t \Psi \left(D^{E}_t\right) - P^{I}_t i_t - P_t \Upsilon^{P}_t \left(p_{i,t-1},p_{i,t},Y_t \right) \right] \\
    & \qquad + Q_{i,t} \left[\left(1-\delta\right) k_{t-1} + \Phi \left(i_{t-1},i_t\right) - k_t \right] \\
    & \qquad + \zeta_t \left( \frac{p_{i,t}}{P_t} - D_t \right) \\
    & \qquad + \mu_t \left[\theta^{IC} \left(F_t - \frac{W_t}{P_t} n_t \right) - \frac{\tilde{r}_t \times \left(B_t+L_t\right)}{P_t}\right]\bigg\}
\end{align*}
\end{small}
The first order condition with respect to $D^{E}_t$ is given by
\begin{align*}
    \Lambda_t &= \Lambda_t \lambda^{i}_t P_t \Psi_{d,t} \left(D^{E}_t\right) \\
    1 &= \lambda^{i}_t P_t \Psi_{d,t} \left(D^{E}_t\right)
\end{align*}
It is useful to define
\begin{align*}
    \lambda^{i}_t &= \frac{1}{P_t \Psi_{d,t} \left(D^{E}_t\right)}
\end{align*}
Variable $\lambda^{i}_t$ represents the shadow price of an additional dollar for the intermediate firm.

\noindent
The first order condition with respect to $p_{i,t}$
\begin{align*}
    - \Lambda_{t} \lambda^{i}_t \phi \left(\frac{p_{i,t}}{p_{i,t-1}} - 1 \right) \frac{P_{t}}{p_{i,t-1}} Y_t + \Lambda_t \frac{\zeta_{i,t}}{P_t} + \Lambda_{t+1} \lambda^{i}_{t+1} \phi \left(\frac{p_{i,t+1}}{p_{i,t}} -1 \right) \frac{p_{i,t+1}}{p^2_{i,t}} P_{t+1}  Y_{t+1} = 0
\end{align*}
Dividing by $\Lambda_t$ we obtain
\begin{align*}
    -\lambda^{i}_{t} \phi \left(\frac{p_{i,t}}{p_{i,t-1}}-1 \right) \frac{P_t}{p_{i,t-1}} Y_t + \frac{\zeta_t}{P_t} + \mathbb{E}_t m_{t+1} \lambda^{i}_{t+1} \phi \left(\frac{p_{i,t+1}}{p_{i,t}}-1\right) P_{t+1} Y_{t+1} \frac{p_{i,t+1}}{p^2_{i,t+1}} = 0
\end{align*}
We replace $\lambda^{i}_{t}$ and $\lambda^{i}_{t+1}$ using the following expressions
\begin{align*}
    \lambda^{i}_{t} &= \frac{1}{P_t \Psi_{d,t}} \\
    \lambda^{i}_{t+1} &= \frac{1}{P_{t+1} \Psi_{d,t+1}}
\end{align*}
we obtain
\begin{align*}
    - \frac{1}{P_t \Psi_{d,t}} \phi \left(\frac{p_{i,t}}{p_{i,t-1}}-1 \right) \frac{P_t}{p_{i,t-1}} Y_t + \frac{\zeta_t}{P_t} + \mathbb{E}_t m_{t+1} \frac{1}{P_{t+1} \Psi_{d,t+1}} \phi \left(\frac{p_{i,t+1}}{p_{i,t}}-1\right) P_{t+1} Y_{t+1} \frac{p_{i,t+1}}{p^2_{i,t+1}} = 0
\end{align*}
Imposing a symmetric equilibrium, i.e. $p_{i,t} = P_t$ and multiplying by $P_t$
\begin{align*}
    - \frac{1}{\Psi_{d,t}} \phi \left(\frac{P_{t}}{P_{t-1}}-1 \right) \frac{P_t}{P_{i,t-1}} Y_t + \zeta_t + \mathbb{E}_t m_{t+1} \frac{P_t}{P_{t+1} \Psi_{d,t+1}} \phi \left(\frac{P_{i,t+1}}{P_{i,t}}-1\right) P_{t+1} Y_{t+1} \frac{P_{t+1}}{P^2_{t}} = 0
\end{align*}
Canceling terms we obtain
\begin{align*}
    - \frac{1}{\Psi_{d,t}} \phi \left(\frac{P_{t}}{P_{t-1}}-1 \right) \frac{P_t}{P_{i,t-1}} Y_t + \zeta_t + \mathbb{E}_t m_{t+1} \frac{1}{\Psi_{d,t+1}} \phi \left(\frac{P_{i,t+1}}{P_{i,t}}-1\right) Y_{t+1} \frac{P_{t+1}}{P_{t}} = 0
\end{align*}
Next, we can define the inflation rates as $\pi_t = P_{t}/P_{t-1}$ and $\pi_{t+1} = P_{t+1}/P_t$. Replacing for this expressions
\begin{align*}
    - \frac{1}{\Psi_{d,t}} \phi \left(\pi_t-1 \right) \pi_t Y_t + \zeta_t + \mathbb{E}_t m_{t+1} \frac{1}{\Psi_{d,t+1}} \phi \left(\pi_{t+1}-1\right) \pi_{t+1} Y_{t+1}  = 0
\end{align*}
Finally, we can multiply by $\lambda^{i}_{t}$, we obtain
\begin{align*}
    - \phi \left(\pi_t-1 \right) \pi_t Y_t + \Psi_{d,t} \zeta_t + \mathbb{E}_t m_{t+1} \frac{\Psi_{d,t}}{\Psi_{d,t+1}} \phi \left(\pi_{t+1}-1\right) \pi_{t+1} Y_{t+1}  = 0
\end{align*}

\noindent
The first order condition with respect to $i_t$
\begin{align*}
    \Lambda_t \lambda^{i}_t P^{I}_t + \Lambda_t Q_{i,t} \Phi_{1,t,t} + \Lambda_{t+1} Q_{i,t+1} \Phi_{2,t,t+1} = 0
\end{align*}
Replacing by $\lambda^{i}_{t} = \frac{1}{P_t \Psi_{d,t}}$ and dividing by $\Lambda_t$ we obtain
\begin{align*}
    \frac{p^{I}_t}{\Psi_{d,t}} = Q_{i,t} \Phi_{1,t,t} + \mathbb{E}_t m_{t+1} Q_{i,t+1} \Phi_{2,t,t+1}
\end{align*}

\noindent
The first order condition with respect to $B_t$ is
\begin{align*}
    \Lambda_t \lambda^{i}_t - \Lambda_{t+1} \lambda^{i}_{t+1} \tilde{R}_{t+1} - \frac{\Lambda_t \times \tilde{r}_t \mu_{t}}{P_t} = 0
\end{align*}
Dividing by $\Lambda_t$ we obtain
\begin{align*}
    \lambda^{i}_t = \mathbb{E}_t m_{t+1} \lambda^{i}_{t+1} \tilde{R}_{t+1} + \frac{\tilde{r}_t \mu_t}{P_t}
\end{align*}
We replace $\lambda^{i}_t = \frac{1}{P_t \Psi_{d,t}}$ and $\lambda^{i}_{t+1} = \frac{1}{P_{t+1} \Psi_{d,t+1}}$ to obtain
\begin{align*}
    \frac{1}{P_t \Psi_{d,t}} = \mathbb{E}_t m_{t+1} \frac{\tilde{R}_{t+1}}{P_{t+1} \Psi_{d,t+1}} + \frac{\mu_t}{P_t}
\end{align*}
Multiplying both sides of the equation by $P_t$ we obtain
\begin{align*}
    \frac{1}{\Psi_{d,t}} &= \mathbb{E}_t m_{t+1} \frac{\tilde{R}_{t+1}}{\pi_{t+1} \Psi_{d,+1}} + \tilde{r}_t \times \mu_t
\end{align*}
Multiplying both sides by $\Psi_{d,t}$ we obtain
\begin{align*}
    1 &= \mathbb{E}_t m_{t+1} \frac{\tilde{R}_{t+1}}{\pi_{t+1}} \frac{\Psi_{d,t}}{\Psi_{d,t+1}} + \Psi_{d,t} \times  \tilde{r}_t \times \mu_t
\end{align*}

\noindent
We compute the first order condition with respect to $n_t$ as
\begin{align*}
    \lambda^{i}_t \left(P_t F_{n,t} - \left(1-\phi + \phi \tilde{R}_t \right) W_t \right) - \zeta_t D_{n,t} + \mu_{t} \theta^{IC} \left(F_{n,t} - \frac{W_t}{P_t} \right) - \mu_t \tilde{r}_t \phi w_t = 0
\end{align*}
Replacing $\lambda^{i}_t = \frac{1}{P_t \Psi_{d,t}}$ we obtain
\begin{align*}
    \left(\frac{P_t F_{n,t} - \left(1-\phi + \phi \tilde{R}_t \right) W_t}{P_t} \right) \frac{1}{\Psi_{d,t}} + \mu_t \theta^{IC} \left(F_{n,t} - \frac{\left(1-\phi + \phi \tilde{R}_t \right) W_t}{P_t} \right) &= \zeta_t D_{n,t} + \mu_t \tilde{r}_t \phi w_t \\
    \left(F_{n,t} - \left(1-\phi + \phi \tilde{R}_t \right) w_t  \right) \left(\frac{1}{\Psi_{d,t}} + \mu_t \theta^{IC} \right) &= \zeta_t D_{n,t} + \mu_t \tilde{r}_t \phi w_t 
\end{align*}
We can re-write this expression to obtain
\begin{align*}
    F_{n,t} \left(\frac{1}{\Psi_{d,t}} + \mu_t \theta^{IC} \right) &= w_t \left(1-\phi+\phi \tilde{R}_t \right) \left(\frac{1}{\Psi_{d,t}} + \mu_t \theta^{IC} \right) + \zeta_t D_{n,t} + \mu_t \phi \tilde{r}_t w_t \\
    F_{n,t} \left(\frac{1}{\Psi_{d,t}} + \mu_t \theta^{IC} \right) &= w_t \left[ \left(1-\phi+\phi \tilde{R}_t \right) \left(\frac{1}{\Psi_{d,t}} + \mu_t \theta^{IC} \right) + \mu_t \phi \tilde{r}_t  \right] + \zeta_t D_{n,t}
\end{align*}
In steady state, this condition becomes
\begin{align*}
    F_{n,t} \left(\frac{1}{\Psi_{d,t}} + \mu_t \theta^{IC} \right) &= w_t \left[ \left(1-\phi+\phi \tilde{R}_t \right) \left(\frac{1}{\Psi_{d,t}} + \mu_t \theta^{IC} \right) + \mu_t \phi \tilde{r}_t  \right] + \zeta_t D_{n,t} \\
    F_{n} \left(1 + \mu \theta^{IC} \right) &= w \left[ \left(1-\phi+\phi \tilde{R} \right) \left(1 + \mu \theta^{IC} \right) + \mu \phi \tilde{r}  \right] \\
    F_{n}  &= w \left[ \left(1-\phi+\phi \tilde{R} \right) \left(1 + \mu \theta^{IC} \right) + \mu \phi \tilde{r}  \right] \left(1 + \mu \theta^{IC} \right)^{-1}
\end{align*}

\noindent
We compute the first order condition with respect to $m_t$ as
\begin{align*}
    \lambda^{i}_t \left(P_t F_{m,t} - \left(1-\phi + \phi \tilde{R}_t \right) P^{m}_t \right) - \zeta_t D_{m,t} + \mu_{t} \theta^{IC} \left(F_{m,t} - \frac{P^{m}_t}{P_t} \right) - \mu_t \tilde{r}_t \phi p^{m}_t = 0
\end{align*}
Replacing $\lambda^{i}_t = \frac{1}{P_t \Psi_{d,t}}$ we obtain
\begin{align*}
    \left(\frac{P_t F_{m,t} - \left(1-\phi + \phi \tilde{R}_t \right) P^{m}_t}{P_t} \right) \frac{1}{\Psi_{d,t}} + \mu_t \theta^{IC} \left(F_{m,t} - \frac{\left(1-\phi + \phi \tilde{R}_t \right) P^{m}_t}{P_t} \right) &= \zeta_t D_{m,t} + \mu_t \tilde{r}_t \phi p^{m}_t \\
    \left(F_{m,t} - \left(1-\phi + \phi \tilde{R}_t \right) p^{m}_t  \right) \left(\frac{1}{\Psi_{d,t}} + \mu_t \theta^{IC} \right) &= \zeta_t D_{n,t} + \mu_t \tilde{r}_t \phi p^{m}_t 
\end{align*}
Once again, we can re-write this expression as
\begin{align*}
    F_{m,t} \left(\frac{1}{\Psi_{d,t}} + \mu_t \theta^{IC} \right) &= p^{m}_t \left(1-\phi + \phi \tilde{R}_t \right)   \left(\frac{1}{\Psi_{d,t}} + \mu_t \theta^{IC} \right) + \zeta_t D_{m,t} + \mu_t \phi \tilde{r}_t p_{m,t} \\
    F_{m,t} \left(\frac{1}{\Psi_{d,t}} + \mu_t \theta^{IC} \right) &= p_{m,t} \left[\left(1-\phi + \phi \tilde{R}_t \right)   \left(\frac{1}{\Psi_{d,t}} + \mu_t \theta^{IC} \right) + \mu_t \phi \tilde{r}_t \right] + \zeta_t D_{m,t} 
\end{align*}
In steady state, this condition becomes
\begin{align*}
    F_{m,t} \left(\frac{1}{\Psi_{d,t}} + \mu_t \theta^{IC} \right) &= p_{m,t} \left[\left(1-\phi + \phi \tilde{R}_t \right)   \left(\frac{1}{\Psi_{d,t}} + \mu_t \theta^{IC} \right) + \mu_t \phi \tilde{r}_t \right] + \zeta_t D_{m,t} \\
    F_{m} \left(1 + \mu_t \theta^{IC} \right) &= p_{m} \left[\left(1-\phi + \phi \tilde{R} \right)   \left(1 + \mu \theta^{IC} \right) + \mu \phi \tilde{r} \right]
\end{align*}

\noindent
Lastly, we compute the first order condition with respect to $k_t$.
\begin{align*}
    - \Lambda_t Q_t + \Lambda_{t+1} \lambda^{i}_{t+1} F_{k,t+1} P_{t+1} + \Lambda_{t+1}Q_{t+1} \left(1-\delta\right) - \Lambda_{t+1} \zeta_{t+1} D_{k,t+1} + \Lambda_{t+1} \mu_{t+1} \theta^{IC} F_{k,t+1} = 0
\end{align*}
Dividing by $\Lambda_t$ and collecting terms we obtain
\begin{align*}
    - Q_t + \frac{\Lambda_{t+1}}{\Lambda_t} \lambda^{i}_{t+1} F_{k,t+1} P_{t+1} + \frac{\Lambda_{t+1}}{\Lambda_t} Q_{t+1} \left(1-\delta\right) - \frac{\Lambda_{t+1}}{\Lambda_t} \zeta_{t+1} D_{k,t+1} + \frac{\Lambda_{t+1}}{\Lambda_t} \mu_{t+1} \theta^{IC} F_{k,t+1} &= 0 \\
    \mathbb{E}_t m_{t+1} \left[F_{k,t+1} \left(\lambda^{i}_{t+1} P_{t+1} +\mu_{t+1} \theta^{IC}\right) - \zeta_t D_{k,t+1} + Q_{t+1} \left(1-\delta\right) \right] &= Q_t
\end{align*}
Replacing $\lambda^{i}_{t+1} = \frac{1}{P_{t+1} \Psi_{d,t+1}}$ we obtain
\begin{align*}
    \mathbb{E}_t m_{t+1} \left[F_{k,t+1} \left( \frac{1}{\Psi_{d,t+1}} +\mu_{t+1} \theta^{IC}\right) - \zeta_t D_{k,t+1} + Q_{t+1} \left(1-\delta\right) \right] &= Q_t
\end{align*}

\noindent
Lastly, we compute the derivatives of functions $D_t$ and $F_t$ with respect to $n_t$, $m_t$ and $k_t$.  Imposing a symmetric equilibrium, such that $Y_t = y_{i,t}$ the condition becomes
\begin{align*}
    D_{n,t} &= \frac{1-\eta}{\eta} \left(1-\alpha-\gamma\right)\frac{1}{n_t} \\
    D_{m,t} &= \frac{1-\eta}{\eta}\gamma\frac{1}{m_t} \\
    D_{k,t+1} &= \frac{1-\eta}{\eta} \alpha \frac{1}{k_t}     
\end{align*}
Next, we turn to the derivatives of function $F$ with respect to $n_t$, $m_t$ and $k_t$
\begin{align*}
    F_{n,t} &= \frac{1}{\eta} \left(1-\alpha-\gamma\right) \frac{y_t}{n_t} \\
    F_{m,t} &= \frac{1}{\eta} \gamma \frac{y_t}{m_t} \\
    F_{k,t+1} &= \frac{1}{\eta} \alpha \frac{y_{t+1}}{k_t}
\end{align*}

\subsubsection{Final Goods} \label{subsubsec:final_goods}

\noindent
\textbf{Investment goods.} There is a large number of identical investment producing firms, with entry/exit prohibited. Because of their large number, each takes input and output prices as given.

To produce investment goods, $I_t$, the capital good firm combines domestic and foreign investment goods using the following production function:
\begin{align*}
    I_t \left(I_{d,t},I_{m,t}\right) = \left[ \gamma^{\frac{1}{\nu_I}}_I I^{\frac{\nu_I-1}{\nu_I}}_{d,t} + \left(1-\gamma_I\right)^{\frac{1}{\nu_I}} I^{\frac{\nu_I-1}{\nu_I}}_{m,t} \right]^{\frac{\nu_I}{\nu_I-1}}
\end{align*}
Clearly, the production function is linear homogeneous.

The firm's cost minimization problem is:
\begin{align*}
    C\left(I_t\right) = \min_{I_{d,t},I_{m,t}} P_{d,t} I_{d,t} + P_{m,t} I_{m,t} + \lambda_{t} \bigg\{I_t - \left[ \gamma^{\frac{1}{\nu_I}}_I I^{\frac{\nu_I-1}{\nu_I}}_{d,t} + \left(1-\gamma_I\right)^{\frac{1}{\nu_I}} I^{\frac{\nu_I-1}{\nu_I-1}}_{m,t} \right]^{\frac{\nu_I}{\nu_I}}\bigg\}
\end{align*}
We assume that the domestic inputs into investment are the domestic homogeneous good, so its price is 
\begin{align*}
    P_{d,t} = P_t
\end{align*}
Also, the imported good is homogeneous so
\begin{align*}
    P_{m,t} = S_t P^{f}_t
\end{align*}
Thus, we can express the price of the investment good as
\begin{align*}
    P_{i,t} = \left[\gamma_I P^{1-\nu_I}_t + \left(1-\gamma_I\right) \left(S_tP^{f}_{t}\right)^{1-\nu_I} \right]^{\frac{1}{1-\nu_I}}
\end{align*}
Scaling by $P_t$
\begin{align*}
    p_{I,t} = \left[\gamma_I + \left(1-\gamma_I\right) \left( p^{c}_t q_t \right)^{1-\nu_I} \right]^{\frac{1}{1-\nu_I}}
\end{align*}
where
\begin{align*}
    q_t = \frac{S_t P^{f}_t}{P^{c}_t}
\end{align*}
The demand functions for inputs are given by
\begin{align*}
    I_{d,t} &= \gamma_I I_t \left(\frac{P_t}{P^{I}_t} \right)^{-\nu_I} \\
    I_{m,t} &= \left(1-\gamma_I\right) I_t \left(\frac{S_tP^{f}_t}{P^{I}_t} \right)^{-\nu_I}
\end{align*}
which becomes
\begin{align*}
    I_{d,t} &= \gamma_I I_t \left(p_{I,t} \right)^{\nu_I} \\
    I_{m,t} &= \left(1-\gamma_I\right) I_t \left(\frac{p_{I,t}}{p^{m}_t} \right)^{\nu_I}
\end{align*}

\noindent
\textbf{Consumption good.} These are the consumption goods purchased and consumed by domestic households. They are produced by a representative, competitive firm using the following production function:
\begin{align*}
    C_t = \left[\left(1-\omega_c\right)^{\frac{1}{\eta_c}} \left(C_{d,t}\right)^{\frac{\eta_c-1}{\eta_c}}  + \omega_c^{\frac{1}{\eta_c}} \left(C_{m,t}\right)^{\frac{\eta_c-1}{\eta_c}} \right]^{\frac{\eta_c}{\eta_c-1}}
\end{align*}
where $C_{d,t}$ is a domestic homogeneous output good with price $P_t$, $C_{m,t}$ is an imported good with price $P^{m}_t = S_t P^{f}_t$, $C_t$ is the final consumption good and $\eta_c$ is the elasticity of substitution between domestic and imported goods.

Profit maximization by the representative firm becomes
\begin{align*}
    \max_{C_t,C_{d,t},C_{m,t}} P^{c}_t C_t - P^{m}_t C_{m,t} - P_t C_{d,t}
\end{align*}
subject to the production function. The first order conditions lead to the following demand functions
\begin{align*}
    C_{m,t} &= \omega_c \left(\frac{P^{c}_t}{P^{m}_t} \right)^{\eta_c} C_t \\
    C_{d,t} &= \left(1-\omega_c\right) \left(\frac{P^{c}_t}{P_t} \right)^{\eta_c} C_t
\end{align*}
The relative price of the consumption good is
\begin{align*}
    p^{c}_t &= \left[\left(1-\omega_c\right) + \omega_c \left(p^{m}_t\right)^{1-\eta_c} \right]^{\frac{1}{1-\eta_c}}
\end{align*}
were 
\begin{align*}
    p^{m}_t = \frac{P^{m}_t}{P_t} = \frac{S_t P^{f}_t}{P_t}
\end{align*}
We can define the consumer price inflation as
\begin{align*}
    \pi^{c}_t \equiv \frac{P^{c}_{t}}{P^{c}_{t-1}} = \frac{P_t p^{c}_t}{P_{t-1} p^{c}_{t-1}} = \left[\frac{\left(1-\omega_c\right) + \omega_c \left(p^{m}_t\right)^{1-\eta_c} }{\left(1-\omega_c\right) + \omega_c \left(p^{m}_{t-1}\right)^{1-\eta_c} }\right]^{\frac{1}{1-\eta_c}}
\end{align*}

\noindent
\textbf{Export goods.} We assume that the domestic homogeneous good can be directly exported. We denote the amount of the good exported by $X_t$. Foreign demand for the domestic good
\begin{align*}
    X_t = \left(\frac{P^{x}_t}{P^{f}_t}\right)^{-\eta_f} Y^{f}_t
\end{align*}
where $Y^{f}_t$ is a foreign demand shifter, $P^{f}_t$ is the foreign currency price of the foreign good, $P^{x}_t$ is the foreign currency price of the export good. 

We assume that there is a perfectly competitive exporter which purchases the domestic homogeneous good at price $P_t$. It sells the good at dollar price $P^{x}_t$, which translates into domestic currency units, $S_t P^{x}_t$. Competition implies that price, $S_t P^{x}_t$, equals marginal cost, $P_t$, so that
\begin{align*}
    S_t P^{x}_t = P_t
\end{align*}
We can express the foreign demand as
\begin{align*}
    X_t &= \left(\frac{P^{x}_t}{P^{f}_t}\right)^{-\eta_f} Y^{f}_t \\
    X_t &= \left(p^{x}_t\right)^{-\eta_f} Y^{f}_t
\end{align*}
where $p^{x}_t = P^{x}_t/P^{f}_t$. Note that we can express $p^{x}_t$ as
\begin{align*}
    p^{x}_t = \frac{P^{x}_t}{P^{f}_t} = \frac{P_t}{S_t P^{f}_t} = \frac{P_t P^{c}_t}{S_t P^{f}_t P^{c}_t} = \frac{1}{p^{c}_t q_t} = \frac{1}{p^{m}_t}
\end{align*}
where $q_t$ denotes the real exchange rate, $q_t = S_t P^{f}_t / P^{c}_t$.

\subsection{Government Policy} \label{subsec:policy}

The government's budget constraint becomes
\begin{align*}
    P_t T_t &= R_tB_t - \tilde{R}_t B_t   \\
    T_t &= R_t b_t - b_t\tilde{R}_t
\end{align*}
where $T_t$ are real lump sum taxes levied on households, and the term 
\begin{align*}
    b_t R_t - b_t \tilde{R}_t
\end{align*}
reflects the tax subsidy given to firms that the government needs to finance.

The monetary authority follows an interest rate rule specified as
\begin{align*}
    \log \left(\frac{R_{d,t}}{R_d}\right) = \rho_R \log \left(\frac{R_{d,t-1}}{R_d}\right) + \left(1-\rho_R\right)\left[r_{\pi} \log \left(\frac{\pi_t}{\bar{\pi}} \right) \right] + \epsilon^R_t
\end{align*}

\subsection{Market Clearing} \label{subsec:market_clearing}

The supply of peso loans comes from households, $D_t$. The demand for pesos comes from intermediate good firms, $B^{\text{peso}}_t$. We assume that foreigners do not participate in the local currency market, so that market clearing implies:
\begin{align*}
    D_t = B_t
\end{align*}
Dividing by $P_t$, it comes
\begin{align*}
    d_t = b_t
\end{align*}
The supply of dollars comes from households $D^{*}_t$ and foreigners $F^{o}_t$. The demand for dollar financing comes from the domestic intermediate good firms, $B^{\text{dollar}}_t$. Market clearing implies
\begin{align*}
    D^{*}_t + F^{o}_t = B^{\text{dollar}}_t
\end{align*}
Recall that this are \textit{dollar} terms. Multiplying by $S_t$ and dividing by $P_t$ we obtain
\begin{align*}
    d^{*}_t + f^{o}_t &=  b^{\text{dollar}}_t
\end{align*}

According to the balance of payments, expressed in dollars, the trade surplus equals the net accumulation of foreign assets:
\begin{align*}
    \frac{P_t}{S_t} X_t - P^{f}_t \left(I_{m,t} + C_{m,t}  \right) = D^{*}_t - R^{*}_{t-1} D^{*}_{t-1} - \left(B^{\text{dollar}}_t - B^{\text{dollar}}_{t-1} R^{*}_{t-1}  \right)= - \left( F^{o}_t - R^{*}_{t-1} F^{o}_{t-1}  \right)
\end{align*}
The first equality is the balance of payments and the second equality is the dollar market clearing.

The demand for exports, expressed in dollars, is:
\begin{align*}
    \frac{P_t X_t}{S_t} &= \frac{P_t}{S_t} \left(\frac{P^{x}_t}{P^{f}_t}\right)^{-\eta_f} Y^{f}_t  \\
    &= \frac{P_t}{S_t} \left(\frac{S_t P^{f}_t}{ S_t P^{x}_t}\right)^{\eta_f} Y^{f}_t \\
    &= \left(p^{c}_t q_t \right)^{\eta_f} Y^{f}_t
\end{align*}
We can express the balance of payments in terms of the domestic homogeneous good as
\begin{align*}
    \left(p^{c}_t q_t \right)^{\eta_f} Y^{f}_t - p^{m}_t \left[ \left(1-\gamma_I\right) \left(\frac{p_{i,t}}{p^{m}_t} \right)^{\nu_I} i_t + \omega_c \left(\frac{p^{c}_t}{p_{m,t}} \right)^{\eta_c} c_t \right] &= - \frac{S_t \left(F^{o}_t-R^{*}_{t-1}F^{o}_{t-1}\right)}{P_t} \\
    &= - \left(f^{o}_t - \frac{s_t R^{*}_{d,t-1}}{\pi_t} f^{o}_{t-1} \right)
\end{align*}

\newpage
\section{Additional Details on Empirical Model} \label{sec:appendix_details_empirical_model}

\subsection{SVAR Model Details} \label{subsec:appendix_details_empirical_SVAR_model}

In this section of the appendix we provide additional details on the estimation of the Structural VAR model presented in Section \ref{subsec:spillover_puzzle_empirical}. The model is estimated using Bayesian methods. In order to carry out the estimation of this model we first re-write the model. In particular, the model can be reformulated in compact form as

\begin{align} \label{eq:model_compact_form}
\underbrace{\begin{pmatrix}
y_{1,t}' \\
y_{2,t}' \\
\vdots  \\
y_{N,t}'
\end{pmatrix}}_{Y_t, \quad N \times n}
&=
\underbrace{\begin{pmatrix}
y_{1,t-1}' \ldots y_{1,t-p}' \\
y_{2,t-1}' \ldots y_{2,t-p}' \\
\vdots  \ddots \vdots \\
y_{N,t-1}' \ldots y_{N,t-p}'
\end{pmatrix}}_{\mathcal{B}, \quad N \times np}
\underbrace{\begin{pmatrix}
\left(A^{1}\right)' \\
\left(A^{2}\right)' \\
\vdots \\
\left(A^{N}\right)'
\end{pmatrix}}_{X_t, \quad np \times n}
+
\underbrace{\begin{pmatrix}
\epsilon_{1,t}' \\
\epsilon_{2,t}' \\
\vdots \\
\epsilon_{N,t}'
\end{pmatrix}}_{\mathcal{E}_t, \quad N \times n}
\end{align}
or
\begin{align}
    Y_t = X_t \mathcal{B} + \mathcal{E}_t
\end{align}
Even more, the model can be written in vectorised form by stacking over the $T$ time periods 
\begin{align}
    \underbrace{vec\left(Y\right)}_{NnT \times 1} = \underbrace{\left(I_n \otimes X \right)}_{NnT \times n np} \quad \underbrace{vec\left(\mathcal{B}\right)}_{n np \times 1} \quad + \quad \underbrace{vec\left(\mathcal{E}\right)}_{NnT \times 1}
\end{align}
or
\begin{align}
    y = \Bar{X} \beta + \epsilon
\end{align}
where $\epsilon \sim \mathcal{N}\left(0, \Bar{\Sigma}\right)$, with $\Bar{\Sigma} = \Sigma_c \otimes I_{NT}$.

The model described above is just a conventional VAR model. Thus, the traditional Normal-Wishart identification strategy is carried out to estimate it. The likelihood function is given by
\begin{align}
    f\left(y | \Bar{X} \right) \propto |\Bar{\Sigma}|^{-\frac{1}{2}} \exp \left(-\frac{1}{2} \left(y - \Bar{X}\beta\right)' \Bar{\Sigma}^{-1} \left(y - \Bar{X}\beta\right) \right)
\end{align}
As for the Normal-Wishart, the prior of $\beta$ is assumed to be multivariate normal and the prior for $\Sigma_c$ is inverse Wishart. For further details, see \cite{dieppe2016bear}. All of the panel SVAR model computations are carried out using the BEAR Toolbox version 5.1.

\subsection{Identification Strategy \& Shock Recovery} \label{subsec:appendix_model_details_identification}

In this section of the appendix, we present additional details on the sign-restriction identification strategy presented in Section \ref{subsec:spillover_puzzle_empirical} which recovers the two FOMC structural shocks. As stated in Section \ref{subsec:spillover_puzzle_empirical}, this identification strategy follows \cite{jarocinski2020deconstructing} and \cite{jarocinski2022central}.

The identification strategy introduced by \cite{jarocinski2022central} exploit the high-frequency surprises of multiple financial instruments to recover two distinct FOMC shocks: a pure monetary policy (MP) shock and information disclosure (ID) shock. In particular, the authors impose sign restrictions conditions on the co-movement of the high-frequency surprises of interest rates and the S\&P 500 around FOMC meetings. This co-movement is informative as standard theory unambiguously predicts that a monetary policy tightening shock should lead to lower stock market valuation. This is because a monetary policy tightening decreases the present value of future dividends by increasing the discount rate and by deteriorating present and future firm's profits and dividends. Thus, MP shocks are identified as those innovations that produce a negative co-movement between these high-frequency financial variables. On the contrary, innovations generating a positive co-movement between interest rates and the S\&P 500 correspond to ID shocks. The co-movement of these high-frequency surprises is remarkably different across FOMC meetings. Although a majority of FOMC meetings exhibit a negative co-movement between the interest rates and S\&P 500 surprises, a significant share of observations exhibit a positive co-movement. The authors' way to account for the positive co-movement is to attribute it to a shock that occurs systematically at the same time the FOMC announces its policy decisions, but that is different from a standard monetary policy shock. In particular, this additional shock is the disclosure of the FOMC's information about the present and future state of the US economy. Hence, by combining both the high-frequency surprises and imposing sign-restriction in their co-movements, the authors separately identify two structural FOMC shocks: a pure US monetary policy (MP) shock (which exhibits a negative co-movement between the interest rates and S\&P high-frequency surprises) and an information disclosure (ID) shock (which exhibits a positive co-movement between the interest rates and S\&P high-frequency surprises).

The high frequency surprise in the policy interest rate, $i^{Total}$, can be decomposed as
\begin{align}
    i^{\text{Total}} = i^{\text{MP}} + i^{\text{ID}}
\end{align}
where $i^{\text{MP}}$ is negatively correlated with the high frequency surprise of the $S\&P 500$ ``$s$'', and $i^{\text{ID}}$ is positively correlated with the ``$s$''. As shown by \cite{jarocinski2022central}, the sign restriction recovery of the structural shocks must satisfy the following decomposition
\begin{align}
    M = UC 
\end{align}
where $U'U$ is a diagonal matrix, $C$ takes the form of
\begin{align} \label{eq:matrix_rotation_appendix}
C =
\begin{pmatrix}
    1 & c^{MP}<0 \\
    1 & c^{ID}>0 \\
\end{pmatrix}
\end{align}
where $M=(i^{Total},s)$ is a $T \times 2$ matrix with $i^{Total}$ in the first column, $s$ in the second; $U = (i^{\text{MP}},i^{\text{ID}})$ is a $T \times 2$ matrix with $i^{\text{MP}}$ in the first column and $i^{\text{ID}}$ in the second column; and $T$ denoting the time length of the sample. By construction, $i^{\text{MP}}$ and $i^{\text{ID}}$ are mutually orthogonal. Matrix $C$ captures how $i^{\text{MP}}$ and $i^{\text{ID}}$ translates into financial market surprises.

The decomposition in \ref{eq:matrix_rotation_appendix} is not unique. In terms of \cite{jarocinski2022central} there is a range of rotations of matrices $U$ and $C$ that satisfy the sign restrictions $c^{MP}<0$ and $c^{ID}>0$.

The matrices $U$ and $C$ are computed as
\begin{align}
    U &= QPD \\
    C &= D^{-1} P' R
\end{align}
where the matrices $Q,P,D,R$ are obtained in three steps. 

\begin{itemize}
    \item[1.] Decompose matrix $M = UC$ into two orthogonal components using a QR decomposition such that
\end{itemize}
\begin{align}
    M &= QR \\
    Q'Q &=\begin{pmatrix}
        1 & 0 \\
        0 & 1 \\        
    \end{pmatrix} \\
    R&=\begin{pmatrix}
        r_{1,1}>0 & r_{1,2} \\
        0 & r_{2,2}>0 \\        
    \end{pmatrix}
\end{align}

\begin{itemize}
    \item [2.] Rotate these orthogonal components using the rotation matrix
    \begin{align}
    P &= \begin{pmatrix}
       \cos \left(\alpha\right) & \sin \left(\alpha\right) \\
       - \sin \left(\alpha\right) & \cos \left(\alpha\right)
    \end{pmatrix}
    \end{align}
    To satisfy the sign restrictions use any angle $\alpha$ in the following range
    \begin{align*}
        & \alpha \in \left( \left(1-w\right) \times \arctan \frac{r_{1,2}}{r_{2,2}} , \frac{w \times \pi}{2} \right) \qquad \text{if } r_{1,2} >0 \\
        & \alpha \in \left( 0 , w \times \arctan \frac{-r_{2,2}}{r_{1,2}} \right) \qquad \text{if } r_{1,2} \leq 0 \\
    \end{align*}
    where $w$ is weight, between 0 and 1, scaling the rotation angle. Setting $w = 0.5$ implies the median rotation angle, assumption used under the benchmark specification
\end{itemize}
\begin{itemize}
    \item[3.] Re-scale the resulting orthogonal components with a diagonal matrix $D$ to ensure that they add up to the interest rate surprises $i^{\text{Total}}$. It is straightforward to show that
    \begin{align}
        D = \begin{pmatrix}
            r_{1,1} \cos \left(\alpha\right) & 0 \\
            0 & r_{1,1} \sin \left(\alpha\right)
        \end{pmatrix}
    \end{align}
\end{itemize}

In \cite{jarocinski2022central}, the angle of rotation $\alpha$ of matrix $P$ is pinned down following these steps
\begin{itemize}
    \item[1.] Construct ``poor man's sign restrictions'' shocks such that
    
    \begin{align*}
        i^{\text{Total}} &= i^{MP} \quad \& \quad i^{\text{ID}} = 0 \qquad \text{if } i^{Total} \times s \leq 0 \\
        i^{\text{Total}} &= i^{ID} \quad \& \quad i^{\text{MP}} = 0 \qquad \text{if } i^{Total} \times s > 0 \\
    \end{align*}
\end{itemize}
In the data, \cite{jarocinski2022central} finds that the poor man's monetary policy shocks account for 88\% of the variance of the Federal Reserve's total interest rate surprises, i.e,
\begin{align*}
    \frac{var \left(i^{\text{MP}}\right)}{var \left(i^{\text{MP}}\right)} = 0.88
\end{align*}
To pin down the decomposition,  \cite{jarocinski2022central} impose that, as in the ``poor man's sign restrictions'' case, $var \left(i^{\text{MP}}\right)/ var \left(i^{\text{MP}}\right) = 0.88$. Additionally, \cite{jarocinski2022central} shows that the angle $\alpha$ can meets this condition can be recovered as 
\begin{align}
    \alpha = \arccos \sqrt{\frac{var \left(i^{\text{MP}}\right)}{var \left(i^{\text{MP}}\right)}}
\end{align}
In particular, the $\alpha$ that meets this condition is set equal to $\alpha = 0.8702$.


\subsection{Data Details} \label{subsec:appendix_data_details}

In this section of the appendix we provide additional details on the construction of the sample used across the paper. The source of the macroeconomic and financial data used for the construction of the variables in the benchmark variable specification is the IMF's ``International Financial Statistics''.\footnote{To access the IMF's IFS datasets go to \url{https://data.imf.org/?sk=4c514d48-b6ba-49ed-8ab9-52b0c1a0179b}.}

First, the benchmark specification is comprised of five variables:
\begin{enumerate}
    \item Nominal Exchange Rate
    
    \item Industrial Production index
    
    \item Consumer Price Index
    
    \item Lending Rate

    \item Equity Index
\end{enumerate}
Next, we present additional details for the construction of each of the variables
\begin{itemize}
    \item \underline{Nominal Exchange Rate:} The variable's full name at the IMF IFS data set is ``Exchange Rates, National Currency Per U.S. Dollar, Period Average, Rate''. 
    
    \item \underline{Industrial Production Index:} In order to construct countries' ``Industrial Production Index'' we rely on three variables of the IMF IFS' dataset:
    
    \begin{itemize}
        \item Economic Activity, Industrial Production, Index
        
        \item Economic Activity, Industrial Production, Seasonally Adjusted, Index
        
        \item Economic Activity, Industrial Production, Manufacturing, Index
    \end{itemize}
    Ideally, we would  construct the variable ``Industrial Production Index'' by choosing only one of the variables mentioned above. However, this is impossible as countries do not report to the IMF all three of these variables for our time sample, January 2004 to December 2016. For instance, Peru provides neither the ``Economic Activity, Industrial Production, Index'' nor the ``Economic Activity, Industrial Production, Seasonally Adjusted, Index'', but does provide the ``Economic Activity, Industrial Production, Manufacturing, Index''. Visiting Peru's Central Bank statistics website, there is no ``Industrial Production Index'', but there is an ``Industrial Production, Manufacturing Index'', which coincides with the variable reported as ``Economic Activity, Industrial Production, Manufacturing, Index'' to the IMF.
    
    In order to deal with this, we establish the following priority between the three IMF IFS variables: (i) ``Economic Activity, Industrial Production, Seasonally Adjusted, Index'' (ii) ``Economic Activity, Industrial Production, Index'', (iii) ``Economic Activity, Industrial Production, Manufacturing, Index''. Table \ref{tab:data_details_industrial} below presents the IMF IFS variable used for each country.
    \begin{table}[ht]
        \centering
        \caption{Construction of Industrial Production Index}
        \footnotesize
        \label{tab:data_details_industrial}
        \begin{tabular}{l l}
\multicolumn{2}{c}{Emerging Markets} \\ \hline \hline
Brazil 	&	Economic Activity, Industrial Production, Seasonally Adjusted, Index	\\
Chile	&	Economic Activity, Industrial Production, Seasonally Adjusted, Index	\\
Colombia	&	Economic Activity, Industrial Production, Seasonally Adjusted, Index	\\
Indonesia	&	Economic Activity, Industrial Production, Manufacturing, Index	\\
Mexico	&	Economic Activity, Industrial Production, Seasonally Adjusted, Index	\\
Peru	&	Economic Activity, Industrial Production, Manufacturing, Index	\\
Philippines	&	Economic Activity, Industrial Production, Manufacturing, Index	\\
South Africa	&	Economic Activity, Industrial Production, Manufacturing, Index	\\
			\\
\multicolumn{2}{c}{Advanced Economies} \\ \hline \hline
Australia	&	Economic Activity, Industrial Production, Seasonally Adjusted, Index	\\
Canada	&	Economic Activity, Industrial Production, Seasonally Adjusted, Index	\\
Japan	&	Economic Activity, Industrial Production, Seasonally Adjusted, Index	\\
South Korea	&	Economic Activity, Industrial Production, Seasonally Adjusted, Index	\\
        \end{tabular}
    \end{table}
I believe that every one of the three variables considered reflects the actual industrial production index. From Table \ref{tab:data_details_industrial}, it is clear that when ``Economic Activity, Industrial Production, Seasonally Adjusted, Index'' is not available, the non-seasonally adjusted is also not available. 

\item \underline{Consumer Price Index:} Data for all countries except Australia is constructed using the variable ``Prices, Consumer Price Index, All items, Index'' from IMF IFS data set. Australia does not report a monthly CPI series to the IMF-IFS data set. Furthermore, the Australian Bureau of Statistics provides only quarterly data on their consumer price index.\footnote{See \url{https://www.abs.gov.au/statistics/economy/price-indexes-and-inflation/consumer-price-index-australia/jun-2022}.} Thus, for the case of Australia we proxy the monthly consumer price index by using the ``Prices, Producer Price Index, All Commodities, Index''. Once again, given that the paper's main results are robust to the different exercises that partition the sample, we believe that using this proxy variable does not guide any of the of the results presented in the paper. 

\item \underline{Lending Rate:} For all countries we use the IMF-IFS' ``Monetary and Financial Accounts, Interest Rates, Other Depository Corporations Rates, Lending Rates, Lending Rate, Percent per Annum'' variable.

\item \underline{Equity Index:} In order to construct countries' ``Equity Index'' we rely on two variables of the IMF IFS' dataset:
    
    \begin{itemize}
        \item Monetary and Financial Accounts, Financial Market Prices, Equities, Index
        
        \item Monetary and Financial Accounts, Financial Market Prices, Equities, End of Period, Index
    \end{itemize}
    
   We establish a priority: (i) ``Monetary and Financial Accounts, Financial Market Prices, Equities, Index'' (ii) ``Monetary and Financial Accounts, Financial Market Prices, Equities, End of Period, Index''. Again, the data coverage is not complete for all countries for the full sample period of January 2004 to December 2016. Table \ref{tab:data_details_equity} presents index used for every country.
    \begin{table}[ht]
        \centering
        \caption{Construction of Equity Index}
        \footnotesize
        \label{tab:data_details_equity}
        \begin{tabular}{l l}
\multicolumn{2}{c}{Emerging Markets} \\ \hline \hline
Brazil 	&	Equities, End of Period, Index	\\
Chile	&	Equities, Index	\\
Colombia	&	Equities, End of Period, Index	\\
Indonesia	&	Equities, Index	\\
Mexico	&	Equities, End of Period, Index	\\
Peru	&	Equities, End of Period, Index	\\
Philippines	&	Equities, Index	\\
South Africa	&	Equities, Index	\\
			\\
\multicolumn{2}{c}{Advanced Economies} \\ \hline \hline
Australia	&	Equities, End of Period, Index	\\
Canada	&	Equities, Index	\\
Japan	&	Equities, Index	\\
Korea, Rep. of	&	Equities, Index	\\
	\\
        \end{tabular}
    \end{table}
\end{itemize}

\end{document}